\documentclass[11pt,tightenlines,preprintnumbers,superscriptaddress,nofootinbib,fleqn]{revtex4-2}
\pdfoutput=1
\makeatletter
\renewcommand{\p@subsection}{}
\renewcommand{\p@subsubsection}{}
\renewcommand{\p@paragraph}{}
\makeatother
\usepackage{verbatim}
\usepackage{array,colortbl}
\usepackage{float}

\usepackage{xcolor}
\definecolor{darkblue}{rgb}{0.0,0,0.5}
\definecolor{darkgreen}{rgb}{0.0,0.3,0.0}
\definecolor{redish}{rgb}{0.675,0,0.2}
\definecolor{red}{rgb}{0.8,0,0}
\definecolor{green}{rgb}{0,0.6,0}
\definecolor{blue}{rgb}{0,0,0.8}
\definecolor{raspberry}{rgb}{0.8,0.,0.5}

\usepackage[pdftitle={Snowmass 2022: Precision QCD, Hadronic Structure \& Forward QCD, Heavy Ions},
  pdfauthor={},unicode=true, bookmarks=false, linkcolor = darkblue, citecolor = redish, breaklinks=false, colorlinks=true, hyperfootnotes=true]{hyperref}

\usepackage{amsmath,amssymb,xspace,graphicx}
\usepackage{scalefnt,listings,multirow}
\usepackage[caption=false]{subfig}

\newcommand{\NLO}[1]{N${}^{#1}$LO\xspace}

\newcommand{\NLOgen}{NLO\xspace}
\newcommand{\NNLOgen}{NNLO\xspace}
\newcommand{\NNNLOgen}{\NLO3}

\newcommand{\NLOH}[1]{N${}^{#1}$LO${}_{\rm HTL}$\xspace}
\newcommand{\NLOHone}{NLO${}_{\rm HTL}$\xspace}

\newcommand{\NNLOHTL}{NNLO${}_{\rm HTL}$\xspace}
\newcommand{\NNNLOHTL}{\NLOH3}

\newcommand{\NLOQ}[1]{N${}^{#1}$LO${}_{\rm QCD}$\xspace}

\newcommand{\LOQCD}{LO${}_{\rm QCD}$\xspace}
\newcommand{\NLOQCD}{NLO${}_{\rm QCD}$\xspace}
\newcommand{\NNLOQCD}{NNLO${}_{\rm QCD}$\xspace}
\newcommand{\NNNLOQCD}{\NLOQ3}

\newcommand{\NLOE}[1]{N${}^{#1}$LO${}_{\rm EW}$\xspace}

\newcommand{\NLOEW}{NLO${}_{\rm EW}$\xspace}

\newcommand{\NLOQE}[2]{N${}^{(#1,#2)}$LO${}_{{\rm QCD}\otimes{\rm EW}}$\xspace}

\newcommand{\NLOHE}[2]{N${}^{(#1,#2)}$LO${}^{\rm (HTL)}_{{\rm QCD}\otimes{\rm EW}}$\xspace}

\newcommand{\NLOQCDVBF}{NLO${}_{\rm QCD}^{(\rm VBF)}$\xspace}
\newcommand{\NNLOQCDVBF}{NNLO${}_{\rm QCD}^{(\rm VBF)}$\xspace}

\newcommand{\NLOEWVBF}{NLO${}_{\rm EW}^{(\rm VBF)}$\xspace}

\newcommand{\NLOQVBFstar}[1]{N${}^{#1}$LO${}_{\rm QCD}^{(\rm VBF^{*})}$\xspace}

\newcommand{\NNLOQCDVBFstar}{NNLO${}_{\rm QCD}^{(\rm VBF^{*})}$\xspace}
\newcommand{\NNNLOQCDVBFstar}{\NLOQVBFstar3}

\newcommand{\NLOggHVtb}[1]{N${}^{#1}$LO${}_{gg\to HZ}^{(t,b)}$\xspace}
\newcommand{\NNLOQCDT}{NNLO${}_{\rm QCD}^{(t)}$\xspace}
\newcommand{\NNLOQCDBC}{NNLO${}_{\rm QCD}^{(b,c)}$\xspace}

\newcommand{\tb}{\bar{t}}

\newcommand{\wodecays}{(w/o decays)}
\newcommand{\wdecays}{}
\newcommand{\wleptdecays}{}

\newcommand{\eg}{\emph{e.g.}\xspace}

\newcommand{\alphas}{\alpha_{s}}
\newcommand{\alphasmZ}{\alphas(m_\mathrm{Z})}

\newcommand{\so}{\sigma_\mathrm{Z}^\mathrm{had}}

\providecommand{\ccbar}{c\overline{c}}
\providecommand{\bbbar}{b\overline{b}}
\providecommand{\ttbar}{t\overline{t}}
\newcommand{\pT}{\ensuremath{p_\mathrm{T}}}

\newcommand{\sqrts}{\sqrt{s}}

\newcommand{\epem}{e^+e^-}

\newcommand\snowmass{\begin{center}\rule[-0.2in]{\hsize}{0.01in}\\\rule{\hsize}{0.01in}\\
\vskip 0.1in Report of Energy Frontier Topical Groups 5, 6, 7  submitted to the US Community Study\\ 
on the Future of Particle Physics (Snowmass 2021)
\rule{\hsize}{0.01in}\\\rule[+0.2in]{\hsize}{0.01in} \end{center}}

\begin{document}
\pagestyle{empty}

\preprint{FERMILAB-CONF-22-733-SCD-T, SMU-HEP-22-06}
\quad \\ \vspace*{-40pt}
\snowmass
\centerline{\large\bf 
 Precision QCD, Hadronic Structure \& Forward QCD, Heavy Ions}

\newcommand{\affBNL}{Physics Department, Brookhaven National Laboratory, Upton, NY 11973, USA}
\newcommand{\affBrown}{Physics Department, Brown University, Providence, RI, 02912, USA}
\newcommand{\affCERN}{CERN, 1211 Geneva 23, Switzerland}
\newcommand{\affFermilab}{Fermi National Accelerator Laboratory, Batavia, IL 60510, USA}
\newcommand{\affFSU}{Department of Physics, Florida State University, Tallahassee, FL 32306-4350, USA }
\newcommand{\affJLAB}{Thomas Jefferson National Accelerator Facility, Newport News, VA 23606, USA}
\newcommand{\affLBL}{Lawrence Berkeley National Laboratory, Berkeley CA 94720 USA}
\newcommand{\affMIT}{The Massachusetts Institute of Technology, Department of Physics, Cambridge, MA 02139, USA}
\newcommand{\affOxford}{Department of Physics, University of Oxford, UK}
\newcommand{\affSMU}{Department of Physics, Southern Methodist University, Dallas, TX 75275, USA}
\newcommand{\affMSU}{Department of Physics \& Astronomy, Michigan State University, MI 48824, USA}
\newcommand{\affMSUb}{Department of Computational Mathematics,
  Science and Engineering, Michigan State University, East Lansing, MI 48824}
\newcommand{\affKansasU}{Department of Physics \& Astronomy, The University of Kansas, Lawrence, KS 66045, USA}
\newcommand{\affNorthwesternU}{Department of Physics \& Astronomy, Northwestern University, Evanston, IL 60208, USA}
\newcommand{\affPSU}{Department of Physics, Pennsylvania State University, University Park, PA 16802, USA}
\newcommand{\affIowa}{Department of Physics \& Astronomy, University of Iowa, Iowa City, IA 52242-1479, USA}
\newcommand{\affDESYHamburg}{Deutsches Elektronen-Synchrotron DESY, Notkestr. 85, 22607 Hamburg, Germany}
\newcommand{\affSUNYBuffalo}{University at Buffalo, State University of New York, Amherst, NY 14221 USA}
\newcommand{\affIIT}{Department of Physics, Illinois Institute of Technology, Chicago, IL 60616, USA}
\newcommand{\affDAMTP}{DAMTP, University of Cambridge, Cambridge, CB3 0WA, UK}
\newcommand{\affPitt}{Pittsburgh Particle Physics,
  Astrophysics, and Cosmology Center,\\ Department of Physics and Astronomy, University of Pittsburgh, Pittsburgh, PA 15260, USA}
\newcommand{\affANL}{High Energy Physics Division, Argonne National Laboratory, Argonne, IL 60439, USA}
\newcommand{\affSLAC}{SLAC National Accelerator Laboratory, Stanford University, Stanford, CA 94039, USA}
\newcommand{\affDukeU}{Department of Physics, Duke University, Durham, North Carolina 27708, USA}
\newcommand{\affPeierls}{Rudolf Peierls Centre for Theoretical Physics, University of Oxford, Oxford, OX1 3PU, UK}
\newcommand{\affAllSouls}{All Souls College, University of Oxford, Oxford OX1 4AL, United Kingdom}
\newcommand{\affYale}{Department of Physics, Yale University, New Haven, CT 06511, USA}
\newcommand{\affECT}{European Centre for Theoretical Studies in Nuclear Physics and Related Areas (ECT*), I-38123 Villazzano, Trento, Italy}
\newcommand{\affFBK}{Fondazione Bruno Kessler (FBK), I-38123 Povo, Trento, Italy}
\newcommand{\affTIFPA}{INFN-TIFPA Trento Institute of Fundamental Physics and Applications, I-38123 Povo, Trento, Italy}
\newcommand{\affEcolePoly} {\'Ecole Polytechnique, Laboratoire Leprince-Ringuet, Av. Chasles, 91120 Palaiseau, France}

\newcommand{\affCalabria}{Dipartimento di Fisica, Universit\'a della Calabria, I-87036 Arcavacata di Rende, Cosenza, Italy}

\newcommand{\affNucleareCalabria}{Istituto Nazionale di Fisica Nucleare, Gruppo collegato di Cosenza, I-87036 Arcavacata di Rende, Cosenza, Italy}
\newcommand{\affSaclay}{Universit\'e Paris-Saclay, CNRS, IJCLab, 91405 Orsay, France}

\newcommand{\affINPKrakow}{Institute of Nuclear Physics, Polish Academy of Sciences, ul. Radzikowskiego 152, 31-342,Krak\'ow, Poland}

\newcommand{\effive}{\thanks{Convener, EF05 topical group}}
\newcommand{\efsix}{\thanks{Convener, EF06 topical group}}
\newcommand{\efseven}{\thanks{Convener, EF07 topical group}}

\author{M.~Begel}\effive\affiliation{\affBNL}
\author{S.~H{\"o}che}\effive\affiliation{\affFermilab}
\author{M.~Schmitt} \effive\affiliation{\affNorthwesternU}
\author{H.-W.~Lin} \efsix \affiliation{\affMSU}\affiliation{\affMSUb}
\author{P.M.~Nadolsky}\efsix\affiliation{\affFermilab}
\affiliation{\affSMU}
\author{C.~Royon}\efsix\affiliation{\affKansasU}
\author{Y-J.~Lee}\efseven\affiliation{\affMIT}
\author{S.~Mukherjee}\efseven\affiliation{\affBNL}

\author{C.~Baldenegro}\affiliation{\affEcolePoly}
\author{J.~Campbell}\affiliation{\affFermilab}
\author{G.~Chachamis}
\affiliation{Laborat\'orio de Instrumenta c\~ao e F\'isica Experimental de Part\'iculas, Av. Prof. Gama Pinto, 2, P-1649-003 Lisboa, Portugal}
\author{F.G.~Celiberto}   \affiliation{\affECT}\affiliation{\affFBK}\affiliation{\affTIFPA}  
\author{A.\ M.~Cooper-Sarkar  }   \affiliation{\affOxford} 
\author{D.~d'Enterria}\affiliation{\affCERN}
\author{M.~Diefenthaler}
\affiliation{\affJLAB}

\author{M.~Fucilla}\affiliation{\affCalabria} \affiliation{\affNucleareCalabria}\affiliation{\affSaclay}

\author{M.~V.~Garzelli}\affiliation{Universit\"at Hamburg, II Institut für Theoretische Physik, 22761 Hamburg, Germany}

\author{M.~Guzzi}
\affiliation{Department of Physics, Kennesaw State University, Kennesaw, GA 30144, USA}

\author{M.~Hentschinski}
\affiliation{Departamento de Actuaria, F\'isica y Matem\'aticas, Universidad de las Americas Puebla, San Andr\'es Cholula, 72820 Puebla, Mexico}

\author{T.\ J.~Hobbs          }   \affiliation{\affANL}\affiliation{\affIIT}     \author{J.~Huston             }   \affiliation{\affMSU}                          \author{J.~Isaacson}\affiliation{\affFermilab}
\author{S.R.~Klein}\affiliation{\affLBL}
\author{F.~Kling}\affiliation{\affDESYHamburg}

\author{P.~Kotko}
\affiliation{AGH University of Science and Technology, Physics Faculty, Mickiewicza 30, 30-059 Krak\'ow, Poland}

\author{Yu. V. Kovchegov} \affiliation{Department of Physics, The Ohio State University, Columbus, OH 43210, USA}

\author{G.~Krintiras} \affiliation{\affKansasU}

\author{A.S.~Kronfeld}\affiliation{\affFermilab}
\author{K.~Kutak} \affiliation{\affINPKrakow}

\author{B.~Mistlberger}\affiliation{\affSLAC}
\author{I.~Moult}\affiliation{\affYale}
\author{S.~Mrenna}\affiliation{\affFermilab}
\author{B.~Nachman}\affiliation{\affLBL}
\author{M.~Narain}\affiliation{\affBrown}
\author{F.I.~Olness}\affiliation{\affSMU} 
\author{A.~Papa} \affiliation{\affCalabria} \affiliation{\affNucleareCalabria}

\author{R.D. Pisarski}\affiliation{\affBNL}

\author{M.~Pitt} \affiliation{\affCERN}

\author{S.~Rappoccio}\affiliation{\affSUNYBuffalo}
\author{L.~Reina}\affiliation{\affFSU}
\author{J.~Reuter}\affiliation{\affDESYHamburg}
\author{M.H.~Reno}\affiliation{\affIowa}
\author{G.P.~Salam}\affiliation{\affPeierls}\affiliation{\affAllSouls}
\author{M.~Strikman}\affiliation{\affPSU}
\author{N.~Tran}\affiliation{\affFermilab}
\author{A.~Tricoli}\affiliation{\affBNL}
\author{M.~Ubiali}\affiliation{\affDAMTP}
\author{A.~van Hameren} \affiliation{\affINPKrakow}
\author{A.~Vossen}\affiliation{\affDukeU}
\author{K.~Xie}\affiliation{\affPitt}

\date{\today}
\maketitle

\clearpage\newpage
\noindent
\centerline{ABSTRACT} 
\\ \bigskip

\noindent This report was prepared on behalf of three Energy Frontier Topical Groups of the Snowmass 2021 Community Planning Exercise. It summarizes the status and implications of studies of strong interactions in high-energy experiments and QCD theory. 
We emphasize the rich landscape and broad impact of these studies in the decade ahead. Hadronic interactions play a central role in the high-luminosity Large Hadron Collider (LHC) physics program, and strong synergies exist between the (HL-)LHC and planned or proposed experiments at the U.S. Electron-Ion Collider, CERN forward physics experiments, high-intensity facilities, and future TeV-range lepton and hadron colliders. Prospects for precision determinations of the strong coupling and a variety of nonperturbative distribution and fragmentation functions are examined. We also review the potential of envisioned tests of new dynamical regimes of QCD in high-energy and high-density scattering processes with nucleon, ion, and photon initial states. The important role of the high-energy heavy-ion program in studies of nuclear structure and the nuclear medium, and its connections with QCD involving nucleons are summarized. We address ongoing and future theoretical advancements in multi-loop QCD computations, lattice QCD, jet substructure, and event generators. Cross-cutting connections between experimental measurements, theoretical predictions, large-scale data analysis, and high-performance computing are emphasized.  
\\ \bigskip

The report is based on many contributions to the EF05, EF06, and EF07 Topical Groups that were presented at the topical group meetings, in feedback on the white papers and the report, and through other communications. We express our gratitude to all Snowmass participants whose strong involvement in the groups' activities guided the vision expressed here. 
The following Snowmass white papers were instrumental in developing the report:
\begin{itemize}\itemsep-0.25em\small
   \item Physics with the Phase-2 ATLAS and CMS Detectors~\cite{ATLAS:2022hsp}
    \item The Forward Physics Facility: Sites, Experiments, and Physics Potential~\cite{Anchordoqui:2021ghd}
    \item The Forward Physics Facility at the High-Luminosity LHC~\cite{Feng:2022inv}
    \item Electron Ion Collider for High Energy Physics~\cite{AbdulKhalek:2022erw}
    \item Some aspects of the impact of the Electron Ion Collider on particle physics at the Energy Frontier~\cite{Chekanov:2022sax}
    \item Heavy Neutral Lepton Searches at the Electron-Ion Collider~\cite{Batell:2022ubw}
    \item Opportunities for precision QCD physics in hadronization at Belle II – a Snowmass whitepaper~\cite{Accardi:2022oog}
    \item The Future Circular Collider: a Summary for the US 2021 Snowmass Process~\cite{Bernardi:2022hny}
    \item The International Linear Collider~\cite{ILCInternationalDevelopmentTeam:2022izu}
    \item The Potential of a TeV-Scale Muon-Ion Collider~\cite{Acosta:2022ejc}
   \item Event Generators for High-Energy Physics  Experiments~\cite{Campbell:2022qmc}
    \item The strong coupling constant: State of the art and the decade ahead~\cite{dEnterria:2022hzv}
    \item Proton structure at the precision frontier~\cite{Amoroso:2022eow}    
    \item Lattice QCD Calculations of Parton Physics~\cite{Constantinou:2022yye}
    \item Impact of lattice $s(x)-\bar s(x)$ data in the CTEQ-TEA global analysis~\cite{Hou:2022sdf}
    \item Forward Physics, BFKL, Saturation Physics and Diffraction~\cite{Hentschinski:2022xnd}
    \item Prompt electron and tau neutrinos and antineutrinos in the forward region at the LHC~\cite{Bai:2022jcs}
    \item Jets and Jet Substructure at Future Colliders~\cite{Nachman:2022emq}
    \item \texttt{xFitter}: An Open Source QCD Analysis Framework~\cite{XFitterSnowmass}
    \item Opportunities for new physics searches with heavy ions at colliders~\cite{dEnterria:2022sut}
\end{itemize}

\clearpage
\centerline{EXECUTIVE SUMMARY} 
\bigskip

Quantum Chromodynamics (QCD), the fundamental theory of strong interactions, plays a unique role in the Standard Model. Being a confining gauge theory, it is an interesting quantum field theory to study in its own right. It is also a crucial tool to enable discovery at virtually every high-energy collider. QCD predicts a rich panoply of phenomena associated with both perturbative and nonperturbative dynamics of the strong interactions. Continued success of the high-energy and nuclear physics research program hinges on an improved understanding of both regimes, as well as the dynamical transition between them.

Future SM measurements and new physics searches will allow the exploration of new kinematic regions, such as very high 
transverse momentum and very forward rapidities, where large scale hierarchies may induce hitherto unseen QCD effects.
The upcoming era --- featuring the HL-LHC, Belle~II, the EIC, new advances in theory including in lattice QCD, and potentially a Higgs factory --- will be a new golden age for QCD easily rivaling the 1990's when the Tevatron, HERA, and LEP were all operating.

To fully exploit the wealth of expected data, precision calculations in QCD perturbation theory are required at one- and two-loop accuracy for many processes at hadron colliders, and in some cases at even higher accuracy. Monte-Carlo event generators serve as the backbone of the majority of collider simulations and must be able to reach similar precision. The advancement of QCD theory is therefore critically important to the physics program at the energy frontier.

Many QCD effects are universal and can be understood through factorization and perturbation theory.
The related systematic uncertainties are often a limiting factor in Standard Model measurements 
and searches for new phenomena. A major goal of the QCD research program has therefore been
to increase the precision of both the experimental measurements and the theoretical predictions.

The strong coupling is the least well measured coupling of the Standard Model,
and substantial progress in its determination is expected from both lattice QCD and future colliders, 
particularly $e^+e^-$ facilities, with a projected reduction in uncertainty of
almost an order of magnitude, leading to a precision on $\alpha_s$ in the permille range.
The RG evolution of the strong coupling will be testable at high precision over a 
large dynamic range with the help of lepton-hadron or hadron-hadron colliders.
Measurements of charm, bottom, and top quark masses at various energy scales test both the QCD dynamics and the mass parameters of the Standard Model, with subpercent precision for bottom quark mass expected at future Higgs factories.

Parton distribution functions (PDFs) and fragmentation functions (FFs) will play a prominent role in future precision experiments and new physics searches. New opportunities emerge for precise determination of these functions in many observations and to predict their behavior in lattice QCD.
Taking advantage of these opportunities requires implementation of two- and three-loop computations of radiative contributions and methodological advances in large-scale phenomenological analyses of QCD data. 

Detection of the decay products of far-forward hadrons at the proposed Forward Physics Facility (FPF) at the HL-LHC  would offer
an unprecedented opportunity for deeper tests of QCD in a novel high-energy regime. Neutrino production of all flavors as well as new particle production could be explored both by the FPF detectors alone and perhaps even in coincidence with ATLAS, leading to improved understanding of small-$x$ dynamics, forward heavy flavor --- particularly charm --- production, neutrino scattering in the TeV range, and hadronization inside nuclear matter. 

Jet substructure has emerged as a powerful framework and tool set for probing the highest energy scales to explore the structure of the strong force in final-state radiation on small angular scales, and to identify Lorentz-boosted massive particles including $W$, $Z$, \& $H$ bosons, top quarks, and exotica. Increased precision is expected to emerge from improved detector capabilities and from the theoretical understanding of particle flow observables, as well as from grooming techniques and other methods to reduce the impact of universal soft gluon and hadronization effects on observables.
Better theoretical understanding of energy correlation functions may enable new opportunities for measurements of fundamental properties of QCD.

Various domains of QCD are strongly interconnected. Progress in one area depends on the other domains. Going forward, the dialogue between experimentalists, theorists and tool developers will become ever more important to achieve precision measurements with reliable systematic uncertainty estimates.
\bigskip\bigskip
\clearpage

\tableofcontents

\clearpage \newpage
\pagestyle{myheadings}
\setcounter{page}{1}
\section{Experimental context}
\label{sec:intro}

Quantum Chromodynamics (QCD) predicts a rich panoply of phenomena associated with both perturbative and nonperturbative dynamics of the strong interactions. QCD affects every analysis at current and future experimental facilities operating with nucleons, heavy nuclei, and hadrons in general. While measurements of QCD parameters and studies of hadron structure are not always the primary goal at these facilities, related systematic uncertainties are often a limiting factor in the extraction of Standard Model (SM) parameters and the reach of new physics searches. QCD effects must therefore be known as precisely as possible. Predictions for QCD phenomena at the modern precision level stimulate developments in quantum field theory and computational methods. They are needed to identify phenomena beyond the Standard Model (BSM) and to understand the dynamics of the BSM physics once it is discovered.

This contribution to the Snowmass Community Study will discuss various aspects of QCD in the context of experimental facilities of interest to the domestic and international high-energy physics research program.
We emphasize that the dialogue between QCD experimentalists, theorists and tool developers will become ever more important to achieve precision measurements with reliable systematic uncertainty estimates. We start with an overview of the QCD role at future Energy Frontier facilities. 

\subsection{The high-luminosity LHC (HL-LHC)}
\label{sec:hl-lhc}

In the coming decade, experiments at the Large Hadron Collider (LHC) 
will make precise measurements of SM parameters,
such as the $W$ mass and Higgs boson couplings. Both the extraction of these parameters and their interpretation
will be limited primarily by the precision of perturbative QCD calculations and their faithful implementation
in Monte-Carlo simulations. Measurements of jet, photon, and heavy-quark cross-sections (including top quarks) will test perturbative QCD at higher orders and constrain parton distribution and fragmentation functions as well as
the running of the strong coupling $\alpha_s$. These analyses are also needed for understanding backgrounds for many 
other interesting processes. 

The HL-LHC provides an opportunity to test QCD with improved precision, particularly 
at high energies where uncertainties are dominated by statistics. A prime example are angular correlation measurements, such as the 
one described in Sec.~\ref{sec:dphi_hl-lhc}.
The average pileup at the LHC is around 25 events, and it is expected to reach values of around 150--200 during the HL-LHC operation. 
This will result in significant degradation in the physics object reconstruction performance and hence on the physics outcome 
without dedicated mitigations. While not an experimental or theoretical consideration of the LHC experiments' original designs,
jet substructure is now being widely used to minimize the impact of pileup, to probe fundamental and emergent properties 
of the strong force, to enhance the precision of measurements of highly-Lorentz-boosted SM particles, and to extend the sensitivity
of searches for new particles.
Novel types of event shape observables also provide new opportunities for precision tests of QCD.

A detailed description of the opportunities for QCD measurements at the HL-LHC is given in Ref.~\cite{ATLAS:2022hsp}.

\subsection{Forward QCD experiments at the HL-LHC}
\label{sec:ForwardExperiments}
\label{sec:fpf}

Detection of the far-forward decay products of hadrons at the Forward Physics Facility (FPF) at the HL-LHC  offers an unprecedented opportunity for deeper tests of QCD in the high-energy regime \cite{Feng:2022inv,Anchordoqui:2021ghd}. The FPF program would expand the physics reach of the ongoing FASER$\nu$ and SND@LHC experiments. 
One class of reactions that can be investigated at the FPF are single-inclusive forward emissions, where a neutrino with rapidity $y \gtrsim 6$ is identified. Both inclusive and exclusive processes can be measured and tested by considering several kinds of final states, such as charged light hadrons, vector mesons (extensively studied at HERA~\cite{Anikin:2011sa,Besse:2013muy,Bolognino:2018rhb,Celiberto:2019slj,Bolognino:2019pba,Bautista:2016xnp,Garcia:2019tne,Hentschinski:2020yfm}), and mesons with open charm or beauty, accompanied by their decay producing at least a forward neutrino.

These kinds of studies can be performed by the FPF detectors alone. Requiring coincidence with the ATLAS detector may allow identification of states with large invariant masses, whose decay products are not entirely captured by the FPF, but fall partly into the FPF's and partly into ATLAS's coverage areas. 
These measurements will lead to an improved understanding of forward heavy flavor---particularly charm production, neutrino scattering in the TeV range, and hadronization inside nuclear matter. 
The possibility of combining information from the ATLAS detector and a forward detector at the FPF relies on the ability to use an FPF event to trigger ATLAS.  This requires very precise timing and has consequences for the design of the forward detector. An additional issue is the extreme pileup expected at the HL-LHC as events with multiple hard-scattering processes within the same bunch crossing. 
Section~\ref{sec:fpf2} continues the discussion of the FPF.

\subsection{The Electron Ion Collider}
\label{sec:eic}
It is vital to understand how the properties of nucleons and nuclei emerge from their constituents: quarks and gluons. To advance this goal, a high-energy and high-luminosity polarized electron-ion collider (EIC) is being designed and constructed by Brookhaven National Laboratory and Jefferson Lab over the course of the next decade \cite{NAP25171} as a high priority on the agenda by the US nuclear physics community. 
The versatile EIC physics program \cite{AbdulKhalek:2021gbh} dedicated to exploration of hadronic matter over a wide-range of center-of-mass energies has significant synergies with exploration of QCD at the HL-LHC (Sec.~\ref{sec:hl-lhc}), Forward Physics Facility (Sec.~\ref{sec:fpf}) and other HEP experiments. Throughout this document, we provide many examples of the positive impact that the EIC program will have on high-energy QCD. The Snowmass EIC whitepapers \cite{AbdulKhalek:2022erw, Chekanov:2022sax,Batell:2022ubw} include many more examples.

Indeed, in addition to performing spin-dependent three-dimensional tomography of nucleons and various ion species, the EIC is capable of obtaining new precise measurements of hadronic structure
in deep inelastic electron-proton and electron-nucleus scattering (DIS)  over a wide range of CM energies, $\sqrt{s}\!=\!20-140$ GeV, and with high instantaneous luminosity of up to $\mathcal{L}=$ \(10^{34}\,\mathrm{cm}^{-2}\cdot \mathrm{s}^{-1}\). With its variable CM energy and excellent detection of final hadronic states, the EIC can precisely probe the unpolarized proton PDFs and their flavor composition at momentum fractions $x>0.1$, in the kinematic region of relevance for BSM searches at the HL-LHC, but at QCD scales of only a few (tens of) GeV where no deviations from the Standard Model are expected. The EIC can therefore elevate the accuracy of DIS experiments to a new level in the large-$x$ region that is currently covered by fixed-target experiments from more than 20 years ago. 

Furthermore, in addition to neutral-current DIS, parity-violating charged-current (CC) reactions can be employed to study nucleon and nuclear structure with highly-polarized beams for electrons and light ions, as well as unpolarized beams for heavy ions. Charge-current DIS on heavy-nucleus targets has large uncertainties, and this in turn is an issue for accelerator-based neutrino oscillation experiments, particularly those at DUNE, that have a significant share of events from neutrino CC DIS. 
Studies of CC DIS at the higher energies typical to the EIC, possibly complemented by studies at the FPF, will benefit HEP objectives in neutrino-nuclear scattering measurements at lower energies. These advancements will depend on new QCD-based theory ingredients for CC DIS simulations, 
as discussed in Sec.~\ref{sec:mcsimulations}.

The EIC will establish a new QCD frontier to address key open questions, such as the origin of nucleon spin, mass, and the emergence of QCD many-body phenomena at extreme parton densities.  There will be ample cross-pollination between studies of the hadron structure in the EIC experiments and using lattice QCD (cf.\ Sec.~\ref{sec:lattice}) and other {\it ab initio} approaches. Precise measurements of most accessible combinations of phenomenological PDFs at the EIC will provide useful benchmarks for lattice QCD calculations, and the latter in turn can make predictions for nonperturbative QCD functions that are not readily accessible in experiments. 

As the EIC will push the DIS experiments into an entirely novel territory, its success will critically depend on accurate and precise theoretical modeling, including theoretical advances in multiloop perturbative QCD+EW calculations, implementation of spin-dependent and transverse momentum dependent QCD evolution, development of new event generators for lepton-hadron scattering, and, just as importantly, the infrastructure for all-out accuracy control in phenomenological analyses of the EIC large data samples. 

\subsection{The Belle II Experiment}
\label{sec:belleii}
The theory uncertainty on the SM prediction for the anomalous magnetic moment of the muon 
is dominated by the leading-order hadronic contribution, which can be calculated from the cross section for
$e^+e^-\to\text{hadrons}$, measured in $e^+e^-$ experiments. The corresponding experimental result is dominated 
by BABAR and KLOE measurements of two pion production, but the two differ notably~\cite{Accardi:2022oog}. Belle II will perform these
measurements with larger data sets, and at least comparable systematic uncertainty, aiming to resolve the discrepancy. 
Furthermore, Belle II’s multi-$\mbox{ab}^{-1}$ data set will facilitate new approaches to suppress systematic uncertainties.
The low-background environment of $e^+e^-$ annihilation exploited at unprecedented statistical precision will also enable
highly impactful tests of transverse-momentum-dependent QCD evolution and factorization in jet and hadron production.
Measurements of multidimensional correlations of momenta and polarizations of final-state hadrons during hadronization 
will further our understanding of soft QCD and will enable refinement and tuning of Monte-Carlo event generators at levels
that may be instrumental to reach the precision needed to accomplish the LHC program. The lever arm in collision energy
with respect to Large Electron-Positron Collider (LEP) data offers a robust basis for extrapolation to LHC energies. The Belle II data set size will enable unique fully multidimensional measurements that capture the fuller picture of hadronization dynamics.

A detailed description of the opportunities of QCD measurements at Belle II is given in Ref.~\cite{Accardi:2022oog}.

\subsection{Future Electron-Positron Colliders}
\label{sec:fccee}
Due to their QCD neutral initial state, $e^+e^-$ colliders are the simplest setting in which to study dynamics in QCD, enabling precision measurements well beyond what is possible at hadron colliders.
At high luminosity, the clean environment will also provide enormous (multi)jet data samples to improve our understanding of parton showers, higher-order logarithmic resummations, as well as hadronization and nonperturbative phenomena.
QCD final states at high-energy electron-positron colliders will generally be more complex compared to earlier experiments.
This complexity is comparable to that already observed in the LHC experiments. For instance, the hadronic Higgs-strahlung 
analysis at a Higgs factory requires excellent jet clustering performance in four-jet final states~\cite{Thomson:2015jda, Lai:2021rko}. At higher energy, di-Higgs, top-quark pair, and $t\bar{t}H$ production lead to six-jet and even eight-jet final states; 
such that jet clustering becomes the dominant experimental limitation~\cite{Boronat:2016tgd}.

There has been much progress since LEP in understanding QCD final states, driven by a renewed interest in studying jet substructure at the LHC. The techniques developed have enabled a variety of new ways of understanding QCD phenomena with increasing sophistication~\cite{Larkoski:2017jix,Marzani:2019hun}. 
Precision determinations of event shapes have also enabled precision extractions of the strong coupling, $\alpha_s$~\cite{Abbate:2010xh,Hoang:2015hka}.
A wide variety of event shapes were measured at LEP in events that always contained two quark-initiated jets. With nonperturbative effects tuned against this rich data set, parton-shower Monte Carlo programs model quark-jet shapes extremely well. The MC programs are less confident in modeling gluon jets, which were not produced that often at LEP, but which are copiously produced at the LHC. A particular advantage of future lepton colliders is the availability of pure samples of gluon jets through the process $e^+e^- \to HZ$, with $Z$ decaying to leptons, and the Higgs boson decaying to $gg$~\cite{dEnterria:2015mgr,Gras:2017jty,Gao:2019mlt}. Understanding of $b$-quark showering and hadronization will also be improved; these are leading sources of systematic uncertainty in the measurement of the forward-backward asymmetry of $b$ quarks in $e^+e^-$ collisions at LEP~\cite{dEnterria:2020cgt}.

The optimization of detector concepts is mainly driven by improving jet energy resolution using particle flow.
The ILD~\cite{ILD:2019kmq} and SiD~\cite{Breidenbach:2021sdo} experiments at the ILC, as well as the CLIC detector~\cite{CLICdp:2018vnx} and the CLD design~\cite{Bacchetta:2019fmz} for the FCC-$ee$, are engineered to efficiently associate tracks and calorimeter energy deposits and, together with improvements in software~\cite{CALICE:2012eac}, might reach jet energy resolutions around 5-20\%. These concepts also offer excellent substructure performance~\cite{Strom:2020wrm}. 
Machine-induced backgrounds at $e^+e^-$ colliders are generally benign compared to the pile-up levels encountered at the LHC, but can have a non-negligible impact on jet reconstruction, especially at higher energy~\cite{Boronat:2014hva, Boronat:2016tgd, Lai:2021rko, Stewart:2015waa}.

\subsection{Future Muon-Muon Colliders}
\label{sec:mumu}

Jet algorithms developed for electron-positron colliders should also apply well to muon colliders. 
Proposed muon colliders offer a tremendous physics reach for discoveries, while maintaining appealing experimental aspects of lepton collider environments such as a lack of pileup and underlying event. 
Muon colliders will produce final states that are generally more complicated than $e^+e^-$ machines due to the higher energies; boosted topologies will also tend to be more prevalent.

A critical difference between muon and electron accelerators is the presence of large beam-induced background (BIB) processes for muon machines~\cite{Collamati:2021sbv,Ally:2022rgk,MuonCollider:2022ded}, which arise due to muons in the beam decaying via $\mu \rightarrow e \nu \bar{\nu}$ before colliding. The resultant electrons interact with experimental elements along the beamline, creating electromagnetic showers of soft photons and neutral particles that can interact with detectors.
Detectors at future muon colliders will need to incorporate specifically-designed shielding and subsystems to mitigate BIB processes. The exact characteristics of the BIB depend strongly on the machine centre-of-mass energy and accelerator lattice, and must be studied in-detail for different scenarios. 
Advanced pileup mitigation techniques studied at the LHC could provide versatile handles to remove beam-induced background contamination during reconstruction~\cite{Thaler:2010tr,Thaler:2011gf,Larkoski:2014gra}.

\subsection{Future Lepton-Hadron Colliders}
\label{sec:LHeCMuIC}
Lepton-hadron deep inelastic scattering (DIS) is a cornerstone process to determine nonperturbative QCD functions, such as PDFs, describing the hadron structure in high-energy collisions. Recent studies \cite{AbdulKhalek:2021gbh} demonstrate that the HL-LHC physics potential in electroweak precision studies and BSM searches can be greatly enhanced by concurrent DIS experiments providing complementary and competitive constraints on the PDFs. This is just one aspect of a versatile physics program, with a variety of QCD observables and final states, at any $\ell h$ collider. One such DIS experiment is the planned EIC that is being designed and constructed by BNL and Jefferson Lab; it was already discussed in Sec.~\ref{sec:eic}. The EIC will provide valuable measurements of PDFs at momentum fractions above 0.1 and relatively low $Q$. An Electron-Ion Collider in China (EIcC) accesses the hadron structure, including spin effects, at a lower energy ($\sqrts=15-20$ GeV), where nonperturbative effects are more pronounced \cite{Anderle:2021wcy}. The Large Hadron Electron Collider (LHeC) at CERN \cite{Jacob:1984vnf,LHeCStudyGroup:2012zhm,LHeC:2020van,Andre:2022xeh}, on the other hand, would extend DIS into the TeV energy range that considerably overlaps with the LHC kinematic region. The LHeC would be able to explore the PDFs, and QCD dynamics in general, at $x$ down to $\sim 10^{-6}$ at  $Q \sim 1$ GeV  and up to $0.5-0.8$ at $Q\sim 0.5-1$  TeV. It can therefore investigate QCD dynamics in the $x$ regions relevant for forward production at the LHC and FPF and to perform unique electroweak measurements and BSM searches. 

An appealing and highly innovative collider configuration for QCD studies is a muon-hadron collider with $\sqrts$ up to 1-6.5 TeV and instantaneous luminosity of up to $\mathcal{L}= 10^{34}\,\mathrm{cm}^{-2}\cdot \mathrm{s}^{-1}$ \cite{Acosta:2022ejc}. It can be constructed through a staged program that involves development of the core muon beam technology followed by installation of {\it one} muon beam at an existing facility (Fermilab, CERN,...) that has a high-energy hadron beam. A muon-hadron collider therefore can have an energy reach in DIS comparable to the LHeC or even FCC-eh, while at the same time serving as a technology demonstrator for $\mu^+\mu^-$ colliders.
A Muon-Ion Collider (MuIC) at the Brookhaven National Laboratory would be a natural successor of the EIC program in 2040's. Key merits of the MuIC proposal are the strong synergy with existing accelerator HEP and nuclear physics facilities in the US and expansion of QCD studies at the EIC into a new range of energies. On the one hand, the MuIC energy would be high enough to conclusively study small-$x$ partonic saturation with multiple types of ions and even beam polarization. On the other hand, at such energies parity-violating processes with weak bosons can be incisively employed to probe the flavor and spin properties of various nucleon and nuclear targets. The MuIC would also be a discovery machine, as it would produce a variety of final states with Higgs bosons and top quarks and perform unique searches for compositeness, leptoquarks, parity-violating BSM interactions. Due to its unique kinematics stipulated by the initial beam energies, the MuIC would require development of a very forward muon spectrometer to operate at pseudorapidities $\eta \approx - 7$ and muon energies up to 1 TeV \cite[Appendix in Ref.][]{Acosta:2022ejc}. 

Figure 1 in Ref.~\cite{Acosta:2022ejc} compares the CM energies and instantaneous luminosities of the past and proposed future electron-hadron and muon-hadron colliders.  Figure 4 of the same whitepaper shows possible timelines for the construction of muon-ion colliders at BNL and CERN. 

\subsection{Future Hadron Colliders}
\label{sec:fcchh}

High-energy hadron colliders provide the best opportunity to make a wide range of precision measurements of perturbative and nonperturbative QCD. Measurements of jet, photon, and top-quark cross-sections test 
higher-order perturbative QCD, and constrain parton distribution and fragmentation functions, and the running of $\alpha_s$. 
There are currently two major future hadron-hadron collider proposals, the FCC-hh at CERN and the SPPC in China, both targeting $pp$ collisions at a center of mass (CM) energy of about 100~TeV. Each machine would deliver an integrated luminosity of around 25 ab$^{-1}$ per experiment, reaching an instantaneous luminosity of $3\times 10^{35}$ cm$^{-2}$ $\mathrm{s}^{-1}$, almost an order of magnitude larger than the HL-LHC. 

These are extremely ambitious projects requiring breakthroughs in accelerator technology, detector design, and physics object reconstruction, and a coherent effort in all aspects is required.
The searches for the heaviest BSM objects in the unprecedented multi-TeV energy regime will observe QCD processes  in which all particles of the Standard Model, including top quarks and electroweak bosons, are emitted within parton showers. If, in addition to reconstructing multi-TeV final states, detectors for a 100 TeV machine are to provide the necessary precision to measure the SM processes, the detector coverage should be extended  with respect to the LHC detectors, since many SM processes are expected to be extremely forward. 
The calorimetry systems must provide excellent energy resolution over a wide range of energies in the central and forward regions, and increased hermetic coverage with respect to the LHC ones (reaching $|\eta|<6$). 

Studies have shown~\cite{calohep} that the granularity of the detector is of particular importance.
For instance,  SM Higgs decays into $ZZ$ pairs would produce two $Z$ bosons with multi-TeV energies, each with p${}_T$ less than 100~GeV, and opening angles between the $Z$ boson decay products of about $0.1$ radian. Detector capabilities to reconstruct these objects are fairly challenging (for instance, an average $Z$ boson would shower mostly within one LHC calorimeter cell). 
This challenge is accentuated by so-called ``hyper-boosted'' jets, whose decay products are collimated into areas the size of single calorimeter cells. Holistic detector designs that integrate tracking, timing, and energy measurements are needed to mitigate for these conditions~\cite{larkoski2015tracking,Chang:2013rca,Elder:2018mcr, Spannowsky:2015eba,ATLAS-CONF-2016-035,Gouskos:2642475}.
The extreme levels of radiation present in a 100~TeV collider pose another challenge for the design. A factor-of-five larger pileup than at the HL-LHC is expected posing stringent criteria on the detector design. 
The energy calibration of calorimeter cells, composite clusters, single particles, and jets is a challenging task at a 100~TeV $pp$ collider~\cite{Aleksa:2019pvl}. 
Detailed studies at higher energies will be needed to achieve the best possible precision at future colliders. 

\section{The strong coupling and tests of RG evolution}
\label{sec:alphas}

The strong coupling, $\alpha_s$, is a fundamental parameter of the SM, and it is also the least well known of its gauge couplings. The uncertainty on $\alpha_s$ will be one of the limiting factors in measurements at the High-Luminosity LHC and other experiments. Detailed summaries of the status of $\alpha_s$ determinations are given in the PDG review on Quantum Chromodynamics \cite{Workman:2022ynf} and a dedicated Snowmass whitepaper~\cite{dEnterria:2022hzv}.
Table~\ref{tab:alphas_prospects} first shows the results of the seven extraction methods that contribute to the PDG world-average combination. The lattice QCD methods are described in more detail in Sec.~\ref{sec:alphas_from_lattice} and involve themselves a variety of techniques. Each
category lists the dominant sources of theoretical and experimental uncertainty that propagate into $\alpha_s(m_\mathrm{Z})$ today, as well as feasible targets for reducing these sources within about 10 years and, in parentheses, in even longer future. The last row shows
the relative uncertainty of the current world average  (0.8\%) and of the one expected within the next decade ($\approx$\,0.4\%).  Section~\ref{sec:alphas_average} briefly reviews the procedures to compute the world average.

The long-term prognoses in the parentheses show that, in principle, $\sim 0.1$\% precision can be ultimately achieved in $\alpha_s(m_\mathrm{Z})$ determinations from at least one lattice QCD method and electroweak fits at a future high-luminosity $\epem$ facility. To translate these advances into the per mil precision of the world-average result in the final row, no large unexplained discrepancies should impact the individual extraction methods. Ruling out such discrepancies presents an emerging challenge for precise analyses of $\alphasmZ$ and PDFs, as the number of contributing systematic factors grows rapidly when the targeted precision increases, and when multiple hard scales are present in the problem. In the latest precision analyses such as the latest PDG $\alpha_s$ combination (cf. Fig.~\ref{fig:alphas_2022}), some experimental determinations show mutual tensions, and those may be further exacerbated when some measurements have a substantially smaller uncertainty than the others. The same complication may occur with theoretical predictions. A significant inconsistency among several $\alpha_s$ extractions hence may be much more likely due to unknown or underestimated systematic errors in experiment or theory than because of BSM physics. Since we will probably continue to face such issues in the future, exhaustive exploration of systematic effects will become critical for interpretation of precise $\alpha_s$ extractions. Agreed-upon protocols for resolution of conflicts among individual determinations could be helpful, as well as cross calibration of common sources of systematics.

\begin{table}[tb]
\centering
\caption{Summary of current and expected future (within the decade ahead or, in parentheses, longer time scales) uncertainties in the $\alpha_s(m_\mathrm{Z})$ extractions used today to derive the world average of $\alpha_s$. Acronyms and symbols: 
CIPT=`contour-improved perturbation theory', FOPT=`fixed-order perturbation theory', NP=`nonperturbative QCD', SF=`structure functions', PS=`Monte Carlo parton shower'.  
Entries of the table are explained in Ref.~\cite{dEnterria:2022hzv}, from which the table is taken. Category 7 is based on $\alpha_s(M_\mathrm{Z})$ determinations from $\ell h$ and $hh$ collider observables performed outside of global PDF fits (category 4)-- see the discussion item (f) in Sec.~10.1 of Ref.~\cite{dEnterria:2022hzv}. 
\label{tab:alphas_prospects}\vspace{0.2cm}}
\renewcommand\arraystretch{1.1}
\resizebox{\textwidth}{!}{%
\begin{tabular}{lcc} \hline\hline
        & \multicolumn{2}{c}{Relative $\alpha_s(m_\mathrm{Z})$ uncertainty}\\ 
Method  & Current &  Near (long-term) future \\
 & theory \& exp.\ uncertainties sources & theory \& experimental progress \\\hline
\multirow{2}{*}{(1) Lattice} & 
\textcolor{red}{$0.7\%$} &  \textcolor{red}{$\approx\,0.3\%~(0.1\%)$} \\
& Finite lattice spacing \& stats. & Reduced latt.\ spacing. Add more observables\\
& N$^{2,3}$LO pQCD truncation      & Add N$^{3,4}$LO, active charm (QED effects)\\
&   & Higher renorm.\ scale via step-scaling to more observ.\\
\hline
\multirow{2}{*}{(2) $\tau$ decays} 
 & \textcolor{red}{$1.6\%$} &  \textcolor{red}{$<1.\%$} \\
 & N$^3$LO CIPT vs.\ FOPT diffs. & Add N$^4$LO terms. Solve CIPT--FOPT diffs. \\
 & Limited $\tau$ spectral data  & Improved $\tau$ spectral functions at Belle~II \\
\hline
\multirow{2}{*}{(3) $Q\bar{Q}$ bound states} & 
 \textcolor{red}{$3.3\%$} &  \textcolor{red}{$\approx\,1.5\%$}  \\
 & N$^{2,3}$LO pQCD truncation & Add N$^{3,4}$LO \& more $(\ccbar)$, $(\bbbar)$ bound states\\
 & $m_\mathrm{c,b}$ uncertainties & Combined $m_\mathrm{c,b}+\alpha_s$ fits\\
\hline
\multirow{2}{*}{(4) DIS \& global PDF fits} 
 & \textcolor{red}{$1.7\%$} & \textcolor{red}{$\approx\,1\%$~(0.2\%)} \\
 & N$^{2,(3)}$LO PDF (SF) fits &  N$^3$LO fits. Add new SF fits: $F^{p,d}_2,\,g_i$ (EIC) \\
 & Span of PDF-based results & Better corr.\ matrices, sampling of PDF solutions. \\
 & & More PDF data (EIC/LHeC/FCC-eh) \\
\hline
\multirow{2}{*}{(5) $e^+e^-$ jets \& evt shapes} 
  & \textcolor{red}{$2.6\%$} & \textcolor{red}{$\approx\,1.5\%$~($<1$\%)} \\
 & \NNLOgen+N$^{(1,2,3)}$LL truncation & Add N$^{2,3}$LO+N$^3$LL, power corrections\\
 & Different NP analytical \& PS corrs.  & Improved NP corrs.\ via: NNLL PS, grooming\\
 & Limited datasets w/ old detectors & New improved data at B factories (FCC-ee)\\
\hline
\multirow{2}{*}{(6) Electroweak fits} 
 & \textcolor{red}{$2.3\%$}  & \textcolor{red}{($\approx\,0.1\%$)} \\
 & N$^3$LO truncation & N$^4$LO, reduced param.\ uncerts.\ ($m_\mathrm{W,Z},\,\alpha,$ CKM) \\
 & Small LEP+SLD datasets & Add $W$ boson. Tera-Z, Oku-W datasets (FCC-ee) \\
 \hline
\multirow{2}{*}{(7) Standalone hadron collider observables}
 & \textcolor{red}{2.4\%} & \textcolor{red}{$\approx\,1.5\%$} \\
  & \NNLOgen(+NNLL) truncation, PDF uncerts. & N$^3$LO+NNLL (for color-singlets), improved PDFs \\
 & Limited data sets ($\ttbar$, $W$, $Z$,e-p jets) & Add more datasets: $Z$ $\pT$, p-p jets, $\sigma_i/\sigma_j$ ratios,...\\
\hline
World average & \textcolor{red}{$0.8\%$}  & \textcolor{red}{$\approx\,0.4\%$~(0.1\%)} \\
 \hline\hline
 \end{tabular}
}
\end{table}

\begin{table}[tp]
\centering
\caption{Values of $\alphasmZ$ determined at N$^3$LO accuracy from Z-boson pseudoobservables ($\Gamma_\mathrm{Z}^\mathrm{tot}$, $R_\mathrm{Z}$, and $\so$) individually, combined, as well as from a global SM fit, with propagated experimental, parametric, and theoretical uncertainties broken down~\cite{dEnterria:2020cpv}. The last two rows list the expected values at the FCC-ee from all $Z$ pseudoobservables 
combined and from the corresponding SM~fit.\vspace{0.2cm}\label{tab:alphas_Z}}
\tabcolsep=4.5mm
\begin{tabular}{lcccc}\hline\hline
Observable & $\alphasmZ$   & \multicolumn{3}{c}{uncertainties}\\
     &     & exp.  &  param. & theor.      \\\hline
$\Gamma_\mathrm{Z}^\mathrm{tot}$ & $0.1192 \pm 0.0047$ & $\pm0.0046$ & $\pm0.0005$ & $\pm0.0008$ \\ 
$R_\mathrm{Z}$ & $0.1207 \pm 0.0041$ & $\pm0.0041$ & $\pm0.0001$ & $\pm0.0009$ \\ 
$\so$ & $0.1206 \pm 0.0068$ & $\pm0.0067$ & $\pm0.0004$ & $\pm0.0012$ \\ 
All above combined & $0.1203 \pm 0.0029$ & $\pm0.0029$ & $\pm0.0002$ & $\pm0.0008$ \\ 
Global SM fit & $0.1202 \pm 0.0028$ & $\pm0.0028$ & $\pm0.0002$ & $\pm0.0008$ \\ \hline
All combined (FCC-ee)  & $0.12030 \pm 0.00026$ & $\pm0.00013$ & $\pm0.00005$ & $\pm 0.00022$\\ 
Global SM fit (FCC-ee) & $0.12020 \pm 0.00026$ & $\pm0.00013$ & $\pm0.00005$ & $\pm 0.00022$\\\hline\hline
\end{tabular}
\end{table}

\begin{minipage}{0.48\textwidth}
\begin{figure}[H]
\centering
\includegraphics[width=0.82\textwidth]{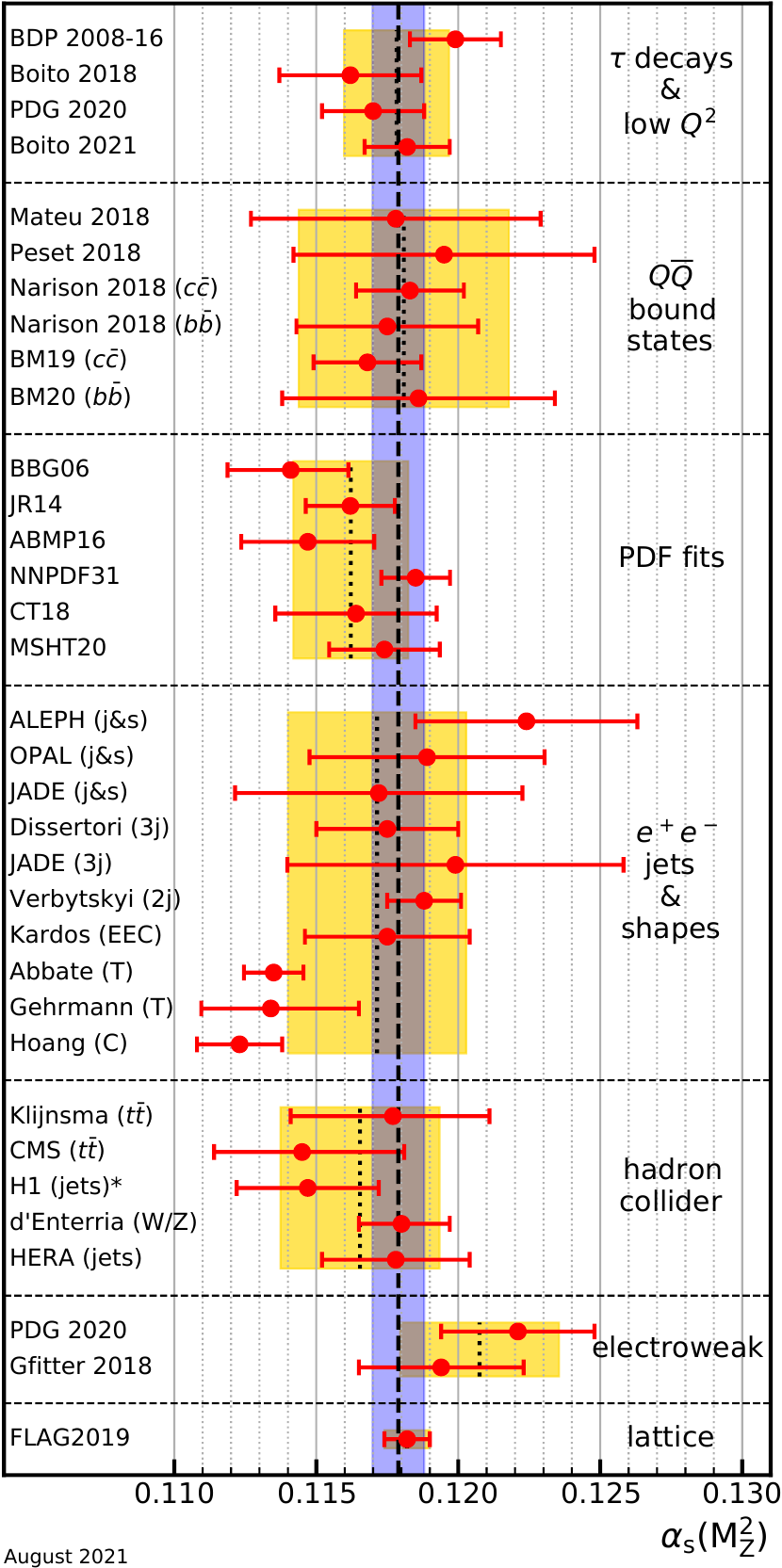}
\caption{Summary of latest determinations of $\alphasmZ$ from seven subfields. 
The yellow (light shaded) bands and dotted lines indicate the pre-average values of each subfield. The dashed line and blue (dark shaded) band represent the final $\alphasmZ$ world average [March'22 update of the PDG'21 results~\cite{ParticleDataGroup:2020ssz}].}
\label{fig:alphas_2022}
\end{figure}
\end{minipage}
\quad\quad
\begin{minipage}{0.5\textwidth}
\begin{figure}[H]
\includegraphics[width=1\textwidth]{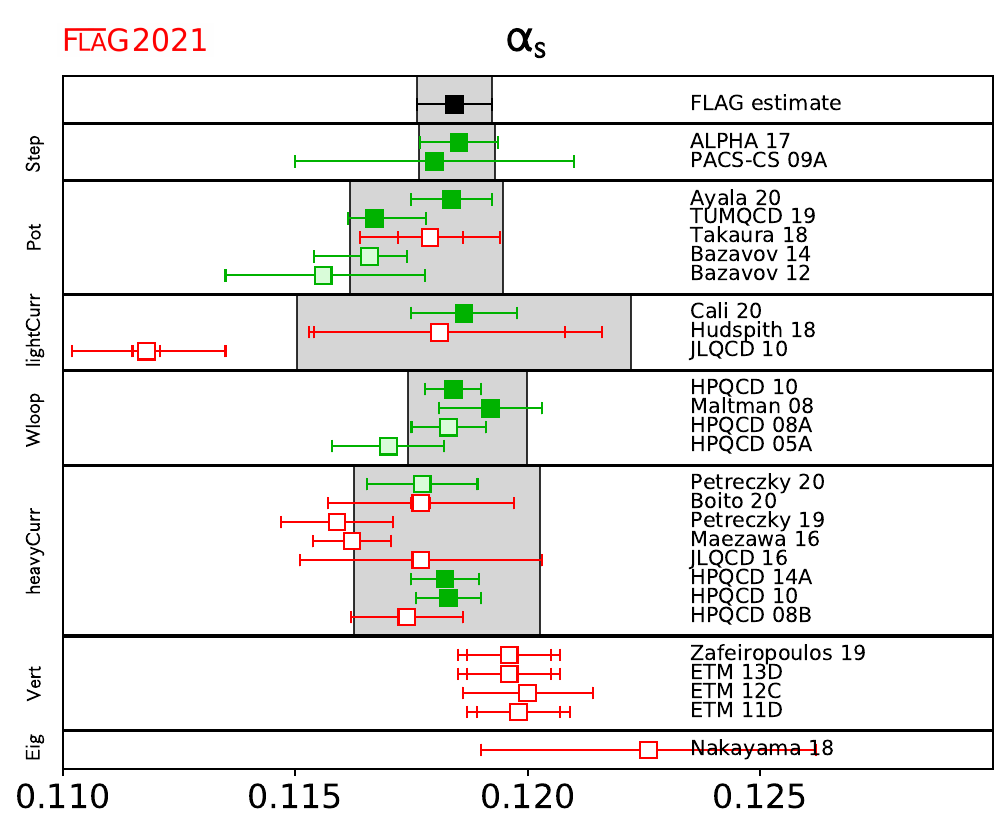}
\includegraphics[width=1\textwidth,clip,trim=0 3mm 0 0]{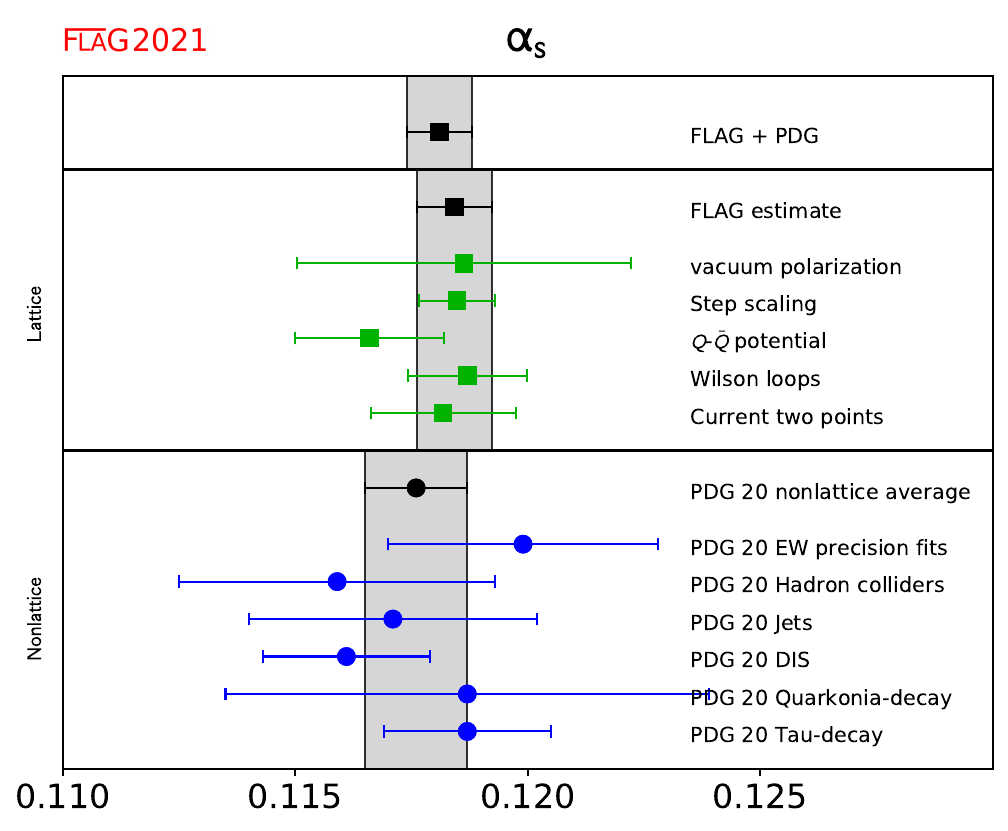}
\caption{Lattice determinations of the strong coupling, pre-averages from different calculational methods and final averaging.
Figures taken from~\cite{Aoki:2021kgd}.\label{fig:alphas_flag}}
\end{figure}
\end{minipage}

\subsection{Extraction of \texorpdfstring{$\alpha_s$}{QCD coupling} from \texorpdfstring{e$^\text{+}$e$^-$}{ep-em} data}

At the FCC-ee, combining the $3\times10^{12}$ $Z$ bosons decaying hadronically 
at the $Z$ pole, and the $\sqrts$ calibration to tens of keV accuracy obtained using resonant depolarization~\cite{Blondel:2021zix}, will provide measurements with unparalleled precision. The statistical uncertainties in the $Z$ mass and width, today of $\pm$1.2~MeV and $\pm$2~MeV (dominated by the LEP beam energy calibration), will be reduced to below $\pm$4~keV and $\pm$7~keV respectively. 
Similarly, the statistical uncertainty in measuring 
$Z$ boson partial widths ($R_\mathrm{\ell, Z}$) will be negligible, and the Z\,$\to\mu^+\mu^-$ decay channel alone, yielding an experimental precision of 0.001 from the knowledge of the detector acceptance, will suffice to reach an absolute (relative) uncertainty of 0.001 ($5\times10^{-5}$) on the ratio of the hadronic-to-leptonic partial $Z$ widths. Thus, accounting for the dominant experimental systematic uncertainties at the FCC-ee, one can expect $\delta m_\mathrm{Z}$~=~0.025--0.1~MeV, $\delta \Gamma_\mathrm{Z} = 0.1$~MeV, $\delta\sigma_\mathrm{Z}^{\rm had}=4.0$~pb, and $\delta R_\mathrm{\ell, Z} = 10^{-3}$~\cite{FCC:2018evy}. In addition, the QED coupling at the $Z$ peak will be measured with a precision of $\delta \alpha  = 3\times10^{-5}$~\cite{Janot:2015gjr}, thereby also reducing the corresponding propagated parametric uncertainties. Implementing the latter uncertainties into GFitter leads to the results listed in the last 
two rows of Table~\ref{tab:alphas_Z}, where the central $\alphasmZ$ value is arbitrarily set at the current SM global fit extraction~\cite{dEnterria:2020cpv}.
The final uncertainties in the QCD coupling constant are reduced to the $\sim$0.1\% level, namely about three times smaller than the propagated theoretical uncertainties today. Theoretical developments in the years to come should further bring down the latter by a factor of four~\cite{Proceedings:2019vxr}. A final QCD coupling constant extraction at the FCC-ee with a 2-per mil total uncertainty is thereby reachable: $\alphasmZ = 0.12030 \pm 0.00013_\mathrm{exp} \pm 0.00005_\mathrm{par} \pm 0.00022_\mathrm{th}$ (Table~\ref{tab:alphas_Z}). 
The large improvement, by more than a factor of ten, in the FCC-ee extraction of $\alphasmZ$ from the $Z$ boson data (and its comparison to the similar extraction from the $W$ boson pseudoobservables) will enable searches for small deviations from the SM predictions that could signal the presence of new physics contributions. 

\subsection{Extraction of \texorpdfstring{$\alpha_s$}{QCD charge} from \texorpdfstring{e${}^\pm$p}{DIS} data}
\begin{figure}[thbp!]                           
  \centering                                      
  \subfloat[][]{\includegraphics[width=0.45\textwidth]{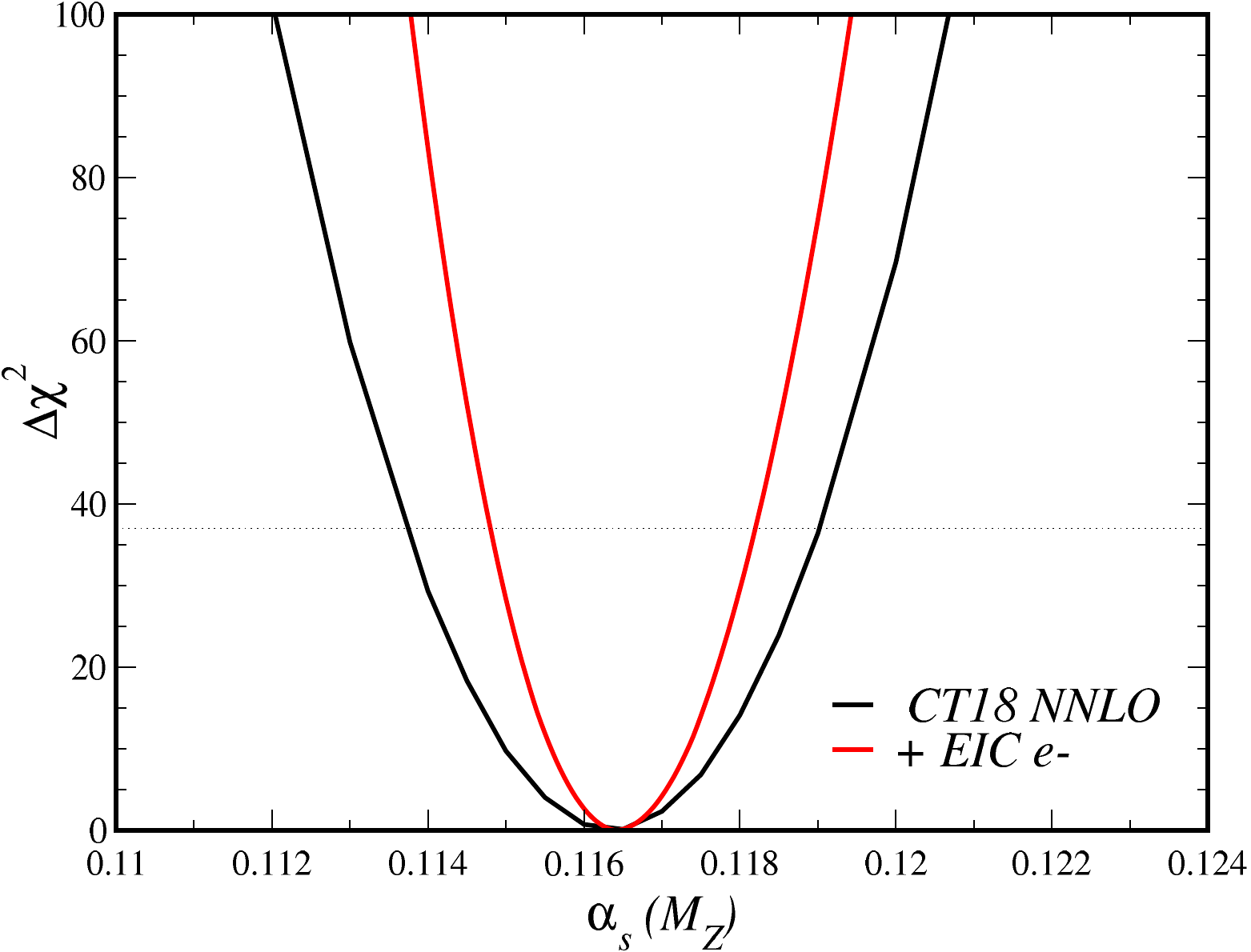}}\\ 
  \subfloat[][]{\includegraphics[width=0.45\textwidth]{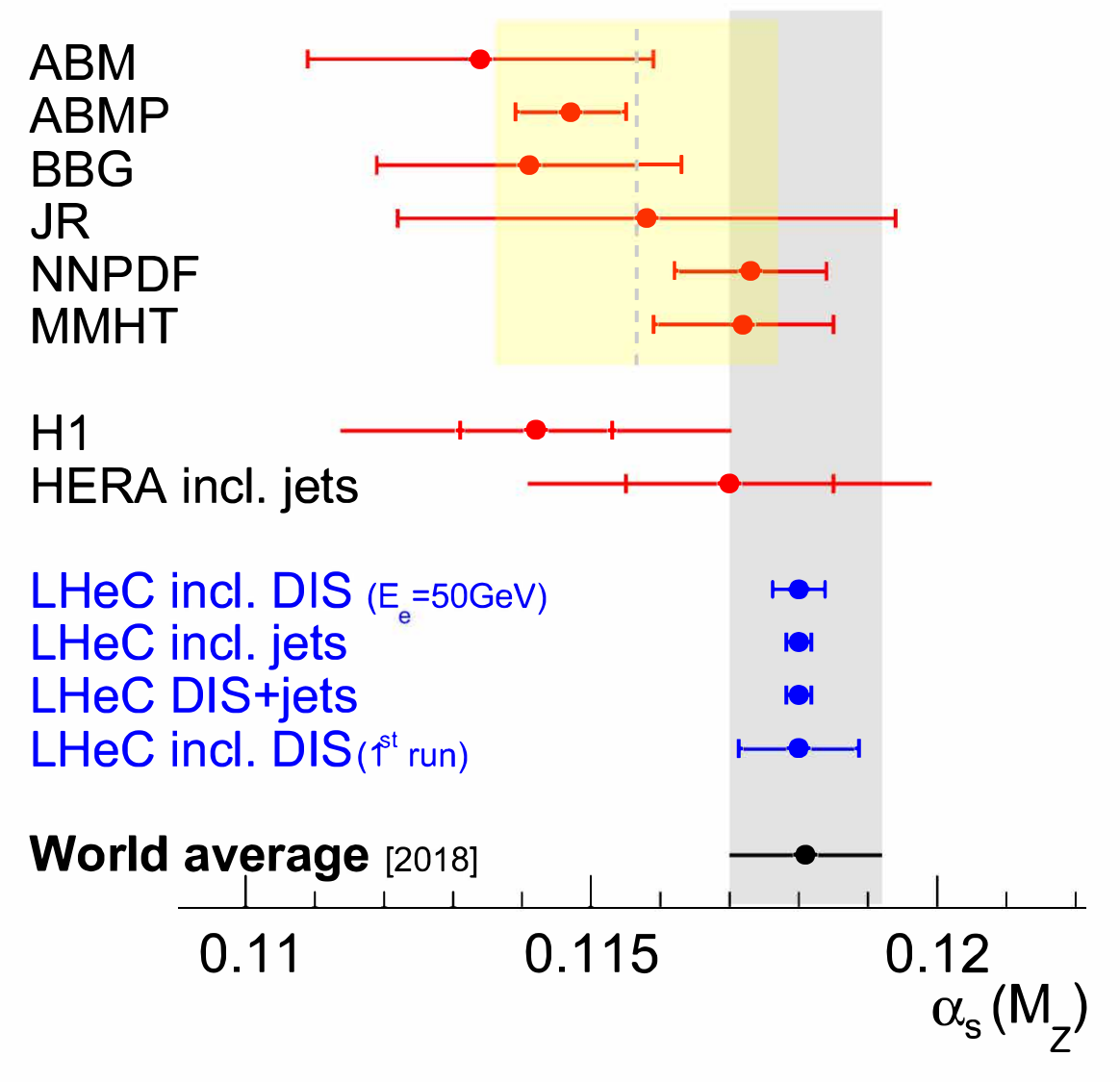}}
  \subfloat[][]{\includegraphics[width=0.45\textwidth]{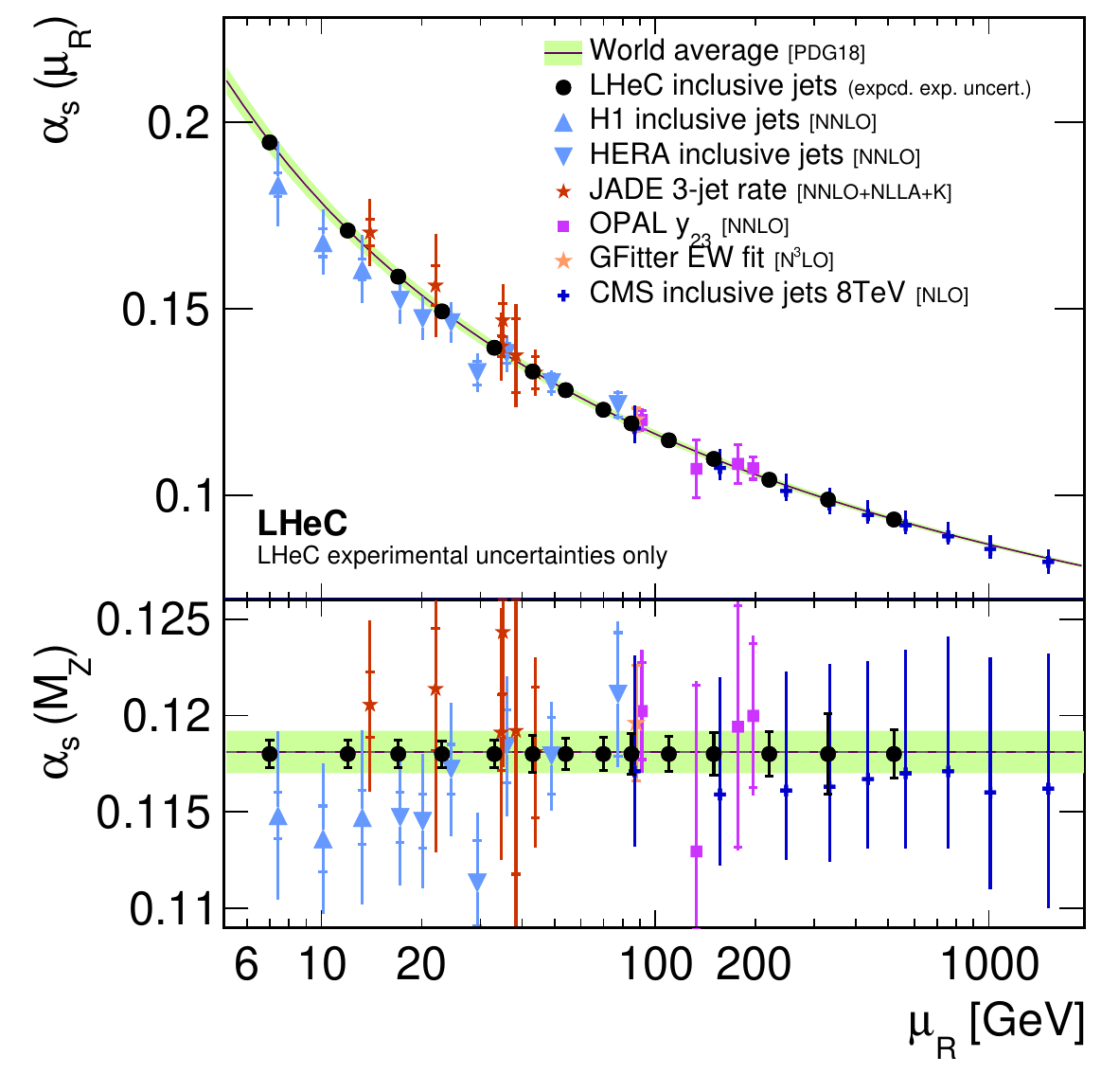}} 
  \caption{
    (a) An estimated precision improvement in the $\alpha_s$ determination in the CT18 NNLO fit \cite{Hou:2019efy}, as quantified by the log-likelihood $\chi^2$, following the inclusion of $100 \mbox{ fb}^{-1}$ of EIC inclusive electron-proton scattering data. From Refs.~\cite{AbdulKhalek:2021gbh,AbdulKhalek:2022erw}.
    (b) Comparison of prospected determination of $\alphasmZ$ from
    inclusive DIS data at the LHeC in comparison to determinations in
    (global) PDF fits.
    (c)
    An illustration of the prospected experimental uncertainties in a
    study of the running of $\alpha_s$ from inclusive jet cross
    sections at the proposed LHeC experiment.
    The LHeC figures are taken from Ref.~\cite{LHeC:2020van}.
     \label{fig:alphasEICLHeC}
  }
 
\end{figure}

Future electron-proton collider experiments provide many opportunities
for precision determinations of $\alphas$. At lower center-of-mass energies, the EIC in the US~\cite{AbdulKhalek:2022erw,Accardi:2012qut,AbdulKhalek:2021gbh} and the EicC in China~\cite{Anderle:2021wcy} would provide new high-luminosity data. 
As an example, Fig.~\ref{fig:alphasEICLHeC}(a) suggests reduction in the $\alpha_s$ uncertainty extracted from the CT18 NNLO global PDF fit \cite{Hou:2019efy} by up to 40\% after a high-statistics sample of the simulated inclusive $ep$ DIS data for the EIC is included. In all $ep$ measurements, the $\alpha_s$ value is correlated with the PDFs, especially the gluon PDF. Hence all provided projections assume that strong constraints on the PDFs will be simultaneously obtained. The EIC will provide also a novel possibility to extract $\alpha_s$ at N$^3$LO accuracy analyzing polarized PDFs, by exploiting the Bjorken sum rule~\cite{Deur:2014vea}, with a few percent precision~\cite{dEnterria:2022hzv}. Also, DIS global event shapes at the EIC on their own, such as 1-jettiness, can determine of $\alphasmZ$ at a level of a few percent~\cite{AbdulKhalek:2021gbh}.

The proposed Large Hadron electron Collider at CERN (LHeC)~\cite{LHeCStudyGroup:2012zhm,LHeC:2020van} would provide
$e^\pm p$ collision data at a center-of-mass energy of 1.3 TeV, and hence its measurements of
hadronic final-state observables would cover a considerably larger kinematic range than at the $ep$ collider HERA. Inclusive neutral-current and charged-current DIS cross sections would be also measured with high precision both in the low-$x$ and  high-$x$ regions, given an excellent detector acceptance and high luminosity. Inclusive DIS data alone would allow one to measure $\alpha_s$ very precisely, again assuming tightly constrained PDFs, and to the extent that was not fully possible with HERA data. An experimental uncertainty of
\begin{equation}
  \delta\alphasmZ = \pm0.00022\,\text{(exp+PDF)}\,,
\end{equation}
could  be possibly achieved in a combined determination of PDFs and $\alphasmZ$ \cite{LHeC:2020van}. These and following estimates assume an idealized uncertainty on the PDFs ("given by $\Delta\chi^2=1$ at 68\% probability"). More realistic PDF uncertainties tend to be larger because of such factors as some inconsistencies between experiments \cite{Kovarik:2019xvh}. As an illustration, Fig.~\ref{fig:alphasEICLHeC}(b) compares the prospected uncertainties after the LHeC using the idealized prescription with recent determinations in global PDF fits. 

A simulation of inclusive jet cross section data, using realistic models of systematic uncertainties, suggests that a determination of $\alphasmZ$ with
uncertainty of
\begin{equation}
  \delta\alphasmZ = \pm0.00013\,\text{(exp)}~\pm0.00010\,\text{(PDF)}
\end{equation}
can be within reach in the HL-LHC + LHeC era. The right-hand side separately shows the experimental and PDF uncertainties (estimated with a disclaimer as above). Similarly as at HERA, the running of $\alpha_s$ could be investigated at the LHeC as a function of the renormalization scale using jet data. Figure~\ref{fig:alphasEICLHeC}(c) displays prospects for such scale-dependent determinations of $\alphasmZ$ (and corresponding values of $\alpha_s(\mu)$).
It is observed that, from LHeC inclusive jet cross sections, the
running could be tested in the range from a few GeV up to about 600\,GeV with per mil precision, resulting in  an indispensable experimental confirmation of validity of renormalized QCD from the low-scale $\alpha_s$ determinations from $\tau$-decays or lattice QCD to TeV scales.

\subsection{Extraction of \texorpdfstring{$\alpha_s$}{QCD charge} from lattice QCD}
\label{sec:alphas_from_lattice}

Multiple lattice QCD methods have been developed to extract the strong 
coupling constant: 
the {step-scaling}~\cite{Luscher:1991wu,Luscher:1993gh,deDivitiis:1994yz,Jansen:1995ck,Symanzik:1981wd,Luscher:1985iu,Luscher:1992an,Sint:1993un,Bode:1999sm,Bruno:2017gxd}, {small Wilson loops}~\cite{Mason:2005zx,Davies:2008sw,Maltman:2008bx,McNeile:2010ji}, 
QCD static energy~\cite{Pineda:2000gza,Brambilla:2004jw,Brambilla:2009bi,Bazavov:2014soa,Takaura:2018lpw,Takaura:2018vcy,Bazavov:2018wmo,Bazavov:2019qoo,Ayala:2020odx,Aoki:2021kgd,Komijani:2020kst}, heavy-quark two-point correlators~\cite{Sturm:2008eb,Kiyo:2009gb,Maier:2009fz,HPQCD:2008kxl,McNeile:2010ji,Chakraborty:2014aca,Maezawa:2016vgv,Petreczky:2019ozv,Petreczky:2020tky}, hadronic vacuum polarization~\cite{Baikov:2008jh,JLQCD:2008bwj,Shintani:2010ph,Hudspith:2018bpz,Cali:2020hrj}, QCD vertex functions~\cite{Alles:1996ka, Boucaud:2001qz,Blossier:2010ky,Blossier:2011tf,Blossier:2012ef,Blossier:2013ioa,Zafeiropoulos:2019flq}, decoupling method~\cite{DallaBrida:2019mqg,dEnterria:2022hzv}, eigenvalues of the Dirac operator~\cite{Chetyrkin:1994ex,Kneur:2015dda} methods. See recent reviews of the lattice methods for $\alpha_s$ in Ref.~\cite{DallaBrida:2020pag,Komijani:2020kst,DelDebbio:2021ryq,dEnterria:2022hzv}.   
The Flavor Lattice Averaging Group (FLAG) report~\cite{Aoki:2013ldr,Aoki:2016frl,FlavourLatticeAveragingGroup:2019iem,Aoki:2021kgd} attempts a global lattice average (cf. Fig.~\ref{fig:alphas_flag}) 
\begin{align}
 &\alpha_s(m_\mathrm{Z}) = 0.1184(8)&&\text{(FLAG global average)~\cite{Aoki:2021kgd}}.
 \label{eq:FLAG average}
\end{align}
FLAG uses a set of quality criteria to decide which determinations to include in the average. This procedure is similar to that adopted by the PDG. Reference~\cite{dEnterria:2022hzv} suggested that the FLAG and PDG procedures should be harmonized as much as possible. Unlike the procedure used for FLAG averages of other quantities, for $\alpha_s$ FLAG applies its own view of the perturbative-truncation uncertainty and inflates the error of some subaverages. Note that there have been updates to individual analyses since the current FLAG 2021 report~\cite{Aoki:2021kgd} was published. 

The current $\pm0.7\%$ precision of the lattice-QCD extraction of $\alphasmZ$ can be reduced by about a factor of two within the next $\sim$10 years. In order to improve the lattice-QCD--based determinations of $\alphas$, it would be important to reach higher renormalization scales by both advancing the lattice simulations and incorporating improved (higher-order) pQCD counterpart calculations. 
Lattice simulations should be run with smaller lattice spacings, allowing even better continuum extrapolations, and they should include charm-quark effects (2+1+1-flavor calculations). Perturbative expansions will require calculating N$^3$LO, N$^4$LO, and/or N$^3$LL contributions, depending on the process under study. In addition, treatment of QED and isospin-breaking effects in both the scale setting and running of $\alphas$ may be needed in some cases (in particular, when aiming at longer-term 0.1\% uncertainties).

To further reduce the error from lattice calculations, sufficient dedicated computing resources are needed to generate state-of-the-art samples for lattice-QCD analyses. 
Enough person-power will be necessary to develop perturbation theory for selected observables in a finite spacetime volume and to compute identified higher-order pQCD corrections to match improved lattice-QCD samples.

\subsection{The world-average combination of \texorpdfstring{$\alpha_s$}{QCD charge}}
\label{sec:alphas_average}
The lower Fig.~\ref{fig:alphas_flag} illustrates the 2021 method to obtain the world-average value of $\alpha_s$ \cite{Workman:2022ynf}. Separate weighted averages of lattice and non-lattice determinations are first computed. Then the final world-average combination of $\alpha_s(m_Z)=0.1179\pm0.0009$ (labeled as "FLAG+PDG" in the figure) is computed as the average of non-lattice and lattice values, with the relative accuracy of 0.8\% shown in Table~\ref{tab:alphas_prospects}.
The future projection of the $\alpha_s$ combination in the rightmost column of Table~\ref{tab:alphas_prospects} is based on a weighted average of the seven categories on equal footing, which  would give $\alpha_s(m_Z)=0.1180\pm0.0006$ (i.e., a smaller uncertainty of 0.5\%) in the case of the current world average.

The world-average combination prescription may evolve as groups of experiments get more precise, possibly revealing currently unseen disagreements. In the future, lattice determinations may be combined with the experimental ones that are affected by systematics of similar origin, such as $\tau$ decays \cite{DelDebbio:2021ryq}. Alternatively, if one group of determinations becomes much more precise than the others, it could be used as a reference. Correlations between different groups of determinations in Table~\ref{tab:alphas_prospects} deserve further scrutiny.  For example, the extractions from the hadron collider category 7 are PDF-dependent and, therefore, must share some degree of correlation with the extraction from the DIS+global-PDF fits category 4.

\subsection{The running bottom quark mass}
Within the Standard Model of particle physics, the masses of quarks are free parameters whose values must be determined experimentally, while their scale dependence is predicted by scheme dependent Renormalization Group Evolution (RGE). These calculations have reached 5-loop accuracy~\cite{Vermaseren:1997fq,Chetyrkin:1997dh,Baikov:2014qja} and have been implemented in public software packages~\cite{Herren:2017osy,Hoang:2021fhn}. 

The most precise extractions of the bottom quark mass~\cite{Narison:2019tym,Peset:2018ria,Kiyo:2015ufa,Penin:2014zaa,Alberti:2014yda,Beneke:2014pta,Dehnadi:2015fra,Lucha:2013gta,Bodenstein:2011fv,Laschka:2011zr,Chetyrkin:2009fv} rely on the measurement of the mass of bottomonium bound states and the $e^+e^- \rightarrow $ hadrons cross section as experimental input, in combination with QCD sum rules and perturbative QCD calculations. Several lattice QCD groups have also published results, the most recent of which reaches a precision of approximately 0.3\%~\cite{Bazavov:2018omf,Colquhoun:2014ica,Bernardoni:2013xba,Lee:2013mla,Dimopoulos:2011gx} (see also the FLAG report~\cite{Aoki:2019cca}). 
The world average provided by the Particle Data Group (PDG)~\cite{ParticleDataGroup:2020ssz} also includes inputs from HERA~\cite{H1:2018flt} and the BaBar and Belle experiments at the B-factories~\cite{Schwanda:2008kw,Aubert:2009qda}. Extractions from $Z$-pole data were performed at LEP~\cite{Abreu:1997ey,Barate:2000ab,Abbiendi:2001tw,Abdallah:2005cv,Abdallah:2008ac} and SLD~\cite{Brandenburg:1999nb,Abe:1998kr}.
Measurements of the bottom quark mass at the scale of the Higgs boson mass were performed in~\cite{Aparisi:2021tym}, based on ATLAS~\cite{ATLAS-CONF-2020-027} and CMS~\cite{Sirunyan:2018koj} experimental data.

In the next decade the study of the ``running'' of the bottom quark mass is expected to turn into a precision test of QCD~\cite{Aparisi:2022yfn}. These investigations will complement analogous studies for the running charm mass, such as in \cite{Gizhko:2017fiu}. Measurements at several energy scales in bottomonium, $Z$, and Higgs production can be used, in a general way, to sense the presence of massive new colored states that may contribute to the quark mass evolution. A dedicated high-luminosity $e^+e^-$ run at the $Z$-pole, i.e.\ the ``GigaZ" program of a linear collider or the ``TeraZ'' run at the circular colliders, yields a sample of $Z$-bosons that exceeds that of the LEP experiments and SLD by orders of magnitude. Ref.~\cite{Fuster:2021ekh} provides an extrapolation under the assumption that the extraction of $m_\mathrm{b}(m_\mathrm{Z})$ from the three-jet rates will be limited by the theory uncertainty and hadronization uncertainties. This requires fixed-order calculations at \NNLOgen accuracy, with full consideration of mass effects.
The Higgs factory program, with several inverse attobarn at a center-of-mass energy of 240-250~GeV can take advantage of radiative-return events. The Lorentz-boost of the $Z$-bosons complicates the selection, reconstruction and interpretation. A dedicated full-simulation study is therefore required to provide a reliable, quantitative projection. However, it is clear that the radiative-return data has the potential to significantly improve the precision of existing LEP/SLC analyses.
Finally, a high-energy electron-positron collider operated at a center-of-mass energy of 250~GeV or above can extend the analysis to higher energies and thus probe the effect of coloured states with masses heavier than that the Higgs boson on the running of the bottom quark mass. The potential of the three-jet rate measurement to determine $m_\mathrm{b}(\mu)$ for $\mu=$ 250~GeV has been studied in Ref.~\cite{Fuster:2021ekh}. A measurement with a precision of 1~GeV was found to be feasible for $\mu=$ 250~GeV

The measurement of $m_\mathrm{b}(m_\mathrm{H})$ from the Higgs decay width to a bottom-antibottom quark pair is expected to increase rapidly in precision as the precision of Higgs coupling measurement improves~\cite{Aparisi:2021tym}. The current theory uncertainty from missing higher orders and parametric uncertainties from $\alpha_s$ and $m_\mathrm{H}$ is estimated to be 60~MeV~\cite{Aparisi:2021tym}, well below the current experimental precision. The theory uncertainty is dominated by the parametric uncertainty from the Higgs boson mass. The current uncertainty on the Higgs mass of 240~MeV leads to an uncertainty of $\sim$40~MeV on $m_\mathrm{b}(m_\mathrm{H})$ and is expected to come down considerably as more precise determinations of $m_\mathrm{H}$ appear. Future prospects for Higgs mass measurements are summarized in Ref.~\cite{deBlas:2019rxi}. Both the HL-LHC~\cite{Cepeda:2019klc} and the Higgs factory~\cite{Yan:2016xyx} are expected to provide a measurement of the Higgs boson mass to 10-20~MeV precision, which is sufficient to reduce the impact of this source of uncertainty on $m_\mathrm{b}(m_\mathrm{H})$ to below 10~MeV. The determination of $m_\mathrm{b}(m_\mathrm{H})$ in $H\rightarrow b\bar{b}$ decay is expected to become the ``golden'' measurement among the high-energy determinations~\cite{Aparisi:2022yfn}.

Ref.~\cite{Cepeda:2019klc} provides the projections for the LHC and its luminosity upgrade, extrapolating the partial run 2 results under the following assumptions: both statistical and systematic uncertainties are envisaged to scale with integrated luminosity $L$ as $1/\sqrt{L}$ up to certain limits, while theory uncertainties are expected to improve by a factor two. This ``S2 scenario'' leads to a projected uncertainty on the Higgs branching ratio to bottom quarks of 4.4\% (1.5\% stat., 1.3\% exp., 4.0\% theo.) and on $\lambda_{bz} = \mu^{bb}/\mu^{ZZ}$ of 3.1\% (1.3\% stat., 1.3\% syst., 2.6\% theo.), an improvement by nearly a factor of ten with respect to the first measurement in Ref.~\cite{Aparisi:2021tym}
 
In the next decades, with the completion of the high luminosity program of the LHC and the construction of a new ``Higgs factory" electron-positron collider, rapid progress is envisaged in the measurement of Higgs coupling measurement. These precise measurements will enable an extraction of the $\overline{\text{MS}}$ bottom quark mass $m_\mathrm{b}(\mu)$ at the scale given by the Higgs boson mass, $m_\mathrm{b}(m_\mathrm{H})$, with a precision of the order of 10~MeV. With a relative precision of 2 per mil, the high-scale measurement can reach a similar precision as $m_\mathrm{b}(m_\mathrm{b})$ based on low-energy measurements.

\begin{figure}[t]
\includegraphics[width=0.8\textwidth]{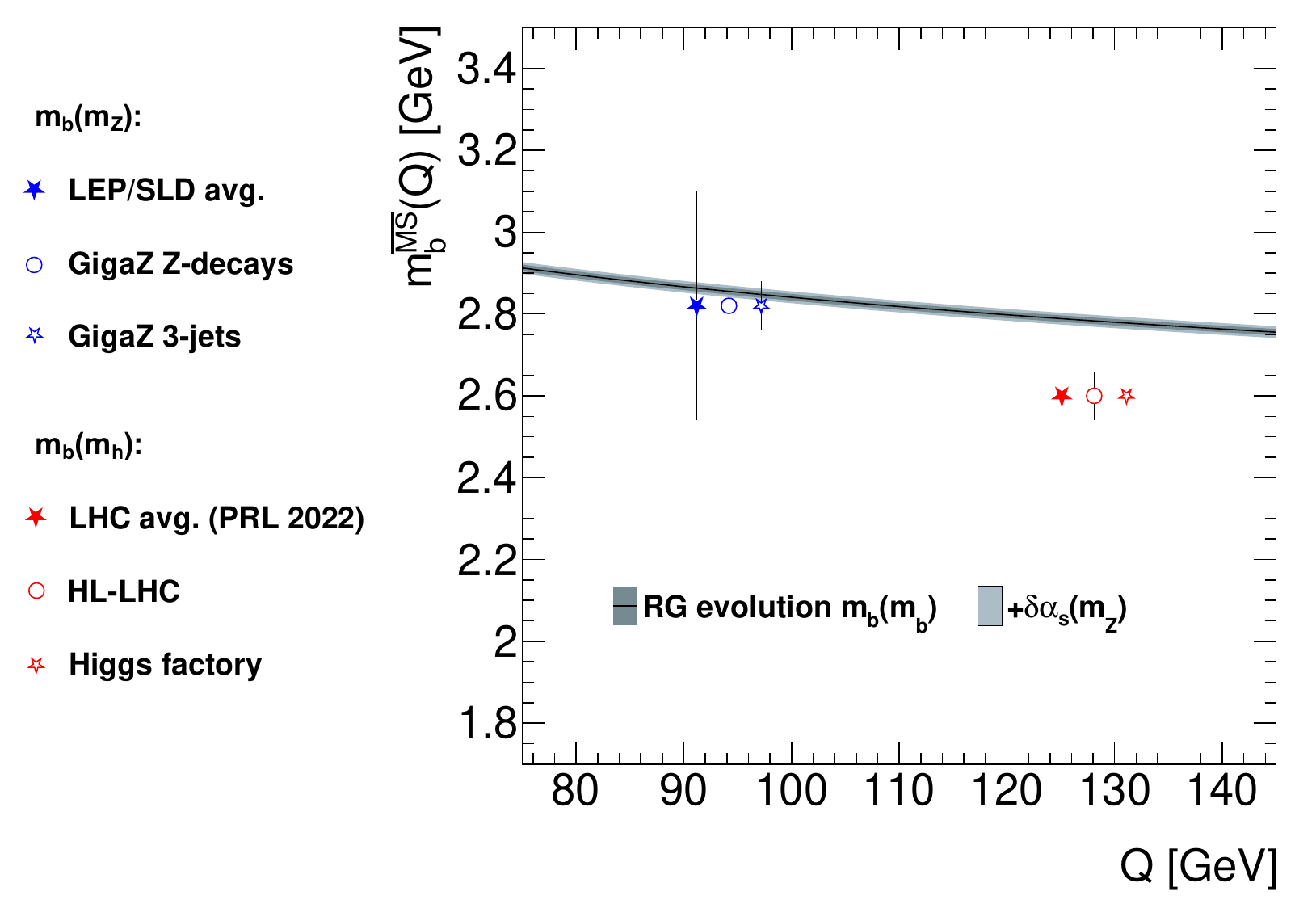}%
\caption{\label{fig:running_mass_projection} Prospects for measurements of the scale evolution of the bottom quark $\overline{\text{MS}}$ mass at future colliders. The markers are projections for $m_\mathrm{b}(m_\mathrm{Z})$ from three-jet rates at the $Z$-pole and for $m_\mathrm{b}(m_\mathrm{H})$ from Higgs boson branching fractions. The RGE evolution of the mass is calculated at five-loop precision with REvolver~\cite{Hoang:2021fhn}. }
\end{figure}
The projections and extrapolations discussed above are summarized in Fig.~\ref{fig:running_mass_projection}. The markers are centered on the current central values for $m_\mathrm{b}(m_\mathrm{Z})$ and $m_\mathrm{b}(m_\mathrm{H})$ and the error bars indicate the projected precision. The solid line indicates the evolution of the PDG world average from $m_\mathrm{b}(m_\mathrm{b})$ to a higher scale using the RGE calculation included in the REvolver code~\cite{Hoang:2021fhn} at five-loop precision. The uncertainty band includes the projected uncertainty of 10~MeV on $m_\mathrm{b}(m_\mathrm{b})$ (dark grey) and an 0.5\% uncertainty on $\alpha_s(m_\mathrm{Z})$.

\section{Parton distribution functions in global QCD analyses}
\label{sec:PDFs}
\subsection{Proton parton distributions} \label{sec:protonPDFs}
\subsubsection{Overview \label{sec:OverviewProtonPDFs}}
The Snowmass whitepaper ``Proton structure at the precision frontier'' \cite{Amoroso:2022eow} summarizes the ubiquitous role of parton distributions functions (PDFs) in future precision measurements. A revolution in computing hard-scattering cross sections in perturbative QCD up to the second and third order in $\alpha_s$ (N$^{2}$LO and N$^{3}$LO, respectively) opens appealing opportunities for precision applications of the PDFs. By knowing the PDFs for the gluon and other quark flavors approximately to 1--2\% accuracy, one greatly reduces the total uncertainties on the Higgs couplings in gluon-gluon fusion and electroweak boson fusion \cite{Cepeda:2019klc}. The energy reach in searches for very heavy new particles at the HL-LHC can be extended to higher masses by better knowing the PDFs at the largest momentum fractions, $x>0.1$, and by pinning down the flavor composition of the partonic sea~\cite{Brady:2011hb}. As interest grows in hadron scattering at very small partonic momentum fractions, $x < 10^{-5}$, at hadron colliders (HL-LHC, LHeC, FCC-hh) as well as in the astrophysics experiments, one must include effects of small-$x$ resummation and saturation the PDFs when warranted \cite{Ball:2017otu}.

\begin{table}[t]
    
    \begin{tabular}{>{\raggedright}p{0.3\textwidth}>{\raggedright}p{0.35\textwidth}>{\raggedright}p{0.35\textwidth}}
\hline 
\textbf{TOPIC} & \textbf{STATUS, Snowmass'2013} & \textbf{STATUS, Snowmass'2021}\tabularnewline
\hline 
\rowcolor{lightgray}Achieved accuracy of PDFs & N$^{2}$LO for evolution, DIS and vector boson production & N$^{2}$LO for all key processes; N$^{3}$LO for some processes\tabularnewline
PDFs with NLO EW contributions & MSTW'04 QED, NNPDF2.3 QED & LuXQED and other photon PDFs from several groups; PDFs with leptons
and massive bosons\tabularnewline
\rowcolor{lightgray} PDFs with resummations & Small x (in progress) & Small-x and threshold resummations implemented in several PDF sets\tabularnewline
Available LHC processes to determine nucleon PDFs & $W/Z$, single-incl. jet, high-$p_{T}$ $Z,$ $t\overline{t}$, $W+c$
production at 7 and 8 TeV & + $t\overline{t}$, single-top, dijet, $\gamma/W/Z+$jet, low-Q Drell
Yan pairs, \ldots{} at 7, 8, 13 TeV\tabularnewline
\rowcolor{lightgray} Current, planned \& proposed experiments to probe PDFs & LHC Run-2\\DIS: LHeC & LHC Run-3, HL-LHC\\DIS: EIC, LHeC, MuIC, \ldots\tabularnewline
Benchmarking of PDFs for the LHC & PDF4LHC'2015 recommendation in preparation & PDF4LHC'21 recommendation issued\tabularnewline
\rowcolor{lightgray} Precision analysis of specialized PDFs &  & Transverse-momentum dependent PDFs, nuclear, meson PDFs \tabularnewline
\hline 
\hline 
\multicolumn{3}{c}{\vspace{6pt} \textbf{NEW TASKS in the HL-LHC ERA} } \tabularnewline
\rowcolor{lightgray} Obtain complete N$^{2}$LO and N$^{3}$LO predictions for PDF-sensitive processes & Improve models for correlated systematic errors & Find ways to constrain large-x PDFs without relying on nuclear targets\tabularnewline
Develop and benchmark fast N$^{2}$LO interfaces & Estimate N$^{2}$LO/N$^{3}$LO theory uncertainties & New methods to combine PDF ensembles, estimate PDF uncertainties,
deliver PDFs for applications\tabularnewline
\hline 
\end{tabular}
    
    \caption{Top part: Some of the PDF-focused topics explored in Snowmass’2013 \cite{Campbell:2013qaa} and '2021 studies. Bottom part: a selection of new critical tasks for the development of a new generation of PDFs that achieve the objectives of the physics program at the high-luminosity LHC.}
    \label{table:Table1}
\end{table}

PDFs for unpolarized protons --- the cornerstone nonperturbative QCD functions ---  are traditionally determined from global analyses of fixed-target and collider data on DIS, production of lepton pairs, jets, top quarks, and increasingly in other processes \cite{Harland-Lang:2014zoa,Dulat:2015mca,H1:2015ubc,Accardi:2016qay,Alekhin:2017kpj,NNPDF:2017mvq,Hou:2019efy,Bailey:2020ooq,NNPDF:2021njg,ATLAS:2021vod}. 
 Table~\ref{table:Table1} illustrates the progress that has been made since the 2013 Snowmass Summer Study \cite{Campbell:2013qaa}.
The bottom part of the table lists new tasks for the PDF analysis that emerge in the HL-LHC era.
While the most precise N$^{2}$LO or even N$^{3}$LO theoretical cross sections should be preferably used in the fit when possible, accuracy of the PDFs also depends on the other commensurate factors that must be properly estimated.  Given the complexity of N$^{2}$LO/N$^{3}$LO calculations, their fast approximate implementations (such as fast \NNLOgen interfaces) must be developed to allow efficient observable computations in the PDF analyses. Control of experimental and theoretical uncertainties requires, in particular, to either fit the experiments that are minimally affected by the unknown factors (for example, to include cross sections only on proton, rather than on nuclear targets to minimize the associated uncertainties in the most precise determinations), or to estimate the associated uncertainty of these unknown factors in the fit.  
The PDFs are provided with uncertainties that must account for acceptable variations in methodology, including the  choice of the functional forms to parametrize PDFs at an initial energy scale and the method for propagation of experimental uncertainties, as well as implementation of physical constraints on the PDFs, such as sum rules, positivity of physical observables, and integrability of relevant PDF combinations. The PDF uncertainties must representatively reflect these factors \cite{Courtoy:2022ocu}. Methodological advances should also include development of practical standards for the delivery of PDFs to a wide range of users. The format of the PDF delivery must optimize for accuracy, versatility, and speed across a broad range of applications---a highly non-trivial task, given the ubiquity of the PDF uses. The PDF4LHC working group \cite{PDF4LHCWG} leads the development of such standards and delivery formats for the LHC community, in particular, by publishing a comprehensive recommendation (PDF4LHC21 \cite{Ball:2022hsh}) on the usage of PDFs and computation of PDF uncertainties at the LHC. The PDF4LHC working group also distributes combined N$^{2}$LO PDF4LHC21 error sets to streamline computations with PDFs across typical LHC studies, such as searches for new physics or theoretical simulations. 

New experimental measurements are essential for constraining the PDFs to the needed accuracy in the HL-LHC era. The large volume of the pre-LHC data, combined with the rapidly growing LHC data, offers a wealth of information about the hadron structure. Yet, the uncertainties on PDFs do not decrease as the square root of the number of data points because of some disagreements among the data sets and systematic uncertainties in many experiments. Constraints on the PDFs can be strengthened by fitting high-luminosity data sets under elevated accuracy control at all stages of the measurements and their analysis. 

\begin{figure}[t!]
    \centering
    \includegraphics[width=0.52\textwidth]{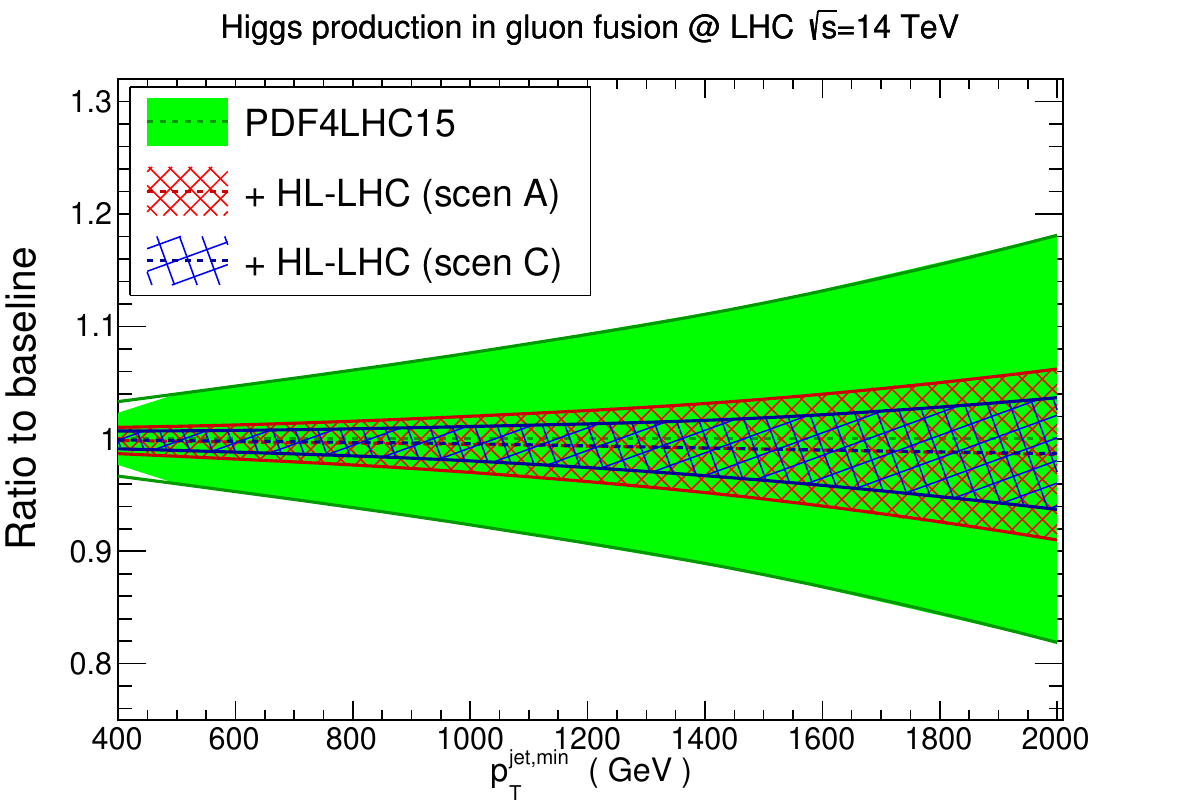}
    \includegraphics[width=0.47\textwidth]{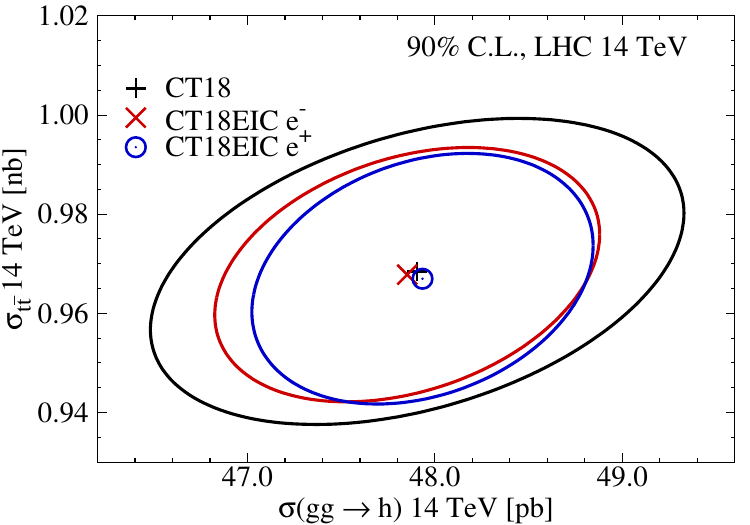}
    \caption{Examples of projections for PDF uncertainties in the HL-LHC era. {\bf Left:} Uncertainties for \NNLOgen Higgs production via gluon fusion at $\sqrt{s} = 14$ TeV obtained with published PDF4LHC15 \NNLOgen PDFs  \cite{Butterworth:2015oua} (green band) and after additional constraints are imposed on these PDFs using simulated HL-LHC data in two scenarios (red and blue bands) \cite{AbdulKhalek:2018rok}. 
    {\bf Right:} 90\% C.L. uncertainty ellipses for \NNLOgen predictions for $gg\to H_{\rm SM}$ and $t\bar t$ production at the LHC 14 TeV obtained using CT18 \NNLOgen PDFs \cite{Hou:2019efy} and after imposing simulated constraints from inclusive DIS at the EIC \cite{AbdulKhalek:2021gbh}. 
    \label{fig:HiggsPDFerrors}
    }
\end{figure}

Recent studies \cite{AbdulKhalek:2018rok,AbdulKhalek:2021gbh} provide projections using various techniques for the reduction of PDF uncertainties under anticipated near-future theoretical and experimental developments. As an illustration, the left panel of Fig.~\ref{fig:HiggsPDFerrors} compares the current PDF uncertainty for $gg\to H_{\rm SM}$ production and its reduction when simulated HL-LHC measurements are included in the conservative (scen A) and optimistic (scen C) scenarios, using PDF4LHC15 \NNLOgen PDFs \cite{Butterworth:2015oua} as the baseline. 

The right panel shows an analogous projection for the reduction of the PDF uncertainty on the SM Higgs and $t\bar t$ cross sections at the LHC upon including the simulated measurements in DIS at the EIC, this time using the CT18 \NNLOgen framework \cite{Hou:2019efy}. The ability of the LHC measurements to reduce the PDF uncertainty critically depends on the control of systematic effects. A lepton-hadron collider such as an EIC (see \cite{AbdulKhalek:2022erw} and below), EIcC, MuIC, or LHeC \cite[][Sec.~3.C]{Amoroso:2022eow} that runs roughly concurrently with the HL-LHC phase would elevate the precision of PDFs in key HL-LHC measurements in a synergistic way that would be unattainable via HL-LHC measurements alone. Section \ref{sec:EICProtonPDFs} includes some examples. A precision QCD program at the EIC is therefore a promising opportunity to obtain  PDF measurements in the  kinematic region of large $x$ and small $Q$ that is currently accessed only in  fixed-target DIS and Drell-Yan experiments. This region is of high relevance to the LHC, as the currently large uncertainties in the PDFs at $x > 0.5$ directly affect the LHC high-mass BSM searches. 
These uncertainties at the largest $x$ (outside of the reach of current experiments) and $Q=1-10$ GeV propagate to smaller $x$ at $Q=100-1000 $ GeV via QCD evolution and affect the LHC precision measurements. 

\subsubsection{HL-LHC experiments to probe PDFs \label{sec:HLLHCProtonPDFs}}

At the HL-LHC, a large range of experiments can either constrain the PDFs or depends on the PDFs \cite[][Sec. 3.A]{Amoroso:2020lgh}. 
There are significant opportunities for constraining the PDFs and general appreciation of their importance. Even so, available estimates of the projected impact on the PDFs may vary considerably even for the same experimental data set, reflecting the methodology of the analysis  and adopted definitions of the PDF uncertainties. These uncertainty estimates generally account for a combination of experimental, theoretical, parametrization, and methodological sources. Just as in the case of world-average QCD coupling determinations discussed in Sec.~\ref{sec:alphas}, the reduction of the PDF uncertainty due to a combination of experiments reflects both the accuracy of individual experiments and mutual consistency of experiments. 

\begin{figure}[thb]
  \centering
   \subfloat[]{\includegraphics[width=0.365\textwidth]{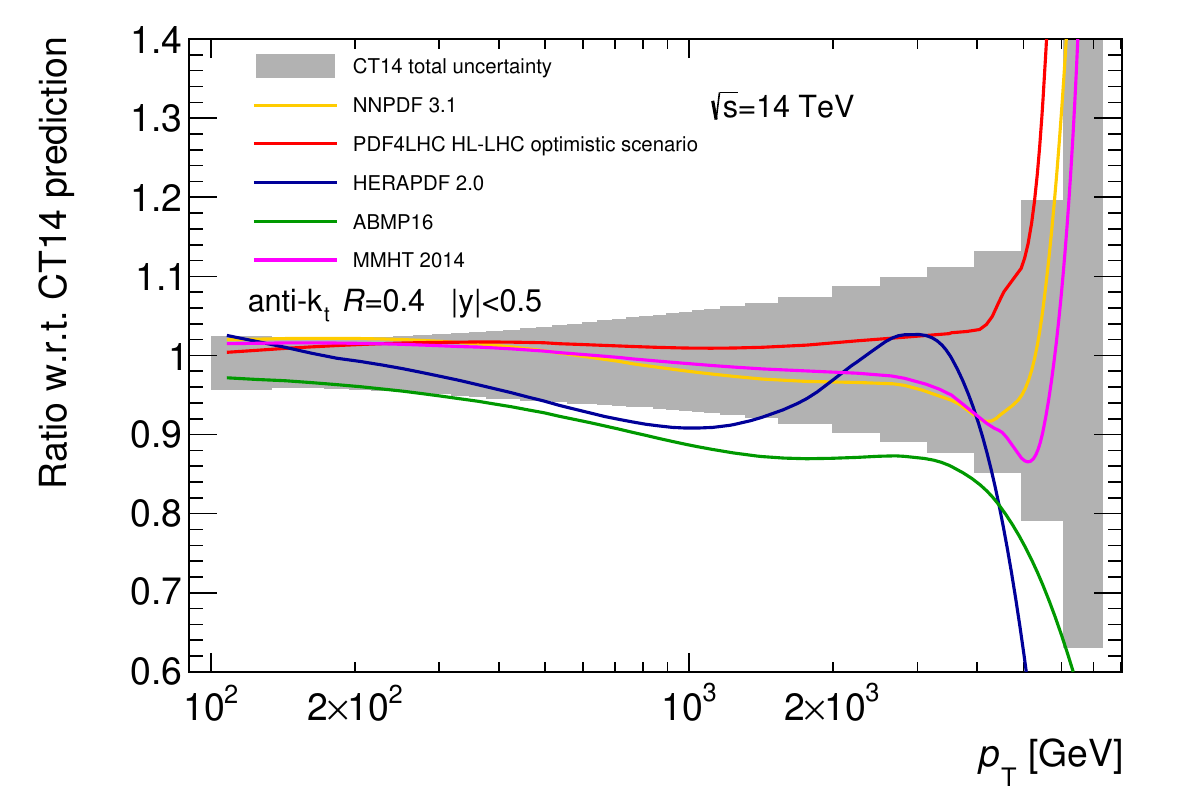} \label{fig:EF05:incljet}}
  \subfloat[]{\includegraphics[width=0.365\textwidth]{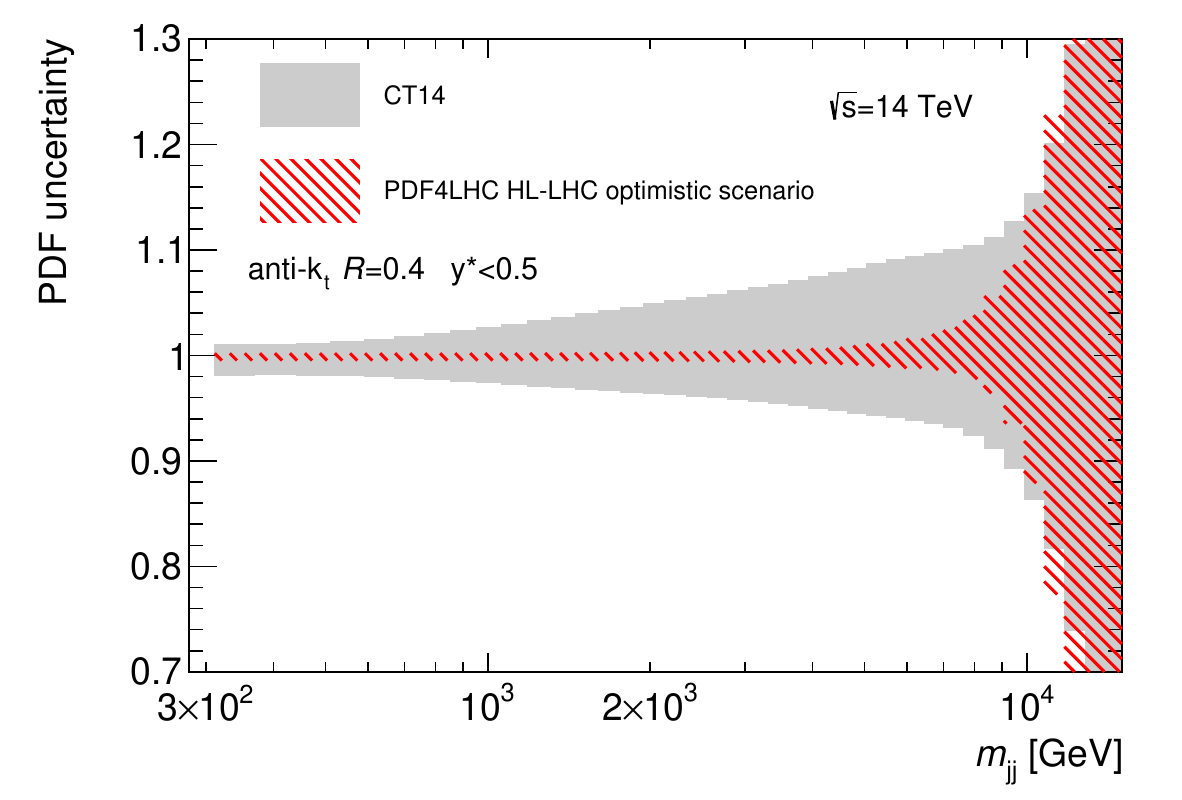}\label{fig:EF05:dijet2}}
   \subfloat[]{}{\includegraphics[width=0.27\textwidth]{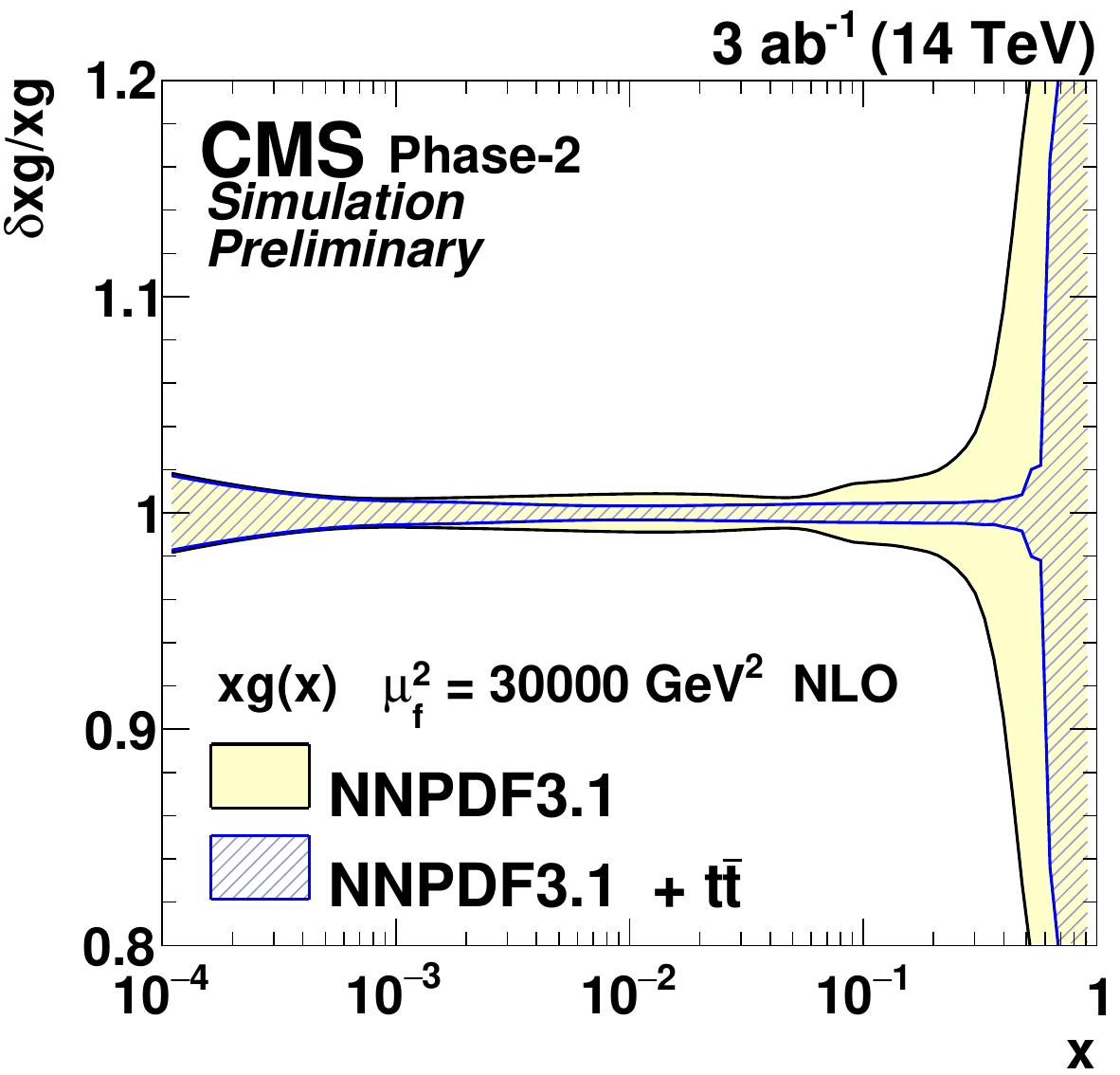}\label{fig:ef03:FTR-18-015_2}} 
  \caption{
  (a) Ratio of cross-sections predictions for several PDF sets to the CT14 PDF prediction and CT14 uncertainty (gray band) for inclusive jet  cross-section at $\sqrt{s} = 14$ TeV \cite{ATL-PHYS-PUB-2018-051}.
   (b) Comparison of the PDF uncertainty in the dijet cross-section
    calculated using the CT14 PDF \cite{Dulat:2015mca} and PDF4LHC HL-LHC sets at $\sqrt{s} = 14$ TeV with 3000~fb$^{-1}$  \cite{ATL-PHYS-PUB-2018-051}.
   (c) An estimated reduction of the relative uncertainty on the gluon PDF by profiling NNPDF3.1 PDFs \cite{NNPDF:2017mvq} using simulated $t\bar t$ measurements at the HL-LHC~\cite{FTR-18-015}. The reduction of the uncertainties depends on the estimation methodology, see the main text. 
   }
  \label{fig:jetxsec:pdfimpact}
\end{figure}

Taking production of hadronic jets as an example, large differences exist between predictions using different PDF sets at the highest jet transverse momenta, $p_{\mathrm{T,j}}$ and dijet invariant masses $m_{\mathrm{jj}}$. These differences are due to sensitivity of the jet cross sections to the gluon density in the proton. As an illustration, Fig.~\ref{fig:jetxsec:pdfimpact}(a) ~\cite{ATLAS:2022hsp} shows the ratio of several PDF sets as a function of $p_{\mathrm{T, j}}$. An LHC data set with a large number of events has high statistical precision in these regions. However, since the shown PDFs already include significant constraints from various jet data sets, the achievable reduction even more depends on the LHC systematic uncertainties, which are insufficiently known. Progress in understanding of the systematic uncertainties is critical for taking full advantage of these promising measurements. This issue is explored in Sec.~5.A of Ref.~\cite{Amoroso:2022eow}.

Figure~\ref{fig:jetxsec:pdfimpact}(b) estimates the impact that inclusion of the HL-LHC measurements into a PDF fit could have on the HL-LHC  dijet production cross section  \cite{AbdulKhalek:2018rok}. 
Under various HL-LHC running scenarios,
the uncertainty estimated with the PDF4LHC15 error PDF ensemble \cite{Butterworth:2015oua} decreases upon adding simulated data in (di)jet, gauge boson, and top quark production. The degree of reduction depends on the various factors mentioned above. In the shown "optimistic" scenario, the reduction of the PDF uncertainty on jet cross sections (and, by extension, on Higgs and other cross sections dominated by gluon scattering) is quite dramatic.  

Figure~\ref{fig:jetxsec:pdfimpact}(c) shows the possible impact on NNPDF3.1 PDFs \cite{NNPDF:2017mvq} upon adding $t\bar t$ cross section measurements at the HL-LHC, also sensitive to the gluon PDF, using the profiling method in the \texttt{xFitter} program \cite{XFitterSnowmass}. The plot is based on estimations of differential $t\bar{t}$ cross-section measurements in the e/$\mu$+jets channels at the HL-LHC with an integrated luminosity of 3000~fb${}^{-1}$ at $\sqrt{s}=14$ TeV by CMS~\cite{FTR-18-015}. 
This final uncertainty can be below 5\%, also reflecting an optimistic projection, as the default profiling in \texttt{xFitter} emphasizes the selected experiment more than the other experiments placing relevant constraints in the fit \cite[Appendix~F in ][]{Hou:2019efy}.
The most significant increase in accuracy is expected to come from an improved jet energy calibration and a reduced uncertainty in the $b$-jet identification---the dominant systematic uncertainties. The precision will profit from the enormous amount of data and the extended $\eta$-coverage of the Phase-2 CMS detector, which enables fine-binned measurements at high rapidity that are not possible with the current detector.

\begin{figure}[t]
  \centering
  \includegraphics[width=0.45\textwidth]{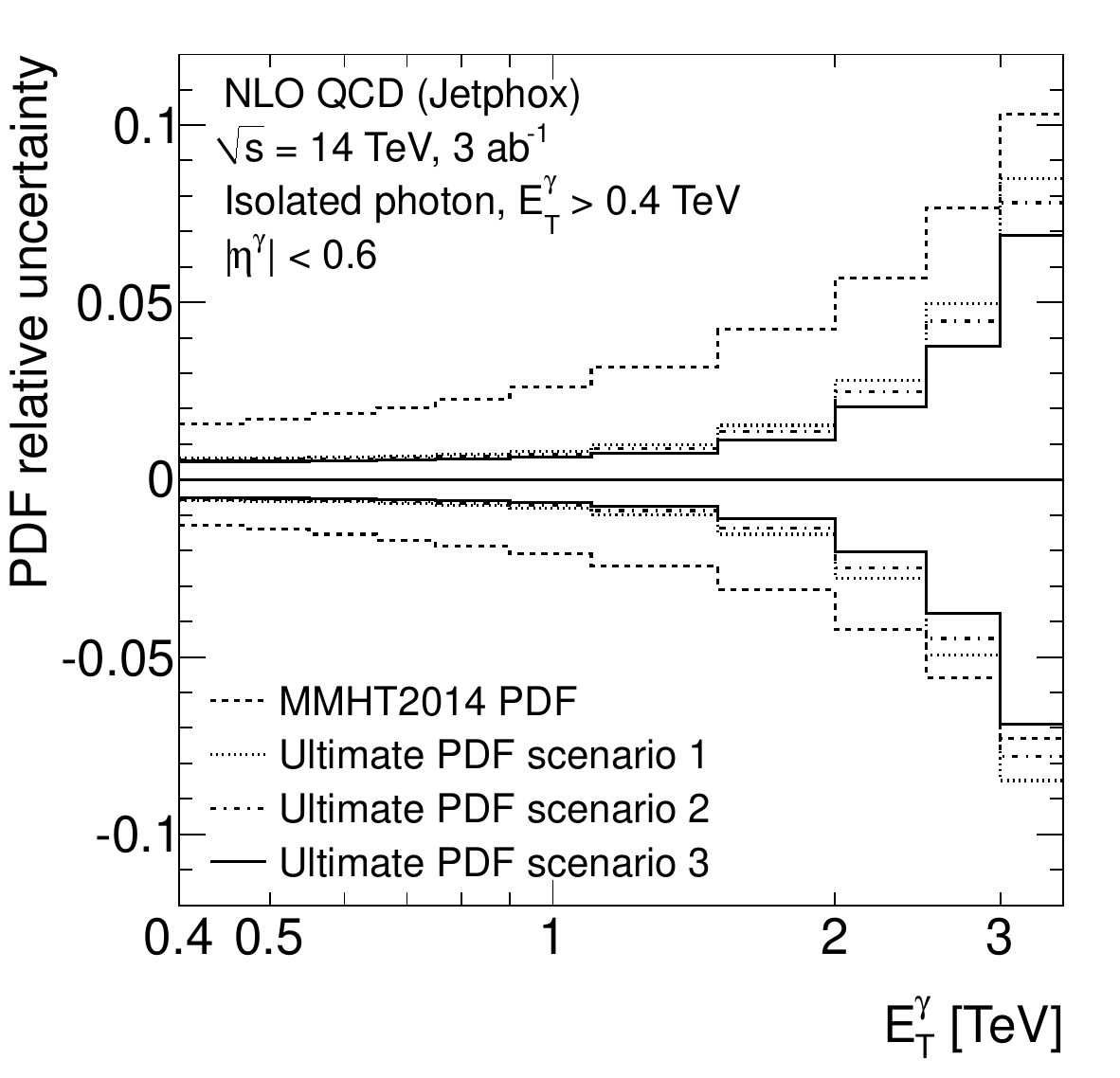}
  \caption{Relative uncertainty in the predicted number of inclusive isolated photon events due to the uncertainties in the PDFs as a function of $E_T^{\gamma}$.~\cite{ATL-PHYS-PUB-2018-051}}
  \label{fig:photon:pdfimpact}
\end{figure}

Many other LHC processes over a wide kinematic region---production of direct photons, massive bosons with jets or heavy quarks, heavy quarks of all three generations---can provide valuable insights about proton PDFs. We refer the reader to the Snowmass PDF whitepaper \cite{Amoroso:2022eow} and recent reviews and textbooks \cite{Gao:2017yyd, AbdulKhalek:2018rok,CampbellHustonKrauss:2018}. It is critical to determine the same (combinations of) PDFs in multiple accurate measurements to pin down systematic uncertainties both in experiment and theory. For example, to further reduce the uncertainty on the critical gluon PDF, one must reconcile occasionally inconsistent pulls on the gluon in the relevant $x$ regions imposed by fixed-target DIS, HERA DIS, jet, and $t\bar t$ production measurements.\footnote{The pulls on the PDFs can be examined by adding individual experiments into a PDF fit or removing them, or directly in a fit to many experiments using a fast sensitivity technique like in \cite{Hobbs:2019gob}, with examples in \cite{Hou:2019efy,Accardi:2021ysh}.} New measurement channels therefore can provide desired independent information. As an illustration, direct photon production studied differentially in $\mathrm{E}_{\mathrm{T}}^{\gamma}$ and $\eta^{\gamma}$ 
is sensitive to the gluon PDF over a large $x$ range \cite{ATL-PHYS-PUB-2018-051}. The photon+jets measurements can be insightful when performed differentially in $\mathrm{E}_{\mathrm{T}}^{\gamma}$, $p_{\mathrm{T, jet}}$, $\mathrm{cos} \, \theta^{*}$, and $m_{\mathrm{\gamma j}}$. With the full 3000~fb${}^{-1}$ dataset, the reach of these measurements will increase, from 3 TeV to 7 TeV in $m_{\mathrm{\gamma j}}$, and from 2.5~TeV to 3.5~TeV for $\mathrm{E}_{\mathrm{T}}^{\gamma}$ and $p_{\mathrm{T}}^{\mathrm{jet}}$. The projected impact of these and other precision measurements at the HL-LHC is shown in Fig.~\ref{fig:photon:pdfimpact}, which compares the PDF uncertainties from the MMHT2014 PDF set to PDF sets derived using projections of measurements from the HL-LHC~\cite{AbdulKhalek:2018rok}. Better control of photon isolation is needed to take full advantage of this channel. 
\subsubsection{PDFs at the EIC \label{sec:EICProtonPDFs}}
An EIC can significantly reduce PDF uncertainties both in HL-LHC EW precision measurements and, as importantly, searches for the heaviest final states, by measuring a range of interactions up to $\sqrt{s}\! =\! 140\,\mathrm{GeV}$ in comparatively clean $ep$ DIS processes. 
Spin-averaged EIC data on inclusive DIS, production of heavy quarks and QCD jets using neutral- and charged-current exchanges will allow for comprehensive flavor separation of (un)polarized PDFs and enhanced-precision determinations of QCD and EW couplings and quark masses. 

\begin{figure}[htb]
\centering
\subfloat{\includegraphics[width=0.5\columnwidth]{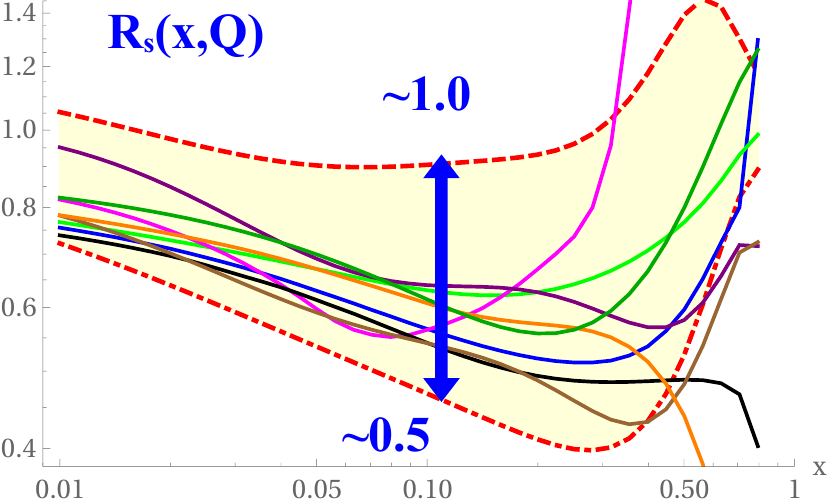}}
\subfloat{\includegraphics[width=0.5\textwidth]{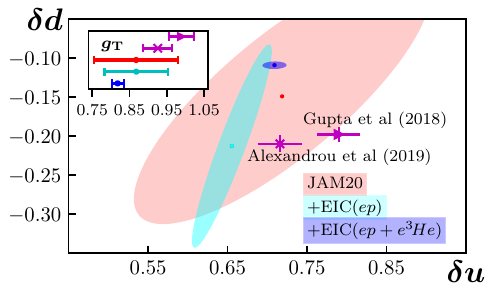} }
\caption{\label{fig:EICrsgT} 
(a) The strange quark ratio $R_s(x,Q)=(s+\bar{s})/(\bar{u}+\bar{d})$  at $Q{=}10$~GeV 
for a selection of PDFs at partonic momentum fractions accesible at the EIC.
(b) Flavor tensor charges $\delta u$, $\delta d$ as well as the isovector charge $g_T$ from the JAM'20 PDF analysis ~\cite{Cammarota:2020qcw} as well as a re-fit that includes EIC Collins effect pion production pseudodata for a proton beam only  and for both proton and $^3$He beams together.  Also shown are the results from two recent lattice-QCD calculations~\cite{Gupta:2018qil,Alexandrou:2019brg} (purple).  All results are at $Q^2=4\,{\rm GeV^2}$ with error bands at $1\sigma$ CL. From \cite{AbdulKhalek:2022erw}.
}
\end{figure}

The EIC can  resolve long-standing questions regarding the precise balance of quark flavors contributing to the proton's structure, in particular the strangeness content of the proton. As Fig.~\ref{fig:EICrsgT}(a) illustrates, 
the ratio $R_s(x,Q)=(s+\bar{s})/(\bar{u}+\bar{d})$ has significant uncertainty in the EIC's kinematical region due to the insufficiently known strangeness PDF.  As explored in Ref.~\cite{Arratia:2020azl}, these uncertainties translate into large event-level shifts in CC DIS charm-jet production at the EIC, implying considerable
potential to constrain the strangeness PDF.

Lattice QCD is making impressive advances in computations of nonperturbative QCD functions at $x>0.1$ and factorization scales $Q$ of a few GeV (Sec.~\ref{sec:lattice}). This is precisely the region covered by the EIC kinematics, which creates ample opportunities for comparing lattice QCD predictions against the EIC data on 3-dimensional hadron structure. Lattice QCD can be compared against precisely known spin-averaged PDFs and make predictions for various spin-dependent PDFs. Not only the $x$ dependence of PDFs can be compared, but also various Mellin moments integrated over the whole $x$ range, as illustrated in Fig.~\ref{fig:EICrsgT}(b) on an example of isovector tensor charges predicted based on the current JAM'20 PDFs,  upon adding the EIC data on the proton and helium beams, or using two lattice QCD calculations.  Section~\ref{sec:LatticeExamples} reviews lattice calculations of collinear PDFs in more depth.

\subsection{Nuclear parton distributions}
\label{sec:EF07:npdf}
The structure of nucleons and nuclei are both key to understanding heavy-ion collisions as well as fundamental  features of the nucleus.  Much less is known about the nuclear PDFs than about the nucleon ones, especially for heavy nuclei at momentum fractions $x$ below 0.1 and above 0.5, where modifications in a nuclear medium rapidly increase. In particular, particle production at small $x$ of the nucleus and the possible QCD effects that may be revealed in it, such as the Color Glass Condensate~\cite{Gelis:2010nm} regime, are poorly known.  Measurements made during the LHC Run 1 and Run 2 in Pb-Pb and especially p-Pb collisions have favored the inclusion of nuclear modification to the parton distribution functions (PDF) extracted for free protons. The p-Pb collision system is an excellent tool to study and constrain these nuclear parton distribution functions (nPDFs), since the asymmetric system allows one to select low-$x$ regions of the nucleus by looking at forward rapidity, namely, the proton-going direction.  

\begin{figure}[!htbp]
\centering
  \subfloat[]{\includegraphics[scale=0.26]{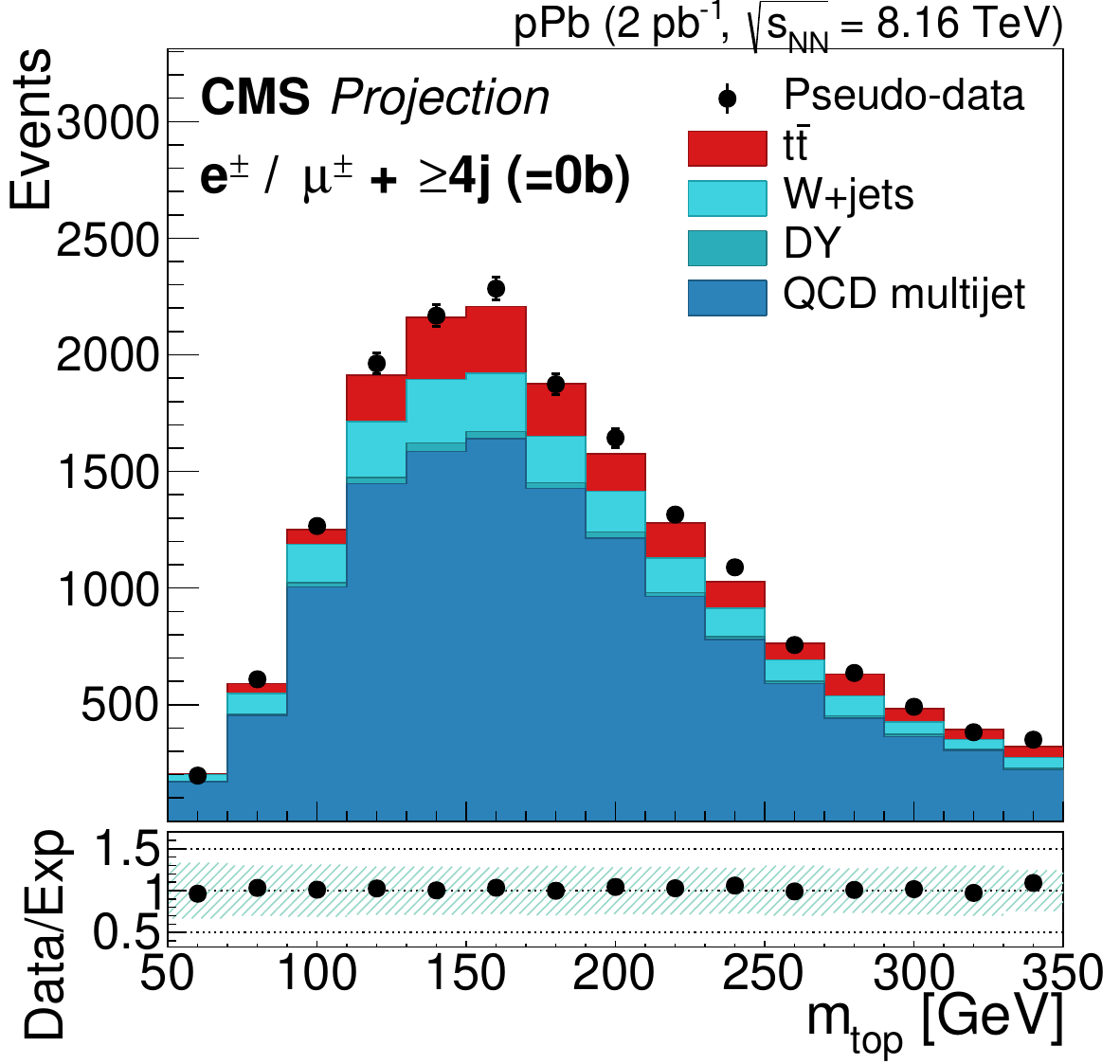}\label{fig:EF07_npdf_ttbar-a}}
  \subfloat[]{\includegraphics[scale=0.26]{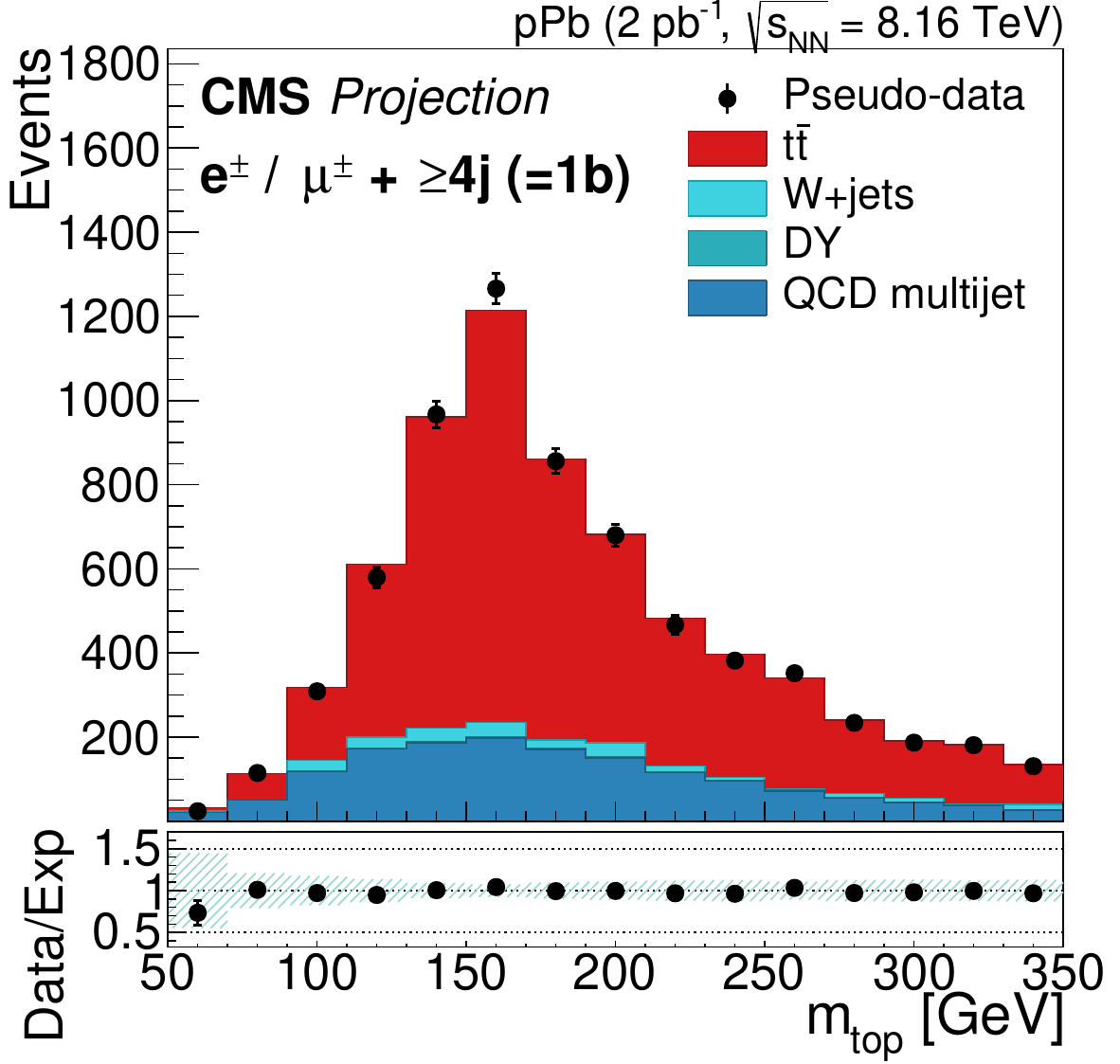}\label{fig:EF07_npdf_ttbar-b}}
  \subfloat[]{\includegraphics[scale=0.26]{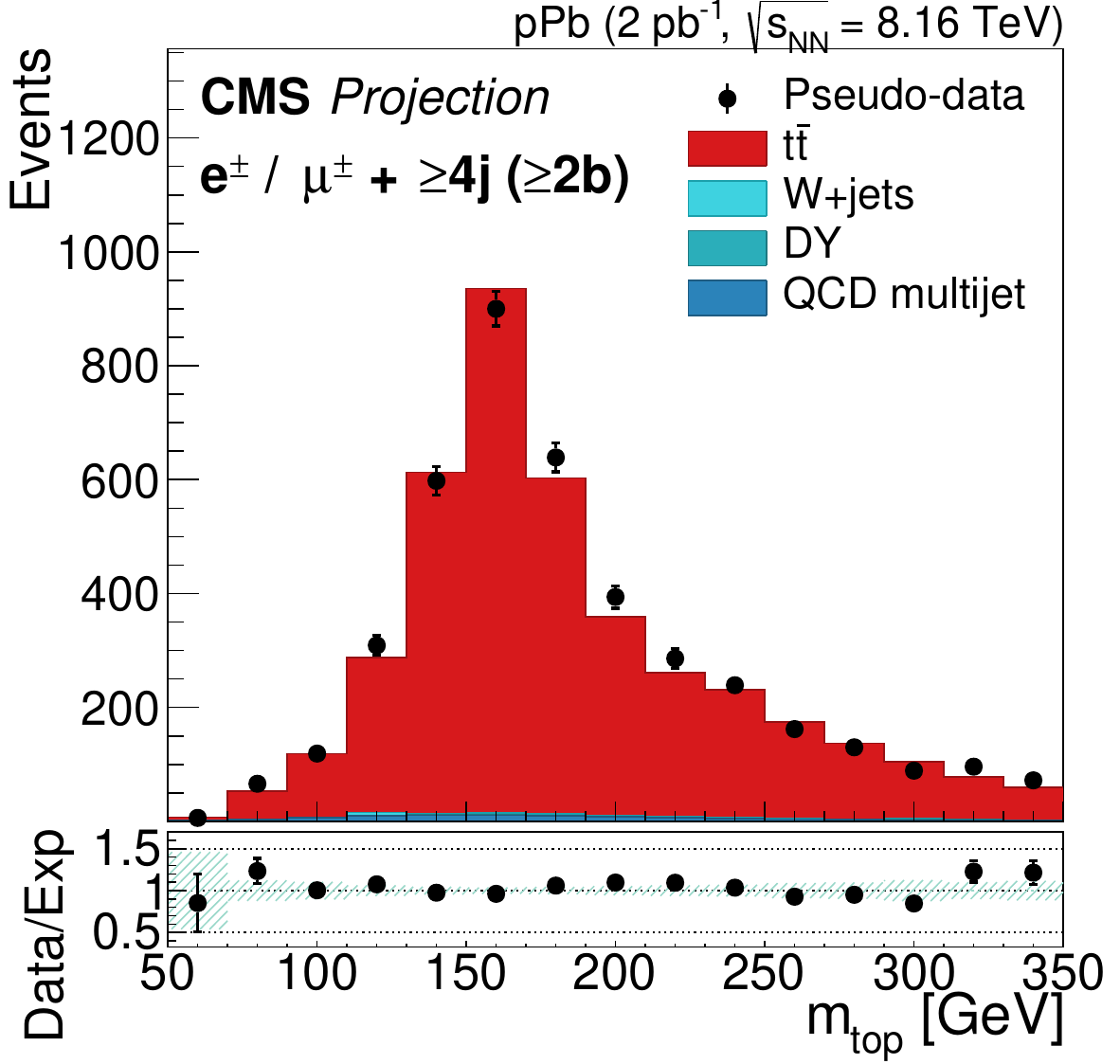}\label{fig:EF07_npdf_ttbar-c}}
  \caption{Distributions of m$_{\mathrm{top}}$ in the \protect\subref{fig:EF07_npdf_ttbar-a} 0, \protect\subref{fig:EF07_npdf_ttbar-b} 1, and \protect\subref{fig:EF07_npdf_ttbar-c} 2 $b$-tagged jet categories. The sum of the predictions for the $t\bar{t}$ signal and background is compared to pseudodata (sampled randomly from the total of the predictions in each category). The bottom panels show the ratio between the pseudodata and the sum of the predictions. The shaded band represents the relative uncertainty due to the limited event count in the simulated samples and the estimate of the normalization of the QCD multijet background~\cite{FTR-18-027}.
  }
  \label{fig:EF07_npdf_ttbar}
\end{figure}

To this end, ATLAS and CMS both intend to measure $W$ and $Z$ boson production from p-Pb collisions to constrain the quark nPDFs~\cite{FTR-18-027,ATL-PHYS-PUB-2018-039}, especially differentially in rapidity of the $Z$ boson or charged lepton pseudorapidity of the $W$ boson. Complementing these measurements, CMS has projected the measurement of dijet pseudorapidity which is sensitive to the gluon nPDF~\cite{FTR-18-027}. The measurement of differential $t\bar{t}$ cross sections in p-Pb collisions is a novel and potentially precise probe of the nuclear gluon density~\cite{dEnterria:2015mgr}.  Figure~\ref{fig:EF07_npdf_ttbar} shows the mass distributions of the top quark and relevant backgrounds projected for p-Pb collisions for three different selections on the number of b-tagged jets~\cite{FTR-18-027}. Additional experimental leverage of the event-by-event sensitivity to $Q^2$ and nuclear $x$ may be obtained by EW boson ($W$ and $Z$) plus jet events, as projected by ATLAS~\cite{ATL-PHYS-PUB-2018-039}.
	 
 \begin{figure}[!htbp]
\centering
  \includegraphics[width=0.55\textwidth]{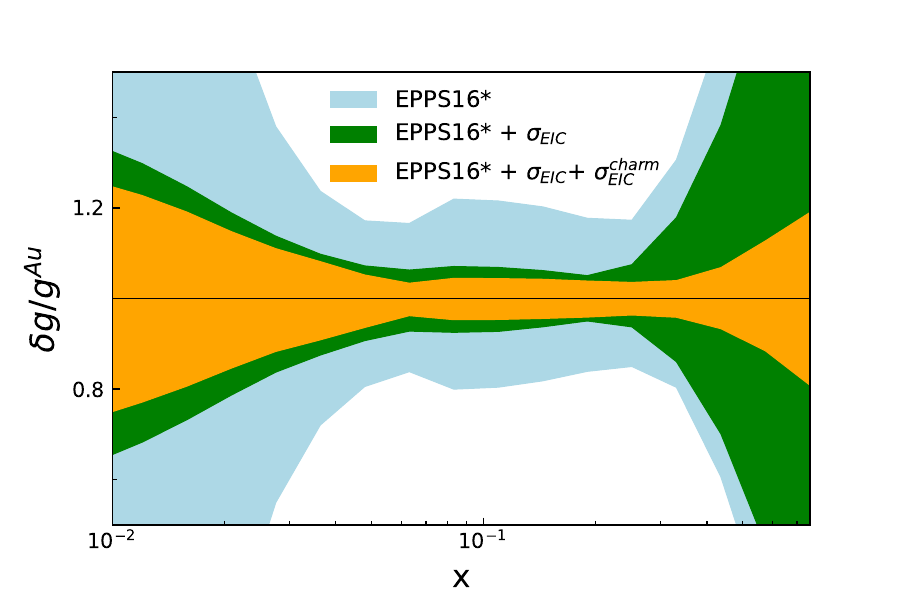}
  \includegraphics[width=0.44\textwidth,clip]{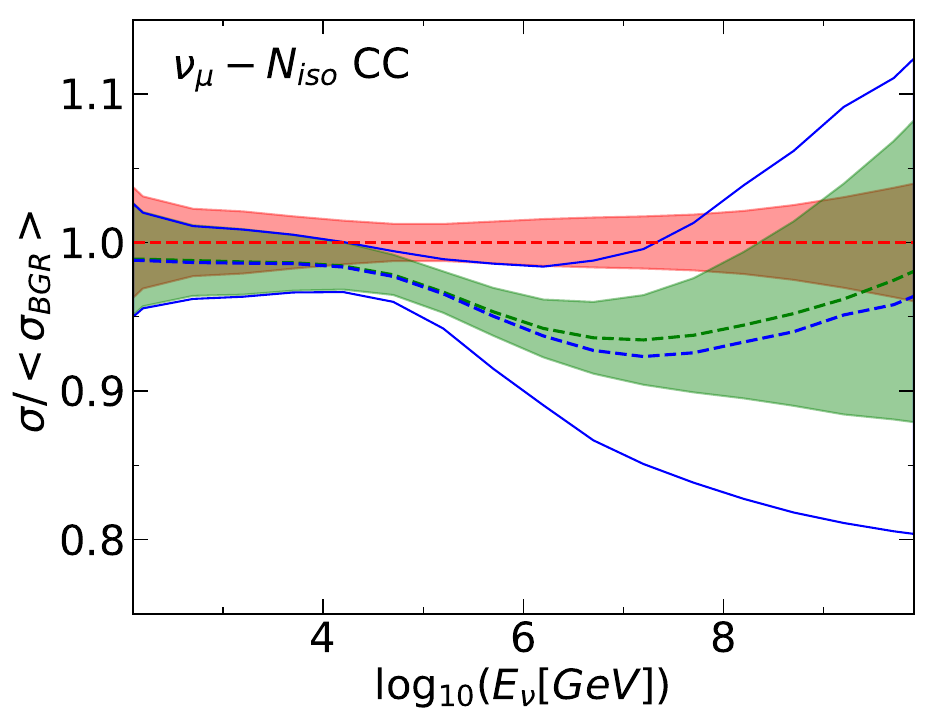}\\

  \caption{Left: Estimated reduction in the EPPS16* \cite{Eskola:2016oht} uncertainty for the gluon distribution in gold upon inclusion of EIC inclusive and charm production data \cite{AbdulKhalek:2021gbh}. Right: Nuclear PDF uncertainties based on nNNPDF2.0 \cite{AbdulKhalek:2019mzd,AbdulKhalek:2020yuc} for CC neutrino-nucleus DIS cross sections, without and with the EIC pseudodata included. Cross sections are plotted as a ratio to predictions for the proton as a function of the neutrino energy $E_\nu$. See details in \cite[][Sec.~3.2.3]{AbdulKhalek:2022erw}.
  \label{fig:EICNuclearPDFs}}
\end{figure}
The EIC can revolutionize understanding of nuclear PDFs for a large span of nuclear mass and charge quantum numbers, $A$ and $Z$. As in the case of nucleon PDFs, studies of nuclear scattering at the EIC and HL-LHC are highly complementary. They will result in the reduction in nuclear PDF uncertainties and flavor separation at previously unaccessible $x$ down to $10^{-2}$, where modifications due to partonic saturation may kick in. As an illustration, the left Fig.~\ref{fig:EICNuclearPDFs} shows the reduction the relative uncertainty on the gluon PDF in gold after fitting inclusive DIS and semi-inclusive charm production pseudodata for the EIC, using the present EPPS16* nuclear PDFs as the baseline. In turn, better knowledge of nuclear PDFs will improve theoretical predictions for neutrino-nucleus scattering at future facilities like DUNE/LBNF. The right Fig.~\ref{fig:EICNuclearPDFs} shows the estimated reduction in the PDF uncertainty for CC neutrino scattering on an isoscalar heavy nucleus with $A=31$ after including the EIC simulated data.

At $x<10^{-2}$, ratios of cross sections of vector meson photoproduction in ultraperipheral collisions (discussed in Sec. \ref{sec:UPCs}) of ions or protons offer a method to probe the small-$x$ nuclear gluon PDF by tying it to the better known gluon PDF in the nucleon. 
First LHC measurements of this kind \cite{ALICE:2021gpt, Duan:2021gzs, CMS:2016itn} all show moderate suppression in lead compared to a proton-target reference, consistently with models predicting moderate shadowing such as \cite{Guzey:2020ntc}. However, this method currently has large uncertainties and  feels contamination from quarks \cite{Eskola:2022vpi}.  Photoproduction of dijets \cite{ATLAS:2017kwa} and heavy quarks will open additional avenues to test universality of nuclear PDFs at small $x$ and look for evidence of partonic saturation. 

Interpretation of many $pA$ and $AA$ collision experiments will require to know nuclear PDFs as a function of the initial parton's impact parameter $b$, in addition to the parton's momentum fraction $x$. Various models predict the $A$ and $x$ dependence of the nuclear modification to the $b$-dependent PDFs and can be tested in the LHC and EIC experiments. At small $x$, a natural assumption is that the nuclear modification primarily depends on  the parton thickness at a given $b$. One could try to determine $b$ dependence of nuclear PDFs in a model in which hard and soft collisions are not correlated \cite{Emelyanov:1999pkc} or using the leading-twist shadowing theory  \cite{Frankfurt:2011cs}.

\subsection{Meson parton distributions \label{sec:MesonPDFs}}
\begin{figure}[htbp]
\includegraphics[width=0.5\textwidth]{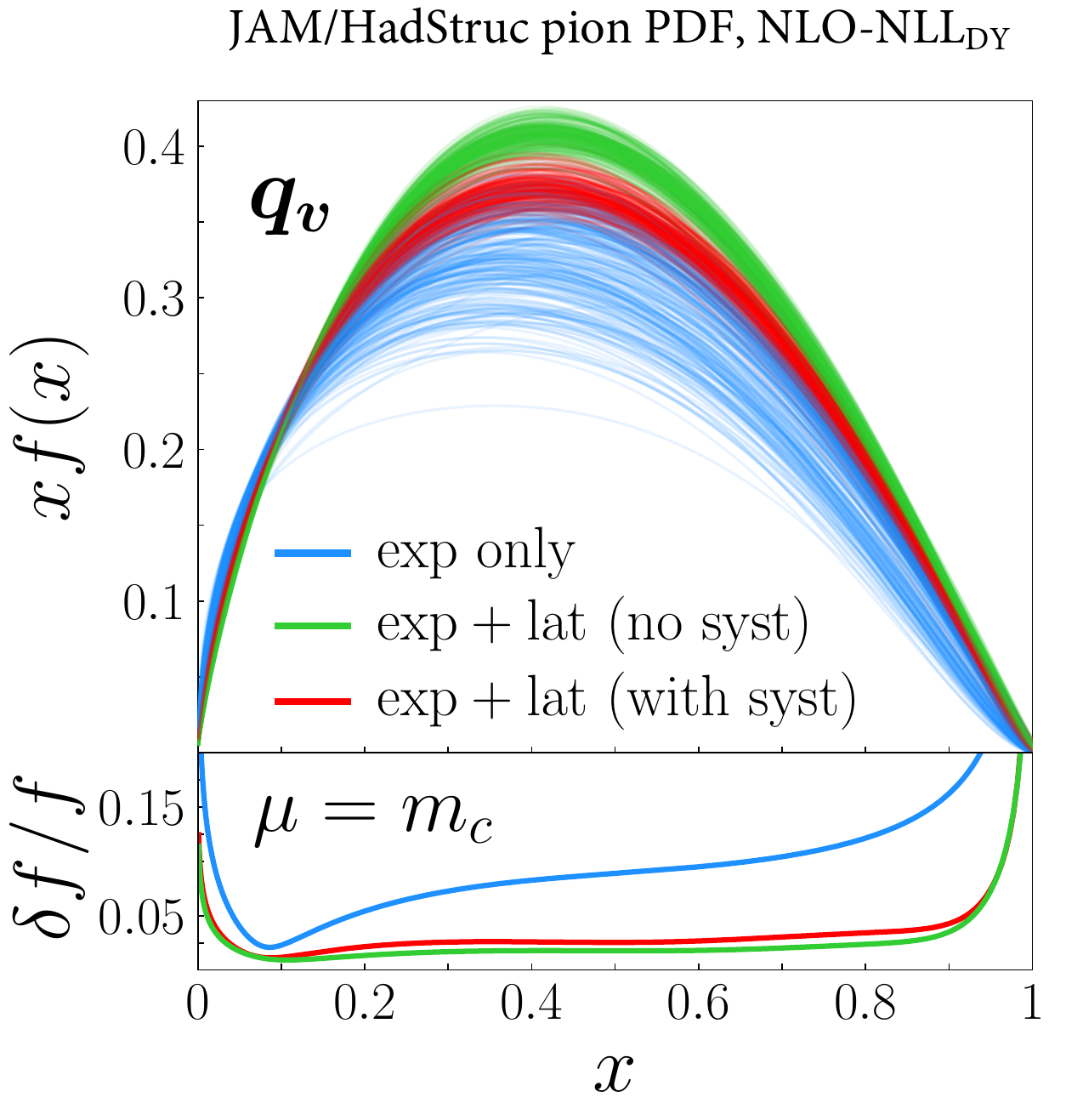}
\caption{
\label{fig:JAMPionPDF} 
Determination of a valence quark PDF in a pion using a combination of experimental and lattice QCD data, and including resummation of threshold radiative contributions \cite{Barry:2022itu}. 
}
\end{figure}

Global fits of PDFs for pions and kaons can clarify mechanisms of formation of hadronic bound states and hadron mass generation -- the central topics in nonperturbative QCD \cite{Aguilar:2019teb, Roberts:2021nhw}. 
New measurements sensitive to meson PDFs are expected to be performed at fixed-target energies, the EIC \cite{AbdulKhalek:2021gbh} and the EIcC \cite{Anderle:2021wcy}, then confronted against predictions from nonperturbative approaches and lattice QCD. As an example of what may be feasible in the near future, Fig.~\ref{fig:JAMPionPDF} shows a phenomenological PDF for a valence quark in the pion extracted from pion-nucleus Drell-Yan and tagged-neutron DIS data in NLO QCD with threshold resummation, and complemented with constraints from lattice QCD \cite{Barry:2022itu}. Such studies can be extended to other processes at \NNLOgen and include more accurate lattice QCD predictions. They offer a window to elucidate the bound-state dynamics and transition to the perturbative QCD regime.

\section{Hadronization and Fragmentation functions}
\label{sec:hadronization}
The process of hadronization describes how detected final-state hadrons are formed from partons. Since hadronization is governed by nonperturbative dynamics, it cannot be calculated analytically and, in contrast to the partonic structure of hadrons, is elusive in lattice calculations~\cite{Metz:2016swz}.
Having an accurate description of hadronization is, however, important for many measurements in high-energy physics and crucial for all measurements at hadron colliders. 

\subsection{Hadronization measurements at Belle II}
A fragmentation function (FF) quantifies the probability for a parton to hadronize into a detected final-state particle of a known momentum. Precision measurements of FFs are instrumental for extracting the spin-averaged and spin-dependent nucleon structure~\cite{Anselmino:2020vlp} in the planned experiments at the EIC and Belle~II. The emphasis of the Belle~II program will be
on investigation of full multidimensional dependency of FFs with complex final states, such as dihadrons
or polarized hyperons. 
These final states are sensitive to spin-orbit correlations in hadronization. Their factorization universality properties and kinematic dependencies are still to be fully mapped out. However, they are important, as tagging on such final-state degrees of freedom allows  more targeted access to the hadron structure in semi-inclusive deep inelastic scattering (SIDIS) experiments, e.g., at JLab and the EIC. One recent example of this is the extraction of the twist-3 PDF $e(x)$ via dihadron correlations, which is sensitive to the force that the gluons exert on a fragmenting quark as it traverses the nucleon remnant~\cite{Burkardt:2008ps,CLAS:2020igs,Hayward:2021psm,Courtoy:2022kca}.
  
Detailed understanding of hadronization is necessary to model background and signal processes for new physics discoveries at B-factories themselves, but also at the LHC. Currently, modeling of backgrounds originating from light-quark fragmentation is mainly performed by Monte-Carlo Event Generators (MCEGs), such as Pythia~\cite{Sjostrand:2014zea}, HERWIG~\cite{Bellm:2019zci} or Sherpa~\cite{Sherpa:2019gpd}. Tuning those generators to a precision needed for discovery science requires a model for correlated production of multiple hadrons that can only be verified with clean semi-inclusive $e^+e^-$ annihilation data. 
Experimental data for this purpose are mostly available from LEP, but, to confidently extrapolate the model to LHC energies, input measurements are also necessary at CM energies an order of magnitude below LEP. The relatively low CM energy at Belle~II, paired with extremely high luminosity, provides a large lever arm when combining Belle~II and LEP/SLD data to probe hadronization effects over a wide energy range.

A comprehensive program with the high-statistics Belle~II data is also needed to reach the precision necessary for the Belle~II analyses themselves. MCEGs are also crucial for inference-based models, e.g.~\cite{Brehmer:2019xox}, which will be applied in the future to extract physical quantities. A recent development has been the extension of MCEGs to include spin-orbit correlations. Belle~II measurements can inform the development of these novel MCEGs by benchmarking against spin dependent di-hadron correlations.
  
Where MCEGs describe full events, and the most common single-hadron FFs integrate over the whole event with the exception of the hadron in question, intermediate representations accounting for more correlations in hadronization gain more recognition in the field. The fragmentation functions for production of hadron pairs mentioned above 
are such an example. Beyond the current factorization theorems, there have been significant recent efforts to define correlation measurements that are sensitive to hadronization dynamics, can be interpreted within hadronization models (e.g., a QCD string model), and, while not yet realized, might be describable in a full QCD calculation with future, extended factorization theorems. These kind of correlation measurements have already been a focus at the LHC (see {\it e.g.}, Ref.~\cite{ATLAS:2020bbn}). At Belle~II, correlations between leading particles  can be precisely measured.
Accurate knowledge of parton (in particular gluon~\cite{,dEnterria:2013sgr}) FFs into hadrons (both inclusively and for individual hadron species) in $\epem$ collisions is also of utmost importance to have an accurate ``QCD vacuum'' baseline to compare with the same objects measured in proton-nucleus and nucleus-nucleus collisions and thereby quantitatively understand final-state (``QCD medium'') modifications of the FFs~\cite{Albino:2008aa,Accardi:2009qv}.

\subsection{Measurements at the Electron Ion Collider}
To capitalize on a new era of experiments like the EIC and HL-LHC, sound predictions for parton dynamics beyond collinear evolution are necessary. 
Transverse-momentum-dependent (TMD) PDFs and FFs will become the primary means to investigate the mechanism of hadronization in a 3D-picture~\cite{AbdulKhalek:2022erw}.
Historically, they have been accessed through Semi-Inclusive DIS (SIDIS) and $e^+e^-$ annihilation with observation of two final-state hadrons.
However, phenomenological extractions based on such processes are complicated by the fact that, in the cross section, the TMD FF does not appear on its own, but it is always convoluted with another TMD (two TMD FFs in $e^+e^-$ annihilations, one TMD PDF and one TMD FF in SIDIS). Disentangling these functions is usually difficult. The problem can be bypassed if one can extend the TMD factorization scheme to cross sections that involve only one TMD FF. In this sense, the cleanest process that accesses one TMD FF is single hadron production in $e^+e^-$ annihilations, $e^+e^- \to h\,X$.

A factorization theorem was recently derived in Ref.~\cite{Boglione:2021wov} for $e^+e^- \to h\,X$, where the transverse momentum $P_T$ of the detected hadron is measured with respect to the thrust axis. Under certain approximations, this cross section can be written as a
convolution of a TMD FF with a coefficient that is totally predicted by perturbative QCD and can be interpreted
as a partonic cross section~\cite{Boglione:2020auc, Boglione:2020cwn}.
Since this process is more inclusive than SIDIS and $e^+e^-$ annihilation into two hadrons, the role of the soft gluons is different. The soft nonperturbative part of the TMD can be disentangled, and it becomes possible to define a phenomenology work plan that involves a much larger number of different processes by dealing with one single unknown at a time.
Within this framework, the future EIC, which will explore a very  broad kinematical region, could provide informative measurements for both TMD PDFs and FFs.

\subsection{Measurements at future $e^+e^-$ colliders}

The reaction of $e^+e^-$ annihilation has always been a method of choice to access hadronization in a clean environment. 
Much of the predictivity of QCD at colliders in fact stems from factorization theorems paired with measurements at
PETRA, PEP, LEP and SLD. There is a class of universal nonperturbative inputs that were not yet defined at the time of LEP and SLD, which could be measured precisely at the ILC and other future $e^+e^-$ machines, and would have a significant impact on the LHC physics program. Modern measurements rely strongly on the use of particle flow and tracking information. However, only observables that are completely inclusive over the spectrum of final states can be computed purely from perturbation theory. The nonperturbative input needed for theoretical predictions of track-based observables is universal and can be parametrized by  so-called ``track functions" \cite{Chang:2013rca,Chang:2013iba}, which describe the fraction of energy carried by charged particles from a fragmenting quark or gluon. Recently it has been shown how to compute jet substructure observables by incorporating track functions \cite{Jaarsma:2022kdd,Li:2021zcf} as a step toward precision jet substructure measurements at the HL-LHC and future colliders.  

\subsection{Hadronization and color reconnection at the HL-LHC and FCC-ee 
\label{sec:dphi_hl-lhc}
\label{sec:HadronizationFCCee}
}
Nonperturbative uncertainties from final-state hadronic effects linked to power-suppressed infrared phenomena, such as color reconnection (CR), hadronization, and multiparticle correlations (in spin, color, space, momenta)---which cannot be currently computed from first-principles QCD and often rely on phenomenological Monte Carlo modeling-- may limit the ultimate accuracy  at hadron-hadron colliders. 
Whereas the perturbative radiation in the process can be in principle theoretically controlled, there is a CR ``cross talk'' among the produced hadronic strings that can only be modelled phenomenologically~\cite{Khoze:1994fu}.
In the $pp$ case, CR can limit precision in the extraction of the top mass, contributing 20--40\% of the net uncertainty~\cite{Argyropoulos:2014zoa}. 
In contrast, 
the FCC-ee would offer a relatively clean radiation environment for systematic study of such effects \cite{Proceedings:2017ocd}. In $e^+e^- \to t\overline{t}$, as the top quarks decay and hadronize closely to one another, their mutual interactions, decays into bottom quarks, and/or gluon radiation rearrange the color flow and thereby the kinematic distributions of the final hadronic state.  To understand CR dynamics that limits the reach of CP-violation searches in 
$H \to W^+ W^-$ hadronic decays~\cite{Christiansen:2015yca}, one can optimally study the process $e^+e^- \to  W^+W^- \to q_1\bar{q}_2 q_3\bar{q}_4$~\cite{Christiansen:2015yca}, where CR can lead to the formation of alternative ``flipped'' singlets $ q_1\bar{q}_4$ and $ q_3\bar{q}_2$, and correspondingly more complicated string topologies~\cite{Sjostrand:1993hi}. The combination of results from all four LEP collaborations excluded the no-CR null hypothesis at 99.5\% CL~\cite{ALEPH:2013dgf}, but the size of the WW data sample was too small for any quantitative studies. At the FCC-ee, if the $W$ mass is determined to better than 1~MeV by a threshold scan, the semileptonic WW measurements (unaffected by CR) can be used to probe the impact of CR in hadronic WW events~\cite{Proceedings:2017ocd,Abada:2019lih}. 

\begin{figure}[thb]
\centering
\includegraphics[width=0.40\textwidth, height=0.41\textwidth]{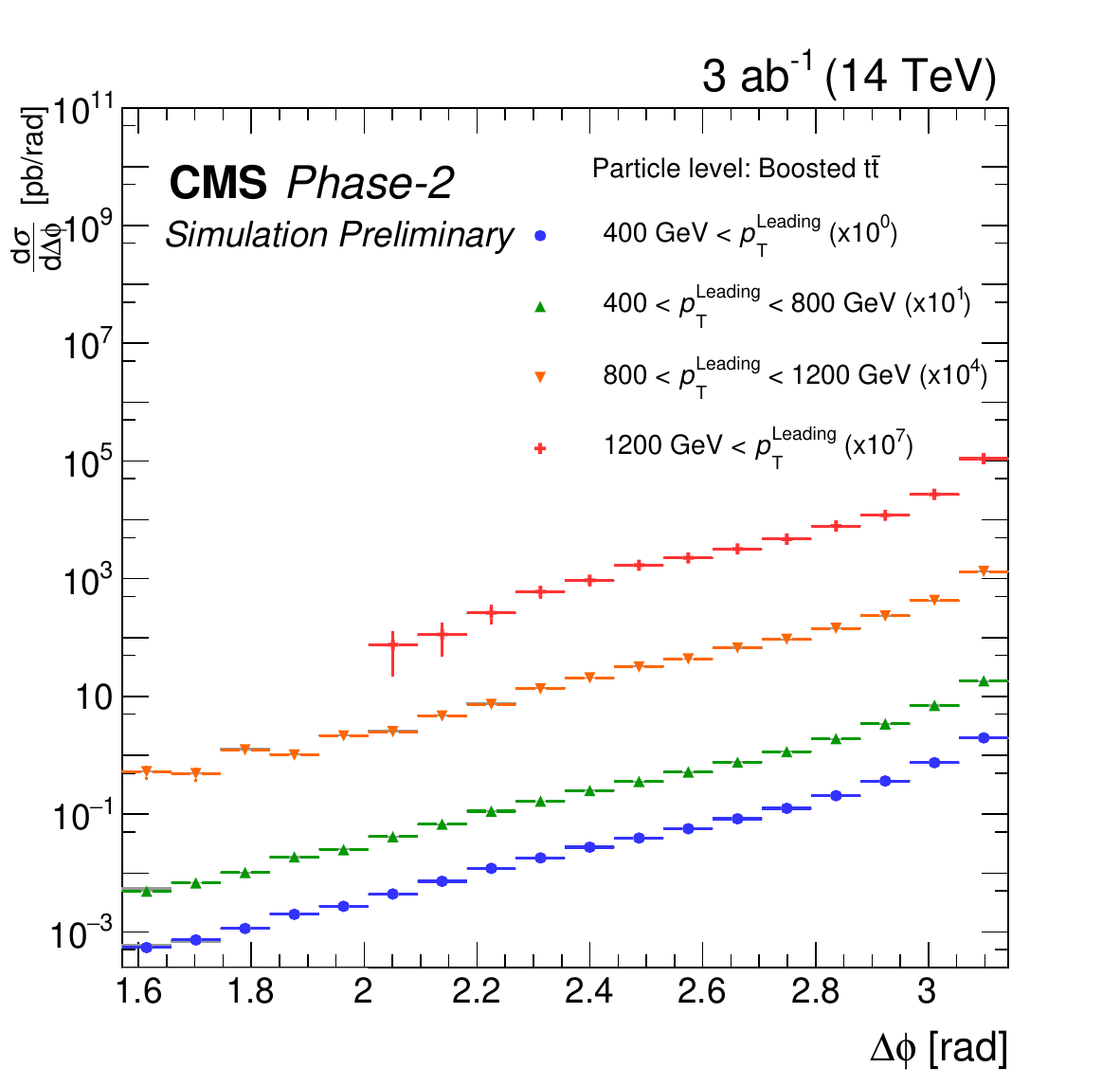}\label{fig:EF05_HighPt-b}
\caption{The particle level cross-section of the $t\bar{t}$ process as a function of  the $\Delta(\phi)$ between the two leading $t\bar{t}$ jets.~\cite{FTR-18-032}}
\label{fig:EF05_HighPt}
\end{figure}

Ultimately, enormous data sets collected at the FCC-ee would lead to negligible statistical and small systematic uncertainties on such observables sensitive to the geometric pattern of soft QCD interference. 
Even at the HL-LHC, these issues can be put under better control in spite of its typically larger uncertainties on particle production due to the underlying event, multiple parton scattering, and dense particle tracks. For example, in production of top-quark pairs, $t$ jets are defined when the top quark decays hadronically, and the decay products are clustered as a single jet. 
Figure~\ref{fig:EF05_HighPt} shows $t\bar{t}$ cross-section as a function of the azimuthal angle separation $\Delta(\phi)$ between the two leading $t\bar{t}$ jets. In this analysis by the CMS Collaboration~\cite{FTR-18-032}, kinematic distributions of jets in inclusive jet production, top quark jets and jets arising out of the hadronic decay of W-bosons have been studied, following previous $\sqrt{s}=$7~TeV analyses~\cite{STDM-2010-15,STDM-2013-03,CMS-FSQ-13-010,CMS-BPH-11-022}.
The azimuthal correlation between the two jets is indicative of the interference effects arising out of the color connection of the jets.    

At the HL-LHC, hadronization uncertainties can also be studied and reduced by taking advantage of better event generators and GEANT detector simulation code, reduced material in trackers, higher reconstruction
efficiency, wider angular and momentum acceptance~\cite{Agostinelli:2002hh}.  During the LHC measurements themselves, hadronization uncertainties can be further reduced thanks to detectors with advanced hadron identification capabilities (e.g., via timing, $dE/dx$, and/or $dN/dx$)
and modern data analysis methods, such as those exploiting the jet substructure and the Lund plane approaches~\cite{Marzani:2019hun,Dreyer:2021hhr}. Better models for hadronization will lead to more accurate experimental measurements and deeper understanding of strong interactions to all orders.

\section{Parton distribution functions in lattice QCD} 
\label{sec:lattice}

Lattice QCD (LQCD) is a theoretical tool that allows us to study the nonperturbative regime of QCD directly with full systematic control.
The approach is based on regularizing QCD on a finite four-dimensional Euclidean spacetime lattice and is often studied using numerical computations of QCD correlation functions in the path-integral formalism using national-scale supercomputers.
To make contact with experimental data, the numerical results are extrapolated to the continuum (with lattice spacing $a \to 0$) and infinite-volume ($L \to \infty$) limits.
When the calculation is done using heavier-than-physical quark masses (to save computational time), one also has to take the $m_\mathrm{q} \to m_\mathrm{q}^\text{phys}$ limit. 
Lattice QCD has been known for great precision in providing flavor-physics inputs, CKM matrix elements, quark masses and more (See Snowmass Rare Precision and Lattice Gauge Theory reports). 
The progress of lattice PDF calculations has long been limited by computational resources, but recent advances in both algorithms and  worldwide investment in pursuing exascale computing have led to exciting progress in LQCD calculations.
Many collaborations have performed hadronic structure calculations directly at physical pion mass with multiple lattice spacings to control lattice artifacts.
Some LQCD calculations have reached a level where they not only complement, but also guide current and forthcoming experimental programs, such as on nucleon tensor charges and the strange-quark contribution to proton spin~\cite{Lin:2017snn,Constantinou:2020hdm}.

There has been rapid progress in new methodology for calculating the momentum fraction ($x$) dependence of PDFs on the lattice.
Here, we will mention a few select examples of this progress; for recent reviews, see Refs.~\cite{Lin:2017snn,Ji:2020ect,Ji:2020byp,Constantinou:2020hdm}.

\begin{itemize}
\item 
Extensive calculations have been carried out for isovector PDFs and distribution amplitudes (DAs). Precision calculations require control over systematics from renormalization, the continuum limit, the
inverse problem in short-distance factorization (SDF) and extrapolation to large lightcone distance in large-momentum effective theory (LaMET). Calibrations can be made against lattice moments and high-precision experimental data. 
Closure tests with artificial data can be used to assess the robustness of the current procedures.  To make an impact on high-energy collider phenomenology, developing calculations at the 5\% level (total systematics) for isovector collinear PDFs and 
improving the precision of the current PDF calculations including sea-quark distributions (as well as large-$x$ quarks and gluons) will require significant increases in computational resources.

\item 
Methods for calculating collinear PDFs, generalized parton distributions (GPDs), TMD distributions and evolution kernels have undergone extensive development. While one-loop matching kernels are widely available, high-precision calculations require two-loop (only available for isovector PDFs) or higher-order matching, as well as quantitative understanding of renormalon uncertainties and higher-twist effects. The key systematic uncertainties that need to be understood arise from inverse problems and coordinate-space extrapolations at large distances. 

\item  While many lattice exploratory studies have been undertaken, extensive high-statistics lattice-QCD data spanning different hadron momenta, quark masses, lattice spacing, volumes, valence and sea quarks, are needed to understand systematic effects. New methods are needed to increase the signal-to-noise ratio for hadronic matrix elements, particularly for large hadron momentum and large spatial correlations. Criteria need to be established for reducing the excited-state contamination, finite-volume effects, and the effects of nonzero lattice spacing.

\item Lattice computations are complementary to phenomenological PDF analyses, and in certain cases they can be used together to obtain hybrid parton distribution sets~\cite{Lin:2017stx,Cichy:2019ebf,Bringewatt:2020ixn,DelDebbio:2020rgv}. This is particularly important 
for three-dimensional nuclear femtography, because extracting the GPDs from experimental data alone can be extremely challenging. The connected-sea and disconnected-sea partons are innately coded in lattice calculations of the PDFs, GPDs and TMDs via the respective insertions. Lattice calculations and phenomenological analyses of the PDFs and their moments can go hand-in-hand in discriminating between the connected-sea and disconnected-sea components of the PDFs. This separation will help to understand the partonic composition of the proton's spin, for example. 

\item 
Besides the collinear PDFs, exploratory calculations have been undertaken for other salient QCD functions, including GPDs, higher-twist distributions, as well as the Collins-Soper rapidity evolution kernel and soft functions introduced by TMD factorization. As some of these functions are not easy to assess in experiment, their controlled lattice calculations can play a prominent role in future studies. 

\end{itemize}

\subsection{New methodologies for PDF computations \label{sec:LatticeMethodology}}

Mellin moments of the collinear PDFs are the simplest quantities to calculate on the lattice. Moments provide ``global'' momentum-space information about partons. It is, however, not easy to connect them directly to a particular experiment in which particles of definite momentum are measured. A more desirable theoretical approach is to directly access dependence of PDFs on the partonic momentum fraction $x$, i.e.,  ``local'' information in momentum space. Toward this goal, two approaches have been developed by lattice QCD in recent years. The first focuses on the
SDF in coordinate space, and the other is based on an expansion in terms of a large hadron momentum, LaMET. Both methods require calculating coordinate-space correlation functions in large-momentum hadron states.

Multiple SDF methods have been developed by the lattice community, based on hadronic tensors~\cite{Liu:1993cv,Liu:1999ak,Liang:2019frk,Liang:2020sqi},
Compton amplitudes or ``OPE without OPE''~\cite{Aglietti:1998ur,Ji:2001wha,Detmold:2005gg,Chambers:2017dov,Detmold:2021uru},
current-current correlators~\cite{Braun:2007wv,Ma:2014jla,Ma:2017pxb,Bali:2018spj,Joo:2020spy,Gao:2020ito,Sufian:2020vzb}
and pseudo-PDFs~\cite{Radyushkin:2017cyf}.
All these approaches provide constraints on collinear PDFs beyond individual moments. To determine the PDFs from the discrete lattice data on a range of coordinate-space correlations (a ``Ioffe-time distribution''), one solves the inverse problem, which may take the form of a fit with either a fixed parametrization or a neural network. The lattice data can be fitted on their own or included into a global analysis of PDFs together with experimental hard-scattering data, as described in Secs.~\ref{sec:EICProtonPDFs} and \ref{sec:MesonPDFs}.

An alternative approach to parton physics on the lattice follows from Feynman's original conception of partons as constituents of hadrons traveling closely to the speed of light~\cite{Bjorken:1969ja}. In this
formulation, PDFs quantify distributions of quarks and gluons in hadrons with large longitudinal momentum,  $P_\infty = P_z\to \infty$. This
approach can compute the collinear PDFs, TMD PDFs and light-front wave functions using standard twist-2 operator definitions in Euclidean space~\cite{Izubuchi:2018srq}. 
The UV behavior of these functions is usually renormalized by (modified) minimal subtraction in $n < 4$ dimensions.
A finite large momentum $P_z$ is used to approximate $P_\infty$,
and a large-momentum expansion is carried out, with systematic power corrections characterized by the expansion parameter $\Lambda_\text{QCD}^2/(x P_z)^2$. This follows from the physical picture of partons, which lose their meaning if their longitudinal momentum reaches the soft
nonperturbative scale
$\Lambda_\text{QCD}$, corresponding to 
zero or soft modes instead of collinear ones. More importantly, the UV cut-off $\Lambda_\text{UV}$ is always taken to
be much larger than $P_z$, and the lattice matrix elements must be matched on the standard lightcone parton distributions to account for different UV behavior. This approach to partons is similar to heavy-quark effective theory, in which heavy-quark masses are taken to infinity, and has been called large-momentum effective theory or LaMET~\cite{Ji:2014gla,Ji:2020ect}.

\begin{figure}[tbp]
\subfloat[]{\includegraphics[width=.32\textwidth]{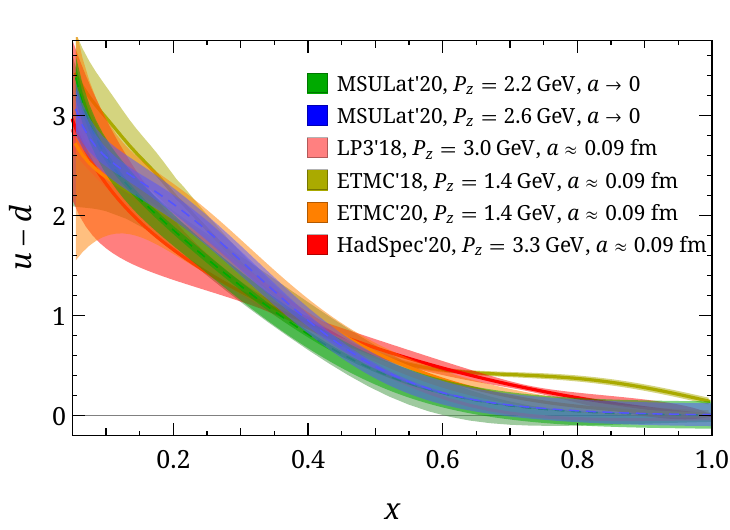}}
\subfloat[]{\includegraphics[width=.32\textwidth]{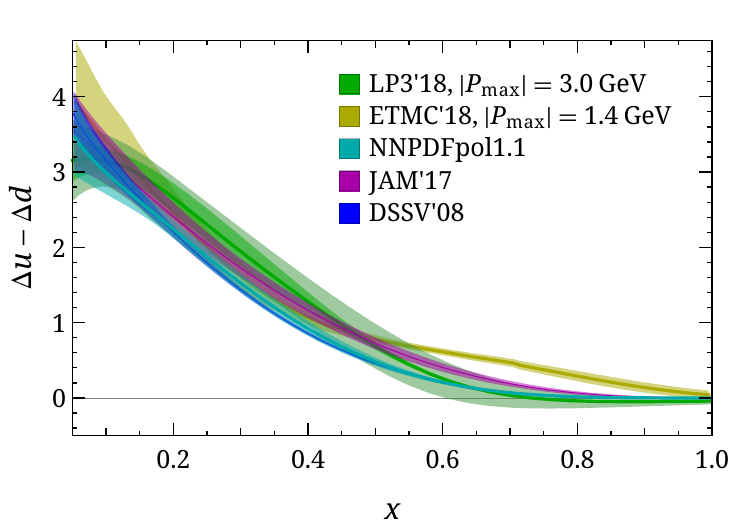}}
\subfloat[]{\includegraphics[width=.32\textwidth]{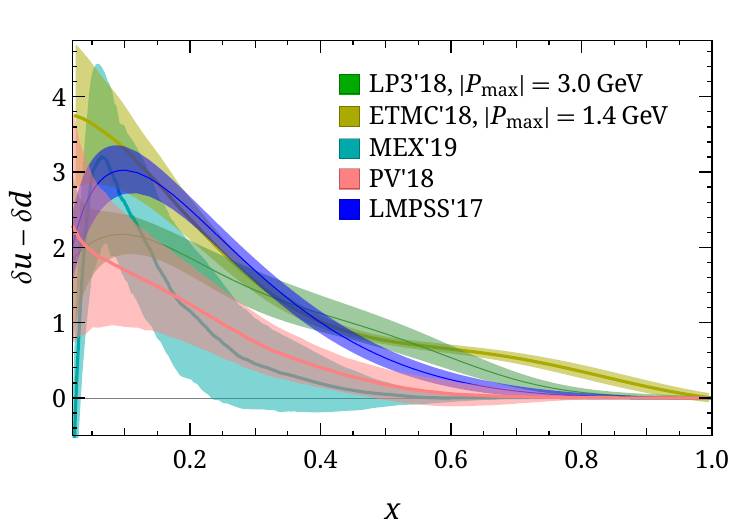}}
\caption{
Comparisons of lattice calculations of the nucleon isovector PDFs. (a) Unpolarized PDFs from the physical-continuum limit, ``MSULat'20''~\cite{Lin:2020fsj}, a single lattice spacing calculation at (or extrapolated to) physical pion mass using LaMET methods, ``LP3'18''~\cite{Chen:2018xof} and ``ETMC'18''~\cite{Alexandrou:2018pbm},  and pseudo-PDF method, ``ETMC'20''~\cite{Bhat:2020ktg} and ``HadSpec'20''\cite{Joo:2020spy}.
(b) Helicity PDFs
from LP318~\cite{Lin:2018pvv,Liu:2018hxv} and ETMC18~\cite{Alexandrou:2018eet,Alexandrou:2019lfo} lattice computations,
and global fits NNPDFpol1.1~\cite{Nocera:2014gqa},
JAM17~\cite{Ethier:2017zbq}, DSSV08~\cite{deFlorian:2009vb}.
(c) Transversity PDFs MEX19~\cite{Benel:2019mcq}, PV18~\cite{Radici:2018iag},
LMPSS17\cite{Lin:2017stx}.
}
\label{fig:PDFresults}
\end{figure}

\subsection{Examples of computations
\label{sec:LatticeExamples}
}
\subsubsection{Nucleon isovector PDFs}
The most studied $x$-dependent structure is the nucleon unpolarized isovector PDF $u(x)-d(x)$.
Multiple collaborations have reported either direct lattice calculations at physical pion mass or extrapolations to physical pion mass using quasi-PDF~\cite{Lin:2017ani,Alexandrou:2018pbm,Chen:2018xof,Lin:2020fsj} and pseudo-PDF methods~\cite{Bhat:2020ktg,Joo:2020spy}.
Figure~\ref{fig:PDFresults}(a) shows the results of lattice calculations using at least one near-physical pion mass, with systematic effects taken into account to varied degrees. 
Overall, there is a reasonable agreement at $x=0.1-0.6$ after scaling up the systematic uncertainties where they may be underestimated. 
The $SU(2)_f$ antiquark asymmetry,  $\overline{d}(x)-\overline{u}(x)$, can be also computed, albeit with a lower accuracy unless an increased value of $P_z$ is used \cite{Lin:2017ani,Chen:2018xof,Lin:2018pvv}.
Increasing the boost momentum of the lattice calculations will be critical for computing these PDF combinations at smaller or larger $x$.

When predicting spin-dependent PDFs, lattice calculations already may provide comparable predictions than phenomenological global analyses. 
Figures~\ref{fig:PDFresults}(b) and (c) summarize lattice predictions for helicity and transversity nucleon isovector PDFs at physical pion mass for the helicity and transversity PDFs~\cite{Alexandrou:2018pbm,Lin:2018pvv,Alexandrou:2018eet,Liu:2018hxv}.The helicity lattice results are compared to two phenomenological fits, NNPDFpol1.1~\cite{Nocera:2014gqa} and JAM17~\cite{Ethier:2017zbq}, exhibiting nice agreement.
The lattice results for the transversity PDFs have better nominal precision than the global analyses by PV18 and LMPSS17~\cite{Lin:2017stx}.
As none of these polarized lattice calculations have taken the continuum limit ($a\rightarrow 0$), and they have remaining lattice artifacts (such as finite-volume effects), further studies will be warranted with more computational resources and multiple lattice spacings and volumes.

\begin{figure}[tbp]
\subfloat[]{\includegraphics[width=0.33\textwidth]{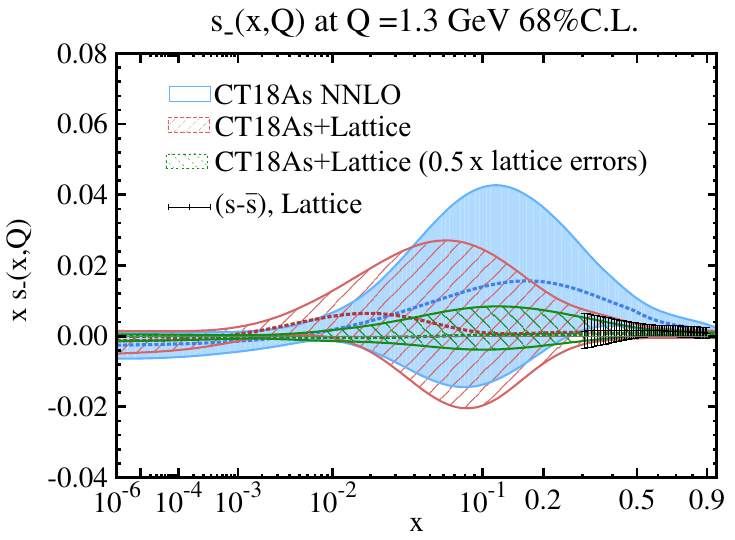}}
\subfloat[]{\includegraphics[width=.32\textwidth]{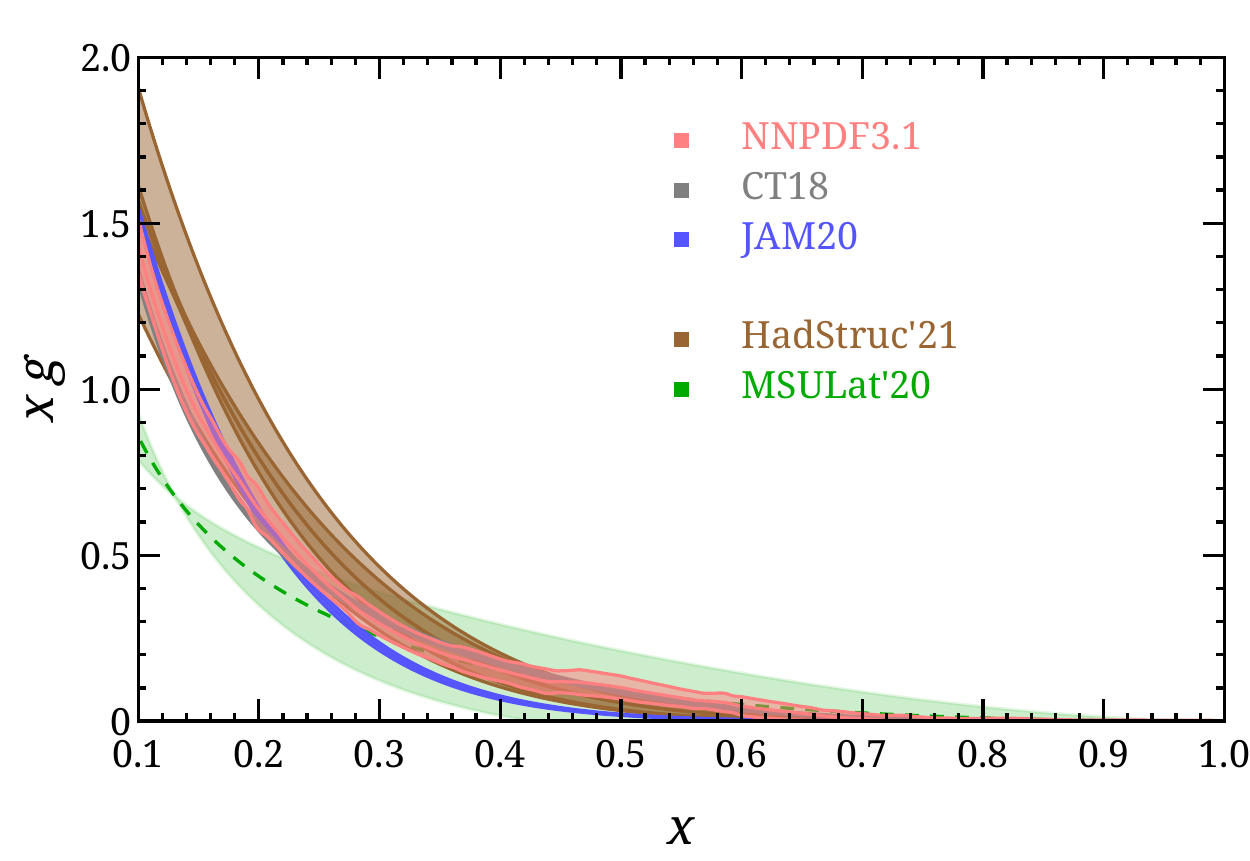}}
\subfloat[]{\includegraphics[width=.3\textwidth]{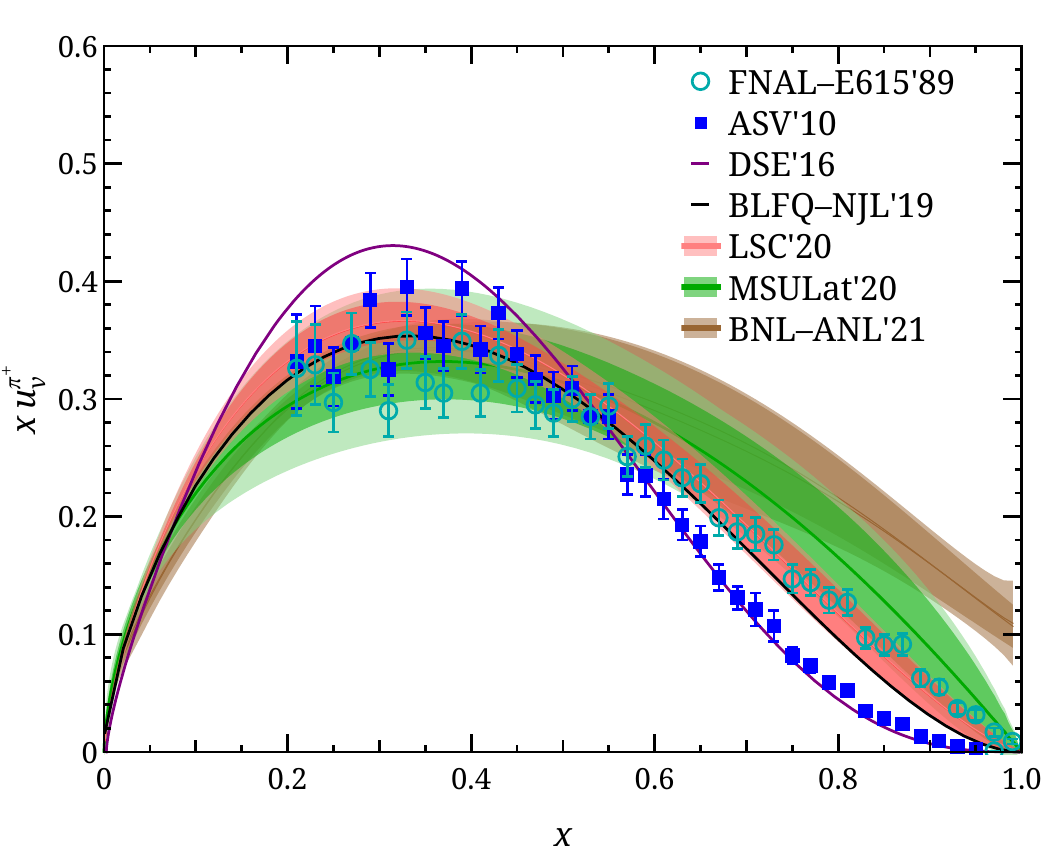}}
\caption{
\label{fig:CT18Aas_Lat_Predicted}
(a) Impact of constraints from lattice QCD (black dashed area) on the difference between strange quark and antiquark PDFs in a recent CT18As NNLO fit~\cite{Hou:2022sdf}. The red (green) error bands are obtained with the current (reduced by 50\%) lattice-QCD errors.
(b) The unpolarized gluon PDF
$xg(x,\mu)$ in $\overline{\text{MS}}$ at $\mu=2$~GeV,
obtained from the fit to the lattice data at pion masses $m_\mathrm{\pi}=135$ (extrapolated), 310 and 690~MeV by MSULat~\cite{Fan:2020cpa} and at pion mass $m_\mathrm{\pi}=380$~MeV by HadStruc21~\cite{HadStruc:2021wmh}, compared with the NNLO CT18
and NNPDF3.1 gluon PDFs.
(c)  Lattice results on the valence-quark distribution of the pion,
BNL~\cite{Gao:2021dbh} and MSULat~\cite{Lin:2020ssv}
lattice groups using LaMET method, JLab and W\&M group~\cite{Sufian:2020vzb}, using LCS method. 
}
\end{figure}

\subsubsection{Strange and anti-strange PDFs}
The experimental uncertainty on the strange quark and antiquark PDFs remains large: in the global fits they contribute in subleading channels or in neutrino-nucleus DIS experiments with substantial uncertainties. DIS and LHC experiments, while not in a certain disagreement on the amount of strangeness in the proton, may exert contradicting pulls on it that depend on the type of the global analysis \cite{CMS:2013pzl,ATLAS:2014jkm,Alekhin:2017olj,Hou:2019efy,Bailey:2020ooq,Faura:2020oom}.
Some of these tensions are relieved by allowing $s(x) \neq \bar{s}(x)$ at the $Q_0$ scale, as is done in some \cite{Bailey:2020ooq,NNPDF:2021njg} but not all analyses. The strangeness contributes a large part of PDF uncertainty in precision $W$ and $Z$ boson measurements at the LHC \cite{Nadolsky:2008zw}, so understanding its behavior is important.
Lattice QCD can already provide competitive constraints on the strangeness asymmetry $s_-(x)\equiv s(x)-\bar s(x)$ and reduce some of the uncertainties that are not constrained in the experiment \cite{Hou:2022sdf}. Figure~\ref{fig:CT18Aas_Lat_Predicted}(a) shows that the uncertainty on $s_-(x)$ in the CT18As NNLO fit is tangibly reduced upon adding the first lattice-QCD constraints on $s_-(x)$ at  $x\in[0.3,0.8]$. In the figure, 
the uncertainty in the lattice data points at $x>0.3$ is quite small compared to the error band of CT18As determined from the global fit, so that including the lattice data in the CT18As\_Lat fit greatly reduces the $s_-$-PDF error band size in the large-$x$ region. The reduction of the CT18As\_Lat error band at $x <0.3$ depends on the chosen parametrization form of $s_-(x)$ at $Q_0=1.3$~GeV. Hence, it is important to have more precise lattice data, extended to smaller $x$ values.
The figure also illustrates the projected reduction in the CT18As\_Lat uncertainty on $s_-(x)$ if the current uncertainties on lattice data are reduced by half. 

\subsubsection{Gluon PDF}
There have been attempts in lattice QCD to constrain the notorious gluon PDFs, which usually require orders-of-magnitude higher statistics than their quark counterparts to get a nonzero result within the statistical uncertainty. 
The first exploratory study of this kind was done by the MSULat group~\cite{Fan:2018dxu}, using $m_\mathrm{\pi}^\text{sea}=330$~MeV.
Up to perturbative matching and power corrections at $O(1/P_z^2)$, the lattice results are compatible with global fits within the statistical uncertainty at large $z$.
Since then, there have been improvements in the operators for the gluon-PDF lattice calculations~\cite{Balitsky:2019krf,Wang:2019tgg,Zhang:2018diq}, which will allow us to take the continuum limit for the gluon PDFs in the future. 
The followup work using the ``pseudo-PDF'' method by MSULat group~\cite{Fan:2020cpa} attempts an extrapolation of the gluon PDF to the physical pion mass at a single lattice spacing. 
HadStruc collaboration~\cite{HadStruc:2021wmh} used a different numerical technique in extracting the gluon PDF with a 360-MeV pion. These results are shown in Fig.~\ref{fig:CT18Aas_Lat_Predicted}(b). While currently these calculations are done at $x$ of order 0.1, the future lattice calculations may be valuable in predicting the gluon PDF at $x>0.5$, where the experimental constraints on the gluon weaken considerably.

\subsubsection{Pion and kaon PDFs}
Lattice QCD predictions are already included in some global fits of pion PDFs \cite{Barry:2022itu}, as reviewed in Sec.~\ref{sec:MesonPDFs}.
Future experiments, such as COMPASS++ and AMBER, will greatly 
improve our knowledge of both the pion and kaon PDFs. Since 2018, there have been several LQCD calculations of the valence-quark dependence of PDFs for pseudoscalar mesons ~\cite{Zhang:2018nsy,Sufian:2019bol,Izubuchi:2019lyk,Joo:2019bzr,Sufian:2020vzb,Shugert:2020tgq,Gao:2020ito,Lin:2020ssv,Gao:2021dbh}.
Reference~\cite{Lin:2020ssv} was the first study of lattice pion and kaon PDFs to take the continuum-physical limit of the matrix elements with a sufficient number of lattice spacings and light pion masses.
There has also been a first next-to-next-to-leading order (N$^2$LO) matching~\cite{Chen:2020ody,Li:2020xml} lattice calculation of the pion valence-quark PDF~\cite{Gao:2021dbh} with 300-MeV pion mass.
Both works take important steps toward precision PDFs from lattice QCD. 
Figure~\ref{fig:CT18Aas_Lat_Predicted}(c) illustrates selected lattice PDF predictions for the valence-quark PDF at $x\to 1$, where it can be compared against various nonperturbative approaches \cite{Farrar:1975yb,Soper:1976jc,Bednar:2018mtf,Zhang:2018nsy,Ding:2019lwe,Novikov:2020snp,Courtoy:2020fex,Gao:2020ito,Alexandrou:2021mmi,Barry:2022itu}.

Gluon PDFs of mesons suffer from significant signal-to-noise issues, so it is harder to get precise signals than for their valence-quark counterparts.
Recently, attempts were made to study the gluon PDFs of the pion~\cite{Fan:2020cpa} and kaon~\cite{Salas-Chavira:2021wui} with lightest pion masses of 220 and 310 MeV, respectively. 
These studies show mild dependence of the pion gluon parton distribution on lattice spacing and pion mass.
The lattice results can be used to improve the large-$x$ region of the global fits with more computational resources and implementation of mixing with the quark PDFs. 

\subsubsection{Other lightcone quantities
\label{sec:LatticeLightconeQuantities}
}

There has been also recent progress made in determining the $x$-dependent  meson distribution amplitudes (DAs)~ \cite{Zhang:2017bzy,Zhang:2017zfe,Bali:2017gfr,Bali:2018spj,RQCD:2019osh,Hua:2020gnw,Zhang:2020gaj,Detmold:2021qln,Juliano:2021hys,Hua:2022kcm,Gao:2022vyh}. 
DAs are important universal quantities to describe exclusive processes at large momentum transfers $Q^2 \gg \Lambda^2_\text{QCD}$ using factorization theorems. Some well-known examples of such processes, which are relevant to measuring fundamental parameters of the Standard Model, include $B \to \pi l \nu$, $\eta l \nu$ giving the CKM matrix element $ |V_{ub}|$, $B \to D \pi$ used for tagging, and $B \to \pi \pi$, $K \pi$, $K\bar{K}$, $\pi \eta$, etc., which are important channels for measuring CP violation (see e.g. Ref.~\cite{Stewart:2003gt}). The lattice studies also help us to 
understand the flavor SU(3) symmetry breaking among light flavors before attributing the effects to enhancement of higher-order amplitudes or even new physics.

New experiments and facilities will explore the three-dimensional  structure of hadrons described by the GPDs and TMDs.
GPDs are hybrid momentum and coordinate-space distributions of partons that bridge the standard nucleon structure observables, form factors and inelastic cross sections. 
Only recently have there been a few lattice calculations made for the pion GPDs with the pion mass of 310 MeV \cite{Chen:2019lcm}, and for nucleon GPDs with the pion masses of 260 MeV~\cite{Alexandrou:2020zbe} and 139 MeV~\cite{Lin:2020rxa,Lin:2021brq}.
These calculations require an increase in computational resources by at least an order of magnitude  relatively to PDF calculations due to the additional boost momenta required. For the best determinations of GPDs, the lattice results can be combined with experimental measurements in a global analysis. 

TMDs depend on the parton's transverse momentum $k_T$, in addition to the longitudinal momentum fraction $x$. They are nonperturbative inputs for processes that follow TMD factorization, such as Drell-Yan process and SIDIS.
Early lattice studies computed selected TMD functions at heavier-than-physical pion masses ranging down to about 300 MeV~\cite{Musch:2010ka,Musch:2011er,Engelhardt:2015xja,Yoon:2015ocs,Yoon:2017qzo}. Recently, there were first extraction of the Collins-Soper kernel, soft function and wavefunctions for TMDs~\cite{Shanahan:2020zxr,LatticeParton:2020uhz,Schlemmer:2021aij,Li:2021wvl,Zhang:2020dbb,Shanahan:2021tst}.
Like for the PDF calculations, lattice precision calculations of TMDs will require large hadron momentum to suppress the power corrections at ${\cal O}(1/(P_zb_T)^2)$.

\subsection{Outlook and challenges
\label{sec:LatticeOutlook}
}
A Snowmass whitepaper~\cite{Constantinou:2022yye} provides more details on the rapid advances in LQCD calculations of PDFs and other QCD functions and has more complete references to relevant work.
Experimental exploration of the three-dimensional structure at the Jefferson Lab, EIC, other facilities will match the ongoing theoretical advancements that open doors to many previously unattainable predictions, from the $x$ dependence of collinear nucleon PDFs to TMDs~\cite{Musch:2010ka,Musch:2011er,Engelhardt:2015xja,Yoon:2016dyh,Yoon:2017qzo} and related functions~\cite{Shanahan:2020zxr,Zhang:2020dbb,Schlemmer:2021aij,Li:2021wvl,Shanahan:2021tst}, GPDs~\cite{Chen:2019lcm, Alexandrou:2020zbe,Lin:2020rxa,Lin:2021brq}, and higher-twist terms, progress that was not envisioned as possible during the 2013 Snowmass study.

There remain challenges to be overcome in the lattice calculations, such as reducing the noise-to-signal ratio, extrapolating to the physical pion mass, and increasing hadronic boosts to suppress systematic uncertainties.
Computational resources place significant limitations on the achievable precision, as sufficiently large and fine lattices are necessary to suppress finite-size and higher-twist contaminating contributions.
New ideas can bypass these limitations.
With sufficient support, lattice QCD can fill in the gaps where the experiments are difficult or not yet available, improve the precision of global fits, and provide better SM inputs to aid new-physics searches across several HEP frontiers.

\section{Forward scattering and saturation}
\label{sec:forward}

Studies of forward and diffractive scattering processes provide rich opportunities to probe QCD dynamics. In this section, we review select examples of ongoing studies and future experiments in this important area. 

\subsection{Low-$x$ physics and BFKL resummation}
\label{sec:LowXBFKL}

In hadron-hadron collisions, final states produced at large rapidities, small partonic momentum fractions $x$, as well as those with the absence of energy in the forward region (with the so-called rapidity gap) in elastic, diffractive, and central exclusive processes offer multiple avenues to learn about QCD dynamics. In these processes, the standard collinear QCD factorization discussed in Secs.~\ref{sec:PDFs} and \ref{sec:perturbativePrecisionCalculations} is generally not applicable. Some of these configurations originate from purely non-perturbative reactions, while others can be explained in terms of multi-parton chains or other extensions of the perturbative QCD parton picture such as the Balitsky-Fadin-Kuraev-Lipatov (BFKL) formalism~\cite{Fadin:1975cb,Kuraev:1976ge,Kuraev:1977fs,Balitsky:1978ic}.

When scattering contributions are evaluated in the high-energy or \emph{Regge} limit, the convergence of the
 perturbative series is spoiled by large logarithms, and an all-order resummation of these
 large logarithms must be carried out. 
The BFKL approach performs this resummation by factorizing QCD cross sections into a
convolution of two process-dependent impact factors
and a process-independent Green's function. This factorization has been proven up to the next-to-leading logarithmic accuracy (NLA).

The BFKL Pomeron predicts a fast powerlike rise of the gluon distribution in the proton in the kinematic regions where $x \to 0$. While such a rise is seen in data, unitarity bounds prohibit such a rise to continue forever: at a certain value of $x$, this rise must slow down to a logarithmic growth. The latter is strongly related with the formation of an over-occupied system of gluons, known as the Color Glass Condensate, whose exploration is one of the central physics goals of the EIC. While at the EIC a dense QCD state will be achieved through scattering of electrons on heavy ions, forward-rapidity experiments at the LHC, including the FPF, allow for a complementary exploration, since high-gluon densities are produced through the low $x$ evolution of the gluon distribution in the proton. 

Closely related to diffractive events are photon-induced reactions at the LHC and other colliders. These reactions are reviewed in the dedicated Section~\ref{sec:UPCs}. In such events, either one or both initial-state protons or ions act as a photon source. The first configuration results in exclusive photon-hadron interaction at highest CM energies and therefore yields another tool to study the highest gluon densities with high precision. Such exclusive reactions are complementary to ion scattering at the EIC, since at the LHC high parton densities are predominately generated by high-energy evolution, while the EIC at its lower CM energy relies on the nuclear enhancement. 
A dedicated Snowmass whitepaper \cite{Hentschinski:2022xnd} elaborates in detail on BFKL and low-$x$ physics, especially Mueller-Navelet jets, jet-gap-jet events, very forward jets, and vector meson production as a possible sign of BFKL resummation and saturation phenomena. 

\subsection{Hard diffraction and sensitivity to the Pomeron}
Hard diffraction corresponds to events when at least one proton is intact after interaction and corresponds to the exchange of a colorless object called the Pomeron. Measurements at the LHC can constrain the Pomeron structure in terms of quarks and gluons that has been previously
derived from QCD fits at HERA and at the Tevatron~\cite{Boonekamp:2011ky,Jung:2009eq}.
One can probe if the Pomeron is universal between
$ep$ and $pp$ colliders.
Tagging both diffractive protons in
allows to probe the QCD evolution of the gluon and quark densities in the Pomeron  and to compare with earlier measurements. In addition, it is possible to assess the gluon and quark densities
using the dijet and $\gamma + \mbox{jet}$ productions~\cite{Marquet:2013rja,Kepka:2007nr,Marquet:2016ulz,Chuinard:2015sva}.
The measurement of the dijet cross section is directly sensitive to the gluon density in the Pomeron and the $\gamma+$jet and $W$ asymmetry measurements~\cite{Chuinard:2015sva} 
are sensitive to the quark densities in the Pomeron. 
However, diffractive measurements are also sensitive to the survival probability which needs to be disentangled from PDF effects, and many different measurements will be needed to distinguish between them.
It is clear that understanding better diffraction and probing different models will be one of the key studies to be performed at the high luminosity LHC, the EIC, and any future hadron collider.

\subsection{Soft diffraction and the Odderon}

Soft diffraction and elastic interactions have been studied for the last 50 years at different colliders. Elastic $pp$ and $p \bar{p}$ scattering at the high energies of the Tevatron and LHC corresponds to the $pp \rightarrow pp$ and $p \bar{p} \rightarrow p \bar{p}$ interactions, where the protons and antiprotons are intact after interaction and scattered at very small angle, and nothing else is produced. 
Many experiments have been looking for evidence of the existence of the Odderon, a $C$-odd counterpart of the Pomeron associated with colorless gluon exchanges~\cite{Lukaszuk:1973nt,Martynov:2018sga}.
At ISR energies~\cite{Breakstone:1985pe, Erhan:1984mv,UA4:1986cgb,UA4:1985oqn,Nagy:1978iw},
there was indication of a possible 3$\sigma$ difference between $pp$ and $p\bar{p}$ interactions.
This was not considered a clean Odderon signal, as elastic scattering at low energies can be due to exchanges of additional particles such as $\rho$, $\omega$, $\phi$ mesons and Reggeons. Distinguishing between all these possible exchanges is not simple and becomes model-dependent.
The advantage of using higher energies such as at the Tevatron or LHC~\cite{D0:2012erd,TOTEM:2018psk, TOTEM:2011vxg,TOTEM:2015oop,TOTEM:2018hki} is that meson and Reggeon exchanges can be neglected. 
Recently, a combination of measurements of elastic cross sections for $p\bar p$ at D0 and for $pp$ by TOTEM was interpreted as a discovery of an Odderon exchange with a significance that ranges from 5.3 to  5.7$\sigma$ (depending on the model) \cite{TOTEM:2020zzr}.
Further measurements of elastic $pp$ cross sections will happen at higher LHC energies, and search for Odderon exchanges will be performed in additional channels, such as production of $\omega$ mesons. It is also clear that the discovery of the Odderon is likely related to the existence of glueballs, and the search for their production will happen at the LHC, RHIC, and the EIC.

\begin{figure}[ht]
\centering
\includegraphics[width=0.48\textwidth]{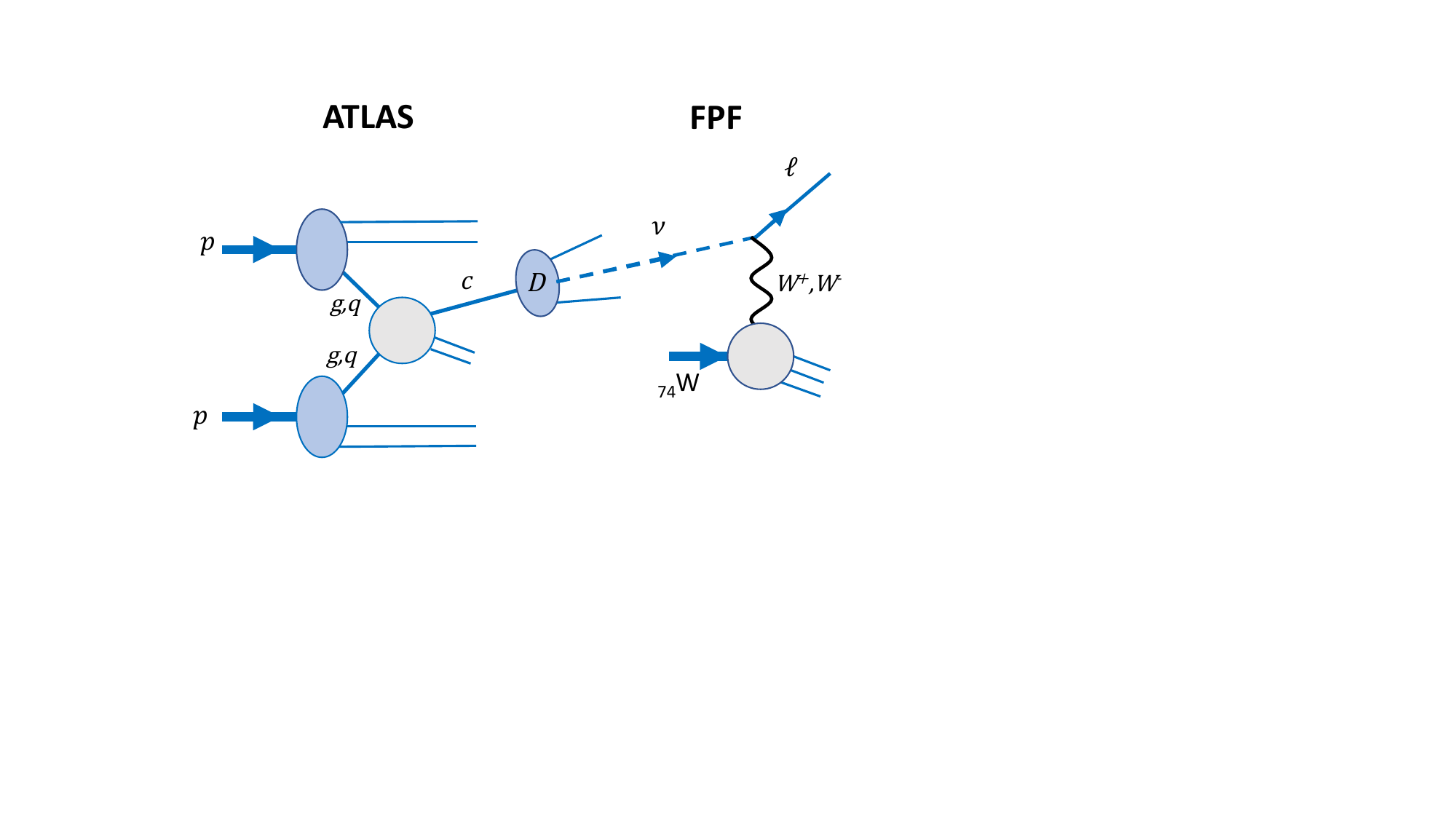}\quad
\includegraphics[width=0.48\textwidth]{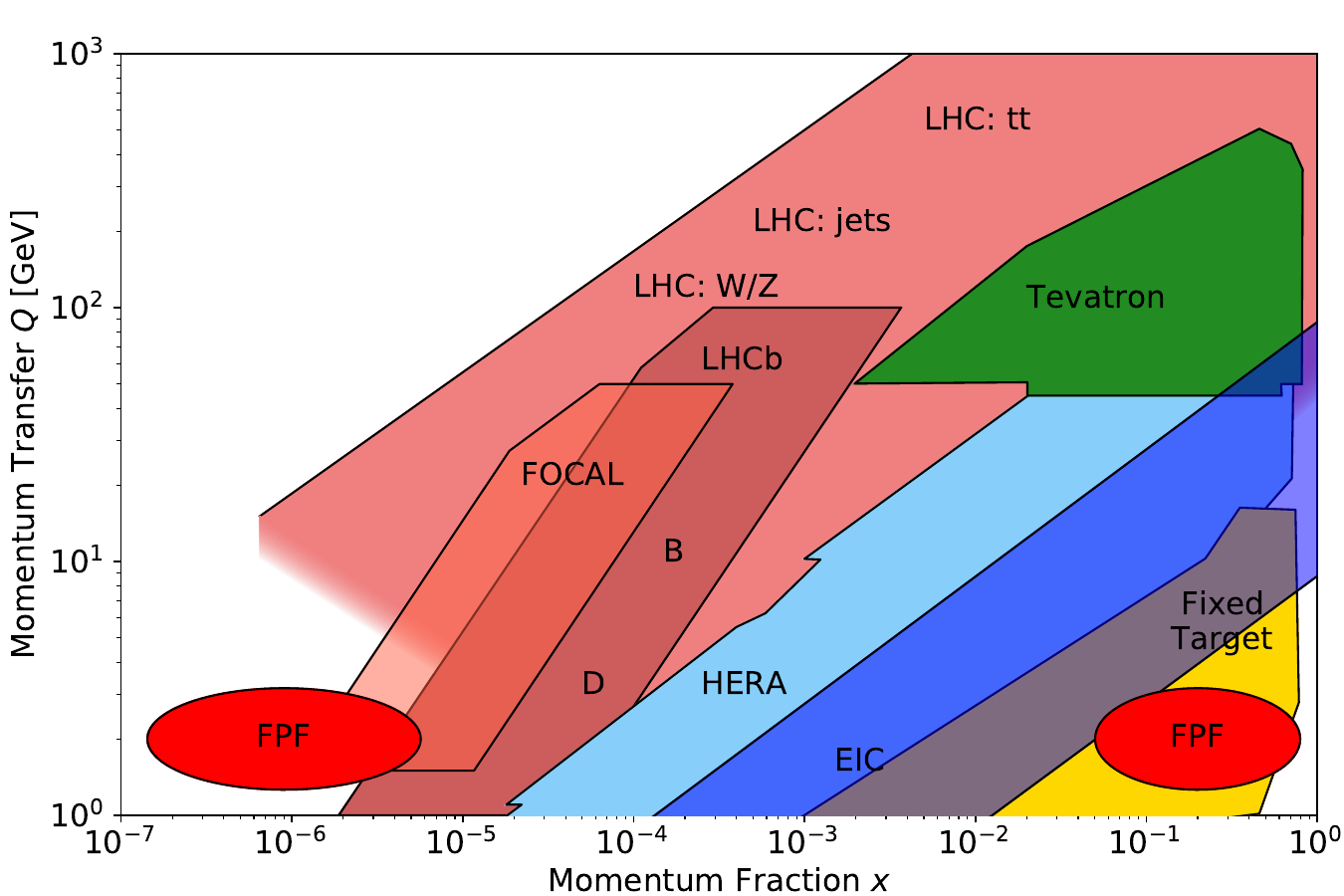}
\caption{(a) The production and detection processes for forward $D$-meson production at the HL-LHC, followed by their decay into neutrinos falling within the FPF acceptance.
(b) The $(x,Q)$ regions (red ovals) that can be accessed at the FPF via this process. 
}
\label{fig:FPF_kin}
\end{figure}

\subsection{Forward Physics Facility}
\label{sec:fpf2}

\subsubsection{Overview}
Experiments at the Forward Physics Facility (FPF) \cite{Feng:2022inv,Anchordoqui:2021ghd} introduced in Sec.~\ref{sec:fpf} offer a unique opportunity to test and study QCD in a new regime. Three FPF experiments, with tungsten or liquid argon targets, will detect far-forward neutrinos that come from decays of hadrons produced in collisions at the LHC ATLAS interaction point, including from decays of charm mesons illustrated in Fig.~\ref{fig:FPF_kin}(a). With its capability of distinguishing neutrinos and antineutrinos of different flavors, the FPF will provide versatile experimental data on both light- and heavy-flavor production at the LHC. 

 The kinematic reach of the FPF in $pp$ collisions is illustrated in Fig.~\ref{fig:FPF_kin}(b). The figure also indicates the approximate kinematic coverage for other experiments that provide inputs for global PDF analyses, including future facilities such as the EIC~\cite{AbdulKhalek:2021gbh} and the ALICE Forward Calorimeter (FOCAL) upgrade~\cite{ALICE:2020mso}.
 The FPF will extend the coverage of the low-$x$ region by almost two orders of magnitude at low-$Q$, reaching down to $x\simeq 10^{-7}$. The FPF will be sensitive to very high-$x$ kinematics as well.

These kinematic ranges open a wide range of opportunities for QCD studies, such as charting the gluon at very low $x$, revealing  BFKL and saturation dynamics, and testing Monte Carlo models for forward hadron production.
Understanding small-$x$ dynamics in $pp$ collisions, already important at the LHC and HL-LHC~\cite{Cepeda:2019klc,Azzi:2019yne}, is crucial for any future higher-energy $pp$ collider such as FCC-hh~\cite{FCC:2018vvp,FCC:2018byv,Mangano:2016jyj,Rojo:2016kwu}, where even standard electroweak processes such as $W$ and $Z$ production become dominated by low--$x$ dynamics, and an accurate calculation of the Higgs production cross section requires accounting for BFKL resummation effects. Therefore, FPF measurements would provide a bridge between the physics program at the HL-LHC and that of an eventual higher-energy $pp$ collider. On the top of that, they will provide improved predictions for DIS in key astroparticle physics processes,
such as improving the extrapolations of the neutrino-nucleus cross section to ultra-high energies and the modeling of showers in cosmic ray interactions in the atmosphere.

By going more differential, e.g., by covering a wide rapidity range either by placing the FPF detectors at different radial distances from the beam collision axis or by making the FPF detectors work in coincidence with the ATLAS detector, the FPF may clarify the transition from collinear to BFKL factorization. Azimuthal-angle distributions, should they be accessible, may allow studies of transverse momentum dependence of production and decay, as well as enhanced searches for BSM signals. As multiple QCD phenomena would contribute at the FPF, disentangling them will strongly benefit from a coordinated program with other facilities, including forward production at LHCb, large-$x$ CC DIS at the EIC~\cite{AbdulKhalek:2021gbh}, and small-$x$ particle production at the HL-LHC and in future DIS experiments  such as the MuIC \cite{Acosta:2022ejc} or LHeC~\cite{LHeC:2020van}.

Heavy-ion collisions at the HL-LHC will produce a large number of hadrons, creating a neutrino flux that can be observed at the FPF. These data could be used to study hadron propagation through nuclear matter or the quark-gluon plasma in different kinematic regimes. In addition, charm production 
in heavy-ion collisions may potentially probe nuclear PDFs for initial-state gluons and test gluon saturation, which is expected to be present at higher $x$ compared to the proton case. For these studies to be feasible, the detection process of neutrino DIS on a nuclear target must be well understood. Again, this makes concurrent studies of CC DIS on heavy ions at the EIC highly beneficial for the success of the FPF program, see Sec.~\ref{sec:CCDIS}. 

\subsubsection{Forward charm production \label{sec:ForwardCharmHybrid}}
\begin{figure}[tbp]
\begin{center}
\includegraphics[width=0.53\textwidth]{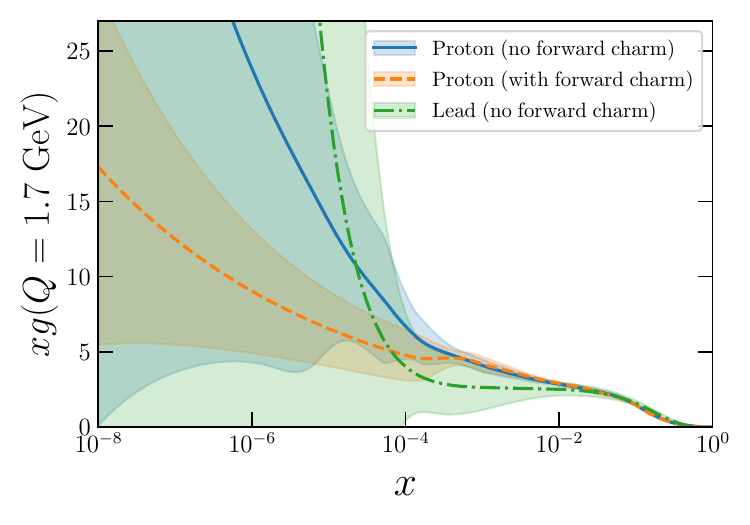}
\includegraphics[width=0.45\textwidth]{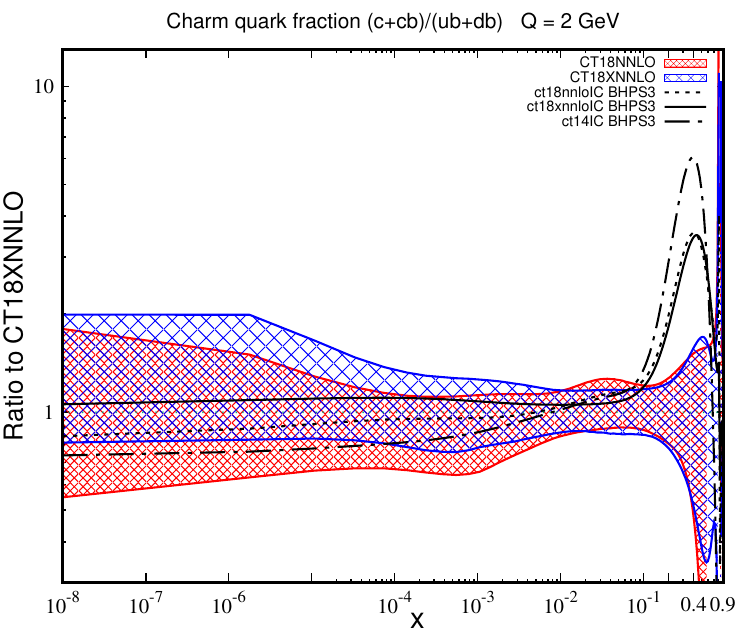}
\caption{(a) The small-$x$ gluon PDF at $Q=1.7$~GeV in the NNPDF3.1 proton fits without and with forward $D$ meson data from LHCb, together with the lead nuclear PDF in the nNNPDF2.0 analysis. (b) The charm-quark fraction ratio $(c(x,Q)+\bar{c}(x,Q))/(\bar{u}(x,Q)+\bar{d}(x,Q))$ for $Q=2$ GeV. The error bands represent the CT18 NNLO (red) and CT18X NNLO (blue) PDF uncertainties at 90\% C.L.~\cite{Hou:2019efy}. From Ref.~\cite{Feng:2022inv}.
\label{fig:PDFs-FPF}}
\end{center}
\end{figure}

Forward charm production at the LHCb is an emerging process to probe the small-$x$ gluon PDF~\cite{PROSA:2015yid,Gauld:2016kpd, Zenaiev:2019ktw, Garzelli:2020fmd}. Figure~\ref{fig:PDFs-FPF}(a) illustrates the impact of the LHCb forward charm production data on the gluon PDF in the nucleon and lead at $x<10^{-2}$ in the recent NNPDF analyses. [The figure does not include large theoretical uncertainties, which will be reduced in upcoming NNLO and resummation calculations.] The FPF measurements will probe this PDF at $x<10^{-5}$, where the current uncertainty is even bigger, and the PDF is reconstructed by extrapolation from large $x$. 

The large-$x$ kinematic region is of interest due to the sensitivity of the FPF experiments to an intrinsic charm component of the proton~\cite{Brodsky:2015fna}. 
The nonperturbative charm component, known as intrinsic charm, may arise due to long-distance interactions of the charm quark with one of the beam remnants. Intrinsic charm results in a characteristic enhancement of the charm PDF at $x>0.1$, as illustrated in Fig.~\ref{fig:PDFs-FPF}(b) on the example of the ratio of charm-to-light flavor PDFs as a function of $x$ for $Q=2$ GeV.
While charm production in $pp$ collisions is dominated  by gluon-gluon scattering, in the presence of the intrinsic charm, the charm-gluon initial state enters and may even be dominant for forward $D$-meson production.
Several  studies have investigated the possible existence of this intrinsic charm,
including measurements by
LHCb~\cite{LHCb:2013xam, LHCb:2015swx, LHCb:2016ikn,LHCb:2021stx}.
FPF measurements would provide a complementary handle on intrinsic charm, which in turn could enhance the expected flux of prompt neutrinos
arising from the decays of charm mesons produced in cosmic-ray collisions in the atmosphere~\cite{Bertone:2018dse,Garcia:2020jwr,Gauld:2015kvh, Garzelli:2016xmx}.
These represent an important background for astrophysical neutrinos at neutrino telescopes such as IceCube~\cite{IceCube:2020wum} and KM3NeT \cite{KM3Net:2016zxf}. See the Snowmass white paper on high-energy and ultra-high-energy neutrinos \cite{Ackermann:2022rqc}. 

\subsubsection{Neutrino-induced deep inelastic scattering \label{sec:CCDIS}}

Among the various detectors proposed for the FPF, the principal detection modes will effectively realize an experiment on high-energy neutrino-induced deep-inelastic scattering (DIS), with event properties being reconstructed from the kinematics of the outgoing charged lepton and hadrons. The ability of neutrino DIS measurements to probe specific quark flavours via the weak current significantly improves global determinations of proton and nuclear PDFs \cite{Gao:2017yyd}.
Neutrino--induced CC DIS structure functions provide access to different quark flavour
combinations compared to charged lepton DIS, and hence FPF data can complement other planned experiments such as the EIC. 
The FPF potential can be further enhanced by using a range of nuclear targets.
The coverage for CC nuclear structure functions at the FPF  broadly overlaps with that for NC charged-lepton expected at the EIC~\cite{AbdulKhalek:2021gbh,Khalek:2021ulf}.
Analogous information from previous neutrino-induced DIS measurements on nuclear targets, such as NuTeV~\cite{NuTeV:2005wsg}, NOMAD~\cite{NOMAD:2013hbk}, CCFR~\cite{NuTeV:2001dfo}, and CHORUS~\cite{CHORUS:1997wxi}, plays a prominent role in many global PDF fits of nucleon and nuclear PDFs (with the two related via nuclear corrections). 

Inclusive CC DIS and especially semi-inclusive charm production in CC DIS are the primary channels to probe the PDFs for strange quarks and antiquarks. Strangeness PDFs offer insights about the nonperturbative proton structure~\cite{Chang:2014jba}, while they are also responsible for a large part of the PDF uncertainty in weak boson mass measurements at the LHC~\cite{Nadolsky:2008zw}. On the experimental side, determination of the (anti-)strangeness PDF has been a hot topic for the PDF community as the fits prefer somewhat different shapes for the strangeness PDFs~\cite{Alekhin:2014sya, Alekhin:2017olj,Hou:2019efy,Faura:2020oom}. The elevated PDF uncertainty from fitting such inconsistent experiments propagates into various pQCD predictions (see e.g. those recently presented in Ref.~\cite{Bevilacqua:2021ovq}).

\subsection{Probing the multidimensional structure of hadrons}
\label{QCD:ssec:HAS_FPF}

Extensions of collinear PDFs can be defined based on alternative factorization theorems to describe strong interactions in special kinematical regions. 
These include transverse-momentum-dependent PDFs (TMDs) introduced in the context of TMD factorization (see e.g.~\cite{Collins:2011zzd}) and the related unintegrated PDFs (uPDFs) in the context of high-energy factorization at small momentum fraction $x$. The connection between the two has been investigated in several papers (see e.g.~\cite{Nefedov:2021vvy,Hentschinski:2021lsh,Celiberto:2019slj}). There is an extended literature on the theory and phenomenology of TMDs, for quarks and gluons, with or without polarization. 
Most recent global fits of quark TMDs~\cite{Bacchetta:2017gcc,Scimemi:2017etj,Scimemi:2019cmh,Bacchetta:2019sam} are based on data from semi-inclusive DIS, Drell-Yan, and $Z$-boson production processes. Studying these and other processes in the low-$x$ and high-$x$ regions opens a window on detailed behavior of QCD radiation. 

Among the TMDs, the gluon unpolarized TMD correlator operator is the least known and has non-trivial spin dependence \cite{Mulders:2000sh}. 
Besides the unpolarized gluon TMD, the counterpart distribution of linearly polarized gluons in an unpolarized nucleon, $h_1^{\perp g}$, gives rise to spin effects even in collisions of unpolarized hadrons~\cite{Nadolsky:2007ba,Catani:2010pd,Boer:2010zf,Sun:2011iw,Boer:2011kf,Pisano:2013cya,Dunnen:2014eta,Lansberg:2017tlc}.
Golden channels for the study and extraction of the gluon TMD and uPDF in hadron-hadron collisions correspond to the inclusive emission of a single particle over forward ranges of rapidity as well as over more central regions in gluon-induced hard scatterings \cite{Chapon:2020heu,Celiberto:2019slj,Celiberto:2020rxb,Celiberto:2021dzy,Bacchetta:2020vty}. 

TMD factorization is violated at high orders in processes with multiple identified hadronic states \cite{Rogers:2010dm}. These violations are likely more pronounced in forward regions and should be experimentally investigated for interpretation of various applications. 

The gluon and quark GPDs characterize the three-dimensional nucleon structure in terms of longitudinal momentum fraction $x$ and momentum transfer to the nucleon, $t$.
The GPDs play an important role in many processes. For instance, the amplitude of onium hard diffractive production in the scaling limit is proportional to the gluon GPD. There are indications, based on HERA data, that the universality limit for $J/\psi$ production is reached already for photoproduction, while in $\rho$ production it is reached only at $Q^2 \gtrsim 15 \mbox{ GeV}^2$ \cite{Frankfurt:2022jns}. 
Investigation of universality of $t$ dependence in $\Upsilon $ and $J/\psi$ production, as well as of connections of GPDs to collinear PDFs \cite{Flett:2019pux,Flett:2020duk}, would be insightful tests of the current GPD models.

Vector meson photoproduction in UPCs (see Ch.~\ref{sec:UPCs}) 
is also an effective way to measure GPDs in ions
 \cite{Klein:2019qfb}, with proof-of-principle measurements of these kinds already available
 \cite{STAR:2017enh}.  
 Multidimensional distributions that go beyond TMDs -- 
the so-called generalized transverse-momentum distributions 
(GTMDs)~\cite{Ji:2003ak,Belitsky:2003nz,Meissner:2009ww} -- 
can be measured in forward processes, for instance exclusive double 
Drell--Yan~\cite{Bhattacharya:2017bvs}, ultraperipheral $pA$ 
collisions~\cite{Hagiwara:2017fye}, and diffractive forward 
production of two quarkonia~\cite{Boussarie:2018zwg}. High-energy studies of TMD's, GPDs, and GTMDs complement 
the program of three-dimensional femtography at the EIC and JLab.


\section{Heavy Ion Physics}
\label{sec:heavyions}

The heavy-ion (HI) program at the LHC has been a successful and important part of the LHC physics program in Runs 1 and 2. The LHC experiments including ALICE, the dedicated detector focuses on HI physics, general purpose detectors CMS, ATLAS and LHCb, have been participated in the data-taking. Its chief aim was the identification and characterization of a Quark Gluon Plasma (QGP) in lead-lead (Pb-Pb) collisions. In addition to QGP studies, the program has included many advances in the understanding of the partonic nuclear structure, collectivity in small collision systems~\cite{CMS:2010ifv,CMS:2012qk,ATLAS:2012cix,ALICE:2012eyl,LHCb:2015coe}, and electromagnetic (EM) interactions~\cite{ATLAS:2017fur,CMS:2018erd}. A detailed plan for the goals and expected measurements at the HL-LHC is presented in Ref.~\cite{Dainese:2703572}. These are summarized below covering results on jet modification and heavy-flavor (HF) hadrons, primarily in Pb-Pb collisions, bulk particle collectivity in both Pb-Pb as well as smaller collision systems, nuclear parton density in proton-lead (p-Pb) and photonuclear collisions, and finally EM interactions from ultraperipheral Pb-Pb collisions.  The latter topic is discussed in the next chapter. 

The nominal expectations for the LHC Run 3 and Run 4 are for: 13~nb${}^{-1}$ of Pb-Pb collisions at $\sqrt{s_{\mathrm{NN}}}=5.5$~TeV and 1.2~pb${}^{-1}$ of p-Pb collisions at $\sqrt{s_{\mathrm{NN}}}=8.8$~TeV.  
In addition to the larger available luminosity, detector upgrades planned for both ATLAS and CMS experiments will benefit the HI program.  In particular, the increased charged particle tracking pseudo-rapidity acceptance will be a boon to bulk particle measurements, the upgraded Zero Degree Calorimeters (ZDC)~\cite{ATLAS-ZDC-LHCC, CMS-ZDC-TDR} will improve triggering and identification for ultraperipheral collisions (UPC), and the addition of time-of-flight particle identification capability enabled by the MIP Timing Detector~\cite{CMS:2667167} will allow to differentiate among low momentum charged hadrons, such as pions, kaons, or protons, improving the HF measurements. The ALICE ITS3 upgrade that is planned for installation before Run 4 will significantly improve the heavy flavor meaurements down to very low transverse momentum. For data-taking in LHC Run 5, the ALICE collaboration planned a major upgrade of its detector referred to as ALICE 3~\cite{ALICE:2803563} which will enable an extensive program to fully exploit the LHC for the study of the properties of the QGP.

At RHIC, the construction of sPHENIX detector~\cite{PHENIX:2015siv}, equipped with calorimeters as well as high precision inner trackers, will be finished and start the heavy-ion data-taking in 2023. The electromagnetic and hadron calorimeter will enable full jet analyses in pp and heavy ion collisions at RHIC energies for the first time. Together with the high precision tracking systems, the collaboration aims to provide high precision heavy flavor meson and quarkonium data in Au-Au collisions.

\subsection{Jet Quenching}

High momentum-transfer interactions between partons in the nuclei produce hard probes of the QGP. One can study the impact of QGP on color charges with fast-moving partons. 
In QGP, medium-induced gluon radiations and elastic scatterings could transfer the parton energy to a large angle from the original mother parton momentum vector. The partons inside the jets could excite a QGP wake. The effect of QGP on color charges can therefore be observed as the attenuation of the jets~\cite{ATLAS:2010isq,CMS:2011iwn,CMS:2012ulu,CMS:2012ytf,CMS:2017ehl,ATLAS:2018dgb,CMS:2017eqd,ATLAS:2012tjt,ATLAS:2014ipv,CMS:2016uxf,ATLAS:2018gwx,CMS:2021vui,ALICE:2015mjv,CMS:2015hkr}, and the modification of their substructure~\cite{CMS:2013lhm,CMS:2014jjt,ATLAS:2014dtd,CMS:2017qlm,CMS:2018jco,ATLAS:2020wmg,ATLAS:2018bvp,CMS:2018mqn,CMS:2018fof,ALICE:2019ykw,ALICE:2018dxf,CMS:2020plq,CMS:2021otx}, often referred to as jet quenching.

\begin{figure}[!htbp]
  \centering
  \subfloat[]{\includegraphics[scale=0.310]{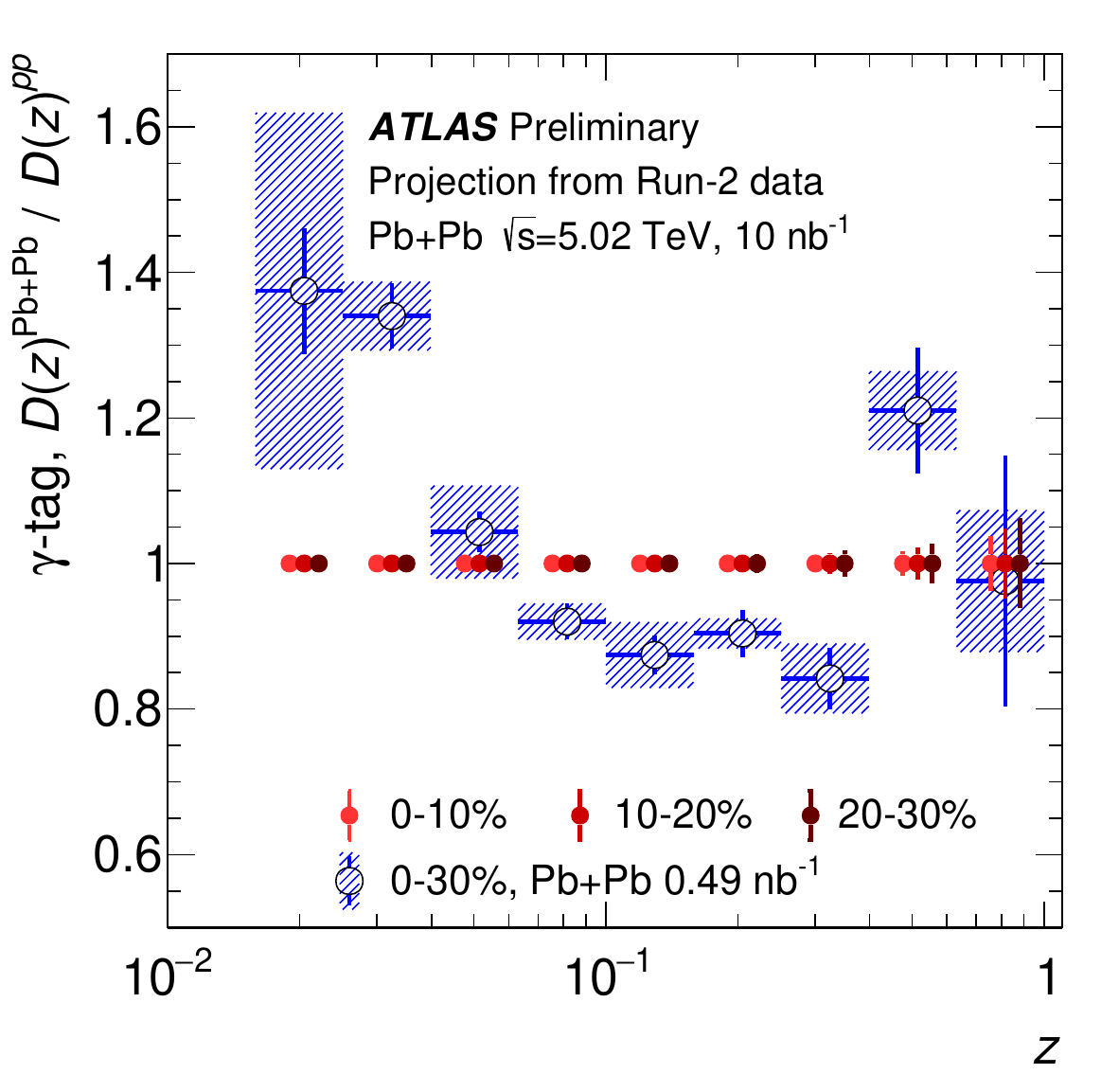}\label{fig:EF07_jet_GammaJet}}
  \subfloat[]{\includegraphics[scale=0.315]{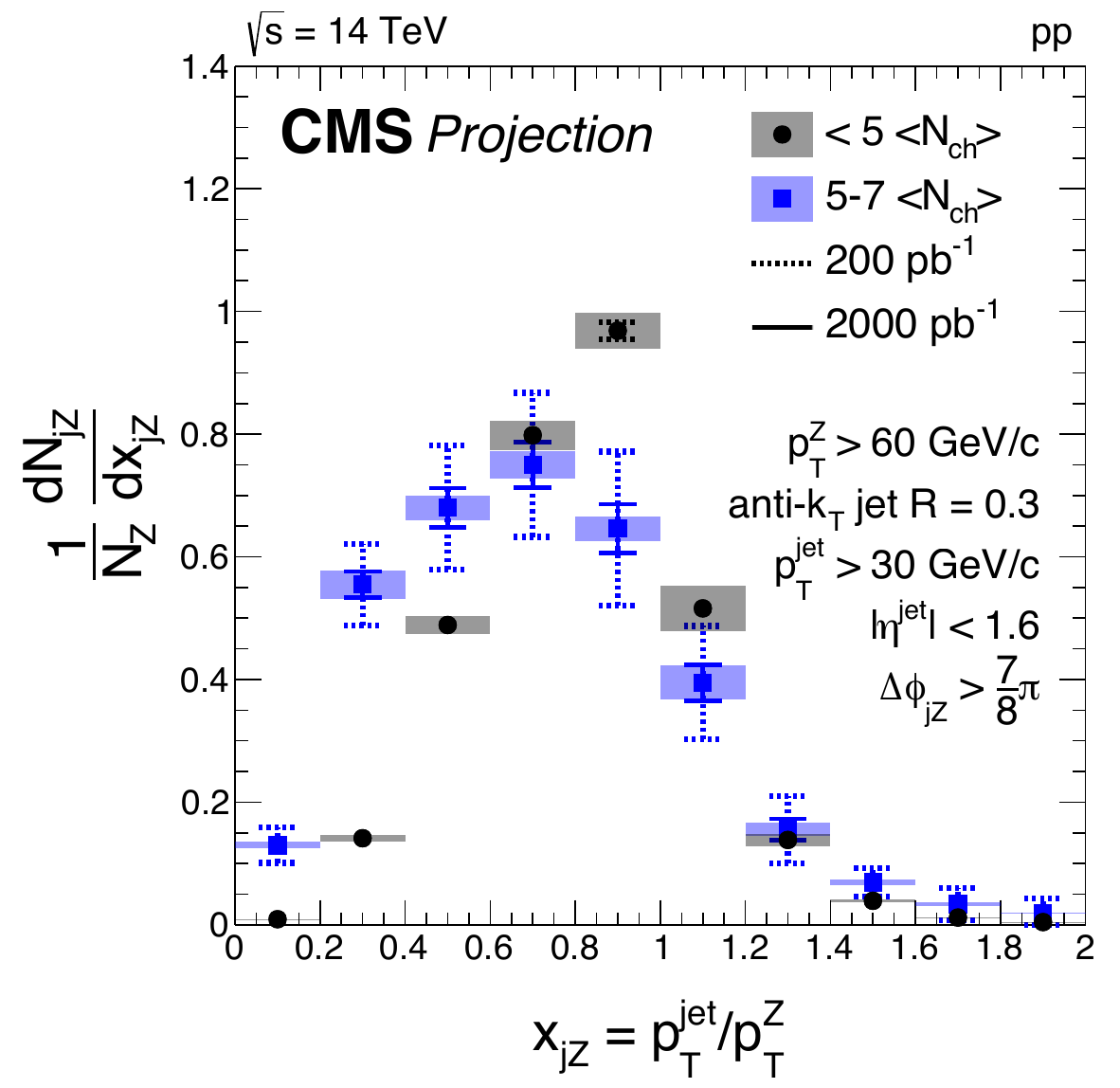}\label{fig:EF07_jet_ZJet}}
  \caption{  \protect\subref{fig:EF07_jet_GammaJet} The statistical precision of the ratio of jet fragmentation functions in Pb–Pb and pp collisions at $\sqrt{s_{\mathrm{NN}}}=5.02$~TeV in ATLAS for jets recoiling from a photon~\cite{ATL-PHYS-PUB-2018-019}.  The error bars at unity for $0-10\%$,  $10-20\%$, and  $20-30\%$ centrality classes show the expected statistical uncertainties at the HL-LHC. \protect\subref{fig:EF07_jet_ZJet} Prospects of $Z$ boson-jet transverse momentum balance distribution in low pileup pp collisions at $\sqrt{s}=$14 TeV in CMS~\cite{FTR-18-025}.
  }
\end{figure}

The large data samples and the improved jet reconstruction due to the upgrade of the tracking system of the CMS and ATLAS detectors will provide significantly reduced statistical and systematic uncertainties for key measurements of medium modification of light (heavy) quark jets using photon/$Z$ ($\mathrm{D^{0}}$-meson) tagged samples~\cite{FTR-18-025, ATL-PHYS-PUB-2018-019}. Moreover, the sPHENIX detector with large acceptance calorimeters~\cite{sPHENIX:2017lqb} will enable the high precision full jet measurements for the first time at RHIC. Since electroweak bosons don't participate in the strong interaction, by measuring the jets tagged against recoiling isolated photons or $Z$ bosons, one can access the energy loss of a parton by using the boson energy as the reference for that of the parton before quenching.  In particular, the jet fragmentation functions and the jet shapes will be measured precisely for high-$z$ region where the difference between pp and Pb-Pb collisions is not resolved yet~\cite{FTR-18-025, ATL-PHYS-PUB-2018-019}, as shown in Fig.~\ref{fig:EF07_jet_GammaJet}, thus providing information on the medium-modified structure of quark-initiated jets. The azimuth and transverse momentum correlations between bosons and jets, measured in $\gamma$+jets and $Z$+jets events presented in Fig.~\ref{fig:EF07_jet_ZJet}, are valuable observables to study parton energy loss in the QGP~\cite{FTR-18-025}. By comparing the LHC and RHIC data, we aim to constrain the temperature dependence of the transport coefficients of QGP.

Another significant improvement expected at the HL-LHC is the measurement of radial distribution of $\mathrm{D^{0}}$ mesons in jets~\cite{FTR-18-025}. By studying the modification of this observable in Pb-Pb compared to pp, one can gain insights into the dynamics of heavy quarks in the QGP. In addition, the large low-PU pp data samples at $\sqrt{s}=$14 TeV can be a great opportunity for precision measurements of the system-size dependence of the jet quenching phenomena.  In this regard,  high multiplicity pp events provide the reference results in small systems, thus completing the system-size dependence of the jet quenching phenomena. 

\subsection{Heavy Quarks}

Heavy quarks provide a unique opportunity to probe the QGP with slow-moving probes. Charm and beauty quarks, as a consequence of their large masses ($m_{\mathrm{c,b}} > \lambda_{\mathrm{QCD}}$), are mostly produced during the early stages of the collision in hard scattering processes. As HF quarks propagate through the medium, they are expected to lose less energy than light flavor through radiation due to the dead-cone effect, i.e., the suppression of small-angle gluon radiation induced by the lower heavy quark velocity at the same kinetic energy as light quarks. These interactions may lead to the thermalization of low-momentum HF quarks, which would then take part in the expansion and hadronization of the medium. In addition, HF mesons, such as quarkonia, can be dissociated in the medium due to Debye color screening or recombined from individual heavy quarks and anti-quarks diffusing through the medium~\cite{Kopeliovich:2014una,Aronson:2017ymv,Du:2017qkv}. Therefore, the measurement of HF hadrons can provide crucial information on the full evolution of the system and allows us to get information on the quark-mass dependence of the medium-induced radiations.

\begin{figure}[!htbp]
  \centering
  \subfloat[]{\includegraphics[width=0.26\textwidth]{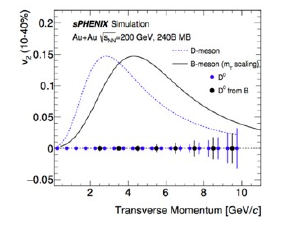}\label{fig:EF07_sPHENIX_HF}}
  \subfloat[]{\includegraphics[width=0.31\textwidth]{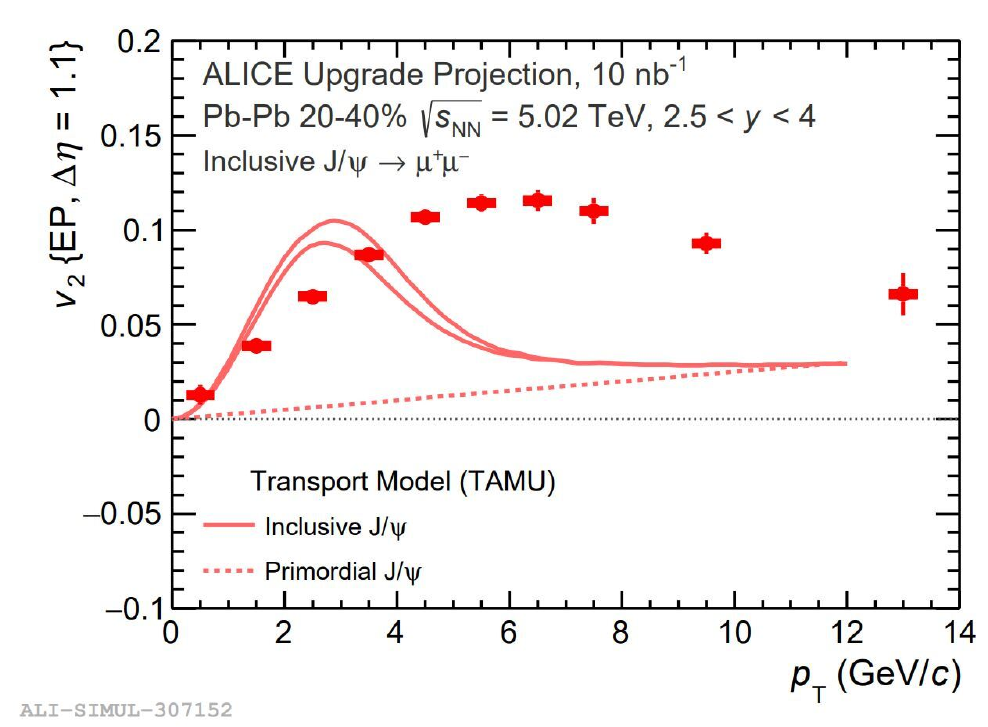}\label{fig:EF07_HF_Jpsi}}
  \subfloat[]{\includegraphics[width=0.34\textwidth]{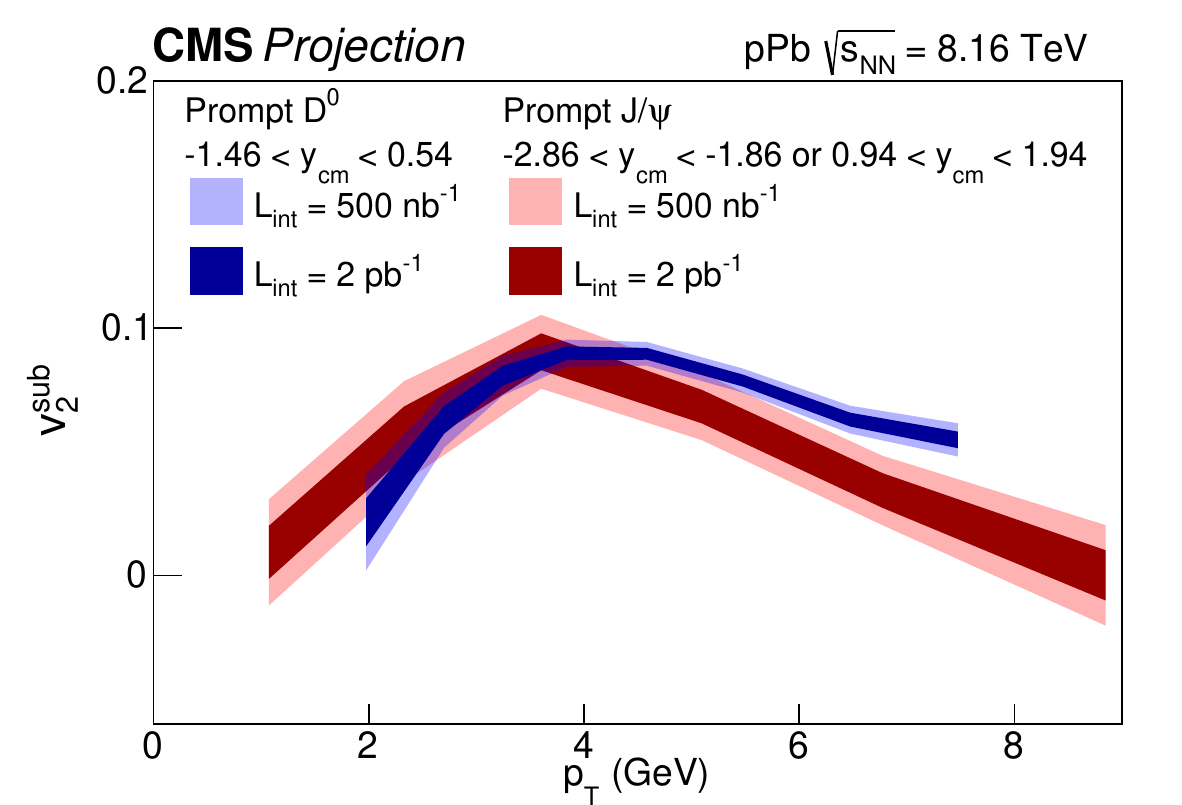}\label{fig:EF07_HF_Charm}}\\
  \subfloat[]{\includegraphics[width=0.39\textwidth]{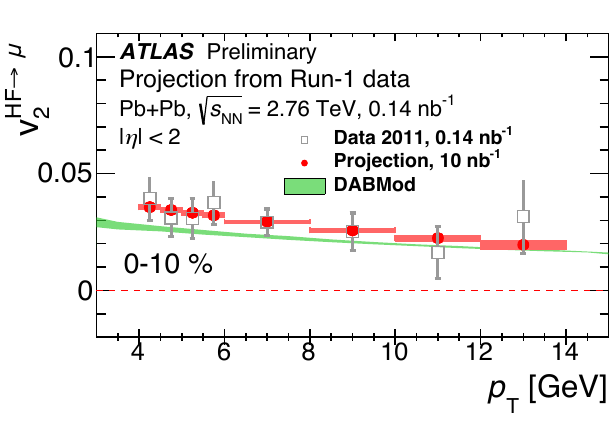}\label{fig:EF07_HF_Muon}}
  \subfloat[]{\includegraphics[width=0.26\textwidth]{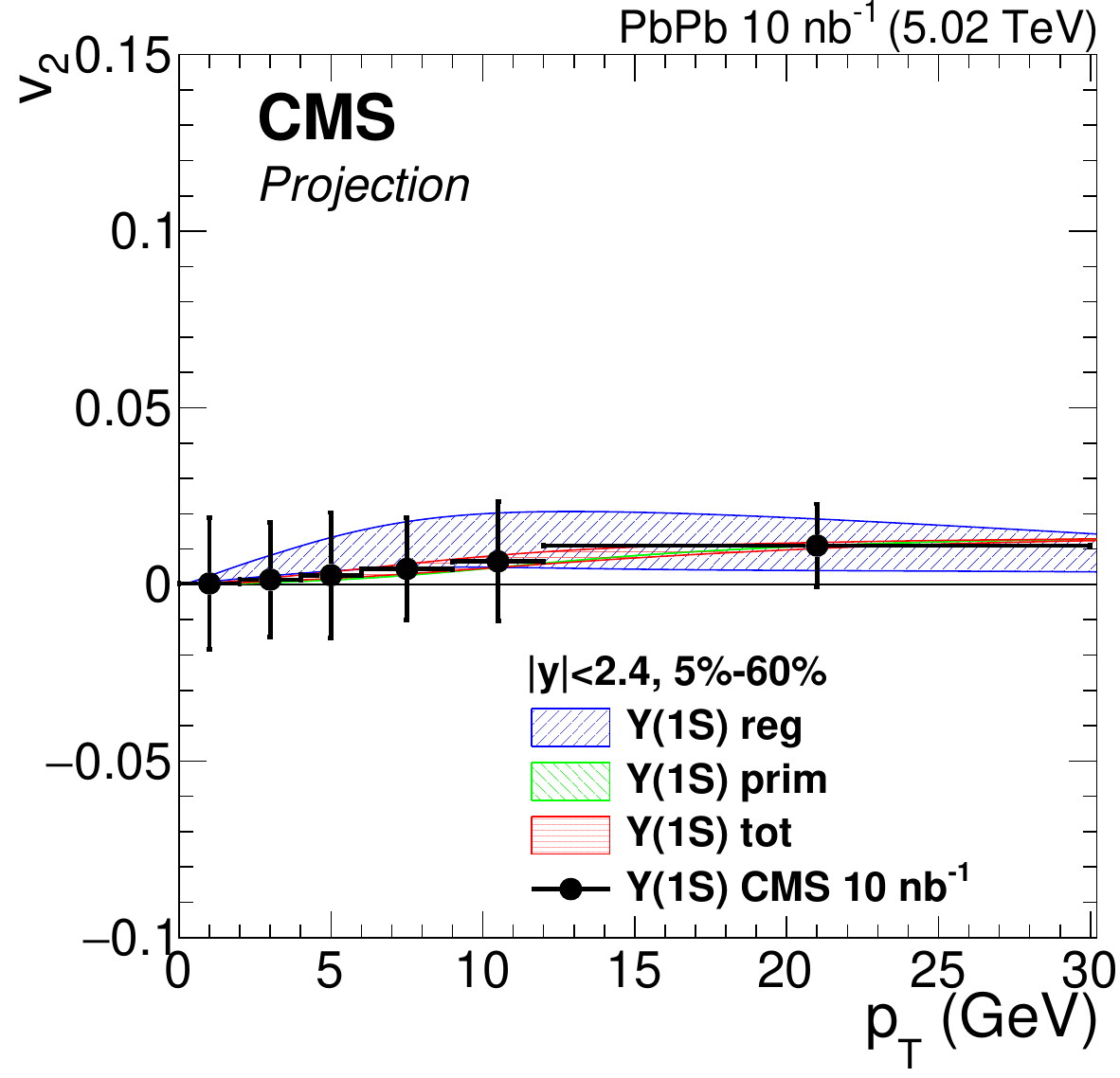}\label{fig:EF07_HF_Upsilon}}
  \caption{Elliptic flow projections as a function of $p_{\mathrm{T}}$ of: \protect\subref{fig:EF07_sPHENIX_HF} sPHENIX performance on D and B mesons at $\sqrt{s_{\mathrm{NN}}}=0.2$~TeV 
    \protect\subref{fig:EF07_HF_Jpsi} J/$\psi$ mesons in ALICE in Pb-Pb collisions at $\sqrt{s_{\mathrm{NN}}}=5.02$~TeV
  \protect\subref{fig:EF07_HF_Charm} prompt D and J/$\psi$ mesons in CMS in  high-multiplicity p-Pb collisions at $\sqrt{s_{\mathrm{NN}}}=8.16$~TeV~\cite{FTR-18-026}, \protect\subref{fig:EF07_HF_Muon} HF decay muons in ATLAS in Pb-Pb collisions at $\sqrt{s_{\mathrm{NN}}}=2.76$~TeV~\cite{ATL-PHYS-PUB-2018-020} and \protect\subref{fig:EF07_HF_Upsilon}  $\Upsilon(1S)$ mesons in CMS in Pb-Pb collisions at $\sqrt{s_{\mathrm{NN}}}=5.02$~TeV~\cite{FTR-18-024}.
  }
  \label{fig:EF07_HF_v2}
\end{figure}

The larger experimental data samples at the HL-LHC, combined with improved detector performance and measurement techniques, will allow the ALICE, CMS, and ATLAS experiments to significantly improve over the current HF hadron~\cite{CMS:2017uoy, CMS:2018eso,ALICE:2018lyv,CMS:2017qjw,ALICE:2015vxz,ALICE:2021mgk,CMS:2018bwt,CMS:2017exb,ATLAS:2018hqe}, and quarkonia~\cite{CMS:2011all,CMS:2012bms,CMS:2012gvv, CMS:2016rpc,ATLAS:2010xzb,ALICE:2012jsl,CMS:2014vjg,ALICE:2015jrl} measurements. The new ALICE ITS2 inner tracking system will give ALICE greatly improved capabilities for charm and bottom \cite{Reidt:2021tvq}. Looking further ahead, the ALICE ITS3, planned for installation before the start of Run 4  will push this even further \cite{Colella:2021stb}. Those upgrade projects aim to measure charm hadrons down to $p_T=0$. The $p_{\mathrm{T}}$ dependence of the quarkonium nuclear modification factor ($R_{\mathrm{AA}}$) will be measured with high precision up to about 80 GeV for prompt J/$\psi$ and 50 GeV for $\Upsilon(1S)$~\cite{FTR-18-024} (compared to 50 and 30 GeV respectively, with the present data), allowing to discern whether quarkonium formation at high $p_{\mathrm{T}}$ is determined by the Debye screening mechanism, or by energy loss of the heavy quark or the quarkonium in the medium. The elliptic flow measurements of charm mesons in p-Pb collisions~\cite{FTR-18-026} and of HF decay muons~\cite{ATL-PHYS-PUB-2018-020} and $\Upsilon(1S)$ mesons~\cite{FTR-18-024} in Pb-Pb collisions will be significantly improved as observed in Fig.~\ref{fig:EF07_HF_v2}, providing insights on the collective expansion and degree of thermalization of HF quarks in the medium at low $p_{\mathrm{T}}$, and on the presence of recombination of bottomonia from deconfined beauty quarks in the QGP. The production of strange B mesons and charm baryons in pp and Pb-Pb collisions~\cite{FTR-18-024} will also be measured with sufficient precision to further investigate the interplay between the predicted enhancement of strange quark production and the quenching mechanism of beauty quarks, and the contribution of recombination of HF quarks with lighter quarks to the hadronization process in HI collisions. Finally, the precise measurements of beauty mesons in p-Pb collisions~\cite{FTR-18-024} will help to elucidate the relative contribution of hadronization and nuclear-matter effects (e.g. nuclear PDF, gluon saturation, and coherent energy loss), as well as serve as a baseline for the understanding of beauty-quark energy loss in Pb-Pb collisions. 

At RHIC, the sPHENIX detector with enhanced capability for the studies of heavy flavor mesons and baryons could provide high precision data at lower collision energy. Together with data from HL-LHC, the measurements of HF hadron spectra, HF particle ratio, and azimuthal anisotropy will provide strong constraints on the heavy quark diffusion coefficient and its temperature dependence.

\subsection{Hadronization and Exotic Hadrons}

Insight into the hadronization process in QGP is provided by the comparison of the $p_T$ distributions of various hadrons in Pb+Pb collision and in pp collisions. In particular, mechanisms such as recombination, involving soft partons close in phase space, could be important for the description of hadron production in high-density environments.

The abundant production of light nuclei and antinuclei measured by ALICE can be greatly improved in HL-LHC. In analogy with the case of light nuclei and of charmonium, the statistical hadronization or coalescence ansatz can be used to gain unique insight into the structure (e.g.\, tetraquark or molecular state) of exotic hadrons, such as the $X(3872)$ studied by LHCb in high-multiplicity pp collisions~\cite{LHCb:2020sey} and by CMS in Pb+Pb collisions~\cite{CMS:2021znk}. Those initial measurements will be followed up with the high statistics data in LHC Run 3 and 4. 

In LHC Run 5, the ALICE 3 detector would provide high precision measurement of multi-charm baryons, expanding the studies of hadronization performed in Run 3 and 4. ALICE 3 would also be the perfect tool for the study of the formation of light nuclei, hyper-nuclei, super-nuclei, and the experimental investigation of exotic states such as $X(3872)$ and the newly discovered $T_{cc}^+$.

\subsection{Particle collectivity in small and large systems}
With the large minimum bias samples of pp, pPb, and PbPb datasets of HL-LHC,
it will be possible to reach an unprecedented experimental precision that will help us to understand the collectivity of small and large systems. The pivotal upgrades of trackers in CMS and ATLAS will enable the measurement of charged particles in the wide pseudo-rapidity range ($|\eta|<4$).
In small systems, a significant statistical improvement is expected for the elaborate flow variables. In particular, the symmetric cumulant observables, SC($m,n$) which are the correlations between Fourier coefficients based on 4-particle correlations will be assessed~\cite{ALICE:2016kpq}.  Since those are very sensitive to the initial state and its fluctuation, precision measurements of them will constrain the current interpretation of the ridge phenomenon in small systems~\cite{CMS-QCD-10-002, CMS-HIN-12-005, ALICE:2012eyl, HION-2012-13}, as well as catching hold of non-flow effects in early stages~\cite{FTR-18-026}. In addition, we expect a crucial improvement in our understanding of the system size of collisions by measuring the Hanbury Brown and Twiss (HBT) radii in small systems~\cite{ATL-PHYS-PUB-2018-020}. With azimuthally sensitive femtoscopy, the spatial ellipticity of the medium at freeze-out can be measured. In particular, the HL-LHC p-Pb data will allow us to unambiguously investigate the normalized second-order Fourier component of the transverse HBT radius as a function of the magnitude of flow.
In Pb-Pb collisions, an interesting observable which can be highly enriched by HL-LHC data is the flow decorrelation~\cite{Gardim:2012im}.
The extended $\eta$ acceptance in Run 4 will lead to significant improvement in characterizing the rapidity dependence of the factorization breaking. A significant improvement of the forward-backward multiplicity correlation and multi-particle cumulants will bring a better understanding of the fluctuations of the medium in early stages~\cite{ATL-PHYS-PUB-2018-020}. 

\section{Photonuclear and photon-photon interactions at colliders}
\label{sec:UPCs}

Ultraperipheral ion collisions (UPCs) are electromagnetic interactions of relativistic heavy ions.  They occur when the nuclei pass by with impact parameter $b>2R_A$, where $R_A$ is the nuclear radius \cite{Bertulani:2005ru,Contreras:2015dqa}.  These photon fields may interact with the opposing nucleus, in a photonuclear interaction, or with each other, in a $\gamma\gamma$ collision. At a lepton-lepton collider, hadroproduction in photon-photon fusion offers complementary ways to study QCD dynamics.  Photonuclear interactions at the LHC (Sec.~\ref{sec:GammaA}) can probe target nuclei down to very small $x$ values, $x\lesssim 10^{-6}$. Meanwhile, photon-photon collisions both at the (HL-)LHC (Sec.~\ref{sec:GammaGammaLHC}) or an $e^+e^-$ collider (Sec.~\ref{sec:GammaGammaEpEm})
can probe high-energy QCD resummations, diffractive interactions, and BSM phenomena.  

In the UPCs, the strongly Lorentz-contracted electromagnetic fields of relativistic nuclei act as nearly-real photon fields that extend up to high energies. The maximum photon energy is $\gamma \hbar c/b$, where $b$ is the transverse distance from the interaction point to the center of the emitting nucleus.   
The LHC can reach $\gamma A$ center-of-mass energies up to 1.5 and 5.4 TeV for proton and lead targets, respectively \cite{Klein:2020fmr}.  For $\gamma\gamma$ interactions, the maximum collision energies are 170 GeV and 4.2 TeV in Pb-Pb and $pp$ collisions, respectively.   

These opportunities were  recognized early \cite{Baltz:2007kq}, and all four LHC experiments have studied UPCs, along with STAR and PHENIX at RHIC. There are extensive plans for UPC studies at the LHC during Runs 3 and 4 \cite{Citron:2018lsq}.   These studies will benefit from the extensive detector upgrades that were implemented during the second long shutdown, such a new streaming DAQ readout in ALICE, which will remove a bottleneck when triggering on low-multiplicity UPC events \cite{Burmasov:2020doi}. 

\subsection{Photonuclear interactions \label{sec:GammaA}}
Photonuclear collisions are an effective tool for the study of the nuclear structure, and several collision observables may be used to constrain the nPDF (similar to the p-Pb studies discussed in Section~\ref{sec:EF07:npdf}).  So far, most UPC studies of photonuclear interactions have involved vector mesons, which are experimentally very clean, but are subject to significant theoretical uncertainty \cite{Klein:2017vua}. 
These uncertainties can be alleviated by going beyond the NLO perturbative approximation that is currently used e.g. in \cite{Eskola:2022vpi}, or by taking the ratio of the cross-sections on proton and ion targets to measure shadowing.

The HL-LHC will allow ALICE and CMS to extend such measurements to the $\psi(2S)$ and potentially the $\Upsilon(1S)$ meson~\cite{FTR-18-027}.
They will also allow for a substantially expanded set of measurements, including photoproduction of dijets (already pursued by ATLAS \cite{ATLAS:2017kwa}) and of open charm, likely bottom and possibly even top quark pairs \cite{Citron:2018lsq,Klein:2000dk,Klein:2002wm,Goncalves:2006xi,Adeluyi:2012sw,Goncalves:2017zdx}.
As was discussed in Sec.~\ref{QCD:ssec:HAS_FPF}, in the limit of high virtualities, exclusive photoproduction of vector mesons provides direct information about the transverse distribution of partons for a given $x$. In the scaling limit, the transverse spread is given by a Fourier transform of photoproduction amplitude $A(x, t)$, which should exhibit universal $t$ dependence. Testing this universality with data will be informative. 

Inelastic hard diffraction at small $t$ provides a unique information  on fluctuations of the gluon field. The ratio of differential cross sections for incoherent and coherent onium production at $t=0$ reflects event-by-event fluctuations of the
square of the  gluon density \cite{Frankfurt:2008vi}, while 
at $t>0$, competition between the fluctuations and parton knockouts takes place \cite{Klein:2019qfb,Frankfurt:2022jns}. 

\begin{figure}[!htbp]
  \centering
\includegraphics[width=0.5\textwidth]{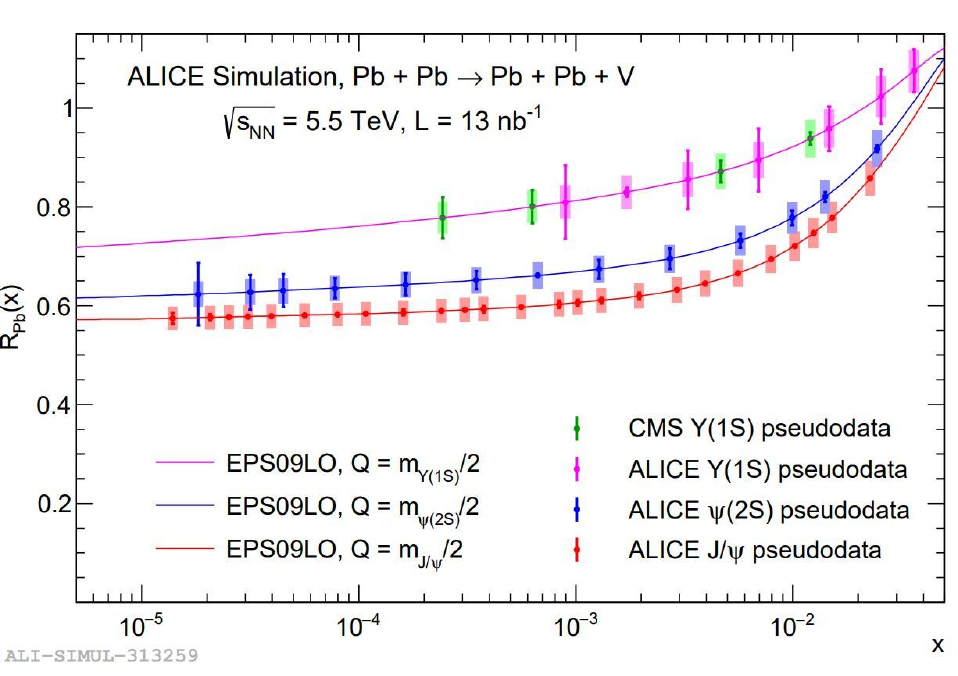}
\caption{Pseudodata projections for nuclear suppression factor by ALICE and CMS measured with the photoproduction of three heavy vector mesons in PbPb UPCs are shown.  From Ref. \cite{Citron:2018lsq}.
  \label{fig:EF07_UPC}}
\end{figure}

Production of multiple vector mesons in UPCs can be also studied, as illustrated in Fig.~\ref{fig:EF07_UPC} showing projections for the nuclear correction in PbPb scattering at ALICE and CMS, including expected uncertainties. For estimated rates for vector meson production in the different LHC experiments during Run 3 and Run 4, refer to \cite{Citron:2018lsq}.

\subsection{Photon-photon scattering}
\label{sec:GammaGammaLHC}
This section considers ultraperipheral events in $pp$ scattering at the LHC, during which the protons remain intact and are measured in special units called roman pots. In these photon-exchange events, quasi-real photons are emitted by the incoming interacting protons.
Both the ATLAS and CMS-TOTEM collaborations have installed roman pot detectors at about 220 meters from the interaction points, the so-called ATLAS Forward Proton (AFP) and CMS-TOTEM Precision Proton Spectrometer (PPS). In standard runs at high luminosity, the ATLAS or CMS detectors with intact protons tagged  in the roman pots can detect final states with typical invariant mass  400-2300 GeV, enabling sensitivity to a wide range of new physics effects. The events are especially clean. As an example, one may observe exclusive pairs of photons or $W$ bosons in the central region. In the same way, one can look for exclusive production of $ZZ$, $\gamma Z$, $t \bar{t}$ events via photon induced processes. The UPCs will be also sensitive to a variety of BSM scenarios, such as those that contain loops of virtual heavy particles coupling to photons that introduce an effective photon quartic compling $\zeta_1$ at a low energy~\cite{Fichet:2014uka,Fichet:2013gsa}.

\subsubsection{Photon-photon collisions at the LHC without tagging intact protons}
\begin{figure}[t]
\centering
  \includegraphics[width=0.48\textwidth]{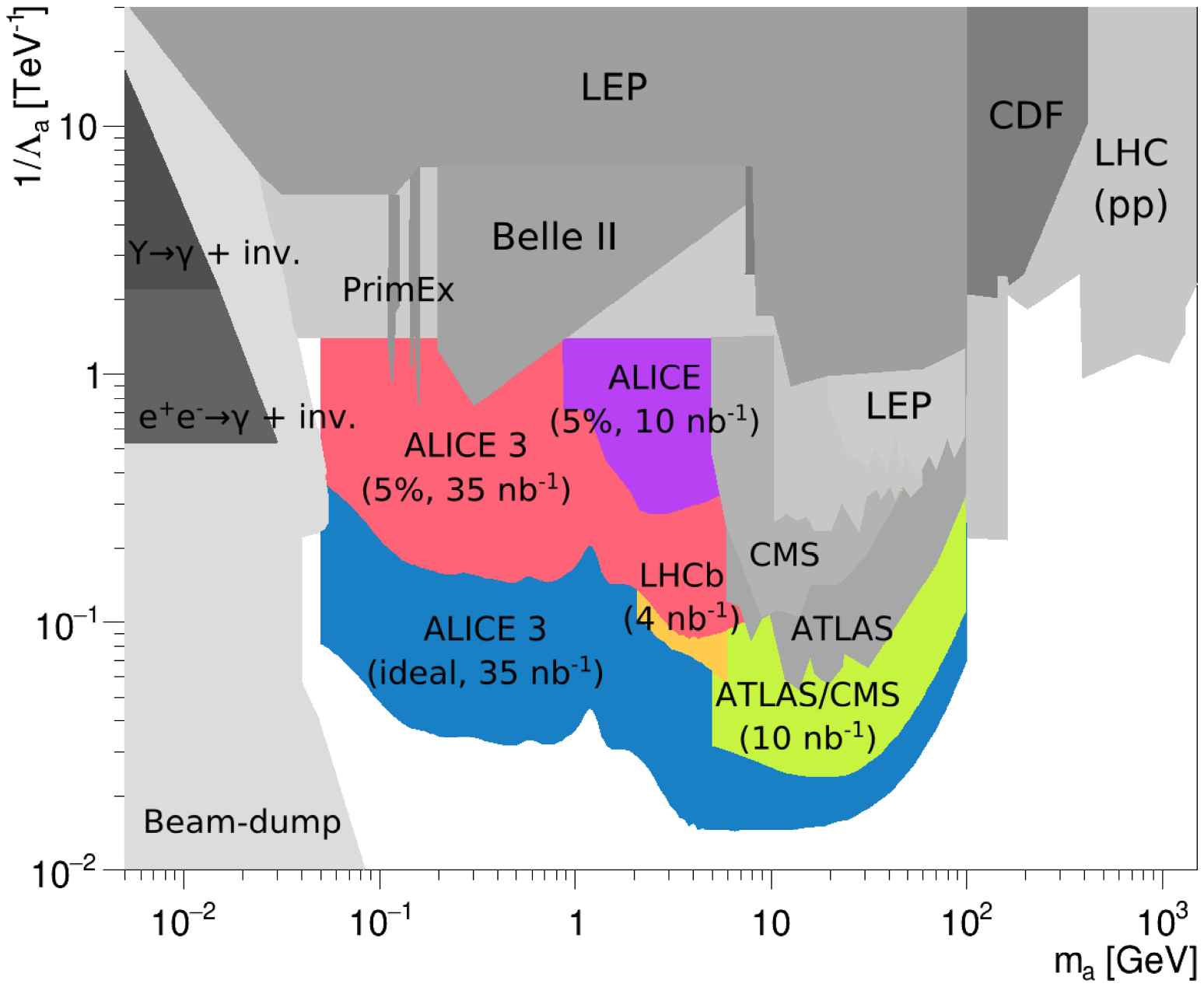}
  \includegraphics[width=0.48\textwidth]{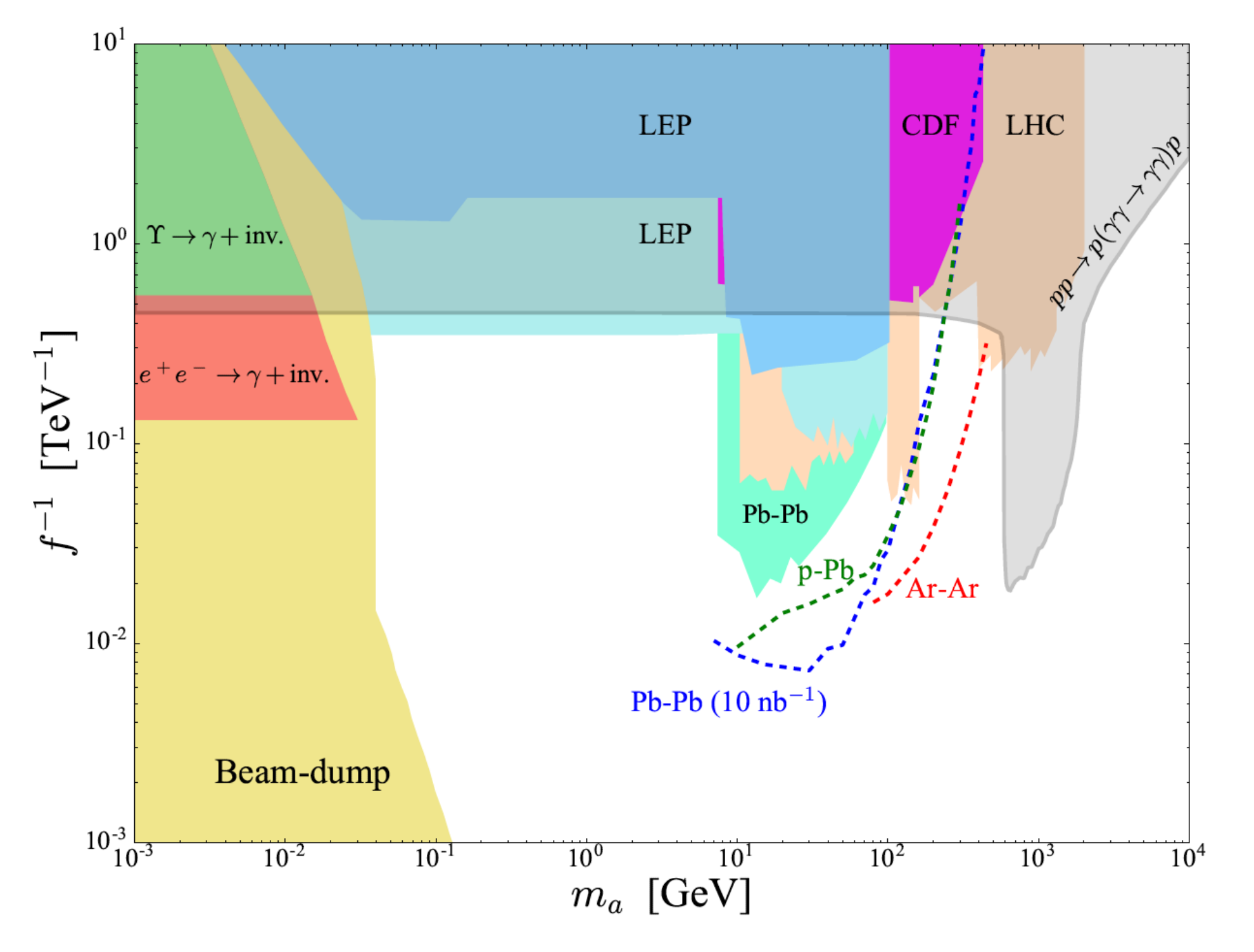}
  \caption{Left: Compilation of exclusion ALPs sensitivities obtained by different studies~\cite{Dainese:2703572, Bauer:2017ris,dEnterria:2022sut}. In light green, the ATLAS $\mathrm{\gamma\gamma\rightarrow\gamma\gamma}$ from $10\;\mathrm{nb^{-1}}$ limit at $\sqrt{s_{\mathrm{NN}}}=5.52$~TeV is presented. The ATLAS $\mathrm{\gamma\gamma\rightarrow\gamma\gamma}$ represents the exclusion limit derived from the LbyL cross section measured in Pb-Pb collisions by ATLAS~\cite{ATL-PHYS-PUB-2018-018}, the CMS $\mathrm{\gamma\gamma\rightarrow\gamma\gamma}$ limit comes from the recent analysis described in Ref.~\cite{CMS-FSQ-16-012} and the projected performance with ALICE 3 detector~\cite{Adamova:2019vkf}. 
  Right: Sensitivity predictions on ALPs using $pp$, p-Pb, Pb-Pb and Ar-Ar interactions at the LHC~\cite{Baldenegro:2018hng,Baldenegro:2019whq}. 
  }
  \label{fig:EF07_lbyl_alp}
\end{figure}

The ATLAS and CMS measurements of exclusive dimuons from $\gamma\gamma\rightarrow\mu^+\mu^-$ will reach a precision that can be used to calibrate the photon flux and reduce the uncertainty on the nuclear charge distribution~\cite{ATL-PHYS-PUB-2018-018}. The rare light-by-light (LbyL) scattering process~\cite{dEnterria:2013zqi}, measured by both ATLAS and CMS, is statistics limited and will greatly benefit from the increased luminosity as well as improvements in triggering capabilities~\cite{ATL-PHYS-PUB-2018-018}. 
Of particular note is the search for BSM axion-like particles, which may be detectable via $\gamma\gamma\rightarrow a \rightarrow \gamma\gamma$~\cite{Knapen:2016moh}, and where LbyL interactions from Pb-Pb measured by both ATLAS and CMS already set the most stringent limits for axion masses between $\sim 5 -100$ GeV and will improve with the new data as shown in Fig.~\ref{fig:EF07_lbyl_alp} where we show the sensitivities in $pp$, p-Pb, Pb-Pb and Ar-Ar collisions. We clearly see the complementarity between heavy ion and proton collisions. Fig.~\ref{fig:EF07_lbyl_alp}, left, does not assume the tagging of intact  protons whereas Fig.~\ref{fig:EF07_lbyl_alp}, right, does, as described in the next section. 
These analyses will benefit from upgrades to forward detectors, see for example~\cite{ATLAS-TDR-24} and~\cite{CMS-TDR-13}.

The $\gamma\gamma \to WW$ process is unique because it occurs at leading-order only via electroweak gauge boson couplings, and is thus an ideal probe for searching for new physics via anomalous gauge boson couplings~\cite{deFavereaudeJeneret:2009db,Pierzchala:2008xc,Chapon:2009hh}. The rare electroweak process $\gamma\gamma \to W^\pm W^\mp \to e^\pm \nu_e \mu^\mp \nu_\mu$ was first observed (rejecting the background-only hypothesis with a significance of $8.4\sigma$) in proton-proton collisions at the LHC by the ATLAS Collaboration in 2020 using 139 fb$^{-1}$ of data collected in Run 2~\cite{STDM-2017-21}.
The increased luminosity at the HL-LHC will allow differential measurements and a large improvement in the statistical precision of this measurement, provided pile-up mitigation and track reconstruction challenges are addressed.
The impact of the HL-LHC environment and planned ATLAS detector upgrade is studied in~\cite{ATL-PHYS-PUB-2021-026}.

\begin{figure}
    \centering
    \subfloat[\label{fig:fig3a}]{\includegraphics[width=.4\linewidth]{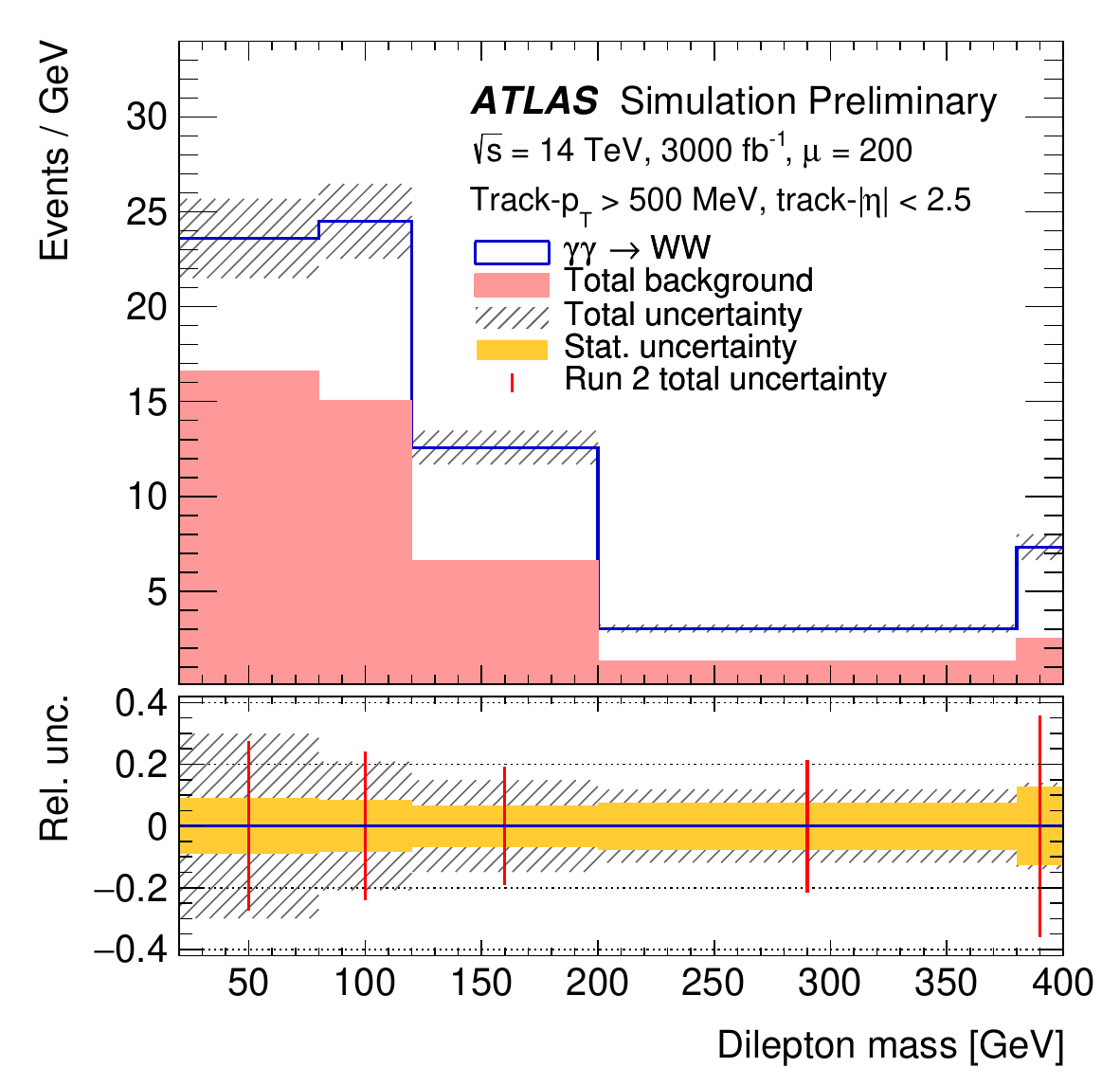}}%
    \subfloat[\label{fig:fig3b}]{\includegraphics[width=.4\linewidth]{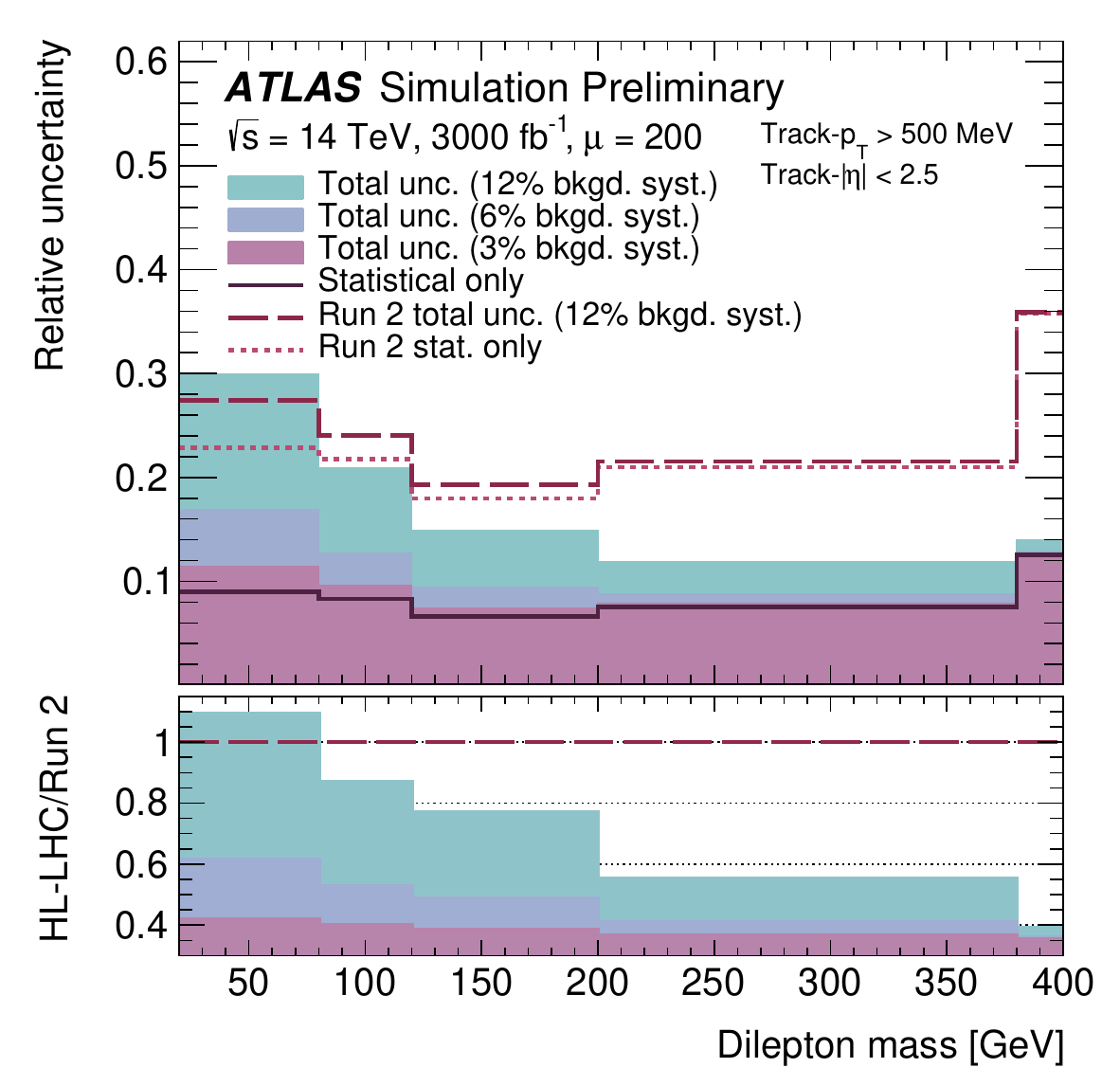}}%
    \caption{
      \protect\subref{fig:fig3a} The expected signal and background yields at the HL-LHC shown stacked and as a function of the dilepton mass, with the relative statistical and total uncertainty on the signal shown in the bottom panel assuming the same background systematic as in Run 2 of 12\%.
      \protect\subref{fig:fig3b} A comparison of the total uncertainty when considering a reduction in the nominal background systematic by a factor of 2 and 4 and compared to the Run 2 relative uncertainties, with the ratio of the total relative uncertainty on the signal yield at the HL-LHC as a ratio to the Run 2 uncertainty shown in the bottom panel. The average pileup in Run 2 was $\langle\mu\rangle$=33.7.~\cite{ATL-PHYS-PUB-2021-026}}
\end{figure}

An example is given in Figs.~\ref{fig:fig3a} and~\ref{fig:fig3b} for a differential measurement in the dilepton invariant mass, $m_{\ell\ell}$, with the aim to probe the sensitivity in the high-$m_{\ell\ell}$ region most sensitive to dimension-8 EFT operators~\cite{Chapon:2009hh}. Figure~\ref{fig:fig3a} shows the expected signal and background yields at the HL-LHC stacked and as a function of $m_\mathrm{\ell\ell}$, illustrating the improvements at high dilepton mass compared to the Run 2 analysis. Figure~\ref{fig:fig3b} compares the main sources of the uncertainty for the HL-LHC and the Run 2 analyses. While the Run 2 analysis is mainly limited by statistical uncertainties, the background systematic uncertainty will be dominant for most of the dilepton mass spectrum at the HL-LHC. It demonstrates the impact of the background modeling uncertainty by considering a reduction in this uncertainty by a factor of 2 and 4 compared to the Run 2 relative uncertainties.

\subsubsection{Photon-photon collisions at the LHC with proton tagging}
Turning now to photon-induced processes at ATLAS and CMS-TOTEM in which scattered protons are detected intact in the roman pots,
we will first concentrate on diphoton exclusive production as an example. The conclusions can be generalized to exclusive $WW$, $ZZ$, $\gamma Z$, and $t \bar{t}$ production via photon exchanges. 
Diphotons can be produced exclusively either via QCD or QED processes, with the cross sections as a function of a diphoton mass available in ~\cite{Chapon:2009hh,Kepka:2008yx,Fichet:2015vvy,Fichet:2016pvq}.
Observing two photons in ATLAS/CMS and two tagged protons means a photon-induced process, with the acceptance of the roman pot detectors starting at about 400 GeV.

\begin{figure}[t]
\centering
\includegraphics[width=0.65\textwidth]{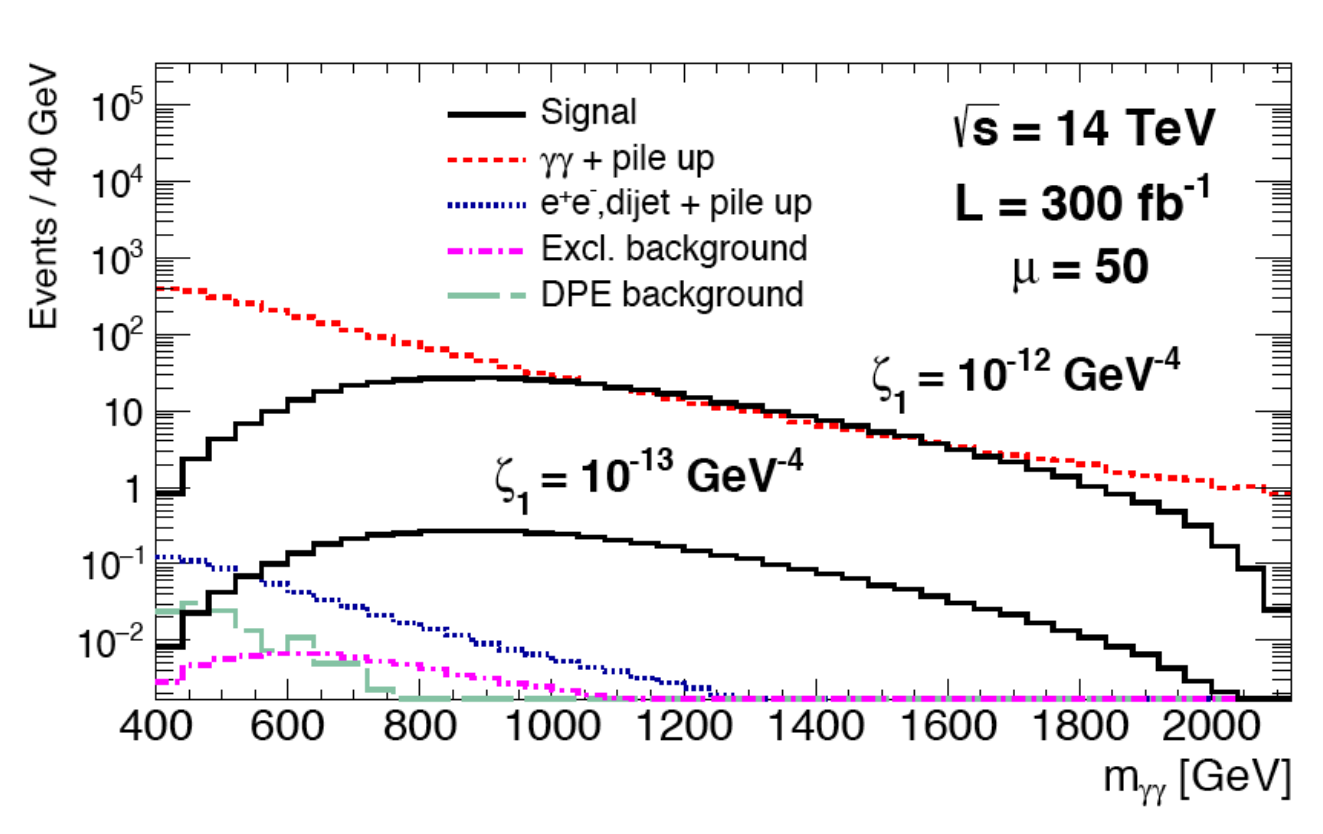}
\caption{Number of events as a function of the diphoton mass for signal and background for exclusive $\gamma \gamma$ production.}
\label{DigammaFig3}
\end{figure}

The number of exclusive diphoton production events for a luminosity of 300 fb$^{-1}$ at the LHC is shown in Fig~\ref{DigammaFig3}.
SM exclusive diphotons (red dashed dotted line) and exclusive dileptons with leptons misidentified as photons (blue dotted line) can be neglected. The only background that matters is shown in red dashed lines and corresponds to  inclusive diphoton production that coincides with detection of intact protons from the pileup. The BSM signal (black lines) is shown for an effective photon quartic complings $\zeta_1$ of order $10^{-12}$ and $10^{-13}\mbox{ GeV}^{-4}$ that are common in BSM models~\cite{Fichet:2014uka,Fichet:2013gsa}.

\begin{figure}[t]
\centering
\includegraphics[width=0.95\textwidth]{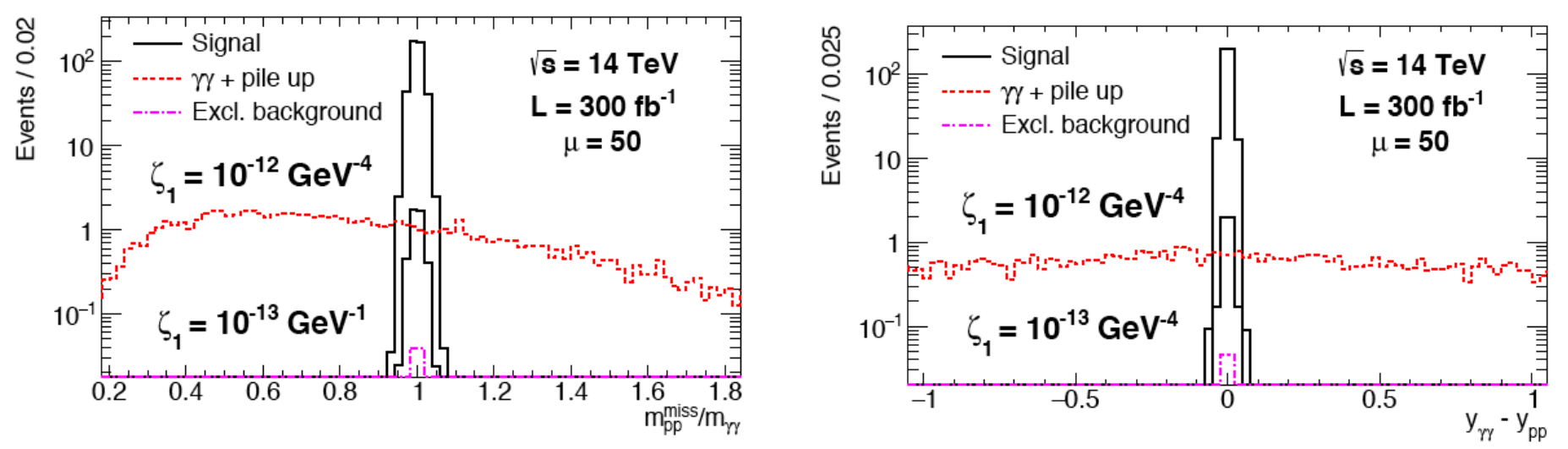}
\caption{Mass ratio and rapidity difference between the $pp$ and $\gamma \gamma$ information for signal (in full line) and pile up background (dashed line).}
\label{DigammaFig4}
\end{figure}

Measuring intact protons is crucial for suppressing the pileup and other background. The method matches the kinematical information, as measured by the two photons, with the one using the two protons. The results are shown in Fig.~\ref{DigammaFig4} for the mass ratio (left) and the rapidity separation (right) between the $pp$ and $\gamma \gamma$ systems. With these exclusivity cuts, one can reduce the background from 80.2 to less than 0.1 events for 300 fb$^{-1}$, lifting the sensitivity to $\zeta_1$ to a few
$10^{-15}$ GeV$^{-4}$ -- better by more than two orders of magnitude compared to ``standard" methods~\cite{Chapon:2009hh,Kepka:2008yx,Fichet:2015vvy,Fichet:2016pvq}. 
This is now becoming a reality, with both CMS-TOTEM and ATLAS reporting observations of QED exclusive dilepton production, and CMS-TOTEM reporting the first limits on $\zeta_1$ with about 9.4 fb$^{-1}$ of data. 

This method can be applied directly to search for axion-like particles (ALPs). For example, if an ALP interacts with the photons, the exclusive $pp\to p(\gamma\gamma \to \gamma\gamma)p$ process at the LHC with 300 fb$^{-1}$ significantly extends the reach of the ALPs search at masses of order 1-2.3 TeV, see the sensitivity plot (coupling vs mass) in Fig.~\ref{fig:EF07_lbyl_alp}, right~\cite{Baldenegro:2018hng,Baldenegro:2019whq}. In the 1 TeV ALP mass range, about two orders of sensitivity are gained by tagging on the protons, as compared to the standard methods. This is complementary with the strong sensitivity of Pb-Pb runs in the region  of lower masses in Fig.~\ref{fig:EF07_lbyl_alp}, $\sim 10-500$ GeV, where the cross section for Pb-Pb runs is enhanced by the fourth power of the large nucleus charge ~\cite{Knapen:2016moh,dEnterria:2022sut}.

In addition to $\gamma\gamma$ final states, tagged $Z\gamma$ and $WW$ states can be also observed. In $Z\gamma$ production, both leptonic and hadronic decays can be studied using the same kinematic techniques as above,  
which may achieve sensitivities to the $\gamma \gamma \gamma Z$ anomalous coupling up to 10$^{-13}\mbox{ GeV}^{-4}$, better by three orders of magnitude~\cite{Baldenegro:2017aen} than with the standard untagged method based on the $Z\to \gamma\gamma\gamma$ decay.

\begin{figure}[ht]
\centering
\includegraphics[width=0.5\textwidth]{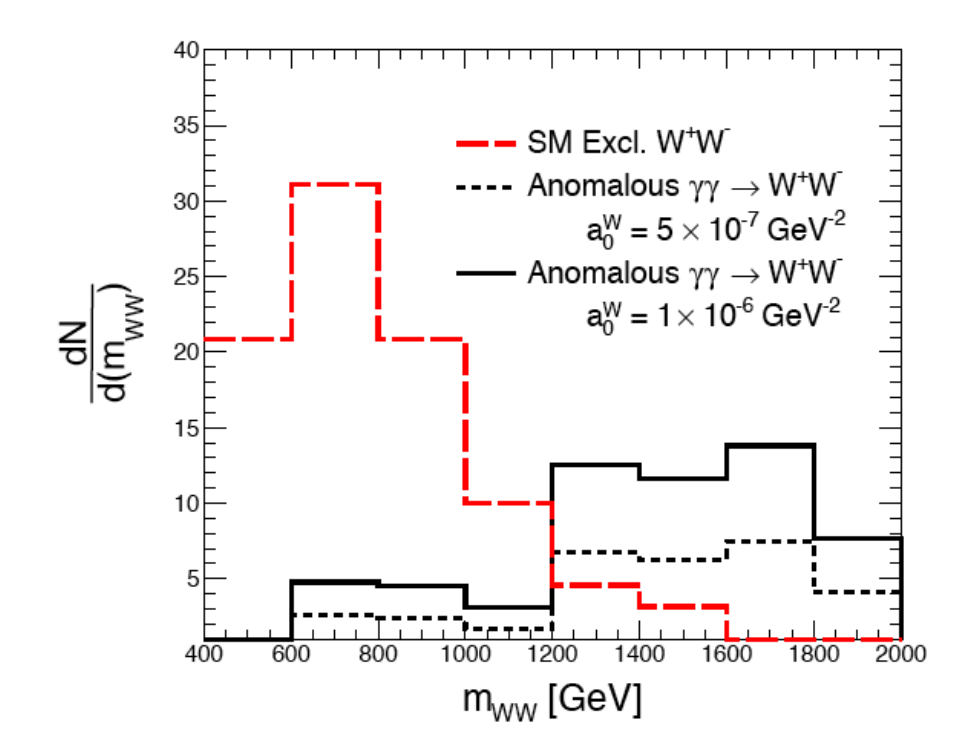}
\caption{$WW$ mass distribution for exclusive $WW$ production (SM is red dashed line and anomalous couplings in full ball line).}
\label{DigammaFig6}
\end{figure}

In the $WW$ exclusive channel, quartic $\gamma \gamma WW$ anomalous couplings can be tested~\cite{Baldenegro:2020qut}. 
Figure~\ref{DigammaFig6} shows that the SM and anomalous (BSM) contributions can be observed at low and high invariant masses of $WW$ pair, respectively.  The best sensitivity to the SM exclusive $WW$ production originates from the leptonic decays of the $W$s, with fast timing detectors desirable to suppress the neutrino background. One strategy to assess the $\gamma \gamma WW$ quartic anomalous couplings is to look for hadronic decays of $W$ bosons at large $m_{WW}$, even if the dijet background is quite high. Advanced jet variables such as subjettiness can help to reject the dijet background.

Following the same ideas, one can also look for exclusive production of $ZZ$ and $t \bar{t}$~\cite{Baldenegro:2022kaa}, as it was recently performed by the CMS and TOTEM collaborations.

\subsection{Photon-photon scattering at $e^+e^-$ colliders}
\label{sec:GammaGammaEpEm}

An $e^+e^-$ collider can function as a photon-photon collider when the two initial-state leptons are detected at small angles to the beam pipe. Since the virtual photons scatter as small transverse-size objects with no
hadronic activity in the initial state, they offer interesting opportunities to test QCD in the Regge regime.  In this  $\gamma^* \gamma^*$ reaction, a future measurement combined with the available LEP2 data would constrain the growth of the total cross section with energy that is predicted by BFKL theory~\cite{Brodsky:1998kn,Brodsky:2002ka,Caporale:2008is,Zheng:2013uja,Chirilli:2014dcb,Ivanov:2014hpa}. 
Diffractive production of two vector mesons ($V$), $\gamma^* \gamma^* \to
VV$, can test NLA predictions for $\rho$ \cite{Ivanov:2005gn,Ivanov:2006gt,Enberg:2005eq} and $J/\psi$ \cite{Kwiecinski:1998sa} mesons. Also, heavy-quark pair production can be observed~\cite{Celiberto:2017nyx,Bolognino:2019pba}. The BFKL resummation -- an important formalism of QCD theory described in Sec.~\ref{sec:LowXBFKL} --  has been 
difficult to confirm in hadronic collisions.  In $\gamma^* \gamma^*$ scattering, dependence on the initial photons' virtualities
gives an additional powerful lever.

\section{Perturbative precision calculations for experiments}
\label{sec:perturbativePrecisionCalculations}

Perturbative precision calculations are crucial for measurements of SM parameters
and a key ingredient for the reliable estimation of SM backgrounds to new physics searches~\cite{TFReport}. 
They also serve as an input to precision simulations in modern MC event generators for collider physics.

\subsection{Fixed-order techniques}
\label{sec:wishlist}
\definecolor{lightgray}{rgb}{0.85,0.85,0.85}
\begin{table}[p!]
\caption{\label{tab:LHWishlist}Summary of the LesHouches precision wishlist for hadron colliders~\cite{Huss:2022ful}.
HTL stands for calculations in heavy top limit, VBF* stands for structure function approximation.}
\renewcommand{\arraystretch}{1.2}
\setlength{\tabcolsep}{5pt}
\scalebox{0.975}{
\centering
  \begin{tabular}{@{}l@{}l@{}l@{}}
    \hline\hline
    \multicolumn{1}{c}{process} & \multicolumn{1}{c}{known} &
    \multicolumn{1}{c}{desired} \\
    \hline\hline
\rowcolor{lightgray}
    $pp\to H$ &
    \NNNLOHTL, \NNLOQCDT, \NLOHE11 &
    \NLOH4 (incl.), \NNLOQCDBC\\
    $pp\to H+j$ &
    \NNLOHTL, \NLOQCD, \NLOQE11  &
    \NNLOHTL$\!\otimes\,$\NLOQCD\!+\,\NLOEW \\
\rowcolor{lightgray}
    $pp\to H+2j$ &
    \NLOHone$\!\otimes\,$\LOQCD &
    \NNLOHTL$\!\otimes\,$\NLOQCD\!+\,\NLOEW,
    \\
\rowcolor{lightgray}
    &     \NNNLOQCDVBFstar (incl.),
    \NNLOQCDVBFstar, 
    \NLOEWVBF &
    \NNLOQCDVBF \\
    $pp\to H+3j$ &
    \NLOHone, \NLOQCDVBF &
    \NLOQCD\!+\,\NLOEW \\
\rowcolor{lightgray}
    $pp\to VH$ &
      \NNLOQCD\!+\,\NLOEW,
      \NLOggHVtb{} & \\
    $pp\to VH + j$ &
      \NNLOQCD  &
      \NNLOQCD + \NLOEW \\
\rowcolor{lightgray}
    $pp\to HH$ &
    \NNNLOHTL$\!\otimes\,$\NLOQCD &
    \NLOEW  \\
    $pp\to HHH$ &
    \NNLOHTL & \\
    \hline
    $pp\to H+t\tb$ &
    \NLOQCD\!+\,\NLOEW,
      \NNLOQCD (off-diag.) &
     \NNLOQCD  \\
\rowcolor{lightgray}
    $pp\to H+t/\tb$ &
      \NLOQCD &
      \NNLOQCD,
      \NLOQCD\!+\,\NLOEW \\
    \hline
    $pp\to V$ &
    \NNNLOQCD,
    \NLOQE11,
    \NLOEW &
      \NNNLOQCD\!+\,\NLOQE11,
      \NLOE2 \\
\rowcolor{lightgray}
    $pp\to VV'$ &
      \NNLOQCD\!+\,\NLOEW{ }\wleptdecays{},
      \!+\,\NLOQCD{ }($gg$) \wleptdecays{} &
      \NLOQCD{ } ($gg$,massive loops)
    \\
\rowcolor{lightgray}
    $pp\to V+j$ &
      \NNLOQCD\!+\,\NLOEW{ }\wleptdecays{} &
      hadronic decays \\
    $pp\to V+2j$ &
      \NLOQCD\!+\,\NLOEW{ }\wleptdecays{},
      \NLOEW{ }\wleptdecays{} &
      \NNLOQCD \wdecays{} \\
\rowcolor{lightgray}
    $pp\to V+b\bar{b}$ &
      \NLOQCD{ }\wleptdecays{} &
      \NNLOQCD \!+\,\NLOEW{ }\wdecays{} \\
    $pp\to VV'+1j$ &
      \NLOQCD\!+\,\NLOEW{ }\wdecays{} &
      \NNLOQCD \\
\rowcolor{lightgray}
    $pp\to VV'+2j$ &
    \NLOQCD \wleptdecays{} (QCD),
    \NLOQCD\!+\,\NLOEW{ }\wleptdecays{}(EW) &
      Full \NLOQCD\!+\,\NLOEW{ }\wdecays{} \\
\rowcolor{lightgray}
    $pp\to W^+W^++2j\quad$ &
      Full \NLOQCD\!+\,\NLOEW{ }\wleptdecays{} &
      \\
\rowcolor{lightgray}
    $pp\to W^+W^-+2j$ &
      \NLOQCD\!+\,\NLOEW{ }\wleptdecays{} (EW component) &
      \\
\rowcolor{lightgray}
    $pp\to W^+Z+2j$ &
      \NLOQCD\!+\,\NLOEW{ }\wleptdecays{} (EW component) &
      \\
\rowcolor{lightgray}
    $pp\to ZZ+2j$ &
      Full \NLOQCD\!+\,\NLOEW{ }\wleptdecays{} &
      \\
   $pp\to VV'V''$ &
      \NLOQCD,
      \NLOEW{ }\wodecays{} &
      \NLOQCD\!+\,\NLOEW \wdecays{} \\
\rowcolor{lightgray}
   $pp\to W^\pm W^+W^-$ &
      \NLOQCD + \NLOEW{ }\wdecays{} &
      \\
    $pp\to \gamma\gamma$ &
      \NNLOQCD\!+\,\NLOEW &
      \NNNLOQCD \\
\rowcolor{lightgray}
    $pp\to \gamma+j$ &
      \NNLOQCD\!+\,\NLOEW &
      \NNNLOQCD \\
    $pp\to \gamma\gamma+j$ &
      \NNLOQCD\!+\,\NLOEW,
      \!+\,\NLOQCD{ }($gg$ channel) & \\
\rowcolor{lightgray}
    $pp\to \gamma\gamma\gamma$ &
      \NNLOQCD &
      \NNLOQCD\!+\,\NLOEW \\
    \hline
    $pp\to 2$\,jets &
      \NNLOQCD, 
      \NLOQCD\!+\,\NLOEW
      &
      \NNNLOQCD\!+\,\NLOEW \\
\rowcolor{lightgray}
    $pp\to 3$\,jets &
      \NNLOQCD\!+\,\NLOEW &
      \\
    \hline
    $pp\to t\tb$ &
    \begin{tabular}{@{}l@{}}
      \NNLOQCD(w/ decays)\!+\,\NLOEW (w/o decays) \\
      \NLOQCD\!+\,\NLOEW{ }(w/ decays, off-shell) \\
      \NNLOQCD{ }
    \end{tabular} &
    \begin{tabular}{@{}l@{}}
      \NNNLOQCD
    \end{tabular} \\
\rowcolor{lightgray}
    $pp\to t\tb+j$ &
    \begin{tabular}{@{}l@{}}
      \NLOQCD{ }(w/ decays, off-shell) \\
      \NLOEW (w/o decays)
    \end{tabular} &
    \begin{tabular}{@{}l@{}}
      \NNLOQCD\!+\,\NLOEW{ }(w/ decays)
    \end{tabular} \\
    $pp\to t\tb+2j$ &
    \begin{tabular}{@{}l@{}}
      \NLOQCD{ }(w/o decays)
    \end{tabular} &
    \begin{tabular}{@{}l@{}}
      \NLOQCD\!+\,\NLOEW{ }(w/ decays)
    \end{tabular} \\
\rowcolor{lightgray}
    $pp\to t\tb+Z$ &
    \begin{tabular}{@{}l@{}}
      \NLOQCD\!+\,\NLOEW{ }(w/o decays) \\
      \NLOQCD{ }(w/ decays, off-shell)
    \end{tabular} &
    \begin{tabular}{@{}l@{}}
      \NNLOQCD\!+\,\NLOEW{ }(w/ decays)
    \end{tabular} \\
    $pp\to t\tb+W$ &
    \begin{tabular}{@{}l@{}}
    \NLOQCD\!+\,\NLOEW{ }(w/ decays, off-shell) \\
    \end{tabular} &
    \begin{tabular}{@{}l@{}}
      \NNLOQCD\!+\,\NLOEW{ }(w/ decays)
    \end{tabular} \\
\rowcolor{lightgray}
    $pp\to t/\tb$ &
    \begin{tabular}{@{}l@{}}
      \NNLOQCD{*}(w/ decays) \\
      \NLOEW{ }(w/o decays)
    \end{tabular} &
    \begin{tabular}{@{}l@{}}
      \NNLOQCD\!+\,\NLOEW{ }(w/ decays)
    \end{tabular} \\
    $pp\to tZj$ &
    \begin{tabular}{@{}l@{}}
      \NLOQCD\!+\,\NLOEW{ }(w/ decays)
    \end{tabular} &
    \begin{tabular}{@{}l@{}}
      \NNLOQCD\!+\,\NLOEW{ } (w/o decays)
    \end{tabular} \\
    \hline\hline
  \end{tabular}}
\end{table}
\renewcommand{\arraystretch}{1.0}
There has been significant recent progress in the computation of QCD radiative corrections~\cite{Amoroso:2020lgh,Heinrich:2020ybq,Cordero:2022gsh,Huss:2022ful}. Several groups have used different approaches to achieve the first 
$2\rightarrow3$ \NNLOgen calculations of hadron collider process. There have also been significant steps forward in the development of improved infrared subtraction schemes including methods to deal with higher-multiplicity processes at \NNLOgen.
There has also been remarkable progress in the area of differential \NNNLOgen calculations,
with first results obtained for $2 \to 1$ benchmark processes.
A summary of the state of the art and targets for future measurements is shown in Tab.~\ref{tab:LHWishlist}.
This Les Houches precision wishlist has served as a summary and repository for the higher-order 
QCD and EW calculations relevant for high-energy colliders, providing a crucial link between 
theory and experiment.

Computing fixed-order amplitudes for scattering processes remains one of the key challenges
and obstacles to producing precise predictions for the LHC. Broadly speaking, one can divide the 
computation into two categories: Obtaining and reducing the amplitudes, and calculating
the integrals which appear. A thorough review of recent formal developments can be found 
in~\cite{Travaglini:2022uwo}. The standard approach is the \textit{projector method}, where
the amplitude is expressed as a linear combination of all potential $D$-dimensional 
tensor structures each multiplied by a scalar coefficient containing scalar integrals.
For on-shell/physical processes, one can restrict the space-time dimension of the external particles
to $D=4$ and directly construct projectors for individual helicity amplitudes~\cite{Chen:2019wyb,Peraro:2019cjj,Peraro:2020sfm}.
The integrals are related through integration-by-parts identities (IBPs)~\cite{Tkachov:1981wb,Chetyrkin:1981qh}
and Lorentz-invariance identities (LIs)~\cite{Gehrmann:1999as}. Redundancies can be eliminated using
the Laporta algorithm~\cite{Laporta:2000dsw}, leading to a set of \textit{master integrals}, 
the evaluation of which is one of the biggest obstacles in obtaining multi-loop/multi-leg amplitudes.
A useful technique for avoiding the intermediate expression swell in the reduction to master integrals
is the use of \textit{finite fields}, typically integers modulo some prime number~\cite{Kant:2013vta,vonManteuffel:2014ixa,Peraro:2016wsq}.
A modern introduction to various techniques for computing multi-loop Feynman integrals can be found in Ref.~\cite{Weinzierl:2022eaz}, 
and further details on recent developments can be found in~\cite{Abreu:2022mfk,Blumlein:2022zkr}.
Another technique that has been recently developed to compute multi-scale two-loop scattering amplitudes is based on numerical unitarity~\cite{Abreu:2017xsl, Abreu:2017hqn, Abreu:2019odu}, which effectively sidesteps tensor integral reduction and directly delivers helicity amplitudes.

The use of the differential equations technique~\cite{Kotikov:1990kg,Gehrmann:1999as},
and particularly Henn's canonical form~\cite{Henn:2013pwa} remains as one of the most important methods for computing Feynman Integrals.
New developments concerning the use of differential equations and their application to cutting edge multi-loop integrals can be found in, 
e.g., Refs.~\cite{Abreu:2020jxa,Frellesvig:2021hkr,Dlapa:2021qsl,Syrrakos:2020kba,Kardos:2022tpo,Henn:2021cyv,Abreu:2021smk}.
In Ref.~\cite{Papadopoulos:2014lla}, a procedure for introducing an auxiliary dimensionless parameter into the kinematics 
of a process and deriving differential equations with respect to this parameter, known as the simplified differential equations 
approach, was described. It has recently been used to compute the 2-loop planar~\cite{Canko:2020ylt} and non-planar~\cite{Kardos:2022tpo} 
5-point functions with one massive leg. The method of differential equations is reviewed in~\cite{Argeri:2007up,Henn:2014qga}.

It is not only the amplitude community that has seen impressive development recently. There have
been significant steps forward on the side of subtraction schemes.
While a full automation of \NLOgen subtractions has been achieved, this is not yet the case at \NNLOgen.
This puts the next frontier in \NNLOgen calculations to $2 \to 3$ processes, as well as revisiting prior 
approximations that could potentially limit the interpretation of theory--data comparisons 
(\eg combination of production and decay subprocesses, flavoured jet definition, photon-jet separation 
and hadron fragmentation, on-shell vs.\ off-shell, etc.).
Antenna subtraction~\cite{Gehrmann-DeRidder:2005btv,Currie:2013vh} is applicable to processes 
with hadronic initial and final states with analytically integrated counterterms. 
It has been extended to cope with identified jet flavours~\cite{Gauld:2019yng,Gauld:2020deh}
and the photon fragmentation function~\cite{Gehrmann:2022cih,Chen:2022gpk}.
Sector-improved residue subtraction~\cite{Czakon:2010td,Czakon:2011ve,Boughezal:2011jf} 
is capable of treating hadronic initial and final states through a fully local subtraction 
that incorporates ideas of the FKS approach at NLO~\cite{Frixione:1995ms,Frederix:2009yq} 
and a sector decomposition~\cite{Binoth:2000ps} approach for real radiation 
singularities~\cite{Heinrich:2002rc,Anastasiou:2003gr,Binoth:2004jv}.
It has been extended to deal with flavoured jets~\cite{Czakon:2020coa} and $B$-hadron production~\cite{Czakon:2021ohs}.
$q_T$-subtraction~\cite{Catani:2007vq} is a slicing approach for processes with a colourless final state
and/or a pair of massive coloured particles and is publicly available in the Matrix program~\cite{Grazzini:2017mhc}
and the MCFM program~\cite{Campbell:2022gdq}.
It has been extended to cope with a pair of massive coloured particles~\cite{Bonciani:2015sha,Angeles-Martinez:2018mqh}
and applied to top-pair production~\cite{Catani:2019iny,Catani:2019hip} and $b\bar{b}$ production~\cite{Catani:2020kkl}.
$N$-jettiness~\cite{Boughezal:2015eha,Boughezal:2015dva,Gaunt:2015pea} is a slicing approach based on 
the resolution variable $\tau_N$ ($N$-jettiness) that is suited for processes beyond the scope of the $q_T$ method,
i.e.\ involving final-state jets and is available in the MCFM program~\cite{Boughezal:2016wmq,Campbell:2019dru}.
ColorFul subtraction~\cite{DelDuca:2015zqa} is a fully local subtraction extending the ideas of the 
Catani--Seymour dipole method at NLO~\cite{Catani:1996vz}.
Nested soft--collinear subtraction~\cite{Caola:2017dug,Caola:2018pxp,Delto:2019asp} is a fully local
subtraction with analytic results for integrated subtraction counterterms. 
Analytic local sector subtraction~\cite{Magnea:2018hab, Magnea:2018ebr} is a local subtraction 
with analytic integration of the counterterms aiming to combine the respective advantages from two 
NLO approaches of FKS subtraction~\cite{Frixione:1995ms,Frederix:2009yq} and dipole subtraction~\cite{Catani:1996vz}.
Finally, projection to Born~\cite{Cacciari:2015jma} is a method based on knowledge of inclusive calculations
that retain the full differential information with respect to Born kinematics.
All these developments will lead to more \NNLOgen precise results becoming available for use by experiments
in the near future.

A separate challenge is to make the \NNLOgen $2\to2$ predictions or complex NLO predictions publicly available 
to experimental analyses. Root nTuples have been a useful tool for complicated final states at NLO 
and allow for very flexible re-weighting and analysis. An extension of APPLgrid~\cite{Carli:2010rw} 
and fastNLO~\cite{Kluge:2006xs} to \NNLOgen offers a convenient method to distribute higher-order predictions.
Despite progress in this direction~\cite{Carrazza:2020gss}, more of these grids should become publicly available.

\subsection{Monte-Carlo simulations}
\label{sec:mcsimulations}
Most QCD experimental programs rely on the modeling of hadronic final states provided by 
particle-level Monte-Carlo event generators. Uncertainties on the results of experimental analyses 
are often dominated by effects associated with these simulations. They arise from the underlying 
physics models and theory, the truncation of perturbative expansions, the PDFs and their implementation, 
the modeling of nonperturbative effects, the tuning of model parameters, and the fundamental parameters 
of the theory.

The types of experiments discussed in Sec.~\ref{sec:intro} span a wide range of energies,
beam particles, targets (collider vs.\ fixed target), temperature
and chemical potential. An event generator employed for a given experimental configuration
may require some dedicated physics models, while other parts can be similar or even identical
to those used for other configurations.
A particular strength of event generators derives from the factorization or assumed factorization 
of physics at different energy scales.   This principle allows some physics models to be universal
and often enables
the modular assembly of (parts of) a generator from existing tools when targeting a new
experiment.   In this manner, previously gained knowledge and experience can be transferred, and a more
comprehensive understanding of the physics models is made possible
by allowing them to be tested against a wealth of data gathered in past and current experiments.
These cross-cutting topics in event generation have been identified as a particular
opportunity for the theory community~\cite{Campbell:2022qmc}. 

A shared feature of all experiments is the hard interaction, which probes the colliding beams
or fixed targets at the shortest distance scales. This component of any reaction is 
often the most interesting, since it is most susceptible to new physics effects. 
It is described in simulations using full quantum mechanical calculations, including 
interference, and typically at the highest order in perturbation theory that is practicable.
Since measurements happen at much larger distance scales, the particles 
produced in hard interactions can radiate a substantial number of additional quanta before being detected.
This radiation is implemented in quasi-classical cascade models, which are matched to the quantum 
mechanical calculation of the hard process to increase precision. If the active degrees of freedom
at the hard scale are the asymptotically free quarks and gluons of QCD, the transition
to color-neutral hadrons must be accounted for, typically through the cluster or string model.
Particles produced in this simulation chain may undergo transport through nuclear matter.
They may also still be unstable and decay on timescales that can 
in some cases be resolved by detectors.
Uncertainties on the results of experimental analyses are often dominated by effects
associated with the simulation of the above effects in event generators.
They arise from the underlying physics models and theory, the truncation of perturbative
expansions, the PDFs and their implementation, the modeling of non-perturbative effects, the tuning of model parameters, 
and the extraction of fundamental parameters of the theory.
The need to address these uncertainties for various facilities and experiments is 
the driving force behind the efforts of the community of Monte-Carlo event generator developers. 
Ref.~\cite{Campbell:2022qmc} discusses these questions in the context of the larger facilities that continue to drive 
the development in the near term.

{\bf High Energy Colliders}:
In the coming decade experiments at the LHC will make precise measurements of Standard Model
parameters, such as the $W$ mass and the Higgs boson couplings. Both the extraction of these parameters and 
their interpretation will be limited primarily by 
the precision of perturbative QCD and EW calculations, both fixed order and resummed. 
The results of some analyses will however also be limited by the number of Monte Carlo events that can be generated,
and computing efficiency will play a crucial role. 
Future highest-energy colliders, including a potential muon collider, may/will require electroweak effects
to be treated on the same footing as QCD and QED effects.

{\bf Neutrino Experiments}:
The next generation neutrino experiments DUNE and HyperK will make precise measurements of the 
CP violating phase, mixing angles and the mass hierarchy. The SBN program will focus on precise measurements
of neutrino cross sections and searches for new physics. None of these experiments will be limited
by statistical uncertainties. Since all running and planned experiments use nuclear targets,
one of the leading systematic uncertainties to the measurements arises from the modeling of neutrino-nucleus interactions.
This requires the use of state-of-the-art nuclear-structure and -reaction theory calculations. 
The implementation of physics models with complete error budgets
will be required to reach the precision goals of these experiments.

{\bf Electron-Ion Collider}:
The EIC  will investigate the structure of nucleons and nuclei at an unprecedented level. This will be accomplished by performing precise measurements of DIS and other processes over the complete relevant kinematic range including the transition region from perturbative to non-perturbative QCD. Highly polarized beams and high luminosity will allow probes of the spatial and spin structure of nucleons,
which will need to be simulated at high precision. This currently not possible with standard event generators and requires
the development of new tools at the interface between particle and nuclear physics. It is expected that measurements at the LHC will greatly benefit from these developments.

{\bf Forward Physics Facility}:
The Forward Physics Facility at the LHC will leverage the intense beam of 
neutrinos, and possibly undiscovered particles, in the far-forward direction to 
search for new physics and calibrate forward particle production. 
These measurements will require an improved description of
forward heavy flavor -- particularly charm -- production,
neutrino scattering in the TeV range, 
and hadronization inside nuclear matter, including uncertainty quantification.

{\bf Lepton Colliders}:
Future lepton colliders would provide per mill level measurements of Higgs boson couplings and $W$ and top-quark masses. The unprecedented experimental precision will require event generators to
cover a much wider range of processes than at previous facilities, both in the Standard Model and beyond. In addition,
predictions for the signal processes must be made with extreme precision.
Some of the methodology is available from the LEP era, while other components will
need to be developed.

Event generation for the above facilities contains many common physics components, such as higher-order QCD and electroweak perturbative corrections, factorization theorems and parton evolution equations, resummation of QCD and QED effects,  hadronization, and final-state modeling. 
Various experiments also require the understanding of heavy-ion collisions and nuclear dynamics at high energies as well as heavy-flavor effects.
In addition to the physics components, there are similar computational 
ingredients, such as interfaces to external tools for analysis, handling of tuning and systematics, and the need for improved computing efficiency.
Many of these aspects may profit from developments in artificial intelligence and machine learning.
A detailed discussion of all these aspects can be found in~\cite{TFReport}.

\section{Analysis techniques}
\subsection{Jet Substructure}
\label{sec:substructure}
Jets produced from high energy quarks and gluons through QCD have a complex composition. This jet substructure has emerged as a powerful framework for studying the SM at particle colliders, and provides a key set of tools for probing nature at the highest energy scales accessible by terrestrial experiments~\cite{Abdesselam:2010pt,Altheimer:2012mn,Altheimer:2013yza,Adams:2015hiv,Larkoski:2017jix,Kogler:2018hem,Marzani:2019hun,Kogler:2021kkw}.
In general, jet substructure techniques are applied to explore the structure of the strong force in final state radiation on small angular scales, and to identify Lorentz-boosted massive particles ($H/W/Z$ bosons, top quarks, and BSM particles). For all of these signatures, there are a variety of physics backgrounds that obfuscate the target signatures. At hadron colliders, this is the result of multiple, nearly simultaneous collisions (pileup) as well as underlying event, and multi-parton interactions.  A variety of \textit{jet grooming} techniques have been developed to mitigate these effects (see e.g., Refs.~\cite{Abdesselam:2010pt,Altheimer:2012mn,Altheimer:2013yza,Adams:2015hiv}). While similar backgrounds in $e^+e^-$ are often much smaller, beam-induced backgrounds in muon colliders~\cite{Collamati:2021sbv} could potentially benefit from similar techniques developed for hadron colliders.

There are several detector technologies that will improve jet substructure and related techniques. These include finer calorimeter granularity~\cite{Yeh:2019xbj,Coleman:2017fiq}, more hermetic coverage of tracking detectors, and precise measurements of timing information. The experience of the LHC has shown that such information can be used to more accurately reconstruct the interaction of hadrons with various detector elements, much of which is used in the `particle flow' (PF) concept already deployed by the LHC experiments. At future muon colliders, `beam background' detectors could also in principle be deployed to reduce the impact on jet substructure.

\paragraph{Light Quark and Gluon Jets}
High-energy quark and gluon jets are important probes of a variety of QCD phenomena. These jets can be used to study perturbative aspects of QCD as well as features of QCD that cannot currently be described with perturbation theory. For the latter case, there are cases where scaling relations can be predicated and tested across a wide range of energies. These final states can be used to measure the strong coupling constant, to extract various universal objects within factorized QCD, to tune parton-shower Monte Carlo generators, as well as other tasks. Quark and gluon jets were also studied at previous colliders, but higher energy machines allow for a suppression of nonperturbative effects as well as a larger lever arm for testing scaling behaviors.

Quark and gluon jets are statistically distinguishable due to their different fragmentation processes.  Quark versus gluon jet tagging has been a standard benchmark for the development of new classical and machine learning-based jet taggers. Many SM and BSM final states of interest are dominated either by quark or gluon jets, in contrast to the dominant background processes.  Quark versus gluon jet tagging can help enhance such signals, although these jets are not as separable as other objects. See also references~\cite{Gras:2017jty,Proceedings:2018jsb,Amoroso:2020lgh} for further details.

Currently, quark vs. gluon tagging has not fulfilled its promise due to large uncertainties in the modelling of gluon jets. Having pure samples of gluon jets in QCD would significantly change this situation and have a major impact on the LHC physics program.
This would be a particular advantage of future lepton colliders for the study of QCD, as they provide pure samples of gluon jets through the process $e^+e^-\to HZ$, with $Z$ decay to leptons and Higgs boson decay to $gg$~\cite{dEnterria:2015mgr,Gras:2017jty,Gao:2019mlt}. 
Although the understanding of gluon jets is quite poor, there in fact exist a wide range of precision perturbative calculations of event shapes on $H\to gg$, which have never been compared to data. Since the perturbative features of gluon jets are well understood, and already available to high accuracies, comparison with data would enable detailed studies of the nonperturbative structure of gluon jets. 

\paragraph{Heavy Flavor Jets}
Bottom quark jets are highly separable from other jets due to the long lifetime of the bottom quark and the heavy mass of bottom-flavored hadrons. In addition to lifetime information, jet substructure can be used to further separate these jets from other jets~\cite{CMS:2017wtu,ATLAS:2019bwq}.  
A similar conclusion holds to a lesser extent for charm quark jets~\cite{CMS:2021scf,ATLAS:2021cxe} and to an even lesser extent for strange quark jets~\cite{Erdmann:2020ovh,Nakai:2020kuu,Erdmann:2019blf}.

\paragraph{Strahlung Jets}
Many future collider scenarios result in $H$, $W$, and $Z$ bosons radiating off of very high energy jets (``Weak-strahlung'').
In the $H\rightarrow b\overline{b}$ and $H\rightarrow c\overline{c}$ final states, flavor and lifetime information can be used in addition to the jet substructure to improve categorization. 
The identification of $W$ and $Z$ bosons is similar to the $H$ boson, however the masses are slightly lower and they often do not decay to bottom or charm quarks, so there are fewer handles to use to identify them.

There may also be top quark production within a jet that originates from light quarks or gluons via gluon splitting to $t\overline{t}$, similar to the case at the Tevatron and LHC for bottom quarks. These types of events will need to be handled separately from events without these gluon splittings. The jet substructure of top quarks is, in some sense, an ideal case, since there are two heavy SM particle masses to utilize (the top quark and $W$ boson), as well as lifetime and flavor information in the final state particles. This provides a strong handle to identify top quarks. Especially at higher-energy future colliders, the analysis of collisions containing top quarks will be reliant on jet substructure and boosted topologies.

\paragraph{Unconventional Jets}
Unconventional signatures include cases where jets are composed of leptons and hadrons, only leptons, only photons, hadrons and missing transverse energy etc.  In addition to the jet kinematics and substructure, the jet timing~\cite{Chiu:2021sgs} information and other information can be used for classification. Examples include jets containing one or more hard leptons~\cite{Chatterjee:2019brg,Mitra:2016kov,Nemevsek:2018bbt,duPlessis:2021xuc,Dube:2017jgo}, displaced vertices~\cite{Nemevsek:2018bbt}, hard photons~\cite{Wang:2021uyb,Sheff:2020jyw}, or significant missing transverse momentum~\cite{Kar:2020bws,Canelli:2021aps}. Some of these anomalous signatures are already started being explored at the LHC~\cite{CMS:2021dzb,ATLAS:2019isd,CMS:2021dzg,ATLAS:2019tkk,CMS:2019qjk}.

\subsection{Event Shapes and Energy Correlations}
\label{sec:eventshapes}

Measurements of the flow of radiation provide one of the most interesting tests of our understanding of QCD. High energy collisions are particularly interesting, since they provide a probe of the dynamics of QCD from asymptotically free quarks and gluons, through the confining phase transition to free hadrons at asymptotic infinity. Energy flow can be studied either using event shapes or using correlation functions. Both approaches have seen significant progress driven by jet substructure at the LHC, giving rise to many interesting new observables that could be measured at future colliders, providing a significantly extended understanding of energy flow in quantum field theory.

\begin{figure}
\begin{center}
\includegraphics[width=0.95\hsize]{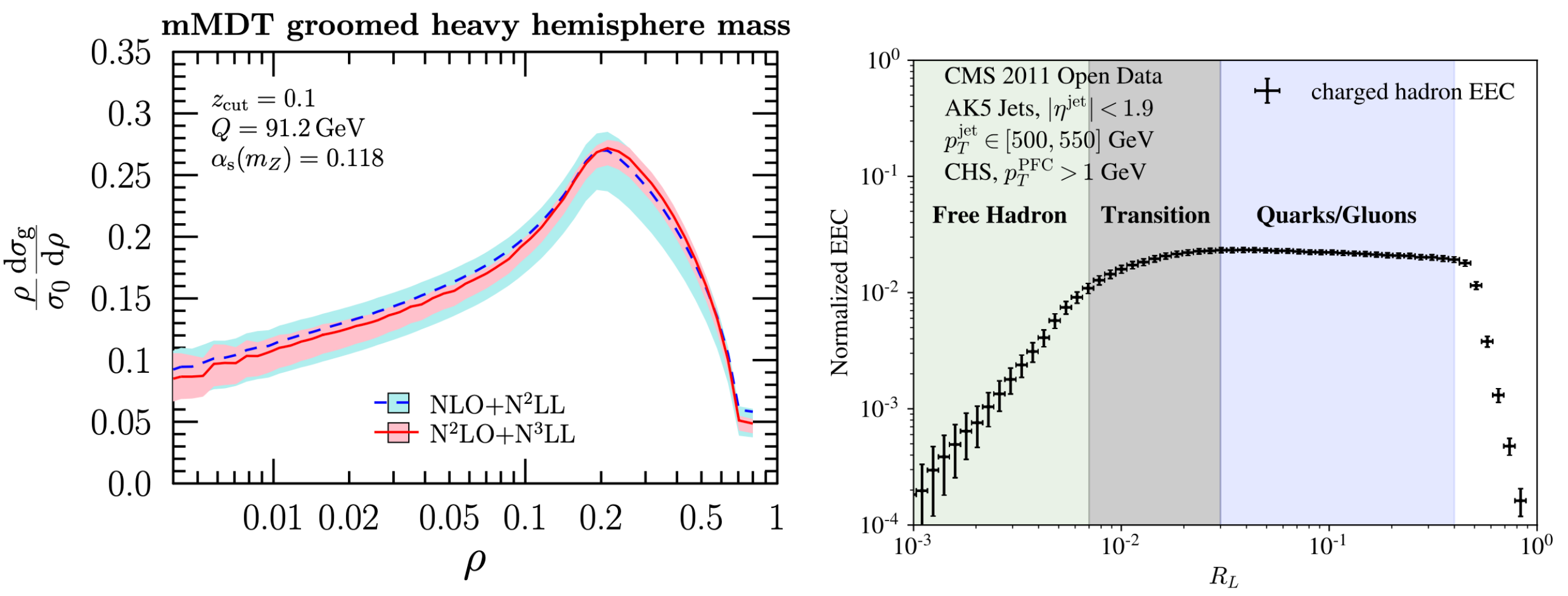} 
\end{center}
  \caption{(a) The groomed mass observable in $e^+e^-$.  The hadronization region is to the left. Figure taken from~\cite{Kardos:2020gty}. (b) A precision measurement of the two-point correlator in the collinear limit at the LHC \cite{Komiske:2022enw}. Both of these new observables provide interesting probes of $\alpha_s$}
  \label{fig:new_obs}
\end{figure}

Energy correlation functions~\cite{Chen:2022jhb,Chen:2021gdk,Chen:2020adz,Chen:2019bpb,Chen:2020vvp,Holguin:2022epo} exhibit simple structures in QCD perturbation theory.
A measurement of the two-point correlator using Open Data from the CMS experiment is shown in Fig.~\ref{fig:new_obs}, illustrating beautiful scaling behavior of weakly coupled quarks and gluons, and a transition to the scaling of free hadrons~\cite{Komiske:2022enw}. Measurements of this quality at future lepton colliders would provide remarkable insights into the dynamics of QCD jets, and the hadronization transition~\cite{Kologlu:2019mfz}.

An insight of the jet substructure program has been the introduction of grooming algorithms that systematically remove low-energy soft radiation~\cite{Dasgupta:2013ihk,Larkoski:2014wba}, thereby reducing traditional double logarithmic observables to single logarithmic observables, and reducing nonperturbative corrections~\cite{dEnterria:2022hzv}. Fig.~\ref{fig:new_obs} shows the groomed hemisphere mass as an example. 
Although the corresponding groomed observables are theoretically cumbersome, they are practically very useful.
The most precise extractions of $\alpha_s$ from event shapes are currently based on thrust and the $C$-parameter~\cite{Abbate:2010xh,Hoang:2015hka}, which are closely related double logarithmic observables. Using the standard methodology, one of the complexities in the measurement is the determination of nonperturbative corrections. These corrections cannot be computed from first principles. Theoretical progress in understanding these power corrections has been made in~\cite{Luisoni:2020efy,Caola:2021kzt,Caola:2022vea}. Nevertheless, they must be simultaneously fit for along with the value of $\alpha_s$. Due to the differing theoretical structure of groomed event shapes, an extraction of $\alpha_s$ from the groomed thrust would provide a relatively independent measurement of the value of the strong coupling. The groomed thrust can be computed to high perturbative accuracy, using a factorization formula, as shown in Fig.~\ref{fig:new_obs}. Nonperturbative corrections to the groomed thrust distribution have been studied in \cite{Hoang:2019ceu}.

While the groomed thrust provides many complementary features to the standard thrust based extraction of $\alpha_s$, it is ultimately based on the same event shape paradigm, and therefore similar assumptions enter in the treatment of nonperturbative effects. Another interesting complementary measurement would be to perform a measurement of the two-point energy correlator (EEC) in the collinear limit \cite{Dixon:2019uzg}. Due to new analytic results at NLO~\cite{Dixon:2018qgp,Luo:2019nig} and the resummation at N$^3$LL$^\prime$ in the back-to-back limit~\cite{Ebert:2020sfi}, the EEC has seen a resurgence of interest. Its collinear limit is described by completely different physics (fixed spin DGLAP) than the Sudakov region, and furthermore, since the energy correlators are not event shape observables, they have a different structure for their nonperturbative effects. Despite being an old observable that was measured at LEP,  extractions of $\alpha_s$ from the collinear limit were never performed at LEP. Comparing the measurement of the two-point correlator at LEP vs. using the modern calorimetry of the LHC shows a completely different understanding of the collinear limit. Achieving a similarly precise measurement in the clean $e^+e^-$ environment of the ILC or FCC-ee would be extremely valuable for precision measurements of $\alpha_s$ (once it reaches NNLO, or beyond, pQCD accuracy), and might resolve the longstanding tensions in the values extracted from event shapes.

\begin{figure}[t]
    \centering
    \includegraphics[width=0.49 \textwidth]{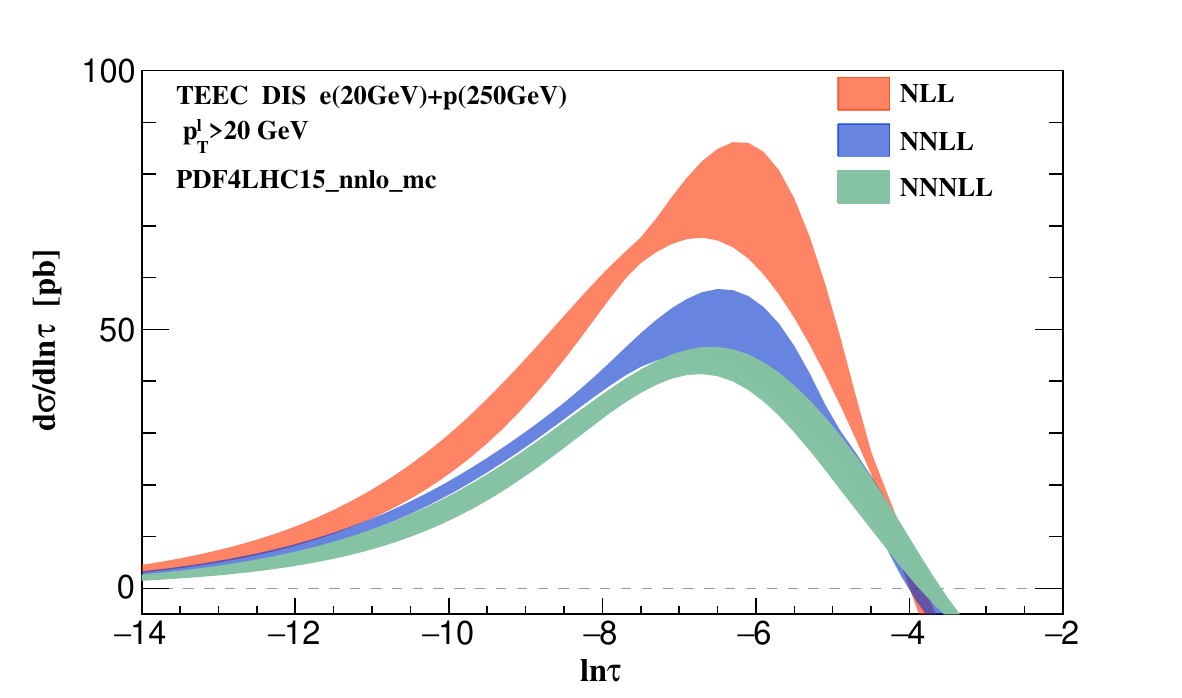}
    \includegraphics[width=0.49 \textwidth]{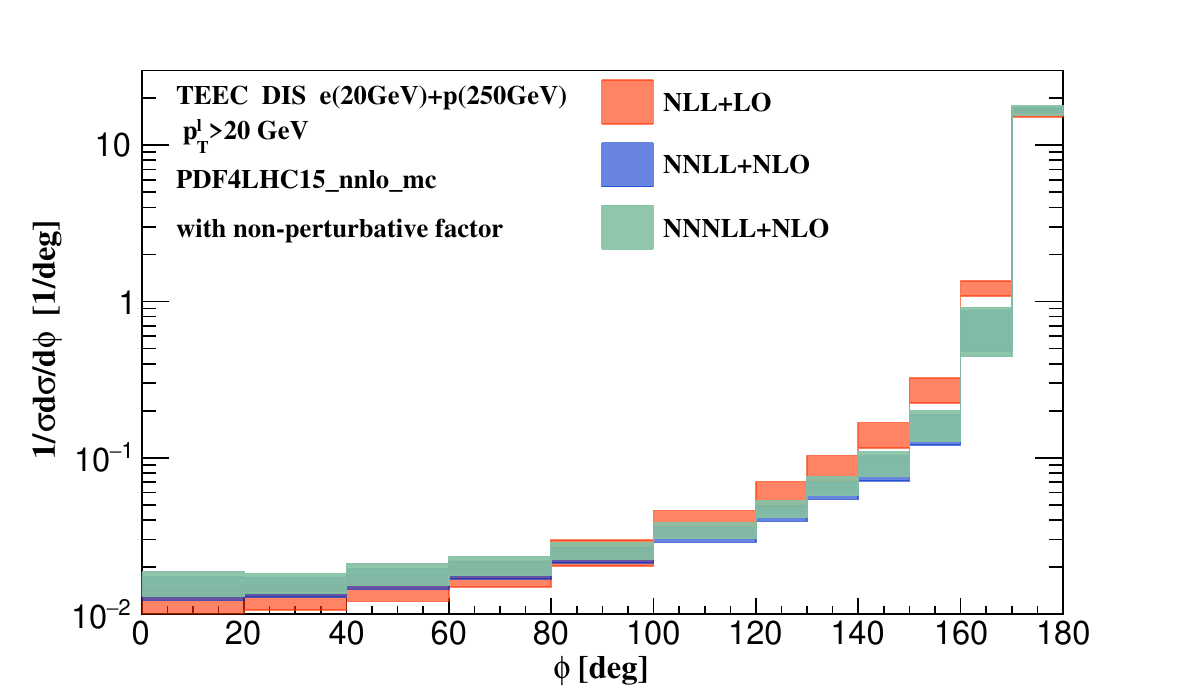}
    \caption{Left: resummed distributions in the back-to-back limit up to N$^3$LL accuracy. Note that results are not normalized by $\sigma$ in the $\tau$ interval shown.
Right: TEEC $\phi$ distribution matched with a nonperturbative model. The orange, blue and green bands are the final predictions with scale uncertainties up to N$^3$LL+NLO. }
    \label{fig:TEEC}
\end{figure}
Moreover, the EEC/TEEC event shape observables can be studied in $e^+e^-$, $ep$ and $pp$ collisions, which provides a way to test the universality of QCD factorization in different colliding systems. These observables can also be used to study TMD physics, which is one of the most important goals of the EIC. 
Hadron colliders present a  more complex environment than $e^+e^-$ or $ep$ colliders, nevertheless, high accuracy can still be achieved. Once EEC/TEEC are computed at NNLO pQCD accuracy, a new avenue will be opened for the rich HL-LHC data to be combined with precision resummed QCD predictions to obtain precision measurement of SM parameters, such as strong coupling constant and various TMD functions. The simplicity in the theoretical structure of EEC/TEEC makes higher-order calculation feasible and can shed light on the structure of perturbation theory. Both of these lead to valuable addition to our understanding of QCD.
Finally, the EEC/TEEC observables can be generalized to $eA$,  $pA$, and $AA$ collisions. They can shed new light on the many-body QCD dynamics in reactions with nuclei, specifically  multi-parton  interactions and the formation of parton showers in matter. In these environments, precise extraction of transport properties of various forms of nuclear matter will greatly benefit from the high perturbative accuracy achieved in the baseline $ep$ and $pp$ reactions.

\section{Cross-Cutting QCD}
\label{sec:CrossCuttingQCD}
QCD interactions play a ubiquitous and multifaceted role in collider phenomenology, and hence successes across many areas depend on future developments at the intersections of QCD and other domains. To take the full advantage of precise perturbative QCD calculations discussed in Sec.~\ref{sec:wishlist}, commensurate advances must be achieved in determinations of long-distance QCD contributions including PDFs, computations of electroweak radiative contributions, event generation,  machine learning, and last but not least, accurate and fast practical implementations. These tasks require collaboration between experimentalists and theorists, model-builders and QCD experts, and, more broadly, support of the {\it QCD infrastructure} that adapts theoretical tools for experimental analyses and provides protocols to validate these tools and assess uncertainties from experimental or theoretical sides. This subsection presents examples of such cross-cutting issues.

\begin{figure}[b]
    \centering
    \includegraphics[width=0.8\textwidth]{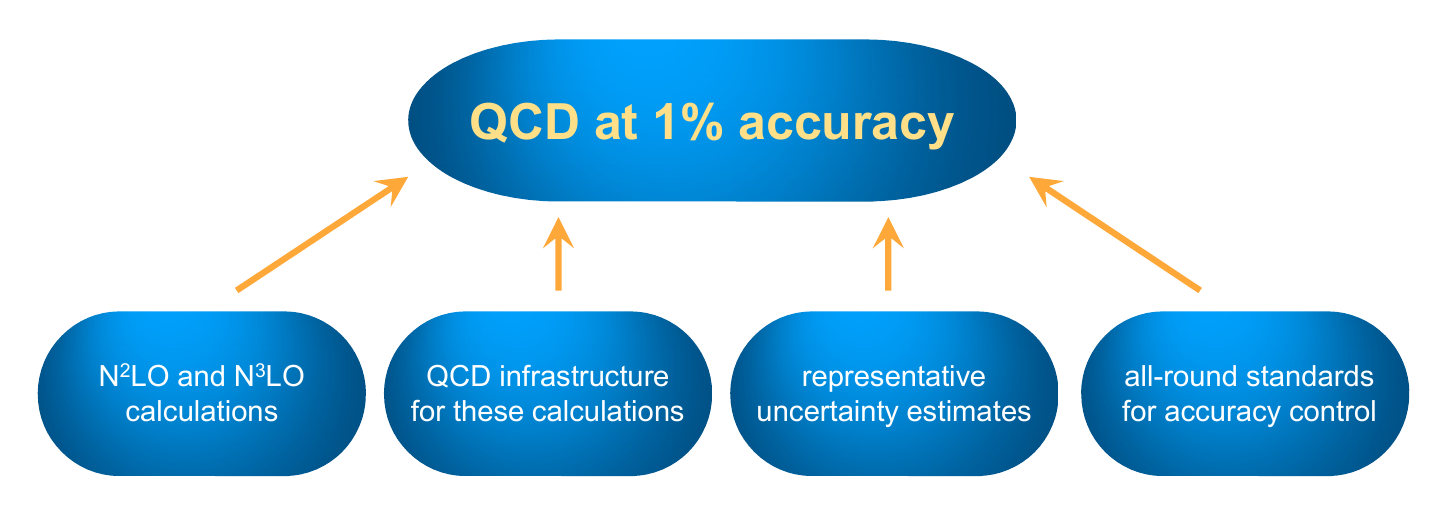}   
    \caption{\label{fig:QCDatOnePercent} Prerequisites for achieving percent-level accuracy in QCD calculations.}
\end{figure}

\subsection{Comprehensive uncertainty estimates}
Achieving the targeted accuracy on the PDFs and key measurements such as $W$ boson mass at the HL-LHC requires better control of systematic uncertainties at all stages \cite{Amoroso:2022eow}, in experimental measurements as well as in numerical computations. This requires a close collaboration between experimentalists and theorists on the consistent usage of QCD predictions and the conversion from parton to particle level (see also Sec.~\ref{sec:mcsimulations}). 
Making higher-order calculations for complex final states available in a form suitable for experimental analyses remains a significant part of this challenge~\cite{Carli:2010rw,Kluge:2006xs,Bern:2013zja}. In addition, more efforts are necessary to present models of systematic uncertainties in the complete form that can be interpreted by external users \cite{Cranmer:2021urp}. New types of complexity issues emerge in comparisons of models with many parameters to very large data samples expected at the LHC Run-3 and HL-LHC.  Such comparisons may be subject to increased risks of undetected biases due to non-representative exploration of contributing systematic factors \cite{MengXL:2018}, as has been recently demonstrated on the example of a PDF global fit \cite{Courtoy:2022ocu}. In short, elevating the accuracy of QCD calculations to one percent requires both individual precise theoretical calculations as well as accurate supporting theoretical infrastructure that would allow, in particular, to explore exhaustively the relevant systematic factors. Reaching this target also requires agreed-upon standards and practices for accuracy control at all stages of the analyses, as is illustrated in Fig.~\ref{fig:QCDatOnePercent}. 

\subsection{AI/ML innovation for QCD}
QCD applications are a fertile ground to explore innovative developments in machine learning (ML) and artificial intelligence (AI) \cite{hepmllivingreview}. ML methods have been proposed to solve a variety of tasks both in experiment (particle reconstruction, event unfolding, anomaly detection) and theory (phase space integration, calculations of scattering amplitudes, simulations of parton showers, and modeling of parton distribution and fragmentation functions) \cite{Butter:2022rso}. ML methods are explored at the same time in the context of lattice QCD calculations \cite{Boyda:2022nmh}. Conceptually, there are significant overlaps between ML and methods of large-scale and multivariate data analysis and fitting that traditionally have been used e.g. for collider particle detection and fitting of nonperturbative functions. ML provides technical tools that can facilitate many advances. Across-the-board outstanding questions also need to be resolved, notably interpretability and faithful estimation of uncertainties of ML/AI results, as detailed in the dedicated Snowmass whitepaper \cite{Shanahan:2022ifi}.

\subsection{QCD in new physics searches and SM EFT fits}
The energy reach of many BSM searches depends on the interplay between precision calculations of matrix elements and global PDF analyses. Examples include searches for new vector bosons, referred to as $Z'$s and $W'$s.
Current LHC bounds on mass disfavor extra vector bosons lighter than approximately 4-5 TeV. BSM searches of $W' / Z'$s with even larger masses are progressively more sensitive to PDFs at large $x$ where uncertainties are still large~\cite{Brady:2011hb} and affected by
nuclear corrections, higher-twist contributions, intrinsic heavy-quark components. Either forward particle production at the LHC or, often more cleanly, DIS at the EIC can constrain PDFs in the large-$x$ region and increase sensitivity of BSM searches in the TeV mass range.

 Search for deviations from SM examined in the language of Effective Field Theories (EFTs)~\cite{Manohar:1996cq} can set lower bounds on the scales in a number of new physics scenarios. Such analyses is an active research area covered in the other EF reports, for example, in a widely adopted EFT expansion of the Standard Model, or SMEFT~\cite{Brivio:2017vri}. SMEFT contributions extracted from collider data may significantly depend on including subleading (dimension-eight) SMEFT operators as well as SM radiative contributions, as examined, e.g., in Refs.~\cite{Brivio:2019ius,Ellis:2020unq,Ethier:2021bye,Corbett:2021eux,Boughezal:2021tih,
Ethier:2021ydt,Miralles:2021dyw,Durieux:2022cvf}. This dependence must be kept in mind when interpreting the SMEFT fits to data from both lepton and hadron colliders. 

\begin{figure}[tb]
    \centering
        \subfloat[]{\includegraphics[width=0.49\textwidth]{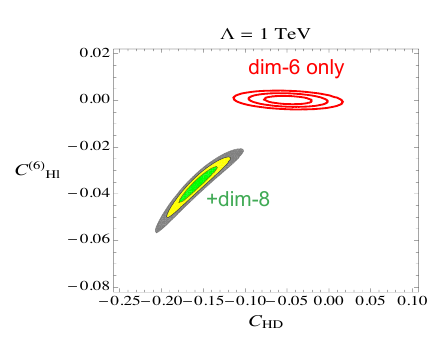}}
    \subfloat[]{\includegraphics[width=0.45\textwidth]{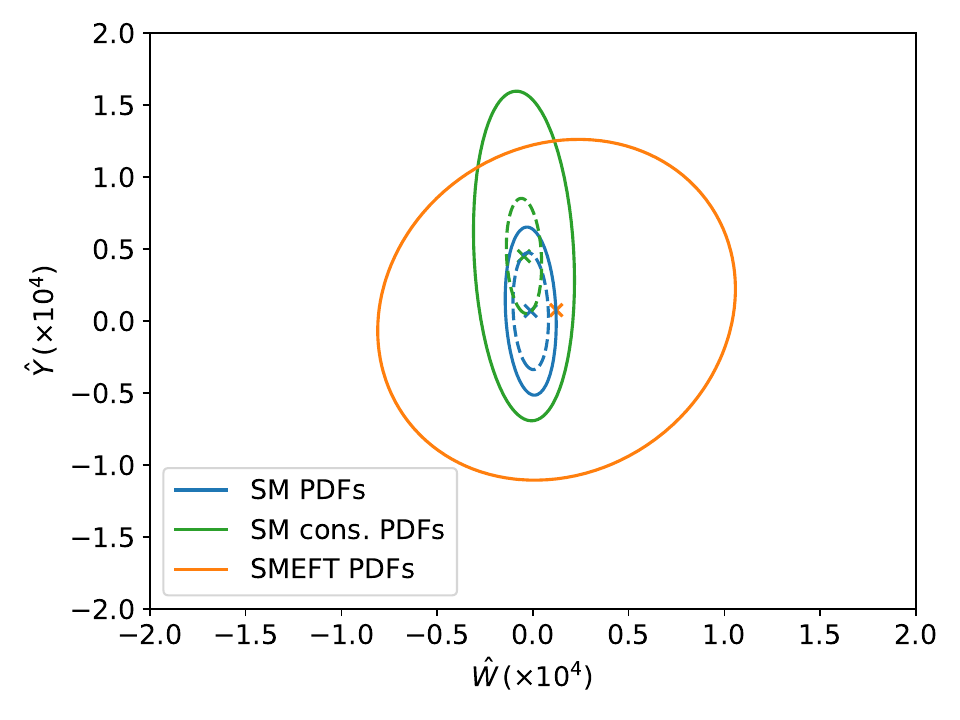}}
    \caption{\label{fig:SMEFT_PDFs} (a) 68\%/95\%/99.9\% exclusion limits on select SMEFT coefficients up to dim-6 and dim-8 in an analysis of LEP electroweak precision data \cite{Corbett:2021eux}.
    (b) The 95\% confidence level bounds on the plane of the Wilson coefficients considered in Ref.~\cite{Greljo:2021kvv} obtained using either fixed SM PDFs (blue) or conservative SM PDFs that do not include high-energy data (green). PDF uncertainties are included in the solid lines and not included in the dashed lines. Results are compared to those obtained in a simultaneous fit of SMEFT and PDFs, when the PDFs are allowed to vary when varying the values of the Wilson coefficients (orange).}
\end{figure}

 As an example, Fig.~\ref{fig:SMEFT_PDFs}(a) illustrates the change in confidence regions for SMEFT coefficients extracted from LEP data upon adding higher-order (dimension-8) operators into the analysis \cite{Corbett:2021eux}.
 As illustrated in Fig.~\ref{fig:SMEFT_PDFs}(b), when including high-mass LHC data in a fit of PDFs and in a fit of SMEFT coefficients, and neglecting the interplay between them, the uncertainties on the EFT parameters may be underestimated.
 Although the proton structure parametrized by PDFs is intrinsically a low-energy input and should in principle be separable from the imprints of SMEFT operators, the complexity of the LHC environment might well intertwine them \cite{Iranipour:2022iak,Liu:2022plj,Greljo:2021kvv,Carrazza:2019sec,ZEUS:2019cou,CMS:2021yzl,Liu:2022plj,Iranipour:2022iak}.  The bounds on the respective Wilson coefficients are relaxed once the are fitted together with the PDFs. Constraints on either the PDFs or EFT operators in low-energy experiments, where at least some new physics contributions are absent, can be crucial for disentangling the SM/BSM degeneracies. In this light, the SM and SMEFT studies at the EIC and other low-energy facilities again are synergistic to those at the (HL-)LHC~\cite{Boughezal:2020uwq,Boughezal:2021kla}, especially for spin-dependent EFT operators.

\subsection{QCD theory for FCC-hh \label{sec:FCChhTheory}}
In 100~TeV $pp$ collisions, in addition to the unprecedented experimental environment reviewed in Sec.~\ref{sec:fcchh}, synergistic developments across multiple areas of QCD theory will be necessary to meet objectives of the FCC-hh experimental program. 
Hadronic and electromagnetic shower components up to several TeV need to be simulated, where extrapolations to these  high energies come with large uncertainties. 
 Differences in the hadronic shower  simulation models in Geant4~\cite{GEANT4:2002zbu} have been reported for pions in the energy range 2--10~GeV~\cite{CALICE:2019vza}, warranting future detailed studies of hadronic showers for the highest-energy regime. Multi-loop perturbative calculations combining both QCD and EW radiative contributions will be combined with PDFs of commensurate precision at multi-TeV energy scales.  A BFKL-like QCD formalism will be necessary to predict parton scattering at momentum fractions as low as $10^{-7}$. Electroweak gauge bosons  $W$ and $Z$, leptons, and top quarks will be copiously produced at the FCC-hh energy and will need to be included into the PDFs together with quarks and gluons \cite{Han:2020uid}. 

\subsection{Legacy data preservation}
Electron-positron and lepton-hadron colliders still provide the most precise data for a number of QCD studies and will likely not be superseded by the LHC dataset.  Legacy data from LEP/SLC and HERA therefore continue to be important for QCD analysis (including event generator tuning and other studies).  In particular, there has been significant theory and experimental innovation (including with machine learning) since these data were collected.  These new insights can be used to extract novel information from these pristine datasets.  A funding model for supporting such analysis does not currently exist, but even a modest investment could produce significant physics results.  

\section{Summary}
\label{sec:summary}
In this report, we reviewed the rich landscape, wide impact, and interconnectedness of QCD studies in 2020's.
During the Snowmass 2021 process, QCD has drawn strong interest in numerous discussions as the only QFT that can be experimentally studied in perturbative and nonperturbative phases, and as the key theory in HEP phenomenology now and in the future. QCD is rich both in data and in ideas. While being the least accurately known fundamental force of the Standard Model, the strong force plays the central role in many measurements. In particular, success of many precision measurements in Higgs, top, electroweak boson production channels relies on both advancements in perturbative QCD calculations and detailed modeling of nonperturbative contributions. Various domains of QCD theory undergo rapid development. 
Completion of the HL-LHC and EIC programs, together with a Higgs factory, will provide unprecedented levels of precision in QCD, in turn impacting most areas in the Energy Frontier.

\section*{Acknowledgments}
We are grateful to all participants of the EF05, 06, and 07 Topical Groups who contributed to many fruitful discussions and Snowmass contributions on which this report is based. 
We gratefully acknowledge comments from  Debasish Das, Michael Engelhardt, Xiaohui Liu, and Jason Nielsen, which helped to improve various sections of this report.

This work was supported by the U.S. Department of Energy, Office of Science, Office of High Energy Physics, and Office of Nuclear Physics under Contracts DE-SC001011, DE-SC0011088, DE-SC0012704, DE-SC0010129, DE-SC0010102, DE-SC0019230, DE-AC02-05CH11231, DE-FG02-93ER40771, DE-AC02-07CH11359, DE-AC05-06OR23177.
A.C.S. acknowledges the support of the Leverhulme Trust.
The work of H.-W.~L. is partially supported by the US National Science Foundation under grant PHY 1653405 and 2209424, and by the  Research  Corporation  for  Science  Advancement through the Cottrell Scholar Award. P.M.N. acknowledges the support by Universities Research Association at Fermilab. 
G.P.S. is supported by a Royal Society Research Professorship
(RP$\backslash$R1$\backslash$180112), by the European Research Council (ERC) under the European Union’s
    Horizon 2020 research and innovation programme (grant agreement No.\
    788223, PanScales), 
    and by the Science and Technology Facilities Council (STFC) under
    grant ST/T000864/1.

\clearpage
\newpage


\begin{thebibliography}{800}

\bibitem{ATLAS:2022hsp}
{\bfseries ATLAS} Collaboration, ``{Snowmass White Paper Contribution: Physics
  with the Phase-2 ATLAS and CMS Detectors}.'' \protect{ATL-PHYS-PUB-2022-018},
  2022.

\bibitem{Anchordoqui:2021ghd}
L.~A. Anchordoqui {\em et~al.}, ``{The Forward Physics Facility: Sites,
  experiments, and physics potential},''
  \href{http://dx.doi.org/10.1016/j.physrep.2022.04.004}{{\em Phys. Rept.}
  {\bfseries 968} (2022) 1--50},
  \href{http://arxiv.org/abs/2109.10905}{{\ttfamily arXiv:2109.10905
  [hep-ph]}}.

\bibitem{Feng:2022inv}
J.~L. Feng {\em et~al.}, ``{The Forward Physics Facility at the High-Luminosity
  LHC},'' \href{http://arxiv.org/abs/2203.05090}{{\ttfamily arXiv:2203.05090
  [hep-ex]}}.

\bibitem{AbdulKhalek:2022erw}
R.~Abdul~Khalek {\em et~al.}, ``{Snowmass 2021 White Paper: Electron Ion
  Collider for High Energy Physics},'' in {\em {2022 Snowmass Summer Study}}.
\newblock 3, 2022.
\newblock \href{http://arxiv.org/abs/2203.13199}{{\ttfamily arXiv:2203.13199
  [hep-ph]}}.

\bibitem{Chekanov:2022sax}
S.~V. Chekanov and S.~Magill, ``{Some aspects of impact of the Electron Ion
  Collider on particle physics at the Energy Frontier},'' in {\em {2022
  Snowmass Summer Study}}.
\newblock 2, 2022.
\newblock \href{http://arxiv.org/abs/2202.11529}{{\ttfamily arXiv:2202.11529
  [hep-ph]}}.

\bibitem{Batell:2022ubw}
{\bfseries $\nu$-Test} Collaboration, B.~Batell, T.~Ghosh, and K.~Xie, ``{Heavy
  Neutral Lepton Searches at the Electron-Ion Collider: A Snowmass
  Whitepaper},'' in {\em {2022 Snowmass Summer Study}}.
\newblock 3, 2022.
\newblock \href{http://arxiv.org/abs/2203.06705}{{\ttfamily arXiv:2203.06705
  [hep-ph]}}.

\bibitem{Accardi:2022oog}
A.~Accardi {\em et~al.}, ``{Opportunities for precision QCD physics in
  hadronization at Belle II -- a snowmass whitepaper},'' in {\em {2022 Snowmass
  Summer Study}}.
\newblock 4, 2022.
\newblock \href{http://arxiv.org/abs/2204.02280}{{\ttfamily arXiv:2204.02280
  [hep-ex]}}.

\bibitem{Bernardi:2022hny}
G.~Bernardi {\em et~al.}, ``{The Future Circular Collider: a Summary for the US
  2021 Snowmass Process},'' \href{http://arxiv.org/abs/2203.06520}{{\ttfamily
  arXiv:2203.06520 [hep-ex]}}.

\bibitem{ILCInternationalDevelopmentTeam:2022izu}
{\bfseries ILC International Development Team} Collaboration, A.~Aryshev {\em
  et~al.}, ``{The International Linear Collider: Report to Snowmass 2021},''
  \href{http://arxiv.org/abs/2203.07622}{{\ttfamily arXiv:2203.07622
  [physics.acc-ph]}}.

\bibitem{Acosta:2022ejc}
D.~Acosta, E.~Barberis, N.~Hurley, W.~Li, O.~M. Colin, D.~Wood, and X.~Zuo,
  ``{The Potential of a TeV-Scale Muon-Ion Collider},'' in {\em {2022 Snowmass
  Summer Study}}.
\newblock 3, 2022.
\newblock \href{http://arxiv.org/abs/2203.06258}{{\ttfamily arXiv:2203.06258
  [hep-ex]}}.

\bibitem{Campbell:2022qmc}
J.~M. Campbell {\em et~al.}, ``{Event Generators for High-Energy Physics
  Experiments},'' \href{http://arxiv.org/abs/2203.11110}{{\ttfamily
  arXiv:2203.11110 [hep-ph]}}.

\bibitem{dEnterria:2022hzv}
D.~d'Enterria {\em et~al.}, ``{The strong coupling constant: State of the art
  and the decade ahead},'' \href{http://arxiv.org/abs/2203.08271}{{\ttfamily
  arXiv:2203.08271 [hep-ph]}}.

\bibitem{Amoroso:2022eow}
S.~Amoroso {\em et~al.}, ``{Snowmass 2021 whitepaper: Proton structure at the
  precision frontier},'' \href{http://arxiv.org/abs/2203.13923}{{\ttfamily
  arXiv:2203.13923 [hep-ph]}}.

\bibitem{Constantinou:2022yye}
M.~Constantinou {\em et~al.}, ``{Lattice QCD Calculations of Parton Physics},''
  \href{http://arxiv.org/abs/2202.07193}{{\ttfamily arXiv:2202.07193
  [hep-lat]}}.

\bibitem{Hou:2022sdf}
T.-J. Hou, H.-W. Lin, M.~Yan, and C.~P. Yuan, ``{Impact of lattice
  $s(x)-\bar{s}(x)$ data in the CTEQ-TEA global analysis},''
  \href{http://arxiv.org/abs/2204.07944}{{\ttfamily arXiv:2204.07944
  [hep-ph]}}.

\bibitem{Hentschinski:2022xnd}
M.~Hentschinski {\em et~al.}, ``{White Paper on Forward Physics, BFKL,
  Saturation Physics and Diffraction},''
  \href{http://arxiv.org/abs/2203.08129}{{\ttfamily arXiv:2203.08129
  [hep-ph]}}.

\bibitem{Bai:2022jcs}
W.~Bai, M.~V. Diwan, M.~V. Garzelli, Y.~S. Jeong, K.~Kumar, and M.~H. Reno,
  ``{Prompt electron and tau neutrinos and antineutrinos in the forward region
  at the LHC},'' \href{http://dx.doi.org/10.1016/j.jheap.2022.05.003}{{\em
  JHEAp} {\bfseries 34} (2022) 212--216},
  \href{http://arxiv.org/abs/2203.07212}{{\ttfamily arXiv:2203.07212
  [hep-ph]}}.

\bibitem{Nachman:2022emq}
B.~Nachman {\em et~al.}, ``{Jets and Jet Substructure at Future Colliders},''
  \href{http://arxiv.org/abs/2203.07462}{{\ttfamily arXiv:2203.07462
  [hep-ph]}}.

\bibitem{XFitterSnowmass}
{The xFitter Collaboration}, ``{xFitter: An Open Source QCD Analysis
  Framework},''. Text available at
  \url{https://tinyurl.com/xFitterSnowmass21WP}; the collaboration website:
  \url{https://www.xfitter.org/}.

\bibitem{dEnterria:2022sut}
D.~d'Enterria {\em et~al.}, ``{Opportunities for new physics searches with
  heavy ions at colliders},'' in {\em {2022 Snowmass Summer Study}}.
\newblock 3, 2022.
\newblock \href{http://arxiv.org/abs/2203.05939}{{\ttfamily arXiv:2203.05939
  [hep-ph]}}.

\bibitem{Anikin:2011sa}
I.~Anikin, A.~Besse, D.~{\relax Yu}. Ivanov, B.~Pire, L.~Szymanowski, and
  S.~Wallon, ``{A phenomenological study of helicity amplitudes of high energy
  exclusive leptoproduction of the rho meson},''
  \href{http://dx.doi.org/10.1103/PhysRevD.84.054004}{{\em Phys. Rev. D}
  {\bfseries 84} (2011) 054004},
  \href{http://arxiv.org/abs/1105.1761}{{\ttfamily arXiv:1105.1761 [hep-ph]}}.

\bibitem{Besse:2013muy}
A.~Besse, L.~Szymanowski, and S.~Wallon, ``{Saturation effects in exclusive
  rhoT, rhoL meson electroproduction},''
  \href{http://dx.doi.org/10.1007/JHEP11(2013)062}{{\em JHEP} {\bfseries 11}
  (2013) 062}, \href{http://arxiv.org/abs/1302.1766}{{\ttfamily arXiv:1302.1766
  [hep-ph]}}.

\bibitem{Bolognino:2018rhb}
A.~D. Bolognino, F.~G. Celiberto, D.~{\relax Yu}. Ivanov, and A.~Papa,
  ``{Unintegrated gluon distribution from forward polarized
  $\rho$-electroproduction},''
  \href{http://dx.doi.org/10.1140/epjc/s10052-018-6493-6}{{\em Eur. Phys. J.}
  {\bfseries C78} no.~12, (2018) 1023},
\href{http://arxiv.org/abs/1808.02395}{{\ttfamily arXiv:1808.02395 [hep-ph]}}.

\bibitem{Celiberto:2019slj}
F.~G. Celiberto, ``{Unraveling the Unintegrated Gluon Distribution in the
  Proton via $\rho$-Meson Leptoproduction},''
  \href{http://dx.doi.org/10.1393/ncc/i2019-19220-9}{{\em Nuovo Cim.}
  {\bfseries C42} (2019) 220},
\href{http://arxiv.org/abs/1912.11313}{{\ttfamily arXiv:1912.11313 [hep-ph]}}.

\bibitem{Bolognino:2019pba}
A.~D. Bolognino, A.~Szczurek, and W.~Schaefer, ``{Exclusive production of
  $\phi$ meson in the $\gamma^*\,p \to \phi\,p$ reaction at large photon
  virtualities within $k_T$-factorization approach},''
  \href{http://dx.doi.org/10.1103/PhysRevD.101.054041}{{\em Phys. Rev. D}
  {\bfseries 101} no.~5, (2020) 054041},
  \href{http://arxiv.org/abs/1912.06507}{{\ttfamily arXiv:1912.06507
  [hep-ph]}}.

\bibitem{Bautista:2016xnp}
I.~Bautista, A.~Fernandez~Tellez, and M.~Hentschinski, ``{BFKL evolution and
  the growth with energy of exclusive $J/\psi$ and $\Upsilon$ photoproduction
  cross sections},'' \href{http://dx.doi.org/10.1103/PhysRevD.94.054002}{{\em
  Phys. Rev. D} {\bfseries 94} no.~5, (2016) 054002},
  \href{http://arxiv.org/abs/1607.05203}{{\ttfamily arXiv:1607.05203
  [hep-ph]}}.

\bibitem{Garcia:2019tne}
A.~Arroyo~Garc\'ia, M.~Hentschinski, and K.~Kutak, ``{QCD evolution based
  evidence for the onset of gluon saturation in exclusive photo-production of
  vector mesons},''
  \href{http://dx.doi.org/10.1016/j.physletb.2019.06.061}{{\em Phys. Lett. B}
  {\bfseries 795} (2019) 569--575},
  \href{http://arxiv.org/abs/1904.04394}{{\ttfamily arXiv:1904.04394
  [hep-ph]}}.

\bibitem{Hentschinski:2020yfm}
M.~Hentschinski and E.~Padr\'on~Molina, ``{Exclusive $J/\Psi$ and $\Psi(2s)$
  photo-production as a probe of QCD low $x$ evolution equations},''
  \href{http://dx.doi.org/10.1103/PhysRevD.103.074008}{{\em Phys. Rev. D}
  {\bfseries 103} no.~7, (2021) 074008},
  \href{http://arxiv.org/abs/2011.02640}{{\ttfamily arXiv:2011.02640
  [hep-ph]}}.

\bibitem{NAP25171}
{National Academies of Sciences, Engineering, and Medicine},
  \href{http://dx.doi.org/10.17226/25171}{{\em An Assessment of U.S.-Based
  Electron-Ion Collider Science}}.
\newblock The National Academies Press, Washington, DC, 2018.

\bibitem{AbdulKhalek:2021gbh}
R.~Abdul~Khalek {\em et~al.}, ``{Science Requirements and Detector Concepts for
  the Electron-Ion Collider: EIC Yellow Report},''
  \href{http://arxiv.org/abs/2103.05419}{{\ttfamily arXiv:2103.05419
  [physics.ins-det]}}.

\bibitem{Thomson:2015jda}
M.~Thomson, ``{Model-independent measurement of the e$^{{+}}$ e$^{-}$
  $\rightarrow $ HZ cross section at a future e$^{{+}}$ e$^{-}$ linear collider
  using hadronic Z decays},''
  \href{http://dx.doi.org/10.1140/epjc/s10052-016-3911-5}{{\em Eur. Phys. J. C}
  {\bfseries 76} no.~2, (2016) 72},
  \href{http://arxiv.org/abs/1509.02853}{{\ttfamily arXiv:1509.02853
  [hep-ex]}}.

\bibitem{Lai:2021rko}
P.-Z. Lai, M.~Ruan, and C.-M. Kuo, ``{Jet performance at the circular
  electron-positron collider},''
  \href{http://dx.doi.org/10.1088/1748-0221/16/07/P07037}{{\em JINST}
  {\bfseries 16} no.~07, (2021) P07037},
  \href{http://arxiv.org/abs/2104.05029}{{\ttfamily arXiv:2104.05029
  [hep-ex]}}.

\bibitem{Boronat:2016tgd}
M.~Boronat, J.~Fuster, I.~Garcia, P.~Roloff, R.~Simoniello, and M.~Vos, ``{Jet
  reconstruction at high-energy electron\textendash{}positron colliders},''
  \href{http://dx.doi.org/10.1140/epjc/s10052-018-5594-6}{{\em Eur. Phys. J. C}
  {\bfseries 78} no.~2, (2018) 144},
  \href{http://arxiv.org/abs/1607.05039}{{\ttfamily arXiv:1607.05039
  [hep-ex]}}.

\bibitem{Larkoski:2017jix}
A.~J. Larkoski, I.~Moult, and B.~Nachman, ``{Jet Substructure at the Large
  Hadron Collider: A Review of Recent Advances in Theory and Machine
  Learning},''
\href{http://arxiv.org/abs/1709.04464}{{\ttfamily arXiv:1709.04464 [hep-ph]}}.

\bibitem{Marzani:2019hun}
S.~Marzani, G.~Soyez, and M.~Spannowsky,
  \href{http://dx.doi.org/10.1007/978-3-030-15709-8}{{\em {Looking inside jets:
  an introduction to jet substructure and boosted-object phenomenology}}},
  vol.~958.
\newblock Springer, 2019.
\newblock \href{http://arxiv.org/abs/1901.10342}{{\ttfamily arXiv:1901.10342
  [hep-ph]}}.

\bibitem{Abbate:2010xh}
R.~Abbate, M.~Fickinger, A.~H. Hoang, V.~Mateu, and I.~W. Stewart, ``{Thrust at
  $N^3LL$ with Power Corrections and a Precision Global Fit for alphas(mZ)},''
  \href{http://dx.doi.org/10.1103/PhysRevD.83.074021}{{\em Phys.Rev.}
  {\bfseries D83} (2011) 074021},
\href{http://arxiv.org/abs/1006.3080}{{\ttfamily arXiv:1006.3080 [hep-ph]}}.

\bibitem{Hoang:2015hka}
A.~H. Hoang, D.~W. Kolodrubetz, V.~Mateu, and I.~W. Stewart, ``{Precise
  determination of $\alpha_s$ from the $C$-parameter distribution},''
  \href{http://dx.doi.org/10.1103/PhysRevD.91.094018}{{\em Phys. Rev. D}
  {\bfseries 91} no.~9, (2015) 094018},
  \href{http://arxiv.org/abs/1501.04111}{{\ttfamily arXiv:1501.04111
  [hep-ph]}}.

\bibitem{dEnterria:2015mgr}
D.~d'Enterria, K.~Krajcz\'ar, and H.~Paukkunen, ``{Top-quark production in
  proton\textendash{}nucleus and nucleus\textendash{}nucleus collisions at LHC
  energies and beyond},''
  \href{http://dx.doi.org/10.1016/j.physletb.2015.04.044}{{\em Phys. Lett. B}
  {\bfseries 746} (2015) 64--72},
  \href{http://arxiv.org/abs/1501.05879}{{\ttfamily arXiv:1501.05879
  [hep-ph]}}.

\bibitem{Gras:2017jty}
P.~Gras, S.~H\"oche, D.~Kar, A.~Larkoski, L.~L\"onnblad, S.~Pl\"atzer,
  A.~Si\'odmok, P.~Skands, G.~Soyez, and J.~Thaler, ``{Systematics of
  quark/gluon tagging},'' \href{http://dx.doi.org/10.1007/JHEP07(2017)091}{{\em
  JHEP} {\bfseries 07} (2017) 091},
  \href{http://arxiv.org/abs/1704.03878}{{\ttfamily arXiv:1704.03878
  [hep-ph]}}.

\bibitem{Gao:2019mlt}
J.~Gao, Y.~Gong, W.-L. Ju, and L.~L. Yang, ``{Thrust distribution in Higgs
  decays at the next-to-leading order and beyond},''
  \href{http://dx.doi.org/10.1007/JHEP03(2019)030}{{\em JHEP} {\bfseries 03}
  (2019) 030}, \href{http://arxiv.org/abs/1901.02253}{{\ttfamily
  arXiv:1901.02253 [hep-ph]}}.

\bibitem{dEnterria:2020cgt}
D.~d'Enterria and C.~Yan, ``{Revised QCD effects on the Z $\to b\bar{b}$
  forward-backward asymmetry},''
  \href{http://arxiv.org/abs/2011.00530}{{\ttfamily arXiv:2011.00530
  [hep-ph]}}.

\bibitem{ILD:2019kmq}
{\bfseries ILD} Collaboration, H.~Abramowicz {\em et~al.}, ``{The ILD detector
  at the ILC},'' \href{http://arxiv.org/abs/1912.04601}{{\ttfamily
  arXiv:1912.04601 [physics.ins-det]}}.

\bibitem{Breidenbach:2021sdo}
M.~Breidenbach, J.~E. Brau, P.~Burrows, T.~Markiewicz, M.~Stanitzki, J.~Strube,
  and A.~P. White, ``{Updating the SiD Detector concept},''
  \href{http://arxiv.org/abs/2110.09965}{{\ttfamily arXiv:2110.09965
  [physics.ins-det]}}.

\bibitem{CLICdp:2018vnx}
{\bfseries CLICdp} Collaboration, D.~Arominski {\em et~al.}, ``{A detector for
  CLIC: main parameters and performance},''
  \href{http://arxiv.org/abs/1812.07337}{{\ttfamily arXiv:1812.07337
  [physics.ins-det]}}.

\bibitem{Bacchetta:2019fmz}
N.~Bacchetta {\em et~al.}, ``{CLD -- A Detector Concept for the FCC-ee},''
  \href{http://arxiv.org/abs/1911.12230}{{\ttfamily arXiv:1911.12230
  [physics.ins-det]}}.

\bibitem{CALICE:2012eac}
{\bfseries CALICE} Collaboration, C.~Adloff {\em et~al.}, ``{Hadronic energy
  resolution of a highly granular scintillator-steel hadron calorimeter using
  software compensation techniques},''
  \href{http://dx.doi.org/10.1088/1748-0221/7/09/P09017}{{\em JINST} {\bfseries
  7} (2012) P09017}, \href{http://arxiv.org/abs/1207.4210}{{\ttfamily
  arXiv:1207.4210 [physics.ins-det]}}.

\bibitem{Strom:2020wrm}
R.~Str\"om and P.~Roloff, ``{Physics potential for boosted topologies in
  top-quark pair production at a multi-TeV Compact Linear Collider},''
  \href{http://arxiv.org/abs/2008.05526}{{\ttfamily arXiv:2008.05526
  [hep-ex]}}.

\bibitem{Boronat:2014hva}
M.~Boronat, J.~Fuster, I.~Garcia, E.~Ros, and M.~Vos, ``{A robust jet
  reconstruction algorithm for high-energy lepton colliders},''
  \href{http://dx.doi.org/10.1016/j.physletb.2015.08.055}{{\em Phys. Lett. B}
  {\bfseries 750} (2015) 95--99},
  \href{http://arxiv.org/abs/1404.4294}{{\ttfamily arXiv:1404.4294 [hep-ex]}}.

\bibitem{Stewart:2015waa}
I.~W. Stewart, F.~J. Tackmann, J.~Thaler, C.~K. Vermilion, and T.~F. Wilkason,
  ``{XCone: N-jettiness as an Exclusive Cone Jet Algorithm},''
  \href{http://dx.doi.org/10.1007/JHEP11(2015)072}{{\em JHEP} {\bfseries 11}
  (2015) 072}, \href{http://arxiv.org/abs/1508.01516}{{\ttfamily
  arXiv:1508.01516 [hep-ph]}}.

\bibitem{Collamati:2021sbv}
F.~Collamati, C.~Curatolo, D.~Lucchesi, A.~Mereghetti, N.~Mokhov, M.~Palmer,
  and P.~Sala, ``{Advanced assessment of beam-induced background at a muon
  collider},'' \href{http://dx.doi.org/10.1088/1748-0221/16/11/P11009}{{\em
  JINST} {\bfseries 16} no.~11, (2021) P11009},
  \href{http://arxiv.org/abs/2105.09116}{{\ttfamily arXiv:2105.09116
  [physics.acc-ph]}}.

\bibitem{Ally:2022rgk}
D.~Ally, L.~Carpenter, T.~Holmes, L.~Lee, and P.~Wagenknecht, ``{Strategies for
  Beam-Induced Background Reduction at Muon Colliders},''
  \href{http://arxiv.org/abs/2203.06773}{{\ttfamily arXiv:2203.06773
  [physics.ins-det]}}.

\bibitem{MuonCollider:2022ded}
{\bfseries Muon Collider} Collaboration, N.~Bartosik {\em et~al.}, ``{Simulated
  Detector Performance at the Muon Collider},'' in {\em {2022 Snowmass Summer
  Study}}.
\newblock 3, 2022.
\newblock \href{http://arxiv.org/abs/2203.07964}{{\ttfamily arXiv:2203.07964
  [hep-ex]}}.

\bibitem{Thaler:2010tr}
J.~Thaler and K.~Van~Tilburg, ``{Identifying Boosted Objects with
  N-subjettiness},'' \href{http://dx.doi.org/10.1007/JHEP03(2011)015}{{\em
  JHEP} {\bfseries 03} (2011) 015},
\href{http://arxiv.org/abs/1011.2268}{{\ttfamily arXiv:1011.2268 [hep-ph]}}.

\bibitem{Thaler:2011gf}
J.~Thaler and K.~Van~Tilburg, ``{Maximizing Boosted Top Identification by
  Minimizing N-subjettiness},''
  \href{http://dx.doi.org/10.1007/JHEP02(2012)093}{{\em JHEP} {\bfseries 02}
  (2012) 093},
\href{http://arxiv.org/abs/1108.2701}{{\ttfamily arXiv:1108.2701 [hep-ph]}}.

\bibitem{Larkoski:2014gra}
A.~J. Larkoski, I.~Moult, and D.~Neill, ``{Power Counting to Better Jet
  Observables},''
\href{http://arxiv.org/abs/1409.6298}{{\ttfamily arXiv:1409.6298 [hep-ph]}}.

\bibitem{Anderle:2021wcy}
D.~P. Anderle {\em et~al.}, ``{Electron-ion collider in China},''
  \href{http://dx.doi.org/10.1007/s11467-021-1062-0}{{\em Front. Phys.
  (Beijing)} {\bfseries 16} no.~6, (2021) 64701},
  \href{http://arxiv.org/abs/2102.09222}{{\ttfamily arXiv:2102.09222
  [nucl-ex]}}.

\bibitem{Jacob:1984vnf}
``{Proceedings, ECFA-CERN Workshop on large hadron collider in the LEP tunnel}:
  {Lausanne and Geneva, Switzerland, March 21-27 March, 1984}.''
  \protect{CERN-84-10-V-2, CERN-84-10-V-1, ECFA-84-85, CERN-YELLOW-84-10-V-1},
  1984.

\bibitem{LHeCStudyGroup:2012zhm}
{\bfseries LHeC Study Group} Collaboration, J.~L. Abelleira~Fernandez {\em
  et~al.}, ``{A Large Hadron Electron Collider at CERN: Report on the Physics
  and Design Concepts for Machine and Detector},''
  \href{http://dx.doi.org/10.1088/0954-3899/39/7/075001}{{\em J. Phys. G}
  {\bfseries 39} (2012) 075001},
  \href{http://arxiv.org/abs/1206.2913}{{\ttfamily arXiv:1206.2913
  [physics.acc-ph]}}.

\bibitem{LHeC:2020van}
{\bfseries LHeC, FCC-he Study Group} Collaboration, P.~Agostini {\em et~al.},
  ``{The Large Hadron-Electron Collider at the HL-LHC},''
  \href{http://dx.doi.org/10.1088/1361-6471/abf3ba}{{\em J. Phys. G} {\bfseries
  48} no.~11, (2021) 110501}, \href{http://arxiv.org/abs/2007.14491}{{\ttfamily
  arXiv:2007.14491 [hep-ex]}}.

\bibitem{Andre:2022xeh}
K.~D.~J. Andr\'e {\em et~al.}, ``{An experiment for electron-hadron scattering
  at the LHC},'' \href{http://dx.doi.org/10.1140/epjc/s10052-021-09967-z}{{\em
  Eur. Phys. J. C} {\bfseries 82} no.~1, (2022) 40},
  \href{http://arxiv.org/abs/2201.02436}{{\ttfamily arXiv:2201.02436
  [hep-ex]}}.

\bibitem{calohep}
E.~Coleman, M.~Freytsis, A.~Hinzmann, M.~Narain, J.~Thaler, N.~Tran, and
  C.~Vernieri, ``The importance of calorimetry for highly-boosted jet
  substructure,'' \href{http://dx.doi.org/10.1088/1748-0221/13/01/t01003}{{\em
  Journal of Instrumentation} {\bfseries 13} no.~01, (Jan, 2018)
  T01003–T01003}. \url{http://dx.doi.org/10.1088/1748-0221/13/01/T01003}.

\bibitem{larkoski2015tracking}
A.~J. Larkoski, F.~Maltoni, and M.~Selvaggi, ``Tracking down hyper-boosted top
  quarks,'' \href{http://arxiv.org/abs/1503.03347}{{\ttfamily arXiv:1503.03347
  [hep-ph]}}.

\bibitem{Chang:2013rca}
H.-M. Chang, M.~Procura, J.~Thaler, and W.~J. Waalewijn, ``{Calculating
  Track-Based Observables for the LHC},''
  \href{http://dx.doi.org/10.1103/PhysRevLett.111.102002}{{\em Phys. Rev.
  Lett.} {\bfseries 111} (2013) 102002},
  \href{http://arxiv.org/abs/1303.6637}{{\ttfamily arXiv:1303.6637 [hep-ph]}}.

\bibitem{Elder:2018mcr}
B.~T. Elder and J.~Thaler, ``{Aspects of Track-Assisted Mass},''
  \href{http://dx.doi.org/10.1007/JHEP03(2019)104}{{\em JHEP} {\bfseries 03}
  (2019) 104}, \href{http://arxiv.org/abs/1805.11109}{{\ttfamily
  arXiv:1805.11109 [hep-ph]}}.

\bibitem{Spannowsky:2015eba}
M.~Spannowsky and M.~Stoll, ``{Tracking New Physics at the LHC and beyond},''
  \href{http://dx.doi.org/10.1103/PhysRevD.92.054033}{{\em Phys. Rev. D}
  {\bfseries 92} no.~5, (2015) 054033},
  \href{http://arxiv.org/abs/1505.01921}{{\ttfamily arXiv:1505.01921
  [hep-ph]}}.

\bibitem{ATLAS-CONF-2016-035}
{\bfseries ATLAS} Collaboration, ``{Jet mass reconstruction with the ATLAS
  Detector in early Run 2 data}.'' \protect{ATLAS-CONF-2016-035}, Jul, 2016.
\newblock \url{http://cds.cern.ch/record/2200211}.

\bibitem{Gouskos:2642475}
L.~Gouskos, A.~Sung, and J.~Incandela, ``{Search for stop scalar quarks at
  FCC-hh},'' tech. rep., CERN, Geneva, Oct, 2018.
\newblock \url{https://cds.cern.ch/record/2642475}.

\bibitem{Aleksa:2019pvl}
M.~Aleksa {\em et~al.}, ``{Calorimeters for the FCC-hh},''
  \href{http://arxiv.org/abs/1912.09962}{{\ttfamily arXiv:1912.09962
  [physics.ins-det]}}.

\bibitem{Workman:2022ynf}
{\bfseries Particle Data Group} Collaboration, R.~L. Workman, ``{Review of
  Particle Physics},'' {\em PTEP} {\bfseries 2022} (2022) 083C01.

\bibitem{dEnterria:2020cpv}
D.~d'Enterria and V.~Jacobsen, ``{Improved strong coupling determinations from
  hadronic decays of electroweak bosons at N$^3$LO accuracy},''
  \href{http://arxiv.org/abs/2005.04545}{{\ttfamily arXiv:2005.04545
  [hep-ph]}}.

\bibitem{ParticleDataGroup:2020ssz}
{\bfseries Particle Data Group} Collaboration, P.~A. Zyla {\em et~al.},
  ``{Review of Particle Physics},''
  \href{http://dx.doi.org/10.1093/ptep/ptaa104}{{\em PTEP} {\bfseries 2020}
  no.~8, (2020) 083C01}.

\bibitem{Aoki:2021kgd}
Y.~Aoki {\em et~al.}, ``{FLAG Review 2021},''
  \href{http://arxiv.org/abs/2111.09849}{{\ttfamily arXiv:2111.09849
  [hep-lat]}}.

\bibitem{Blondel:2021zix}
A.~Blondel and E.~Gianfelice, ``{The challenges of beam polarization and
  keV-scale centre-of-mass energy calibration at the FCC-ee},''
  \href{http://dx.doi.org/10.1140/epjp/s13360-021-02038-y}{{\em Eur. Phys. J.
  Plus} {\bfseries 136} no.~11, (2021) 1103}.

\bibitem{FCC:2018evy}
{\bfseries FCC} Collaboration, A.~Abada {\em et~al.}, ``{FCC-ee: The Lepton
  Collider}: {Future Circular Collider Conceptual Design Report Volume 2},''
  \href{http://dx.doi.org/10.1140/epjst/e2019-900045-4}{{\em Eur. Phys. J. ST}
  {\bfseries 228} no.~2, (2019) 261--623}.

\bibitem{Janot:2015gjr}
P.~Janot, ``{Direct measurement of $\alpha_{QED}(m_{Z}^{2})$ at the FCC-ee},''
  \href{http://dx.doi.org/10.1007/JHEP02(2016)053}{{\em JHEP} {\bfseries 02}
  (2016) 053}, \href{http://arxiv.org/abs/1512.05544}{{\ttfamily
  arXiv:1512.05544 [hep-ph]}}. [Erratum: JHEP 11, 164 (2017)].

\bibitem{Proceedings:2019vxr}
A.~Blondel, J.~Gluza, S.~Jadach, P.~Janot, and T.~Riemann, eds.,
  \href{http://dx.doi.org/10.23731/CYRM-2020-003}{{\em {Theory for the FCC-ee}:
  {Report on the 11th FCC-ee Workshop Theory and Experiments}}}, vol.~3/2020 of
  {\em CERN Yellow Reports: Monographs}.
\newblock CERN, Geneva, May, 2019.
\newblock \href{http://arxiv.org/abs/1905.05078}{{\ttfamily arXiv:1905.05078
  [hep-ph]}}.

\bibitem{Accardi:2012qut}
A.~Accardi {\em et~al.}, ``{Electron Ion Collider: The Next QCD Frontier}:
  {Understanding the glue that binds us all},''
  \href{http://dx.doi.org/10.1140/epja/i2016-16268-9}{{\em Eur. Phys. J. A}
  {\bfseries 52} no.~9, (2016) 268},
  \href{http://arxiv.org/abs/1212.1701}{{\ttfamily arXiv:1212.1701 [nucl-ex]}}.

\bibitem{Hou:2019efy}
T.-J. Hou {\em et~al.}, ``{New CTEQ global analysis of quantum chromodynamics
  with high-precision data from the LHC},''
  \href{http://dx.doi.org/10.1103/PhysRevD.103.014013}{{\em Phys. Rev. D}
  {\bfseries 103} no.~1, (2021) 014013},
  \href{http://arxiv.org/abs/1912.10053}{{\ttfamily arXiv:1912.10053
  [hep-ph]}}.

\bibitem{Deur:2014vea}
A.~Deur, Y.~Prok, V.~Burkert, D.~Crabb, F.~X. Girod, K.~A. Griffioen, N.~Guler,
  S.~E. Kuhn, and N.~Kvaltine, ``{High precision determination of the $Q^2$
  evolution of the Bjorken Sum},''
  \href{http://dx.doi.org/10.1103/PhysRevD.90.012009}{{\em Phys. Rev. D}
  {\bfseries 90} no.~1, (2014) 012009},
  \href{http://arxiv.org/abs/1405.7854}{{\ttfamily arXiv:1405.7854 [nucl-ex]}}.

\bibitem{Kovarik:2019xvh}
K.~Kova\v{r}\'\i{}k, P.~M. Nadolsky, and D.~E. Soper, ``{Hadronic structure in
  high-energy collisions},''
  \href{http://dx.doi.org/10.1103/RevModPhys.92.045003}{{\em Rev. Mod. Phys.}
  {\bfseries 92} no.~4, (2020) 045003},
  \href{http://arxiv.org/abs/1905.06957}{{\ttfamily arXiv:1905.06957
  [hep-ph]}}.

\bibitem{Luscher:1991wu}
M.~Luscher, P.~Weisz, and U.~Wolff, ``{A Numerical method to compute the
  running coupling in asymptotically free theories},''
  \href{http://dx.doi.org/10.1016/0550-3213(91)90298-C}{{\em Nucl. Phys. B}
  {\bfseries 359} (1991) 221--243}.

\bibitem{Luscher:1993gh}
M.~Luscher, R.~Sommer, P.~Weisz, and U.~Wolff, ``{A Precise determination of
  the running coupling in the SU(3) Yang-Mills theory},''
  \href{http://dx.doi.org/10.1016/0550-3213(94)90629-7}{{\em Nucl. Phys. B}
  {\bfseries 413} (1994) 481--502},
  \href{http://arxiv.org/abs/hep-lat/9309005}{{\ttfamily
  arXiv:hep-lat/9309005}}.

\bibitem{deDivitiis:1994yz}
{\bfseries Alpha} Collaboration, G.~de~Divitiis, R.~Frezzotti, M.~Guagnelli,
  M.~Luscher, R.~Petronzio, R.~Sommer, P.~Weisz, and U.~Wolff, ``{Universality
  and the approach to the continuum limit in lattice gauge theory},''
  \href{http://dx.doi.org/10.1016/0550-3213(94)00019-B}{{\em Nucl. Phys. B}
  {\bfseries 437} (1995) 447--470},
  \href{http://arxiv.org/abs/hep-lat/9411017}{{\ttfamily
  arXiv:hep-lat/9411017}}.

\bibitem{Jansen:1995ck}
K.~Jansen, C.~Liu, M.~Luscher, H.~Simma, S.~Sint, R.~Sommer, P.~Weisz, and
  U.~Wolff, ``{Nonperturbative renormalization of lattice QCD at all scales},''
  \href{http://dx.doi.org/10.1016/0370-2693(96)00075-5}{{\em Phys. Lett. B}
  {\bfseries 372} (1996) 275--282},
  \href{http://arxiv.org/abs/hep-lat/9512009}{{\ttfamily
  arXiv:hep-lat/9512009}}.

\bibitem{Symanzik:1981wd}
K.~Symanzik, ``{Schrodinger Representation and Casimir Effect in Renormalizable
  Quantum Field Theory},''
  \href{http://dx.doi.org/10.1016/0550-3213(81)90482-X}{{\em Nucl. Phys. B}
  {\bfseries 190} (1981) 1--44}.

\bibitem{Luscher:1985iu}
M.~Luscher, ``{SCHRODINGER REPRESENTATION IN QUANTUM FIELD THEORY},''
  \href{http://dx.doi.org/10.1016/0550-3213(85)90210-X}{{\em Nucl. Phys. B}
  {\bfseries 254} (1985) 52--57}.

\bibitem{Luscher:1992an}
M.~Luscher, R.~Narayanan, P.~Weisz, and U.~Wolff, ``{The Schr\"odinger
  functional: A Renormalizable probe for nonAbelian gauge theories},''
  \href{http://dx.doi.org/10.1016/0550-3213(92)90466-O}{{\em Nucl. Phys. B}
  {\bfseries 384} (1992) 168--228},
  \href{http://arxiv.org/abs/hep-lat/9207009}{{\ttfamily
  arXiv:hep-lat/9207009}}.

\bibitem{Sint:1993un}
S.~Sint, ``{On the Schrodinger functional in QCD},''
  \href{http://dx.doi.org/10.1016/0550-3213(94)90228-3}{{\em Nucl. Phys. B}
  {\bfseries 421} (1994) 135--158},
  \href{http://arxiv.org/abs/hep-lat/9312079}{{\ttfamily
  arXiv:hep-lat/9312079}}.

\bibitem{Bode:1999sm}
{\bfseries ALPHA} Collaboration, A.~Bode, P.~Weisz, and U.~Wolff, ``{Two loop
  computation of the Schrodinger functional in lattice QCD},''
  \href{http://dx.doi.org/10.1016/S0550-3213(00)00187-5}{{\em Nucl. Phys. B}
  {\bfseries 576} (2000) 517--539},
  \href{http://arxiv.org/abs/hep-lat/9911018}{{\ttfamily
  arXiv:hep-lat/9911018}}. [Erratum: Nucl.Phys.B 608, 481--481 (2001), Erratum:
  Nucl.Phys.B 600, 453--453 (2001)].

\bibitem{Bruno:2017gxd}
{\bfseries ALPHA} Collaboration, M.~Bruno, M.~Dalla~Brida, P.~Fritzsch,
  T.~Korzec, A.~Ramos, S.~Schaefer, H.~Simma, S.~Sint, and R.~Sommer, ``{QCD
  Coupling from a Nonperturbative Determination of the Three-Flavor $\Lambda$
  Parameter},'' \href{http://dx.doi.org/10.1103/PhysRevLett.119.102001}{{\em
  Phys. Rev. Lett.} {\bfseries 119} no.~10, (2017) 102001},
  \href{http://arxiv.org/abs/1706.03821}{{\ttfamily arXiv:1706.03821
  [hep-lat]}}.

\bibitem{Mason:2005zx}
{\bfseries HPQCD, UKQCD} Collaboration, Q.~Mason, H.~D. Trottier, C.~T.~H.
  Davies, K.~Foley, A.~Gray, G.~P. Lepage, M.~Nobes, and J.~Shigemitsu,
  ``{Accurate determinations of alpha(s) from realistic lattice QCD},''
  \href{http://dx.doi.org/10.1103/PhysRevLett.95.052002}{{\em Phys. Rev. Lett.}
  {\bfseries 95} (2005) 052002},
  \href{http://arxiv.org/abs/hep-lat/0503005}{{\ttfamily
  arXiv:hep-lat/0503005}}.

\bibitem{Davies:2008sw}
{\bfseries HPQCD} Collaboration, C.~T.~H. Davies, K.~Hornbostel, I.~D. Kendall,
  G.~P. Lepage, C.~McNeile, J.~Shigemitsu, and H.~Trottier, ``{Update: Accurate
  Determinations of alpha(s) from Realistic Lattice QCD},''
  \href{http://dx.doi.org/10.1103/PhysRevD.78.114507}{{\em Phys. Rev. D}
  {\bfseries 78} (2008) 114507},
  \href{http://arxiv.org/abs/0807.1687}{{\ttfamily arXiv:0807.1687 [hep-lat]}}.

\bibitem{Maltman:2008bx}
K.~Maltman, D.~Leinweber, P.~Moran, and A.~Sternbeck, ``{The Realistic Lattice
  Determination of alpha(s)(M(Z)) Revisited},''
  \href{http://dx.doi.org/10.1103/PhysRevD.78.114504}{{\em Phys. Rev. D}
  {\bfseries 78} (2008) 114504},
  \href{http://arxiv.org/abs/0807.2020}{{\ttfamily arXiv:0807.2020 [hep-lat]}}.

\bibitem{McNeile:2010ji}
C.~McNeile, C.~T.~H. Davies, E.~Follana, K.~Hornbostel, and G.~P. Lepage,
  ``{High-Precision c and b Masses, and QCD Coupling from Current-Current
  Correlators in Lattice and Continuum QCD},''
  \href{http://dx.doi.org/10.1103/PhysRevD.82.034512}{{\em Phys. Rev. D}
  {\bfseries 82} (2010) 034512},
  \href{http://arxiv.org/abs/1004.4285}{{\ttfamily arXiv:1004.4285 [hep-lat]}}.

\bibitem{Pineda:2000gza}
A.~Pineda and J.~Soto, ``{The Renormalization group improvement of the QCD
  static potentials},''
  \href{http://dx.doi.org/10.1016/S0370-2693(00)01261-2}{{\em Phys. Lett. B}
  {\bfseries 495} (2000) 323--328},
  \href{http://arxiv.org/abs/hep-ph/0007197}{{\ttfamily arXiv:hep-ph/0007197}}.

\bibitem{Brambilla:2004jw}
N.~Brambilla, A.~Pineda, J.~Soto, and A.~Vairo, ``{Effective Field Theories for
  Heavy Quarkonium},'' \href{http://dx.doi.org/10.1103/RevModPhys.77.1423}{{\em
  Rev. Mod. Phys.} {\bfseries 77} (2005) 1423},
  \href{http://arxiv.org/abs/hep-ph/0410047}{{\ttfamily arXiv:hep-ph/0410047}}.

\bibitem{Brambilla:2009bi}
N.~Brambilla, A.~Vairo, X.~Garcia~i Tormo, and J.~Soto, ``{The QCD static
  energy at NNNLL},'' \href{http://dx.doi.org/10.1103/PhysRevD.80.034016}{{\em
  Phys. Rev. D} {\bfseries 80} (2009) 034016},
  \href{http://arxiv.org/abs/0906.1390}{{\ttfamily arXiv:0906.1390 [hep-ph]}}.

\bibitem{Bazavov:2014soa}
A.~Bazavov, N.~Brambilla, X.~G. Tormo, I, P.~Petreczky, J.~Soto, and A.~Vairo,
  ``{Determination of $\alpha_s$ from the QCD static energy: An update},''
  \href{http://dx.doi.org/10.1103/PhysRevD.90.074038}{{\em Phys. Rev. D}
  {\bfseries 90} no.~7, (2014) 074038},
  \href{http://arxiv.org/abs/1407.8437}{{\ttfamily arXiv:1407.8437 [hep-ph]}}.
  [Erratum: Phys.Rev.D 101, 119902 (2020)].

\bibitem{Takaura:2018lpw}
H.~Takaura, T.~Kaneko, Y.~Kiyo, and Y.~Sumino, ``{Determination of $\alpha_s$
  from static QCD potential with renormalon subtraction},''
  \href{http://dx.doi.org/10.1016/j.physletb.2018.12.060}{{\em Phys. Lett. B}
  {\bfseries 789} (2019) 598--602},
  \href{http://arxiv.org/abs/1808.01632}{{\ttfamily arXiv:1808.01632
  [hep-ph]}}.

\bibitem{Takaura:2018vcy}
H.~Takaura, T.~Kaneko, Y.~Kiyo, and Y.~Sumino, ``{Determination of $\alpha_s$
  from static QCD potential: OPE with renormalon subtraction and lattice
  QCD},'' \href{http://dx.doi.org/10.1007/JHEP04(2019)155}{{\em JHEP}
  {\bfseries 04} (2019) 155}, \href{http://arxiv.org/abs/1808.01643}{{\ttfamily
  arXiv:1808.01643 [hep-ph]}}.

\bibitem{Bazavov:2018wmo}
{\bfseries TUMQCD} Collaboration, A.~Bazavov, N.~Brambilla, P.~Petreczky,
  A.~Vairo, and J.~H. Weber, ``{Color screening in (2+1)-flavor QCD},''
  \href{http://dx.doi.org/10.1103/PhysRevD.98.054511}{{\em Phys. Rev. D}
  {\bfseries 98} no.~5, (2018) 054511},
  \href{http://arxiv.org/abs/1804.10600}{{\ttfamily arXiv:1804.10600
  [hep-lat]}}.

\bibitem{Bazavov:2019qoo}
{\bfseries TUMQCD} Collaboration, A.~Bazavov, N.~Brambilla, X.~Garcia~i Tormo,
  P.~Petreczky, J.~Soto, A.~Vairo, and J.~H. Weber, ``{Determination of the QCD
  coupling from the static energy and the free energy},''
  \href{http://dx.doi.org/10.1103/PhysRevD.100.114511}{{\em Phys. Rev. D}
  {\bfseries 100} no.~11, (2019) 114511},
  \href{http://arxiv.org/abs/1907.11747}{{\ttfamily arXiv:1907.11747
  [hep-lat]}}.

\bibitem{Ayala:2020odx}
C.~Ayala, X.~Lobregat, and A.~Pineda, ``{Determination of $\alpha(M_z)$ from an
  hyperasymptotic approximation to the energy of a static quark-antiquark
  pair},'' \href{http://dx.doi.org/10.1007/JHEP09(2020)016}{{\em JHEP}
  {\bfseries 09} (2020) 016}, \href{http://arxiv.org/abs/2005.12301}{{\ttfamily
  arXiv:2005.12301 [hep-ph]}}.

\bibitem{Komijani:2020kst}
J.~Komijani, P.~Petreczky, and J.~H. Weber, ``{Strong coupling constant and
  quark masses from lattice QCD},''
  \href{http://dx.doi.org/10.1016/j.ppnp.2020.103788}{{\em Prog. Part. Nucl.
  Phys.} {\bfseries 113} (2020) 103788},
  \href{http://arxiv.org/abs/2003.11703}{{\ttfamily arXiv:2003.11703
  [hep-lat]}}.

\bibitem{Sturm:2008eb}
C.~Sturm, ``{Moments of Heavy Quark Current Correlators at Four-Loop Order in
  Perturbative QCD},''
  \href{http://dx.doi.org/10.1088/1126-6708/2008/09/075}{{\em JHEP} {\bfseries
  09} (2008) 075}, \href{http://arxiv.org/abs/0805.3358}{{\ttfamily
  arXiv:0805.3358 [hep-ph]}}.

\bibitem{Kiyo:2009gb}
Y.~Kiyo, A.~Maier, P.~Maierhofer, and P.~Marquard, ``{Reconstruction of heavy
  quark current correlators at O(alpha(s)**3)},''
  \href{http://dx.doi.org/10.1016/j.nuclphysb.2009.08.010}{{\em Nucl. Phys. B}
  {\bfseries 823} (2009) 269--287},
  \href{http://arxiv.org/abs/0907.2120}{{\ttfamily arXiv:0907.2120 [hep-ph]}}.

\bibitem{Maier:2009fz}
A.~Maier, P.~Maierhofer, P.~Marquard, and A.~V. Smirnov, ``{Low energy moments
  of heavy quark current correlators at four loops},''
  \href{http://dx.doi.org/10.1016/j.nuclphysb.2009.08.011}{{\em Nucl. Phys. B}
  {\bfseries 824} (2010) 1--18},
  \href{http://arxiv.org/abs/0907.2117}{{\ttfamily arXiv:0907.2117 [hep-ph]}}.

\bibitem{HPQCD:2008kxl}
{\bfseries HPQCD} Collaboration, I.~Allison {\em et~al.}, ``{High-Precision
  Charm-Quark Mass from Current-Current Correlators in Lattice and Continuum
  QCD},'' \href{http://dx.doi.org/10.1103/PhysRevD.78.054513}{{\em Phys. Rev.
  D} {\bfseries 78} (2008) 054513},
  \href{http://arxiv.org/abs/0805.2999}{{\ttfamily arXiv:0805.2999 [hep-lat]}}.

\bibitem{Chakraborty:2014aca}
B.~Chakraborty, C.~T.~H. Davies, B.~Galloway, P.~Knecht, J.~Koponen, G.~C.
  Donald, R.~J. Dowdall, G.~P. Lepage, and C.~McNeile, ``{High-precision quark
  masses and QCD coupling from $n_f=4$ lattice QCD},''
  \href{http://dx.doi.org/10.1103/PhysRevD.91.054508}{{\em Phys. Rev. D}
  {\bfseries 91} no.~5, (2015) 054508},
  \href{http://arxiv.org/abs/1408.4169}{{\ttfamily arXiv:1408.4169 [hep-lat]}}.

\bibitem{Maezawa:2016vgv}
Y.~Maezawa and P.~Petreczky, ``{Quark masses and strong coupling constant in
  2+1 flavor QCD},'' \href{http://dx.doi.org/10.1103/PhysRevD.94.034507}{{\em
  Phys. Rev. D} {\bfseries 94} no.~3, (2016) 034507},
  \href{http://arxiv.org/abs/1606.08798}{{\ttfamily arXiv:1606.08798
  [hep-lat]}}.

\bibitem{Petreczky:2019ozv}
P.~Petreczky and J.~H. Weber, ``{Strong coupling constant and heavy quark
  masses in ( 2+1 )-flavor QCD},''
  \href{http://dx.doi.org/10.1103/PhysRevD.100.034519}{{\em Phys. Rev. D}
  {\bfseries 100} no.~3, (2019) 034519},
  \href{http://arxiv.org/abs/1901.06424}{{\ttfamily arXiv:1901.06424
  [hep-lat]}}.

\bibitem{Petreczky:2020tky}
P.~Petreczky and J.~H. Weber, ``{Strong coupling constant from moments of
  quarkonium correlators revisited},''
  \href{http://dx.doi.org/10.1140/epjc/s10052-022-09998-0}{{\em Eur. Phys. J.
  C} {\bfseries 82} no.~1, (2022) 64},
  \href{http://arxiv.org/abs/2012.06193}{{\ttfamily arXiv:2012.06193
  [hep-lat]}}.

\bibitem{Baikov:2008jh}
P.~A. Baikov, K.~G. Chetyrkin, and J.~H. Kuhn, ``{Order alpha**4(s) QCD
  Corrections to Z and tau Decays},''
  \href{http://dx.doi.org/10.1103/PhysRevLett.101.012002}{{\em Phys. Rev.
  Lett.} {\bfseries 101} (2008) 012002},
  \href{http://arxiv.org/abs/0801.1821}{{\ttfamily arXiv:0801.1821 [hep-ph]}}.

\bibitem{JLQCD:2008bwj}
{\bfseries JLQCD, TWQCD} Collaboration, E.~Shintani, S.~Aoki, T.~W. Chiu,
  S.~Hashimoto, T.~H. Hsieh, T.~Kaneko, H.~Matsufuru, J.~Noaki, T.~Onogi, and
  N.~Yamada, ``{Lattice study of the vacuum polarization function and
  determination of the strong coupling constant},''
  \href{http://dx.doi.org/10.1103/PhysRevD.79.074510}{{\em Phys. Rev. D}
  {\bfseries 79} (2009) 074510},
  \href{http://arxiv.org/abs/0807.0556}{{\ttfamily arXiv:0807.0556 [hep-lat]}}.

\bibitem{Shintani:2010ph}
E.~Shintani, S.~Aoki, H.~Fukaya, S.~Hashimoto, T.~Kaneko, T.~Onogi, and
  N.~Yamada, ``{Strong coupling constant from vacuum polarization functions in
  three-flavor lattice QCD with dynamical overlap fermions},''
  \href{http://dx.doi.org/10.1103/PhysRevD.82.074505}{{\em Phys. Rev. D}
  {\bfseries 82} no.~7, (2010) 074505},
  \href{http://arxiv.org/abs/1002.0371}{{\ttfamily arXiv:1002.0371 [hep-lat]}}.
  [Erratum: Phys.Rev.D 89, 099903 (2014)].

\bibitem{Hudspith:2018bpz}
R.~J. Hudspith, R.~Lewis, K.~Maltman, and E.~Shintani, ``{$\alpha_s$ from the
  Lattice Hadronic Vacuum Polarisation},''
  \href{http://arxiv.org/abs/1804.10286}{{\ttfamily arXiv:1804.10286
  [hep-lat]}}.

\bibitem{Cali:2020hrj}
S.~Cali, K.~Cichy, P.~Korcyl, and J.~Simeth, ``{Running coupling constant from
  position-space current-current correlation functions in three-flavor lattice
  QCD},'' \href{http://dx.doi.org/10.1103/PhysRevLett.125.242002}{{\em Phys.
  Rev. Lett.} {\bfseries 125} (2020) 242002},
  \href{http://arxiv.org/abs/2003.05781}{{\ttfamily arXiv:2003.05781
  [hep-lat]}}.

\bibitem{Alles:1996ka}
B.~Alles, D.~Henty, H.~Panagopoulos, C.~Parrinello, C.~Pittori, and D.~G.
  Richards, ``{$\alpha_s$ from the nonperturbatively renormalised lattice three
  gluon vertex},'' \href{http://dx.doi.org/10.1016/S0550-3213(97)00483-5}{{\em
  Nucl. Phys. B} {\bfseries 502} (1997) 325--342},
  \href{http://arxiv.org/abs/hep-lat/9605033}{{\ttfamily
  arXiv:hep-lat/9605033}}.

\bibitem{Boucaud:2001qz}
P.~Boucaud, J.~P. Leroy, H.~Moutarde, J.~Micheli, O.~Pene,
  J.~Rodriguez-Quintero, and C.~Roiesnel, ``{Preliminary calculation of
  alpha(s) from Green functions with dynamical quarks},''
  \href{http://dx.doi.org/10.1088/1126-6708/2002/01/046}{{\em JHEP} {\bfseries
  01} (2002) 046}, \href{http://arxiv.org/abs/hep-ph/0107278}{{\ttfamily
  arXiv:hep-ph/0107278}}.

\bibitem{Blossier:2010ky}
{\bfseries ETM} Collaboration, B.~Blossier, P.~Boucaud, F.~De~soto, V.~Morenas,
  M.~Gravina, O.~Pene, and J.~Rodriguez-Quintero, ``{Ghost-gluon coupling,
  power corrections and $\Lambda_{\overline {\rm MS}}$ from twisted-mass
  lattice QCD at Nf=2},''
  \href{http://dx.doi.org/10.1103/PhysRevD.82.034510}{{\em Phys. Rev. D}
  {\bfseries 82} (2010) 034510},
  \href{http://arxiv.org/abs/1005.5290}{{\ttfamily arXiv:1005.5290 [hep-lat]}}.

\bibitem{Blossier:2011tf}
B.~Blossier, P.~Boucaud, M.~Brinet, F.~De~Soto, X.~Du, M.~Gravina, V.~Morenas,
  O.~Pene, K.~Petrov, and J.~Rodriguez-Quintero, ``{Ghost-gluon coupling, power
  corrections and $\Lambda_{\bar{\rm MS}}$ from lattice QCD with a dynamical
  charm},'' \href{http://dx.doi.org/10.1103/PhysRevD.85.034503}{{\em Phys. Rev.
  D} {\bfseries 85} (2012) 034503},
  \href{http://arxiv.org/abs/1110.5829}{{\ttfamily arXiv:1110.5829 [hep-lat]}}.

\bibitem{Blossier:2012ef}
B.~Blossier, P.~Boucaud, M.~Brinet, F.~De~Soto, X.~Du, V.~Morenas, O.~Pene,
  K.~Petrov, and J.~Rodriguez-Quintero, ``{The Strong running coupling at
  $\tau$ and $Z_0$ mass scales from lattice QCD},''
  \href{http://dx.doi.org/10.1103/PhysRevLett.108.262002}{{\em Phys. Rev.
  Lett.} {\bfseries 108} (2012) 262002},
  \href{http://arxiv.org/abs/1201.5770}{{\ttfamily arXiv:1201.5770 [hep-ph]}}.

\bibitem{Blossier:2013ioa}
{\bfseries ETM} Collaboration, B.~Blossier, P.~Boucaud, M.~Brinet, F.~De~Soto,
  V.~Morenas, O.~Pene, K.~Petrov, and J.~Rodriguez-Quintero, ``{High statistics
  determination of the strong coupling constant in Taylor scheme and its OPE
  Wilson coefficient from lattice QCD with a dynamical charm},''
  \href{http://dx.doi.org/10.1103/PhysRevD.89.014507}{{\em Phys. Rev. D}
  {\bfseries 89} no.~1, (2014) 014507},
  \href{http://arxiv.org/abs/1310.3763}{{\ttfamily arXiv:1310.3763 [hep-ph]}}.

\bibitem{Zafeiropoulos:2019flq}
S.~Zafeiropoulos, P.~Boucaud, F.~De~Soto, J.~Rodr\'\i{}guez-Quintero, and
  J.~Segovia, ``{Strong Running Coupling from the Gauge Sector of Domain Wall
  Lattice QCD with Physical Quark Masses},''
  \href{http://dx.doi.org/10.1103/PhysRevLett.122.162002}{{\em Phys. Rev.
  Lett.} {\bfseries 122} no.~16, (2019) 162002},
  \href{http://arxiv.org/abs/1902.08148}{{\ttfamily arXiv:1902.08148
  [hep-ph]}}.

\bibitem{DallaBrida:2019mqg}
{\bfseries ALPHA} Collaboration, M.~Dalla~Brida, R.~H\"ollwieser, F.~Knechtli,
  T.~Korzec, A.~Ramos, and R.~Sommer, ``{Non-perturbative renormalization by
  decoupling},'' \href{http://dx.doi.org/10.1016/j.physletb.2020.135571}{{\em
  Phys. Lett. B} {\bfseries 807} (2020) 135571},
  \href{http://arxiv.org/abs/1912.06001}{{\ttfamily arXiv:1912.06001
  [hep-lat]}}.

\bibitem{Chetyrkin:1994ex}
K.~G. Chetyrkin and J.~H. Kuhn, ``{Quartic mass corrections to R(had)},''
  \href{http://dx.doi.org/10.1016/0550-3213(94)90605-X}{{\em Nucl. Phys. B}
  {\bfseries 432} (1994) 337--350},
  \href{http://arxiv.org/abs/hep-ph/9406299}{{\ttfamily arXiv:hep-ph/9406299}}.

\bibitem{Kneur:2015dda}
J.-L. Kneur and A.~Neveu, ``{Chiral condensate from renormalization group
  optimized perturbation},''
  \href{http://dx.doi.org/10.1103/PhysRevD.92.074027}{{\em Phys. Rev. D}
  {\bfseries 92} no.~7, (2015) 074027},
  \href{http://arxiv.org/abs/1506.07506}{{\ttfamily arXiv:1506.07506
  [hep-ph]}}.

\bibitem{DallaBrida:2020pag}
M.~Dalla~Brida, ``{Past, present, and future of precision determinations of the
  QCD parameters from lattice QCD},''
  \href{http://dx.doi.org/10.1140/epja/s10050-021-00381-3}{{\em Eur. Phys. J.
  A} {\bfseries 57} no.~2, (2021) 66},
  \href{http://arxiv.org/abs/2012.01232}{{\ttfamily arXiv:2012.01232
  [hep-lat]}}.

\bibitem{DelDebbio:2021ryq}
L.~Del~Debbio and A.~Ramos, ``{Lattice determinations of the strong
  coupling},'' \href{http://arxiv.org/abs/2101.04762}{{\ttfamily
  arXiv:2101.04762 [hep-lat]}}.

\bibitem{Aoki:2013ldr}
S.~Aoki {\em et~al.}, ``{Review of Lattice Results Concerning Low-Energy
  Particle Physics},''
  \href{http://dx.doi.org/10.1140/epjc/s10052-014-2890-7}{{\em Eur. Phys. J. C}
  {\bfseries 74} (2014) 2890}, \href{http://arxiv.org/abs/1310.8555}{{\ttfamily
  arXiv:1310.8555 [hep-lat]}}.

\bibitem{Aoki:2016frl}
S.~Aoki {\em et~al.}, ``{Review of lattice results concerning low-energy
  particle physics},''
  \href{http://dx.doi.org/10.1140/epjc/s10052-016-4509-7}{{\em Eur. Phys. J. C}
  {\bfseries 77} no.~2, (2017) 112},
  \href{http://arxiv.org/abs/1607.00299}{{\ttfamily arXiv:1607.00299
  [hep-lat]}}.

\bibitem{FlavourLatticeAveragingGroup:2019iem}
{\bfseries Flavour Lattice Averaging Group} Collaboration, S.~Aoki {\em
  et~al.}, ``{FLAG Review 2019: Flavour Lattice Averaging Group (FLAG)},''
  \href{http://dx.doi.org/10.1140/epjc/s10052-019-7354-7}{{\em Eur. Phys. J. C}
  {\bfseries 80} no.~2, (2020) 113},
  \href{http://arxiv.org/abs/1902.08191}{{\ttfamily arXiv:1902.08191
  [hep-lat]}}.

\bibitem{Vermaseren:1997fq}
J.~Vermaseren, S.~Larin, and T.~van Ritbergen, ``{The four loop quark mass
  anomalous dimension and the invariant quark mass},''
  \href{http://dx.doi.org/10.1016/S0370-2693(97)00660-6}{{\em Phys.Lett.}
  {\bfseries B405} (1997) 327--333},
\href{http://arxiv.org/abs/hep-ph/9703284}{{\ttfamily arXiv:hep-ph/9703284
  [hep-ph]}}.

\bibitem{Chetyrkin:1997dh}
K.~G. Chetyrkin, ``{Quark mass anomalous dimension to O (alpha-s**4)},''
  \href{http://dx.doi.org/10.1016/S0370-2693(97)00535-2}{{\em Phys. Lett. B}
  {\bfseries 404} (1997) 161--165},
  \href{http://arxiv.org/abs/hep-ph/9703278}{{\ttfamily arXiv:hep-ph/9703278}}.

\bibitem{Baikov:2014qja}
P.~A. Baikov, K.~G. Chetyrkin, and J.~H. K\"uhn, ``{Quark Mass and Field
  Anomalous Dimensions to ${\cal O}(\alpha_s^5)$},''
  \href{http://dx.doi.org/10.1007/JHEP10(2014)076}{{\em JHEP} {\bfseries 10}
  (2014) 076}, \href{http://arxiv.org/abs/1402.6611}{{\ttfamily arXiv:1402.6611
  [hep-ph]}}.

\bibitem{Herren:2017osy}
F.~Herren and M.~Steinhauser, ``{Version 3 of RunDec and CRunDec},''
  \href{http://dx.doi.org/10.1016/j.cpc.2017.11.014}{{\em Comput. Phys.
  Commun.} {\bfseries 224} (2018) 333--345},
  \href{http://arxiv.org/abs/1703.03751}{{\ttfamily arXiv:1703.03751
  [hep-ph]}}.

\bibitem{Hoang:2021fhn}
A.~H. Hoang, C.~Lepenik, and V.~Mateu, ``{REvolver: Automated running and
  matching of couplings and masses in QCD},''
  \href{http://arxiv.org/abs/2102.01085}{{\ttfamily arXiv:2102.01085
  [hep-ph]}}.

\bibitem{Narison:2019tym}
S.~Narison, ``{$\overline m_c$ and $\overline m_b$ from $M_{Bc}$ and improved
  estimates of $f_{Bc}$ and $f_{Bc(2S)}$},''
  \href{http://dx.doi.org/10.1016/j.physletb.2020.135221}{{\em Phys. Lett. B}
  {\bfseries 802} (2020) 135221},
  \href{http://arxiv.org/abs/1906.03614}{{\ttfamily arXiv:1906.03614
  [hep-ph]}}.

\bibitem{Peset:2018ria}
C.~Peset, A.~Pineda, and J.~Segovia, ``{The charm/bottom quark mass from heavy
  quarkonium at N3LO},'' \href{http://dx.doi.org/10.1007/JHEP09(2018)167}{{\em
  JHEP} {\bfseries 09} (2018) 167},
  \href{http://arxiv.org/abs/1806.05197}{{\ttfamily arXiv:1806.05197
  [hep-ph]}}.

\bibitem{Kiyo:2015ufa}
Y.~Kiyo, G.~Mishima, and Y.~Sumino, ``{Determination of m$_c$ and m$_b$ from
  quarkonium 1S energy levels in perturbative QCD},''
  \href{http://dx.doi.org/10.1016/j.physletb.2015.11.040,
  10.1016/j.physletb.2017.09.024}{{\em Phys. Lett.} {\bfseries B752} (2016)
  122--127}, \href{http://arxiv.org/abs/1510.07072}{{\ttfamily arXiv:1510.07072
  [hep-ph]}}.
[Erratum: Phys. Lett.B772,878(2017)].

\bibitem{Penin:2014zaa}
A.~A. Penin and N.~Zerf, ``{Bottom Quark Mass from $\Upsilon$ Sum Rules to
  ${\cal O}(\alpha_s^3)$},''
  \href{http://dx.doi.org/10.1007/JHEP04(2014)120}{{\em JHEP} {\bfseries 04}
  (2014) 120},
\href{http://arxiv.org/abs/1401.7035}{{\ttfamily arXiv:1401.7035 [hep-ph]}}.

\bibitem{Alberti:2014yda}
A.~Alberti, P.~Gambino, K.~J. Healey, and S.~Nandi, ``{Precision Determination
  of the Cabibbo-Kobayashi-Maskawa Element $V_{cb}$},''
  \href{http://dx.doi.org/10.1103/PhysRevLett.114.061802}{{\em Phys. Rev.
  Lett.} {\bfseries 114} no.~6, (2015) 061802},
  \href{http://arxiv.org/abs/1411.6560}{{\ttfamily arXiv:1411.6560 [hep-ph]}}.

\bibitem{Beneke:2014pta}
M.~Beneke, A.~Maier, J.~Piclum, and T.~Rauh, ``{The bottom-quark mass from
  non-relativistic sum rules at NNNLO},''
  \href{http://dx.doi.org/10.1016/j.nuclphysb.2014.12.001}{{\em Nucl. Phys. B}
  {\bfseries 891} (2015) 42--72},
  \href{http://arxiv.org/abs/1411.3132}{{\ttfamily arXiv:1411.3132 [hep-ph]}}.

\bibitem{Dehnadi:2015fra}
B.~Dehnadi, A.~H. Hoang, and V.~Mateu, ``{Bottom and Charm Mass Determinations
  with a Convergence Test},''
  \href{http://dx.doi.org/10.1007/JHEP08(2015)155}{{\em JHEP} {\bfseries 08}
  (2015) 155}, \href{http://arxiv.org/abs/1504.07638}{{\ttfamily
  arXiv:1504.07638 [hep-ph]}}.

\bibitem{Lucha:2013gta}
W.~Lucha, D.~Melikhov, and S.~Simula, ``{Accurate bottom-quark mass from Borel
  QCD sum rules for $f_B$ and $f_{B_s}$},''
  \href{http://dx.doi.org/10.1103/PhysRevD.88.056011}{{\em Phys. Rev. D}
  {\bfseries 88} (2013) 056011},
  \href{http://arxiv.org/abs/1305.7099}{{\ttfamily arXiv:1305.7099 [hep-ph]}}.

\bibitem{Bodenstein:2011fv}
S.~Bodenstein, J.~Bordes, C.~Dominguez, J.~Penarrocha, and K.~Schilcher,
  ``{Bottom-quark mass from finite energy QCD sum rules},''
  \href{http://dx.doi.org/10.1103/PhysRevD.85.034003}{{\em Phys. Rev. D}
  {\bfseries 85} (2012) 034003},
  \href{http://arxiv.org/abs/1111.5742}{{\ttfamily arXiv:1111.5742 [hep-ph]}}.

\bibitem{Laschka:2011zr}
A.~Laschka, N.~Kaiser, and W.~Weise, ``{Quark-antiquark potential to order 1/m
  and heavy quark masses},''
  \href{http://dx.doi.org/10.1103/PhysRevD.83.094002}{{\em Phys. Rev. D}
  {\bfseries 83} (2011) 094002},
  \href{http://arxiv.org/abs/1102.0945}{{\ttfamily arXiv:1102.0945 [hep-ph]}}.

\bibitem{Chetyrkin:2009fv}
K.~Chetyrkin, J.~Kuhn, A.~Maier, P.~Maierhofer, P.~Marquard, M.~Steinhauser,
  and C.~Sturm, ``{Charm and Bottom Quark Masses: An Update},''
  \href{http://dx.doi.org/10.1103/PhysRevD.80.074010}{{\em Phys. Rev. D}
  {\bfseries 80} (2009) 074010},
  \href{http://arxiv.org/abs/0907.2110}{{\ttfamily arXiv:0907.2110 [hep-ph]}}.

\bibitem{Bazavov:2018omf}
{\bfseries Fermilab Lattice, MILC, TUMQCD} Collaboration, A.~Bazavov {\em
  et~al.}, ``{Up-, down-, strange-, charm-, and bottom-quark masses from
  four-flavor lattice QCD},''
  \href{http://dx.doi.org/10.1103/PhysRevD.98.054517}{{\em Phys. Rev. D}
  {\bfseries 98} no.~5, (2018) 054517},
  \href{http://arxiv.org/abs/1802.04248}{{\ttfamily arXiv:1802.04248
  [hep-lat]}}.

\bibitem{Colquhoun:2014ica}
B.~Colquhoun, R.~Dowdall, C.~Davies, K.~Hornbostel, and G.~Lepage,
  ``{$\Upsilon$ and $\Upsilon^{\prime}$ Leptonic Widths, $a_{\mu}^b$ and $m_b$
  from full lattice QCD},''
  \href{http://dx.doi.org/10.1103/PhysRevD.91.074514}{{\em Phys. Rev. D}
  {\bfseries 91} no.~7, (2015) 074514},
  \href{http://arxiv.org/abs/1408.5768}{{\ttfamily arXiv:1408.5768 [hep-lat]}}.

\bibitem{Bernardoni:2013xba}
F.~Bernardoni {\em et~al.}, ``{The b-quark mass from non-perturbative $N_f=2$
  Heavy Quark Effective Theory at $O(1/m_h)$},''
  \href{http://dx.doi.org/10.1016/j.physletb.2014.01.046}{{\em Phys. Lett. B}
  {\bfseries 730} (2014) 171--177},
  \href{http://arxiv.org/abs/1311.5498}{{\ttfamily arXiv:1311.5498 [hep-lat]}}.

\bibitem{Lee:2013mla}
{\bfseries HPQCD} Collaboration, A.~Lee, C.~Monahan, R.~Horgan, C.~Davies,
  R.~Dowdall, and J.~Koponen, ``{Mass of the b quark from lattice NRQCD and
  lattice perturbation theory},''
  \href{http://dx.doi.org/10.1103/PhysRevD.87.074018}{{\em Phys. Rev. D}
  {\bfseries 87} no.~7, (2013) 074018},
  \href{http://arxiv.org/abs/1302.3739}{{\ttfamily arXiv:1302.3739 [hep-lat]}}.

\bibitem{Dimopoulos:2011gx}
{\bfseries ETM} Collaboration, P.~Dimopoulos {\em et~al.}, ``{Lattice QCD
  determination of $m_b, f_B$ and $f_{Bs}$ with twisted mass Wilson
  fermions},'' \href{http://dx.doi.org/10.1007/JHEP01(2012)046}{{\em JHEP}
  {\bfseries 01} (2012) 046}, \href{http://arxiv.org/abs/1107.1441}{{\ttfamily
  arXiv:1107.1441 [hep-lat]}}.

\bibitem{Aoki:2019cca}
{\bfseries Flavour Lattice Averaging Group} Collaboration, S.~Aoki {\em
  et~al.}, ``{FLAG Review 2019: Flavour Lattice Averaging Group (FLAG)},''
  \href{http://dx.doi.org/10.1140/epjc/s10052-019-7354-7}{{\em Eur. Phys. J. C}
  {\bfseries 80} no.~2, (2020) 113},
  \href{http://arxiv.org/abs/1902.08191}{{\ttfamily arXiv:1902.08191
  [hep-lat]}}.

\bibitem{H1:2018flt}
{\bfseries H1, ZEUS} Collaboration, H.~Abramowicz {\em et~al.}, ``{Combination
  and QCD analysis of charm and beauty production cross-section measurements in
  deep inelastic ep scattering at HERA},''
  \href{http://dx.doi.org/10.1140/epjc/s10052-018-5848-3}{{\em Eur. Phys. J. C}
  {\bfseries 78} no.~6, (2018) 473},
  \href{http://arxiv.org/abs/1804.01019}{{\ttfamily arXiv:1804.01019
  [hep-ex]}}.

\bibitem{Schwanda:2008kw}
{\bfseries Belle} Collaboration, C.~Schwanda {\em et~al.}, ``{Measurement of
  the Moments of the Photon Energy Spectrum in B ---\ensuremath{>} X(s) gamma
  Decays and Determination of |V(cb)| and\ m(b) at Belle},''
  \href{http://dx.doi.org/10.1103/PhysRevD.78.032016}{{\em Phys. Rev. D}
  {\bfseries 78} (2008) 032016},
  \href{http://arxiv.org/abs/0803.2158}{{\ttfamily arXiv:0803.2158 [hep-ex]}}.

\bibitem{Aubert:2009qda}
{\bfseries BaBar} Collaboration, B.~Aubert {\em et~al.}, ``{Measurement and
  interpretation of moments in inclusive semileptonic decays anti-B
  ---\ensuremath{>} X(c) l- anti-nu},''
  \href{http://dx.doi.org/10.1103/PhysRevD.81.032003}{{\em Phys. Rev. D}
  {\bfseries 81} (2010) 032003},
  \href{http://arxiv.org/abs/0908.0415}{{\ttfamily arXiv:0908.0415 [hep-ex]}}.

\bibitem{Abreu:1997ey}
{\bfseries DELPHI} Collaboration, P.~Abreu {\em et~al.}, ``{$m_b$ at $M_Z$},''
\href{http://dx.doi.org/10.1016/S0370-2693(97)01442-1}{{\em Phys. Lett.}
  {\bfseries B418} (1998) 430--442}.

\bibitem{Barate:2000ab}
{\bfseries ALEPH} Collaboration, R.~Barate {\em et~al.}, ``{A Measurement of
  the b quark mass from hadronic Z decays},''
  \href{http://dx.doi.org/10.1007/s100520000533}{{\em Eur. Phys. J.} {\bfseries
  C18} (2000) 1--13},
\href{http://arxiv.org/abs/hep-ex/0008013}{{\ttfamily arXiv:hep-ex/0008013
  [hep-ex]}}.

\bibitem{Abbiendi:2001tw}
{\bfseries OPAL} Collaboration, G.~Abbiendi {\em et~al.}, ``{Determination of
  the b quark mass at the Z mass scale},''
  \href{http://dx.doi.org/10.1007/100520100746}{{\em Eur. Phys. J.} {\bfseries
  C21} (2001) 411--422},
\href{http://arxiv.org/abs/hep-ex/0105046}{{\ttfamily arXiv:hep-ex/0105046
  [hep-ex]}}.

\bibitem{Abdallah:2005cv}
{\bfseries DELPHI} Collaboration, J.~Abdallah {\em et~al.}, ``{Determination of
  the b quark mass at the M(Z) scale with the DELPHI detector at LEP},''
  \href{http://dx.doi.org/10.1140/epjc/s2006-02497-6}{{\em Eur. Phys. J.}
  {\bfseries C46} (2006) 569--583},
\href{http://arxiv.org/abs/hep-ex/0603046}{{\ttfamily arXiv:hep-ex/0603046
  [hep-ex]}}.

\bibitem{Abdallah:2008ac}
{\bfseries DELPHI} Collaboration, J.~Abdallah {\em et~al.}, ``{Study of b-quark
  mass effects in multijet topologies with the DELPHI detector at LEP},''
  \href{http://dx.doi.org/10.1140/epjc/s10052-008-0631-5}{{\em Eur. Phys. J.}
  {\bfseries C55} (2008) 525--538},
\href{http://arxiv.org/abs/0804.3883}{{\ttfamily arXiv:0804.3883 [hep-ex]}}.

\bibitem{Brandenburg:1999nb}
A.~Brandenburg, P.~N. Burrows, D.~Muller, N.~Oishi, and P.~Uwer, ``{Measurement
  of the running $b$ quark mass using $e^+ e^- \to b \bar b g$ events},''
  \href{http://dx.doi.org/10.1016/S0370-2693(99)01194-6}{{\em Phys. Lett.}
  {\bfseries B468} (1999) 168--177},
\href{http://arxiv.org/abs/hep-ph/9905495}{{\ttfamily arXiv:hep-ph/9905495
  [hep-ph]}}.

\bibitem{Abe:1998kr}
{\bfseries SLD} Collaboration, K.~Abe {\em et~al.}, ``{An Improved test of the
  flavor independence of strong interactions},''
  \href{http://dx.doi.org/10.1103/PhysRevD.59.012002}{{\em Phys. Rev.}
  {\bfseries D59} (1999) 012002},
\href{http://arxiv.org/abs/hep-ex/9805023}{{\ttfamily arXiv:hep-ex/9805023
  [hep-ex]}}.

\bibitem{Aparisi:2021tym}
J.~Aparisi {\em et~al.}, ``{mb at mH: The Running Bottom Quark Mass and the
  Higgs Boson},'' \href{http://dx.doi.org/10.1103/PhysRevLett.128.122001}{{\em
  Phys. Rev. Lett.} {\bfseries 128} no.~12, (2022) 122001},
  \href{http://arxiv.org/abs/2110.10202}{{\ttfamily arXiv:2110.10202
  [hep-ph]}}.

\bibitem{ATLAS-CONF-2020-027}
{ATLAS Collaboration}, ``{A combination of measurements of Higgs boson
  production and decay using up to \(139\,\text{fb}^{-1}\) of proton--proton
  collision data at \(\sqrt{s} = 13\,\text{TeV}\) collected with the ATLAS
  experiment}.'' {ATLAS-CONF-2020-027}, 2020.
\newblock \url{https://cds.cern.ch/record/2725733}.

\bibitem{Sirunyan:2018koj}
{CMS collaboration}, ``{Combined measurements of Higgs boson couplings in
  proton–proton collisions at $\sqrt{s}=13\,\text {Te}\text {V} $},''
  \href{http://dx.doi.org/10.1140/epjc/s10052-019-6909-y}{{\em Eur. Phys. J.}
  {\bfseries C79} no.~5, (2019) 421},
\href{http://arxiv.org/abs/1809.10733}{{\ttfamily arXiv:1809.10733 [hep-ex]}}.

\bibitem{Aparisi:2022yfn}
J.~Aparisi {\em et~al.}, ``{Snowmass White Paper: prospects for measurements of
  the bottom quark mass},'' \href{http://arxiv.org/abs/2203.16994}{{\ttfamily
  arXiv:2203.16994 [hep-ex]}}.

\bibitem{Gizhko:2017fiu}
A.~Gizhko {\em et~al.}, ``{Running of the Charm-Quark Mass from HERA
  Deep-Inelastic Scattering Data},''
  \href{http://dx.doi.org/10.1016/j.physletb.2017.11.002}{{\em Phys. Lett. B}
  {\bfseries 775} (2017) 233--238},
  \href{http://arxiv.org/abs/1705.08863}{{\ttfamily arXiv:1705.08863
  [hep-ph]}}.

\bibitem{Fuster:2021ekh}
J.~Fuster, A.~Irles, G.~Rodrigo, S.~Tairafune, M.~Vos, H.~Yamamoto, and
  R.~Yonamine, ``{Prospects for the measurement of the $b$-quark mass at the
  ILC},'' in {\em {International Workshop on Future Linear Colliders}}.
\newblock 4, 2021.
\newblock \href{http://arxiv.org/abs/2104.09924}{{\ttfamily arXiv:2104.09924
  [hep-ex]}}.

\bibitem{deBlas:2019rxi}
J.~de~Blas {\em et~al.}, ``{Higgs Boson Studies at Future Particle
  Colliders},'' \href{http://dx.doi.org/10.1007/JHEP01(2020)139}{{\em JHEP}
  {\bfseries 01} (2020) 139}, \href{http://arxiv.org/abs/1905.03764}{{\ttfamily
  arXiv:1905.03764 [hep-ph]}}.

\bibitem{Cepeda:2019klc}
M.~Cepeda {\em et~al.}, ``{Report from Working Group 2}: {Higgs Physics at the
  HL-LHC and HE-LHC},''
  \href{http://dx.doi.org/10.23731/CYRM-2019-007.221}{{\em CERN Yellow Rep.
  Monogr.} {\bfseries 7} (2019) 221--584},
  \href{http://arxiv.org/abs/1902.00134}{{\ttfamily arXiv:1902.00134
  [hep-ph]}}.

\bibitem{Yan:2016xyx}
J.~Yan, S.~Watanuki, K.~Fujii, A.~Ishikawa, D.~Jeans, J.~Strube, J.~Tian, and
  H.~Yamamoto, ``{Measurement of the Higgs boson mass and $e^+e^- \to ZH$ cross
  section using $Z \to \mu^+\mu^-$ and $Z \to e^+ e^-$ at the ILC},''
  \href{http://dx.doi.org/10.1103/PhysRevD.94.113002}{{\em Phys. Rev. D}
  {\bfseries 94} no.~11, (2016) 113002},
  \href{http://arxiv.org/abs/1604.07524}{{\ttfamily arXiv:1604.07524
  [hep-ex]}}. [Erratum: Phys.Rev.D 103, 099903 (2021)].

\bibitem{Brady:2011hb}
L.~T. Brady, A.~Accardi, W.~Melnitchouk, and J.~F. Owens, ``{Impact of PDF
  uncertainties at large x on heavy boson production},''
  \href{http://dx.doi.org/10.1007/JHEP06(2012)019}{{\em JHEP} {\bfseries 06}
  (2012) 019}, \href{http://arxiv.org/abs/1110.5398}{{\ttfamily arXiv:1110.5398
  [hep-ph]}}.

\bibitem{Ball:2017otu}
R.~D. Ball, V.~Bertone, M.~Bonvini, S.~Marzani, J.~Rojo, and L.~Rottoli,
  ``{Parton distributions with small-x resummation: evidence for BFKL dynamics
  in HERA data},'' \href{http://dx.doi.org/10.1140/epjc/s10052-018-5774-4}{{\em
  Eur. Phys. J. C} {\bfseries 78} no.~4, (2018) 321},
  \href{http://arxiv.org/abs/1710.05935}{{\ttfamily arXiv:1710.05935
  [hep-ph]}}.

\bibitem{Campbell:2013qaa}
J.~Campbell {\em et~al.}, ``{Working Group Report: Quantum Chromodynamics},''
  in {\em {Community Summer Study 2013}: {Snowmass on the Mississippi}}.
\newblock 10, 2013.
\newblock \href{http://arxiv.org/abs/1310.5189}{{\ttfamily arXiv:1310.5189
  [hep-ph]}}.

\bibitem{Harland-Lang:2014zoa}
L.~A. Harland-Lang, A.~D. Martin, P.~Motylinski, and R.~S. Thorne, ``{Parton
  distributions in the LHC era: MMHT 2014 PDFs},''
  \href{http://dx.doi.org/10.1140/epjc/s10052-015-3397-6}{{\em Eur. Phys. J. C}
  {\bfseries 75} no.~5, (2015) 204},
  \href{http://arxiv.org/abs/1412.3989}{{\ttfamily arXiv:1412.3989 [hep-ph]}}.

\bibitem{Dulat:2015mca}
S.~Dulat, T.-J. Hou, J.~Gao, M.~Guzzi, J.~Huston, P.~Nadolsky, J.~Pumplin,
  C.~Schmidt, D.~Stump, and C.~P. Yuan, ``{New parton distribution functions
  from a global analysis of quantum chromodynamics},''
  \href{http://dx.doi.org/10.1103/PhysRevD.93.033006}{{\em Phys. Rev.}
  {\bfseries D93} no.~3, (2016) 033006},
\href{http://arxiv.org/abs/1506.07443}{{\ttfamily arXiv:1506.07443 [hep-ph]}}.

\bibitem{H1:2015ubc}
{\bfseries H1, ZEUS} Collaboration, H.~Abramowicz {\em et~al.}, ``{Combination
  of measurements of inclusive deep inelastic ${e^{\pm }p}$ scattering cross
  sections and QCD analysis of HERA data},''
  \href{http://dx.doi.org/10.1140/epjc/s10052-015-3710-4}{{\em Eur. Phys. J. C}
  {\bfseries 75} no.~12, (2015) 580},
  \href{http://arxiv.org/abs/1506.06042}{{\ttfamily arXiv:1506.06042
  [hep-ex]}}.

\bibitem{Accardi:2016qay}
A.~Accardi, L.~T. Brady, W.~Melnitchouk, J.~F. Owens, and N.~Sato,
  ``{Constraints on large-$x$ parton distributions from new weak boson
  production and deep-inelastic scattering data},''
  \href{http://dx.doi.org/10.1103/PhysRevD.93.114017}{{\em Phys. Rev. D}
  {\bfseries 93} no.~11, (2016) 114017},
  \href{http://arxiv.org/abs/1602.03154}{{\ttfamily arXiv:1602.03154
  [hep-ph]}}.

\bibitem{Alekhin:2017kpj}
S.~Alekhin, J.~Bl\"umlein, S.~Moch, and R.~Placakyte, ``{Parton distribution
  functions, $\alpha_s$, and heavy-quark masses for LHC Run II},''
  \href{http://dx.doi.org/10.1103/PhysRevD.96.014011}{{\em Phys. Rev. D}
  {\bfseries 96} no.~1, (2017) 014011},
  \href{http://arxiv.org/abs/1701.05838}{{\ttfamily arXiv:1701.05838
  [hep-ph]}}.

\bibitem{NNPDF:2017mvq}
{\bfseries NNPDF} Collaboration, R.~D. Ball {\em et~al.}, ``{Parton
  distributions from high-precision collider data},''
  \href{http://dx.doi.org/10.1140/epjc/s10052-017-5199-5}{{\em Eur. Phys. J. C}
  {\bfseries 77} no.~10, (2017) 663},
  \href{http://arxiv.org/abs/1706.00428}{{\ttfamily arXiv:1706.00428
  [hep-ph]}}.

\bibitem{Bailey:2020ooq}
S.~Bailey, T.~Cridge, L.~A. Harland-Lang, A.~D. Martin, and R.~S. Thorne,
  ``{Parton distributions from LHC, HERA, Tevatron and fixed target data:
  MSHT20 PDFs},'' \href{http://dx.doi.org/10.1140/epjc/s10052-021-09057-0}{{\em
  Eur. Phys. J. C} {\bfseries 81} no.~4, (2021) 341},
  \href{http://arxiv.org/abs/2012.04684}{{\ttfamily arXiv:2012.04684
  [hep-ph]}}.

\bibitem{NNPDF:2021njg}
{\bfseries NNPDF} Collaboration, R.~D. Ball {\em et~al.}, ``{The path to proton
  structure at 1\% accuracy},''
  \href{http://dx.doi.org/10.1140/epjc/s10052-022-10328-7}{{\em Eur. Phys. J.
  C} {\bfseries 82} no.~5, (2022) 428},
  \href{http://arxiv.org/abs/2109.02653}{{\ttfamily arXiv:2109.02653
  [hep-ph]}}.

\bibitem{ATLAS:2021vod}
{\bfseries ATLAS} Collaboration, G.~Aad {\em et~al.}, ``{Determination of the
  parton distribution functions of the proton using diverse ATLAS data from
  $pp$ collisions at $\sqrt{s} = 7$, 8 and 13~TeV},''
  \href{http://dx.doi.org/10.1140/epjc/s10052-022-10217-z}{{\em Eur. Phys. J.
  C} {\bfseries 82} no.~5, (2022) 438},
  \href{http://arxiv.org/abs/2112.11266}{{\ttfamily arXiv:2112.11266
  [hep-ex]}}.

\bibitem{Courtoy:2022ocu}
A.~Courtoy, J.~Huston, P.~Nadolsky, K.~Xie, M.~Yan, and C.~P. Yuan, ``{Parton
  distributions need representative sampling},''
  \href{http://arxiv.org/abs/2205.10444}{{\ttfamily arXiv:2205.10444
  [hep-ph]}}.

\bibitem{PDF4LHCWG}
\protect{The PDF4LHC working group}. \url{https://www.hep.ucl.ac.uk/pdf4lhc/}.

\bibitem{Ball:2022hsh}
R.~D. Ball {\em et~al.}, ``{The PDF4LHC21 combination of global PDF fits for
  the LHC Run III},'' \href{http://arxiv.org/abs/2203.05506}{{\ttfamily
  arXiv:2203.05506 [hep-ph]}}.

\bibitem{Butterworth:2015oua}
J.~Butterworth {\em et~al.}, ``{PDF4LHC recommendations for LHC Run II},''
  \href{http://dx.doi.org/10.1088/0954-3899/43/2/023001}{{\em J. Phys. G}
  {\bfseries 43} (2016) 023001},
\href{http://arxiv.org/abs/1510.03865}{{\ttfamily arXiv:1510.03865 [hep-ph]}}.

\bibitem{AbdulKhalek:2018rok}
R.~Abdul~Khalek, S.~Bailey, J.~Gao, L.~Harland-Lang, and J.~Rojo, ``{Towards
  Ultimate Parton Distributions at the High-Luminosity LHC},''
  \href{http://dx.doi.org/10.1140/epjc/s10052-018-6448-y}{{\em Eur. Phys. J. C}
  {\bfseries 78} no.~11, (2018) 962},
  \href{http://arxiv.org/abs/1810.03639}{{\ttfamily arXiv:1810.03639
  [hep-ph]}}.

\bibitem{Amoroso:2020lgh}
S.~Amoroso {\em et~al.}, ``{Les Houches 2019: Physics at TeV Colliders:
  Standard Model Working Group Report},'' in {\em {11th Les Houches Workshop on
  Physics at TeV Colliders}: {PhysTeV Les Houches}}.
\newblock 3, 2020.
\newblock \href{http://arxiv.org/abs/2003.01700}{{\ttfamily arXiv:2003.01700
  [hep-ph]}}.

\bibitem{ATL-PHYS-PUB-2018-051}
{ATLAS Collaboration}, ``{Prospects for jet and photon physics at the HL-LHC
  and HE-LHC}.'' {ATL-PHYS-PUB-2018-051}, 2018.
\newblock \url{https://cds.cern.ch/record/2652285}.

\bibitem{FTR-18-015}
{CMS Collaboration}, ``{Projection of measurements of differential ttbar
  production cross sections in the e/u+jets channels in pp collisions at the
  HL-LHC},'' CMS Physics Analysis Summary CMS-PAS-FTR-18-015, CERN, 2018.
\newblock \url{http://cds.cern.ch/record/2651195}.

\bibitem{Gao:2017yyd}
J.~Gao, L.~Harland-Lang, and J.~Rojo, ``{The Structure of the Proton in the LHC
  Precision Era},'' \href{http://dx.doi.org/10.1016/j.physrep.2018.03.002}{{\em
  Phys. Rept.} {\bfseries 742} (2018) 1--121},
  \href{http://arxiv.org/abs/1709.04922}{{\ttfamily arXiv:1709.04922
  [hep-ph]}}.

\bibitem{CampbellHustonKrauss:2018}
J.~Campbell, J.~Huston, and F.~Krauss, {\em The Black Book of Quantum
  Chromodynamics: A Primer for the LHC Era}.
\newblock Oxford University Press, 2018.

\bibitem{Hobbs:2019gob}
T.~J. Hobbs, B.-T. Wang, P.~M. Nadolsky, and F.~I. Olness, ``{Charting the
  coming synergy between lattice QCD and high-energy phenomenology},''
  \href{http://dx.doi.org/10.1103/PhysRevD.100.094040}{{\em Phys. Rev. D}
  {\bfseries 100} no.~9, (2019) 094040},
  \href{http://arxiv.org/abs/1904.00022}{{\ttfamily arXiv:1904.00022
  [hep-ph]}}.

\bibitem{Accardi:2021ysh}
A.~Accardi, T.~J. Hobbs, X.~Jing, and P.~M. Nadolsky, ``{Deuterium scattering
  experiments in CTEQ global QCD analyses: a comparative investigation},''
  \href{http://dx.doi.org/10.1140/epjc/s10052-021-09318-y}{{\em Eur. Phys. J.
  C} {\bfseries 81} no.~7, (2021) 603},
  \href{http://arxiv.org/abs/2102.01107}{{\ttfamily arXiv:2102.01107
  [hep-ph]}}.

\bibitem{Cammarota:2020qcw}
{\bfseries Jefferson Lab Angular Momentum} Collaboration, J.~Cammarota,
  L.~Gamberg, Z.-B. Kang, J.~A. Miller, D.~Pitonyak, A.~Prokudin, T.~C. Rogers,
  and N.~Sato, ``{Origin of single transverse-spin asymmetries in high-energy
  collisions},'' \href{http://dx.doi.org/10.1103/PhysRevD.102.054002}{{\em
  Phys. Rev. D} {\bfseries 102} no.~5, (2020) 054002},
  \href{http://arxiv.org/abs/2002.08384}{{\ttfamily arXiv:2002.08384
  [hep-ph]}}.

\bibitem{Gupta:2018qil}
R.~Gupta, Y.-C. Jang, B.~Yoon, H.-W. Lin, V.~Cirigliano, and T.~Bhattacharya,
  ``{Isovector Charges of the Nucleon from 2+1+1-flavor Lattice QCD},''
  \href{http://dx.doi.org/10.1103/PhysRevD.98.034503}{{\em Phys. Rev.}
  {\bfseries D98} (2018) 034503},
  \href{http://arxiv.org/abs/1806.09006}{{\ttfamily arXiv:1806.09006
  [hep-lat]}}.

\bibitem{Alexandrou:2019brg}
C.~Alexandrou, S.~Bacchio, M.~Constantinou, J.~Finkenrath, K.~Hadjiyiannakou,
  K.~Jansen, G.~Koutsou, and A.~Vaquero Aviles-Casco, ``{Nucleon axial, tensor,
  and scalar charges and $\sigma$-terms in lattice QCD},''
  \href{http://dx.doi.org/10.1103/PhysRevD.102.054517}{{\em Phys. Rev. D}
  {\bfseries 102} no.~5, (2020) 054517},
  \href{http://arxiv.org/abs/1909.00485}{{\ttfamily arXiv:1909.00485
  [hep-lat]}}.

\bibitem{Arratia:2020azl}
M.~Arratia, Y.~Furletova, T.~J. Hobbs, F.~Olness, and S.~J. Sekula, ``{Charm
  jets as a probe for strangeness at the future Electron-Ion Collider},''
  \href{http://dx.doi.org/10.1103/PhysRevD.103.074023}{{\em Phys. Rev. D}
  {\bfseries 103} no.~7, (2021) 074023},
  \href{http://arxiv.org/abs/2006.12520}{{\ttfamily arXiv:2006.12520
  [hep-ph]}}.

\bibitem{Gelis:2010nm}
F.~Gelis, E.~Iancu, J.~Jalilian-Marian, and R.~Venugopalan, ``{The Color Glass
  Condensate},''
  \href{http://dx.doi.org/10.1146/annurev.nucl.010909.083629}{{\em Ann. Rev.
  Nucl. Part. Sci.} {\bfseries 60} (2010) 463--489},
  \href{http://arxiv.org/abs/1002.0333}{{\ttfamily arXiv:1002.0333 [hep-ph]}}.

\bibitem{FTR-18-027}
{CMS Collaboration}, ``{Constraining nuclear parton distributions with heavy
  ion collisions at the HL-LHC with the CMS experiment},'' CMS Physics Analysis
  Summary CMS-PAS-FTR-18-027, CERN, 2018.
\newblock \url{http://cds.cern.ch/record/2652030}.

\bibitem{ATL-PHYS-PUB-2018-039}
{ATLAS Collaboration}, ``{Expected ATLAS Measurement Capabilities of
  Observables Sensitive to Nuclear Parton Distributions}.''
  {ATL-PHYS-PUB-2018-039}, 2018.
\newblock \url{https://cds.cern.ch/record/2649445}.

\bibitem{Eskola:2016oht}
K.~J. Eskola, P.~Paakkinen, H.~Paukkunen, and C.~A. Salgado, ``{EPPS16: Nuclear
  parton distributions with LHC data},''
  \href{http://dx.doi.org/10.1140/epjc/s10052-017-4725-9}{{\em Eur. Phys. J. C}
  {\bfseries 77} no.~3, (2017) 163},
  \href{http://arxiv.org/abs/1612.05741}{{\ttfamily arXiv:1612.05741
  [hep-ph]}}.

\bibitem{AbdulKhalek:2019mzd}
{\bfseries NNPDF} Collaboration, R.~Abdul~Khalek, J.~J. Ethier, and J.~Rojo,
  ``{Nuclear parton distributions from lepton-nucleus scattering and the impact
  of an electron-ion collider},''
  \href{http://dx.doi.org/10.1140/epjc/s10052-019-6983-1}{{\em Eur. Phys. J. C}
  {\bfseries 79} no.~6, (2019) 471},
  \href{http://arxiv.org/abs/1904.00018}{{\ttfamily arXiv:1904.00018
  [hep-ph]}}.

\bibitem{AbdulKhalek:2020yuc}
R.~Abdul~Khalek, J.~J. Ethier, J.~Rojo, and G.~van Weelden, ``{nNNPDF2.0: quark
  flavor separation in nuclei from LHC data},''
  \href{http://dx.doi.org/10.1007/JHEP09(2020)183}{{\em JHEP} {\bfseries 09}
  (2020) 183}, \href{http://arxiv.org/abs/2006.14629}{{\ttfamily
  arXiv:2006.14629 [hep-ph]}}.

\bibitem{ALICE:2021gpt}
{\bfseries ALICE} Collaboration, S.~Acharya {\em et~al.}, ``{Coherent $J/\psi$
  and $\psi'$ photoproduction at midrapidity in ultra-peripheral Pb-Pb
  collisions at $\sqrt{s_{\mathrm{NN}}}~=~5.02$ TeV}''
  \href{http://dx.doi.org/10.1140/epjc/s10052-021-09437-6}{{\em Eur. Phys. J.
  C} {\bfseries 81} no.~8, (2021) 712},
  \href{http://arxiv.org/abs/2101.04577}{{\ttfamily arXiv:2101.04577
  [nucl-ex]}}.

\bibitem{Duan:2021gzs}
{\bfseries LHCb} Collaboration, W.~Duan, ``{Charmonia photo-production in
  ultra-peripheral and peripheral PbPb collisions with LHCb},'' in {\em {Low-x
  Workshop 2021}}.
\newblock 12, 2021.
\newblock \href{http://arxiv.org/abs/2112.10300}{{\ttfamily arXiv:2112.10300
  [nucl-ex]}}.

\bibitem{CMS:2016itn}
{\bfseries CMS} Collaboration, V.~Khachatryan {\em et~al.}, ``{Coherent
  $J/\psi$ photoproduction in ultra-peripheral PbPb collisions at $\sqrt
  {s_{NN}} =$ 2.76 TeV with the CMS experiment},''
  \href{http://dx.doi.org/10.1016/j.physletb.2017.07.001}{{\em Phys. Lett. B}
  {\bfseries 772} (2017) 489--511},
  \href{http://arxiv.org/abs/1605.06966}{{\ttfamily arXiv:1605.06966
  [nucl-ex]}}.

\bibitem{Guzey:2020ntc}
V.~Guzey, E.~Kryshen, M.~Strikman, and M.~Zhalov, ``{Nuclear suppression from
  coherent $J /\psi$ photoproduction at the Large Hadron Collider},''
  \href{http://dx.doi.org/10.1016/j.physletb.2021.136202}{{\em Phys. Lett. B}
  {\bfseries 816} (2021) 136202},
  \href{http://arxiv.org/abs/2008.10891}{{\ttfamily arXiv:2008.10891
  [hep-ph]}}.

\bibitem{Eskola:2022vpi}
K.~J. Eskola, C.~A. Flett, V.~Guzey, T.~L\"oyt\"ainen, and H.~Paukkunen,
  ``{Exclusive $J/\psi$ photoproduction in ultraperipheral Pb+Pb collisions at
  the LHC to next-to-leading order perturbative QCD},''
  \href{http://arxiv.org/abs/2203.11613}{{\ttfamily arXiv:2203.11613
  [hep-ph]}}.

\bibitem{ATLAS:2017kwa}
{\bfseries ATLAS} Collaboration, ``{Photo-nuclear dijet production in
  ultra-peripheral Pb+Pb collisions}.'' \protect{ATLAS-CONF-2017-011}, 2017.

\bibitem{Emelyanov:1999pkc}
V.~Emel'yanov, A.~Khodinov, S.~R. Klein, and R.~Vogt, ``{The Effect of
  shadowing on initial conditions, transverse energy and hard probes in
  ultrarelativistic heavy ion collisions},''
  \href{http://dx.doi.org/10.1103/PhysRevC.61.044904}{{\em Phys. Rev. C}
  {\bfseries 61} (2000) 044904},
  \href{http://arxiv.org/abs/hep-ph/9909427}{{\ttfamily arXiv:hep-ph/9909427}}.

\bibitem{Frankfurt:2011cs}
L.~Frankfurt, V.~Guzey, and M.~Strikman, ``{Leading Twist Nuclear Shadowing
  Phenomena in Hard Processes with Nuclei},''
  \href{http://dx.doi.org/10.1016/j.physrep.2011.12.002}{{\em Phys. Rept.}
  {\bfseries 512} (2012) 255--393},
  \href{http://arxiv.org/abs/1106.2091}{{\ttfamily arXiv:1106.2091 [hep-ph]}}.

\bibitem{Aguilar:2019teb}
A.~C. Aguilar {\em et~al.}, ``{Pion and Kaon Structure at the Electron-Ion
  Collider},'' \href{http://dx.doi.org/10.1140/epja/i2019-12885-0}{{\em Eur.
  Phys. J. A} {\bfseries 55} no.~10, (2019) 190},
  \href{http://arxiv.org/abs/1907.08218}{{\ttfamily arXiv:1907.08218
  [nucl-ex]}}.

\bibitem{Roberts:2021nhw}
C.~D. Roberts, D.~G. Richards, T.~Horn, and L.~Chang, ``{Insights into the
  emergence of mass from studies of pion and kaon structure},''
  \href{http://dx.doi.org/10.1016/j.ppnp.2021.103883}{{\em Prog. Part. Nucl.
  Phys.} {\bfseries 120} (2021) 103883},
  \href{http://arxiv.org/abs/2102.01765}{{\ttfamily arXiv:2102.01765
  [hep-ph]}}.

\bibitem{Barry:2022itu}
P.~C. Barry {\em et~al.}, ``{Complementarity of experimental and lattice QCD
  data on pion parton distributions},''
  \href{http://arxiv.org/abs/2204.00543}{{\ttfamily arXiv:2204.00543
  [hep-ph]}}.

\bibitem{Metz:2016swz}
A.~Metz and A.~Vossen, ``{Parton Fragmentation Functions},''
  \href{http://dx.doi.org/10.1016/j.ppnp.2016.08.003}{{\em Prog. Part. Nucl.
  Phys.} {\bfseries 91} (2016) 136--202},
  \href{http://arxiv.org/abs/1607.02521}{{\ttfamily arXiv:1607.02521
  [hep-ex]}}.

\bibitem{Anselmino:2020vlp}
M.~Anselmino, A.~Mukherjee, and A.~Vossen, ``{Transverse spin effects in hard
  semi-inclusive collisions},''
  \href{http://dx.doi.org/10.1016/j.ppnp.2020.103806}{{\em Prog. Part. Nucl.
  Phys.} {\bfseries 114} (2020) 103806},
  \href{http://arxiv.org/abs/2001.05415}{{\ttfamily arXiv:2001.05415
  [hep-ph]}}.

\bibitem{Burkardt:2008ps}
M.~Burkardt, ``{Transverse force on quarks in deep-inelastic scattering},''
  \href{http://dx.doi.org/10.1103/PhysRevD.88.114502}{{\em Phys. Rev. D}
  {\bfseries 88} (2013) 114502},
  \href{http://arxiv.org/abs/0810.3589}{{\ttfamily arXiv:0810.3589 [hep-ph]}}.

\bibitem{CLAS:2020igs}
{\bfseries CLAS} Collaboration, M.~Mirazita {\em et~al.}, ``{Beam Spin
  Asymmetry in Semi-Inclusive Electroproduction of Hadron Pairs},''
  \href{http://dx.doi.org/10.1103/PhysRevLett.126.062002}{{\em Phys. Rev.
  Lett.} {\bfseries 126} no.~6, (2021) 062002},
  \href{http://arxiv.org/abs/2010.09544}{{\ttfamily arXiv:2010.09544
  [hep-ex]}}.

\bibitem{Hayward:2021psm}
T.~B. Hayward {\em et~al.}, ``{Observation of Beam Spin Asymmetries in the
  Process $ep\rightarrow{e}^{'}{\pi}^{+}{\pi}^{-}X$ with CLAS12},''
  \href{http://dx.doi.org/10.1103/PhysRevLett.126.152501}{{\em Phys. Rev.
  Lett.} {\bfseries 126} (2021) 152501},
  \href{http://arxiv.org/abs/2101.04842}{{\ttfamily arXiv:2101.04842
  [hep-ex]}}.

\bibitem{Courtoy:2022kca}
A.~Courtoy, A.~S. Miramontes, H.~Avakian, M.~Mirazita, and S.~Pisano,
  ``{Extraction of the higher-twist parton distribution e(x) from CLAS data},''
  \href{http://dx.doi.org/10.1103/PhysRevD.106.014027}{{\em Phys. Rev. D}
  {\bfseries 106} no.~1, (2022) 014027},
  \href{http://arxiv.org/abs/2203.14975}{{\ttfamily arXiv:2203.14975
  [hep-ph]}}.

\bibitem{Sjostrand:2014zea}
T.~Sj{\"o}strand, S.~Ask, J.~R. Christiansen, R.~Corke, N.~Desai, P.~Ilten,
  S.~Mrenna, S.~Prestel, C.~O. Rasmussen, and P.~Z. Skands, ``{An Introduction
  to PYTHIA 8.2}'' \href{http://dx.doi.org/10.1016/j.cpc.2015.01.024}{{\em
  Comput. Phys. Commun.} {\bfseries 191} (2015) 159--177},
\href{http://arxiv.org/abs/1410.3012}{{\ttfamily arXiv:1410.3012 [hep-ph]}}.

\bibitem{Bellm:2019zci}
J.~Bellm {\em et~al.}, ``{Herwig 7.2 release note},''
  \href{http://dx.doi.org/10.1140/epjc/s10052-020-8011-x}{{\em Eur. Phys. J. C}
  {\bfseries 80} no.~5, (2020) 452},
  \href{http://arxiv.org/abs/1912.06509}{{\ttfamily arXiv:1912.06509
  [hep-ph]}}.

\bibitem{Sherpa:2019gpd}
{\bfseries Sherpa} Collaboration, E.~Bothmann {\em et~al.}, ``{Event Generation
  with Sherpa 2.2}'' \href{http://dx.doi.org/10.21468/SciPostPhys.7.3.034}{{\em
  SciPost Phys.} {\bfseries 7} no.~3, (2019) 034},
  \href{http://arxiv.org/abs/1905.09127}{{\ttfamily arXiv:1905.09127
  [hep-ph]}}.

\bibitem{Brehmer:2019xox}
J.~Brehmer, F.~Kling, I.~Espejo, and K.~Cranmer, ``{MadMiner: Machine
  learning-based inference for particle physics},''
  \href{http://dx.doi.org/10.1007/s41781-020-0035-2}{{\em Comput. Softw. Big
  Sci.} {\bfseries 4} no.~1, (2020) 3},
  \href{http://arxiv.org/abs/1907.10621}{{\ttfamily arXiv:1907.10621
  [hep-ph]}}.

\bibitem{ATLAS:2020bbn}
{\bfseries ATLAS} Collaboration, G.~Aad {\em et~al.}, ``{Measurement of the
  Lund Jet Plane Using Charged Particles in 13 TeV Proton-Proton Collisions
  with the ATLAS Detector},''
  \href{http://dx.doi.org/10.1103/PhysRevLett.124.222002}{{\em Phys. Rev.
  Lett.} {\bfseries 124} no.~22, (2020) 222002},
  \href{http://arxiv.org/abs/2004.03540}{{\ttfamily arXiv:2004.03540
  [hep-ex]}}.

\bibitem{dEnterria:2013sgr}
D.~d'Enterria, K.~J. Eskola, I.~Helenius, and H.~Paukkunen, ``{Confronting
  current NLO parton fragmentation functions with inclusive charged-particle
  spectra at hadron colliders},''
  \href{http://dx.doi.org/10.1016/j.nuclphysb.2014.04.006}{{\em Nucl. Phys. B}
  {\bfseries 883} (2014) 615--628},
  \href{http://arxiv.org/abs/1311.1415}{{\ttfamily arXiv:1311.1415 [hep-ph]}}.

\bibitem{Albino:2008aa}
S.~Albino {\em et~al.}, ``{Parton fragmentation in the vacuum and in the
  medium},'' \href{http://arxiv.org/abs/0804.2021}{{\ttfamily arXiv:0804.2021
  [hep-ph]}}.

\bibitem{Accardi:2009qv}
A.~Accardi, F.~Arleo, W.~K. Brooks, D.~D'Enterria, and V.~Muccifora, ``{Parton
  Propagation and Fragmentation in QCD Matter},''
  \href{http://dx.doi.org/10.1393/ncr/i2009-10048-0}{{\em Riv. Nuovo Cim.}
  {\bfseries 32} no.~9-10, (2009) 439--554},
  \href{http://arxiv.org/abs/0907.3534}{{\ttfamily arXiv:0907.3534 [nucl-th]}}.

\bibitem{Boglione:2021wov}
M.~Boglione and A.~Simonelli, ``{Kinematic regions in the $e^+e^- \to h \, X$
  factorized cross section in a $2$-jet topology with thrust},''
  \href{http://dx.doi.org/10.1007/JHEP02(2022)013}{{\em JHEP} {\bfseries 02}
  (2022) 013}, \href{http://arxiv.org/abs/2109.11497}{{\ttfamily
  arXiv:2109.11497 [hep-ph]}}.

\bibitem{Boglione:2020auc}
M.~Boglione and A.~Simonelli, ``{Factorization of $e^+e^- \to H \, X$ cross
  section, differential in $z_h$, $P_T$ and thrust, in the $2$-jet limit},''
  \href{http://dx.doi.org/10.1007/JHEP02(2021)076}{{\em JHEP} {\bfseries 02}
  (2021) 076}, \href{http://arxiv.org/abs/2011.07366}{{\ttfamily
  arXiv:2011.07366 [hep-ph]}}.

\bibitem{Boglione:2020cwn}
M.~Boglione and A.~Simonelli, ``{Universality-breaking effects in $e^+e^-$
  hadronic production processes},''
  \href{http://dx.doi.org/10.1140/epjc/s10052-020-08821-y}{{\em Eur. Phys. J.
  C} {\bfseries 81} no.~1, (2021) 96},
  \href{http://arxiv.org/abs/2007.13674}{{\ttfamily arXiv:2007.13674
  [hep-ph]}}.

\bibitem{Chang:2013iba}
H.-M. Chang, M.~Procura, J.~Thaler, and W.~J. Waalewijn, ``{Calculating Track
  Thrust with Track Functions},''
  \href{http://dx.doi.org/10.1103/PhysRevD.88.034030}{{\em Phys. Rev. D}
  {\bfseries 88} (2013) 034030},
  \href{http://arxiv.org/abs/1306.6630}{{\ttfamily arXiv:1306.6630 [hep-ph]}}.

\bibitem{Jaarsma:2022kdd}
M.~Jaarsma, Y.~Li, I.~Moult, W.~Waalewijn, and H.~X. Zhu, ``{Renormalization
  Group Flows for Track Function Moments},''
  \href{http://arxiv.org/abs/2201.05166}{{\ttfamily arXiv:2201.05166
  [hep-ph]}}.

\bibitem{Li:2021zcf}
Y.~Li, I.~Moult, S.~S. van Velzen, W.~J. Waalewijn, and H.~X. Zhu, ``{Extending
  Precision Perturbative QCD with Track Functions},''
  \href{http://dx.doi.org/10.1103/PhysRevLett.128.182001}{{\em Phys. Rev.
  Lett.} {\bfseries 128} no.~18, (2022) 182001},
  \href{http://arxiv.org/abs/2108.01674}{{\ttfamily arXiv:2108.01674
  [hep-ph]}}.

\bibitem{Khoze:1994fu}
V.~A. Khoze and T.~Sjostrand, ``{Color correlations and multiplicities in top
  events},'' \href{http://dx.doi.org/10.1016/0370-2693(94)91506-7}{{\em Phys.
  Lett. B} {\bfseries 328} (1994) 466},
  \href{http://arxiv.org/abs/hep-ph/9403394}{{\ttfamily arXiv:hep-ph/9403394}}.

\bibitem{Argyropoulos:2014zoa}
S.~Argyropoulos and T.~Sj\"ostrand, ``{Effects of color reconnection on
  $t\bar{t}$ final states at the LHC},''
  \href{http://dx.doi.org/10.1007/JHEP11(2014)043}{{\em JHEP} {\bfseries 11}
  (2014) 043}, \href{http://arxiv.org/abs/1407.6653}{{\ttfamily arXiv:1407.6653
  [hep-ph]}}.

\bibitem{Proceedings:2017ocd}
D.~d'Enterria, P.~Z. Skands, {\em et~al.}, ``Parton radiation and fragmentation
  from {LHC} to {FCC-ee},'' \href{http://arxiv.org/abs/1702.01329}{{\ttfamily
  arXiv:1702.01329 [hep-ph]}}.

\bibitem{Christiansen:2015yca}
J.~R. Christiansen and T.~Sj\"ostrand, ``{Color reconnection at future e$^{+}$
  e$^{-}$ colliders},''
  \href{http://dx.doi.org/10.1140/epjc/s10052-015-3674-4}{{\em Eur. Phys. J. C}
  {\bfseries 75} no.~9, (2015) 441},
  \href{http://arxiv.org/abs/1506.09085}{{\ttfamily arXiv:1506.09085
  [hep-ph]}}.

\bibitem{Sjostrand:1993hi}
T.~Sjostrand and V.~A. Khoze, ``{On Color rearrangement in hadronic W+ W-
  events},'' \href{http://dx.doi.org/10.1007/BF01560244}{{\em Z. Phys. C}
  {\bfseries 62} (1994) 281--310},
  \href{http://arxiv.org/abs/hep-ph/9310242}{{\ttfamily hep-ph/9310242}}.

\bibitem{ALEPH:2013dgf}
{\bfseries ALEPH, DELPHI, L3, OPAL, LEP Electroweak} Collaboration, S.~Schael
  {\em et~al.}, ``{Electroweak Measurements in Electron-Positron Collisions at
  W-Boson-Pair Energies at LEP},''
  \href{http://dx.doi.org/10.1016/j.physrep.2013.07.004}{{\em Phys. Rept.}
  {\bfseries 532} (2013) 119--244},
  \href{http://arxiv.org/abs/1302.3415}{{\ttfamily arXiv:1302.3415 [hep-ex]}}.

\bibitem{Abada:2019lih}
{\bfseries FCC} Collaboration, A.~Abada {\em et~al.}, ``{FCC Physics
  Opportunities},''
\href{http://dx.doi.org/10.1140/epjc/s10052-019-6904-3}{{\em Eur. Phys. J.}
  {\bfseries C79} no.~6, (2019) 474}.

\bibitem{FTR-18-032}
{CMS Collaboration}, ``{High-$p_T$ jet measurements at the HL-LHC},'' CMS
  Physics Analysis Summary CMS-PAS-FTR-18-032, CERN, 2018.
\newblock \url{http://cds.cern.ch/record/2651219}.

\bibitem{STDM-2010-15}
{ATLAS Collaboration}, ``{Measurement of the inclusive and dijet cross-sections
  of \(b\)-jets in \(pp\) collisions at \(\sqrt{s} = 7\,\text{TeV}\) with the
  ATLAS detector},''
  \href{http://dx.doi.org/10.1140/epjc/s10052-011-1846-4}{{\em Eur. Phys. J. C}
  {\bfseries 71} (2011) 1846}, \href{http://arxiv.org/abs/1109.6833}{{\ttfamily
  arXiv:1109.6833 [hep-ex]}}.

\bibitem{STDM-2013-03}
{ATLAS Collaboration}, ``{Measurement of the \(b\bar{b}\) dijet cross section
  in \(pp\) collisions at \(\sqrt{s} = 7\,\text{TeV}\) with the ATLAS
  detector},'' \href{http://dx.doi.org/10.1140/epjc/s10052-016-4521-y}{{\em
  Eur. Phys. J. C} {\bfseries 76} (2016) 670},
  \href{http://arxiv.org/abs/1607.08430}{{\ttfamily arXiv:1607.08430
  [hep-ex]}}.

\bibitem{CMS-FSQ-13-010}
{CMS Collaboration}, ``{Studies of inclusive four-jet production with two
  \(b\)-tagged jets in proton--proton collisions at \(7\,\text{TeV}\)},''
  \href{http://dx.doi.org/10.1103/PhysRevD.94.112005}{{\em Phys. Rev. D}
  {\bfseries 94} (2016) 112005},
  \href{http://arxiv.org/abs/1609.03489}{{\ttfamily arXiv:1609.03489
  [hep-ex]}}.

\bibitem{CMS-BPH-11-022}
{CMS Collaboration}, ``{Inclusive \(b\)-jet production in \(pp\) collisions at
  \(\sqrt{s} = 7\,\text{TeV}\)},''
  \href{http://dx.doi.org/10.1007/JHEP04(2012)084}{{\em JHEP} {\bfseries 04}
  (2012) 084}, \href{http://arxiv.org/abs/1202.4617}{{\ttfamily arXiv:1202.4617
  [hep-ex]}}.

\bibitem{Agostinelli:2002hh}
{GEANT4 Collaboration}, S.~Agostinelli, {\em et~al.}, ``{\textsc{Geant4} -- a
  simulation toolkit},''
\href{http://dx.doi.org/10.1016/S0168-9002(03)01368-8}{{\em Nucl. Instrum.
  Meth. A} {\bfseries 506} (2003) 250}.

\bibitem{Dreyer:2021hhr}
F.~Dreyer, G.~Soyez, and A.~Takacs, ``Quarks and gluons in the {Lund} plane,''
  \href{http://arxiv.org/abs/2112.09140}{{\ttfamily arXiv:2112.09140
  [hep-ph]}}.

\bibitem{Lin:2017snn}
H.-W. Lin {\em et~al.}, ``{Parton distributions and lattice QCD calculations: a
  community white paper},''
  \href{http://dx.doi.org/10.1016/j.ppnp.2018.01.007}{{\em Prog. Part. Nucl.
  Phys.} {\bfseries 100} (2018) 107--160},
  \href{http://arxiv.org/abs/1711.07916}{{\ttfamily arXiv:1711.07916
  [hep-ph]}}.

\bibitem{Constantinou:2020hdm}
M.~Constantinou {\em et~al.}, ``{Parton distributions and lattice-QCD
  calculations: Toward 3D structure},''
  \href{http://dx.doi.org/10.1016/j.ppnp.2021.103908}{{\em Prog. Part. Nucl.
  Phys.} {\bfseries 121} (2021) 103908},
  \href{http://arxiv.org/abs/2006.08636}{{\ttfamily arXiv:2006.08636
  [hep-ph]}}.

\bibitem{Ji:2020ect}
X.~Ji, Y.-S. Liu, Y.~Liu, J.-H. Zhang, and Y.~Zhao, ``{Large-momentum effective
  theory},'' \href{http://dx.doi.org/10.1103/RevModPhys.93.035005}{{\em Rev.
  Mod. Phys.} {\bfseries 93} no.~3, (2021) 035005},
  \href{http://arxiv.org/abs/2004.03543}{{\ttfamily arXiv:2004.03543
  [hep-ph]}}.

\bibitem{Ji:2020byp}
X.~Ji, ``{Why is LaMET an effective field theory for partonic structure?},''
  \href{http://arxiv.org/abs/2007.06613}{{\ttfamily arXiv:2007.06613
  [hep-ph]}}.

\bibitem{Lin:2017stx}
H.-W. Lin, W.~Melnitchouk, A.~Prokudin, N.~Sato, and H.~Shows, ``{First Monte
  Carlo Global Analysis of Nucleon Transversity with Lattice QCD
  Constraints},'' \href{http://dx.doi.org/10.1103/PhysRevLett.120.152502}{{\em
  Phys. Rev. Lett.} {\bfseries 120} no.~15, (2018) 152502},
  \href{http://arxiv.org/abs/1710.09858}{{\ttfamily arXiv:1710.09858
  [hep-ph]}}.

\bibitem{Cichy:2019ebf}
K.~Cichy, L.~Del~Debbio, and T.~Giani, ``{Parton distributions from lattice
  data: the nonsinglet case},''
  \href{http://dx.doi.org/10.1007/JHEP10(2019)137}{{\em JHEP} {\bfseries 10}
  (2019) 137}, \href{http://arxiv.org/abs/1907.06037}{{\ttfamily
  arXiv:1907.06037 [hep-ph]}}.

\bibitem{Bringewatt:2020ixn}
J.~Bringewatt, N.~Sato, W.~Melnitchouk, J.-W. Qiu, F.~Steffens, and
  M.~Constantinou, ``{Confronting lattice parton distributions with global QCD
  analysis},'' \href{http://dx.doi.org/10.1103/PhysRevD.103.016003}{{\em Phys.
  Rev. D} {\bfseries 103} no.~1, (2021) 016003},
  \href{http://arxiv.org/abs/2010.00548}{{\ttfamily arXiv:2010.00548
  [hep-ph]}}.

\bibitem{DelDebbio:2020rgv}
L.~Del~Debbio, T.~Giani, J.~Karpie, K.~Orginos, A.~Radyushkin, and
  S.~Zafeiropoulos, ``{Neural-network analysis of Parton Distribution Functions
  from Ioffe-time pseudodistributions},''
  \href{http://dx.doi.org/10.1007/JHEP02(2021)138}{{\em JHEP} {\bfseries 02}
  (2021) 138}, \href{http://arxiv.org/abs/2010.03996}{{\ttfamily
  arXiv:2010.03996 [hep-ph]}}.

\bibitem{Liu:1993cv}
K.-F. Liu and S.-J. Dong, ``{Origin of difference between anti-d and anti-u
  partons in the nucleon},''
  \href{http://dx.doi.org/10.1103/PhysRevLett.72.1790}{{\em Phys. Rev. Lett.}
  {\bfseries 72} (1994) 1790--1793},
  \href{http://arxiv.org/abs/hep-ph/9306299}{{\ttfamily arXiv:hep-ph/9306299}}.

\bibitem{Liu:1999ak}
K.-F. Liu, ``{Parton degrees of freedom from the path integral formalism},''
  \href{http://dx.doi.org/10.1103/PhysRevD.62.074501}{{\em Phys. Rev. D}
  {\bfseries 62} (2000) 074501},
  \href{http://arxiv.org/abs/hep-ph/9910306}{{\ttfamily arXiv:hep-ph/9910306}}.

\bibitem{Liang:2019frk}
{\bfseries XQCD} Collaboration, J.~Liang, T.~Draper, K.-F. Liu, A.~Rothkopf,
  and Y.-B. Yang, ``{Towards the nucleon hadronic tensor from lattice QCD},''
  \href{http://dx.doi.org/10.1103/PhysRevD.101.114503}{{\em Phys. Rev. D}
  {\bfseries 101} no.~11, (2020) 114503},
  \href{http://arxiv.org/abs/1906.05312}{{\ttfamily arXiv:1906.05312
  [hep-ph]}}.

\bibitem{Liang:2020sqi}
{\bfseries \ensuremath{\chi}QCD} Collaboration, J.~Liang and K.-F. Liu, ``{PDFs
  and Neutrino-Nucleon Scattering from Hadronic Tensor},''
  \href{http://dx.doi.org/10.22323/1.363.0046}{{\em PoS} {\bfseries
  LATTICE2019} (2020) 046}, \href{http://arxiv.org/abs/2008.12389}{{\ttfamily
  arXiv:2008.12389 [hep-lat]}}.

\bibitem{Aglietti:1998ur}
U.~Aglietti, M.~Ciuchini, G.~Corbo, E.~Franco, G.~Martinelli, and
  L.~Silvestrini, ``{Model independent determination of the light cone wave
  functions for exclusive processes},''
  \href{http://dx.doi.org/10.1016/S0370-2693(98)01138-1}{{\em Phys. Lett. B}
  {\bfseries 441} (1998) 371--375},
  \href{http://arxiv.org/abs/hep-ph/9806277}{{\ttfamily arXiv:hep-ph/9806277}}.

\bibitem{Ji:2001wha}
X.-d. Ji and C.-w. Jung, ``{Studying hadronic structure of the photon in
  lattice QCD},'' \href{http://dx.doi.org/10.1103/PhysRevLett.86.208}{{\em
  Phys. Rev. Lett.} {\bfseries 86} (2001) 208},
  \href{http://arxiv.org/abs/hep-lat/0101014}{{\ttfamily
  arXiv:hep-lat/0101014}}.

\bibitem{Detmold:2005gg}
W.~Detmold and C.~J.~D. Lin, ``{Deep-inelastic scattering and the operator
  product expansion in lattice QCD},''
  \href{http://dx.doi.org/10.1103/PhysRevD.73.014501}{{\em Phys. Rev. D}
  {\bfseries 73} (2006) 014501},
  \href{http://arxiv.org/abs/hep-lat/0507007}{{\ttfamily
  arXiv:hep-lat/0507007}}.

\bibitem{Chambers:2017dov}
A.~J. Chambers, R.~Horsley, Y.~Nakamura, H.~Perlt, P.~E.~L. Rakow,
  G.~Schierholz, A.~Schiller, K.~Somfleth, R.~D. Young, and J.~M. Zanotti,
  ``{Nucleon Structure Functions from Operator Product Expansion on the
  Lattice},'' \href{http://dx.doi.org/10.1103/PhysRevLett.118.242001}{{\em
  Phys. Rev. Lett.} {\bfseries 118} no.~24, (2017) 242001},
  \href{http://arxiv.org/abs/1703.01153}{{\ttfamily arXiv:1703.01153
  [hep-lat]}}.

\bibitem{Detmold:2021uru}
{\bfseries HOPE} Collaboration, W.~Detmold, A.~V. Grebe, I.~Kanamori, C.~J.~D.
  Lin, R.~J. Perry, and Y.~Zhao, ``{Parton physics from a heavy-quark operator
  product expansion: Formalism and Wilson coefficients},''
  \href{http://dx.doi.org/10.1103/PhysRevD.104.074511}{{\em Phys. Rev. D}
  {\bfseries 104} no.~7, (2021) 074511},
  \href{http://arxiv.org/abs/2103.09529}{{\ttfamily arXiv:2103.09529
  [hep-lat]}}.

\bibitem{Braun:2007wv}
V.~Braun and D.~M\"uller, ``{Exclusive processes in position space and the pion
  distribution amplitude},''
  \href{http://dx.doi.org/10.1140/epjc/s10052-008-0608-4}{{\em Eur. Phys. J. C}
  {\bfseries 55} (2008) 349--361},
  \href{http://arxiv.org/abs/0709.1348}{{\ttfamily arXiv:0709.1348 [hep-ph]}}.

\bibitem{Ma:2014jla}
Y.-Q. Ma and J.-W. Qiu, ``{Extracting Parton Distribution Functions from
  Lattice QCD Calculations},''
  \href{http://dx.doi.org/10.1103/PhysRevD.98.074021}{{\em Phys. Rev. D}
  {\bfseries 98} no.~7, (2018) 074021},
  \href{http://arxiv.org/abs/1404.6860}{{\ttfamily arXiv:1404.6860 [hep-ph]}}.

\bibitem{Ma:2017pxb}
Y.-Q. Ma and J.-W. Qiu, ``{Exploring Partonic Structure of Hadrons Using ab
  initio Lattice QCD Calculations},''
  \href{http://dx.doi.org/10.1103/PhysRevLett.120.022003}{{\em Phys. Rev.
  Lett.} {\bfseries 120} no.~2, (2018) 022003},
  \href{http://arxiv.org/abs/1709.03018}{{\ttfamily arXiv:1709.03018
  [hep-ph]}}.

\bibitem{Bali:2018spj}
G.~S. Bali, V.~M. Braun, B.~Gl\"a\ss{}le, M.~G\"ockeler, M.~Gruber, F.~Hutzler,
  P.~Korcyl, A.~Sch\"afer, P.~Wein, and J.-H. Zhang, ``{Pion distribution
  amplitude from Euclidean correlation functions: Exploring universality and
  higher-twist effects},''
  \href{http://dx.doi.org/10.1103/PhysRevD.98.094507}{{\em Phys. Rev. D}
  {\bfseries 98} no.~9, (2018) 094507},
  \href{http://arxiv.org/abs/1807.06671}{{\ttfamily arXiv:1807.06671
  [hep-lat]}}.

\bibitem{Joo:2020spy}
B.~Jo\'o, J.~Karpie, K.~Orginos, A.~V. Radyushkin, D.~G. Richards, and
  S.~Zafeiropoulos, ``{Parton Distribution Functions from Ioffe Time
  Pseudodistributions from Lattice Calculations: Approaching the Physical
  Point},'' \href{http://dx.doi.org/10.1103/PhysRevLett.125.232003}{{\em Phys.
  Rev. Lett.} {\bfseries 125} no.~23, (2020) 232003},
  \href{http://arxiv.org/abs/2004.01687}{{\ttfamily arXiv:2004.01687
  [hep-lat]}}.

\bibitem{Gao:2020ito}
X.~Gao, L.~Jin, C.~Kallidonis, N.~Karthik, S.~Mukherjee, P.~Petreczky,
  C.~Shugert, S.~Syritsyn, and Y.~Zhao, ``{Valence parton distribution of the
  pion from lattice QCD: Approaching the continuum limit},''
  \href{http://dx.doi.org/10.1103/PhysRevD.102.094513}{{\em Phys. Rev. D}
  {\bfseries 102} no.~9, (2020) 094513},
  \href{http://arxiv.org/abs/2007.06590}{{\ttfamily arXiv:2007.06590
  [hep-lat]}}.

\bibitem{Sufian:2020vzb}
R.~S. Sufian, C.~Egerer, J.~Karpie, R.~G. Edwards, B.~Jo\'o, Y.-Q. Ma,
  K.~Orginos, J.-W. Qiu, and D.~G. Richards, ``{Pion Valence Quark Distribution
  from Current-Current Correlation in Lattice QCD},''
  \href{http://dx.doi.org/10.1103/PhysRevD.102.054508}{{\em Phys. Rev. D}
  {\bfseries 102} no.~5, (2020) 054508},
  \href{http://arxiv.org/abs/2001.04960}{{\ttfamily arXiv:2001.04960
  [hep-lat]}}.

\bibitem{Radyushkin:2017cyf}
A.~V. Radyushkin, ``{Quasi-parton distribution functions, momentum
  distributions, and pseudo-parton distribution functions},''
  \href{http://dx.doi.org/10.1103/PhysRevD.96.034025}{{\em Phys. Rev. D}
  {\bfseries 96} no.~3, (2017) 034025},
  \href{http://arxiv.org/abs/1705.01488}{{\ttfamily arXiv:1705.01488
  [hep-ph]}}.

\bibitem{Bjorken:1969ja}
J.~D. Bjorken and E.~A. Paschos, ``{Inelastic Electron Proton and gamma Proton
  Scattering, and the Structure of the Nucleon},''
  \href{http://dx.doi.org/10.1103/PhysRev.185.1975}{{\em Phys. Rev.} {\bfseries
  185} (1969) 1975--1982}.

\bibitem{Izubuchi:2018srq}
T.~Izubuchi, X.~Ji, L.~Jin, I.~W. Stewart, and Y.~Zhao, ``{Factorization
  Theorem Relating Euclidean and Light-Cone Parton Distributions},''
  \href{http://dx.doi.org/10.1103/PhysRevD.98.056004}{{\em Phys. Rev. D}
  {\bfseries 98} no.~5, (2018) 056004},
  \href{http://arxiv.org/abs/1801.03917}{{\ttfamily arXiv:1801.03917
  [hep-ph]}}.

\bibitem{Ji:2014gla}
X.~Ji, ``{Parton Physics from Large-Momentum Effective Field Theory},''
  \href{http://dx.doi.org/10.1007/s11433-014-5492-3}{{\em Sci. China Phys.
  Mech. Astron.} {\bfseries 57} (2014) 1407--1412},
  \href{http://arxiv.org/abs/1404.6680}{{\ttfamily arXiv:1404.6680 [hep-ph]}}.

\bibitem{Lin:2020fsj}
H.-W. Lin, J.-W. Chen, and R.~Zhang, ``{Lattice Nucleon Isovector Unpolarized
  Parton Distribution in the Physical-Continuum Limit},''
  \href{http://arxiv.org/abs/2011.14971}{{\ttfamily arXiv:2011.14971
  [hep-lat]}}.

\bibitem{Chen:2018xof}
J.-W. Chen, L.~Jin, H.-W. Lin, Y.-S. Liu, Y.-B. Yang, J.-H. Zhang, and Y.~Zhao,
  ``{Lattice Calculation of Parton Distribution Function from LaMET at Physical
  Pion Mass with Large Nucleon Momentum},''
  \href{http://arxiv.org/abs/1803.04393}{{\ttfamily arXiv:1803.04393
  [hep-lat]}}.

\bibitem{Alexandrou:2018pbm}
C.~Alexandrou, K.~Cichy, M.~Constantinou, K.~Jansen, A.~Scapellato, and
  F.~Steffens, ``{Light-Cone Parton Distribution Functions from Lattice QCD},''
  \href{http://dx.doi.org/10.1103/PhysRevLett.121.112001}{{\em Phys. Rev.
  Lett.} {\bfseries 121} no.~11, (2018) 112001},
  \href{http://arxiv.org/abs/1803.02685}{{\ttfamily arXiv:1803.02685
  [hep-lat]}}.

\bibitem{Bhat:2020ktg}
M.~Bhat, K.~Cichy, M.~Constantinou, and A.~Scapellato, ``{Flavor nonsinglet
  parton distribution functions from lattice QCD at physical quark masses via
  the pseudodistribution approach},''
  \href{http://dx.doi.org/10.1103/PhysRevD.103.034510}{{\em Phys. Rev. D}
  {\bfseries 103} no.~3, (2021) 034510},
  \href{http://arxiv.org/abs/2005.02102}{{\ttfamily arXiv:2005.02102
  [hep-lat]}}.

\bibitem{Lin:2018pvv}
H.-W. Lin, J.-W. Chen, X.~Ji, L.~Jin, R.~Li, Y.-S. Liu, Y.-B. Yang, J.-H.
  Zhang, and Y.~Zhao, ``{Proton Isovector Helicity Distribution on the Lattice
  at Physical Pion Mass},''
  \href{http://dx.doi.org/10.1103/PhysRevLett.121.242003}{{\em Phys. Rev.
  Lett.} {\bfseries 121} no.~24, (2018) 242003},
  \href{http://arxiv.org/abs/1807.07431}{{\ttfamily arXiv:1807.07431
  [hep-lat]}}.

\bibitem{Liu:2018hxv}
Y.-S. Liu, J.-W. Chen, L.~Jin, R.~Li, H.-W. Lin, Y.-B. Yang, J.-H. Zhang, and
  Y.~Zhao, ``{Nucleon Transversity Distribution at the Physical Pion Mass from
  Lattice QCD},'' \href{http://arxiv.org/abs/1810.05043}{{\ttfamily
  arXiv:1810.05043 [hep-lat]}}.

\bibitem{Alexandrou:2018eet}
C.~Alexandrou, K.~Cichy, M.~Constantinou, K.~Jansen, A.~Scapellato, and
  F.~Steffens, ``{Transversity parton distribution functions from lattice
  QCD},'' \href{http://dx.doi.org/10.1103/PhysRevD.98.091503}{{\em Phys. Rev.
  D} {\bfseries 98} no.~9, (2018) 091503},
  \href{http://arxiv.org/abs/1807.00232}{{\ttfamily arXiv:1807.00232
  [hep-lat]}}.

\bibitem{Alexandrou:2019lfo}
C.~Alexandrou, K.~Cichy, M.~Constantinou, K.~Hadjiyiannakou, K.~Jansen,
  A.~Scapellato, and F.~Steffens, ``{Systematic uncertainties in parton
  distribution functions from lattice QCD simulations at the physical point},''
  \href{http://dx.doi.org/10.1103/PhysRevD.99.114504}{{\em Phys. Rev. D}
  {\bfseries 99} no.~11, (2019) 114504},
  \href{http://arxiv.org/abs/1902.00587}{{\ttfamily arXiv:1902.00587
  [hep-lat]}}.

\bibitem{Nocera:2014gqa}
{\bfseries NNPDF} Collaboration, E.~R. Nocera, R.~D. Ball, S.~Forte,
  G.~Ridolfi, and J.~Rojo, ``{A first unbiased global determination of
  polarized PDFs and their uncertainties},''
  \href{http://dx.doi.org/10.1016/j.nuclphysb.2014.08.008}{{\em Nucl. Phys.}
  {\bfseries B887} (2014) 276--308},
  \href{http://arxiv.org/abs/1406.5539}{{\ttfamily arXiv:1406.5539 [hep-ph]}}.

\bibitem{Ethier:2017zbq}
J.~J. Ethier, N.~Sato, and W.~Melnitchouk, ``{First simultaneous extraction of
  spin-dependent parton distributions and fragmentation functions from a global
  QCD analysis},'' \href{http://dx.doi.org/10.1103/PhysRevLett.119.132001}{{\em
  Phys. Rev. Lett.} {\bfseries 119} no.~13, (2017) 132001},
  \href{http://arxiv.org/abs/1705.05889}{{\ttfamily arXiv:1705.05889
  [hep-ph]}}.

\bibitem{deFlorian:2009vb}
D.~de~Florian, R.~Sassot, M.~Stratmann, and W.~Vogelsang, ``{Extraction of
  Spin-Dependent Parton Densities and Their Uncertainties},''
  \href{http://dx.doi.org/10.1103/PhysRevD.80.034030}{{\em Phys. Rev. D}
  {\bfseries 80} (2009) 034030},
  \href{http://arxiv.org/abs/0904.3821}{{\ttfamily arXiv:0904.3821 [hep-ph]}}.

\bibitem{Benel:2019mcq}
J.~Benel, A.~Courtoy, and R.~Ferro-Hernandez, ``{A constrained fit of the
  valence transversity distributions from dihadron production},''
  \href{http://dx.doi.org/10.1140/epjc/s10052-020-8039-y}{{\em Eur. Phys. J. C}
  {\bfseries 80} no.~5, (2020) 465},
  \href{http://arxiv.org/abs/1912.03289}{{\ttfamily arXiv:1912.03289
  [hep-ph]}}.

\bibitem{Radici:2018iag}
M.~Radici and A.~Bacchetta, ``{First Extraction of Transversity from a Global
  Analysis of Electron-Proton and Proton-Proton Data},''
  \href{http://dx.doi.org/10.1103/PhysRevLett.120.192001}{{\em Phys. Rev.
  Lett.} {\bfseries 120} no.~19, (2018) 192001},
  \href{http://arxiv.org/abs/1802.05212}{{\ttfamily arXiv:1802.05212
  [hep-ph]}}.

\bibitem{Lin:2017ani}
{\bfseries LP3} Collaboration, H.-W. Lin, J.-W. Chen, T.~Ishikawa, and J.-H.
  Zhang, ``{Improved parton distribution functions at the physical pion
  mass},'' \href{http://dx.doi.org/10.1103/PhysRevD.98.054504}{{\em Phys. Rev.
  D} {\bfseries 98} no.~5, (2018) 054504},
  \href{http://arxiv.org/abs/1708.05301}{{\ttfamily arXiv:1708.05301
  [hep-lat]}}.

\bibitem{Fan:2020cpa}
Z.~Fan, R.~Zhang, and H.-W. Lin, ``{Nucleon gluon distribution function from 2
  + 1 + 1-flavor lattice QCD},''
  \href{http://dx.doi.org/10.1142/S0217751X21500809}{{\em Int. J. Mod. Phys. A}
  {\bfseries 36} no.~13, (2021) 2150080},
  \href{http://arxiv.org/abs/2007.16113}{{\ttfamily arXiv:2007.16113
  [hep-lat]}}.

\bibitem{HadStruc:2021wmh}
{\bfseries HadStruc} Collaboration, T.~Khan {\em et~al.}, ``{Unpolarized gluon
  distribution in the nucleon from lattice quantum chromodynamics},''
  \href{http://dx.doi.org/10.1103/PhysRevD.104.094516}{{\em Phys. Rev. D}
  {\bfseries 104} no.~9, (2021) 094516},
  \href{http://arxiv.org/abs/2107.08960}{{\ttfamily arXiv:2107.08960
  [hep-lat]}}.

\bibitem{Gao:2021dbh}
X.~Gao, A.~D. Hanlon, S.~Mukherjee, P.~Petreczky, P.~Scior, S.~Syritsyn, and
  Y.~Zhao, ``{Lattice QCD Determination of the Bjorken-x Dependence of Parton
  Distribution Functions at Next-to-Next-to-Leading Order},''
  \href{http://dx.doi.org/10.1103/PhysRevLett.128.142003}{{\em Phys. Rev.
  Lett.} {\bfseries 128} no.~14, (2022) 142003},
  \href{http://arxiv.org/abs/2112.02208}{{\ttfamily arXiv:2112.02208
  [hep-lat]}}.

\bibitem{Lin:2020ssv}
H.-W. Lin, J.-W. Chen, Z.~Fan, J.-H. Zhang, and R.~Zhang, ``{Valence-Quark
  Distribution of the Kaon and Pion from Lattice QCD},''
  \href{http://dx.doi.org/10.1103/PhysRevD.103.014516}{{\em Phys. Rev. D}
  {\bfseries 103} no.~1, (2021) 014516},
  \href{http://arxiv.org/abs/2003.14128}{{\ttfamily arXiv:2003.14128
  [hep-lat]}}.

\bibitem{CMS:2013pzl}
{\bfseries CMS} Collaboration, S.~Chatrchyan {\em et~al.}, ``{Measurement of
  the Muon Charge Asymmetry in Inclusive $pp \to W+X$ Production at $\sqrt s =$
  7 TeV and an Improved Determination of Light Parton Distribution
  Functions},'' \href{http://dx.doi.org/10.1103/PhysRevD.90.032004}{{\em Phys.
  Rev. D} {\bfseries 90} no.~3, (2014) 032004},
  \href{http://arxiv.org/abs/1312.6283}{{\ttfamily arXiv:1312.6283 [hep-ex]}}.

\bibitem{ATLAS:2014jkm}
{\bfseries ATLAS} Collaboration, G.~Aad {\em et~al.}, ``{Measurement of the
  production of a $W$ boson in association with a charm quark in $pp$
  collisions at $\sqrt{s} =$ 7 TeV with the ATLAS detector},''
  \href{http://dx.doi.org/10.1007/JHEP05(2014)068}{{\em JHEP} {\bfseries 05}
  (2014) 068}, \href{http://arxiv.org/abs/1402.6263}{{\ttfamily arXiv:1402.6263
  [hep-ex]}}.

\bibitem{Alekhin:2017olj}
S.~Alekhin, J.~Bl\"umlein, and S.~Moch, ``{Strange sea determination from
  collider data},''
  \href{http://dx.doi.org/10.1016/j.physletb.2017.12.024}{{\em Phys. Lett. B}
  {\bfseries 777} (2018) 134--140},
  \href{http://arxiv.org/abs/1708.01067}{{\ttfamily arXiv:1708.01067
  [hep-ph]}}.

\bibitem{Faura:2020oom}
F.~Faura, S.~Iranipour, E.~R. Nocera, J.~Rojo, and M.~Ubiali, ``{The Strangest
  Proton?},'' \href{http://dx.doi.org/10.1140/epjc/s10052-020-08749-3}{{\em
  Eur. Phys. J. C} {\bfseries 80} no.~12, (2020) 1168},
  \href{http://arxiv.org/abs/2009.00014}{{\ttfamily arXiv:2009.00014
  [hep-ph]}}.

\bibitem{Nadolsky:2008zw}
P.~M. Nadolsky {\em et~al.}, ``{Implications of CTEQ global analysis for
  collider observables},''
  \href{http://dx.doi.org/10.1103/PhysRevD.78.013004}{{\em Phys. Rev.}
  {\bfseries D78} (2008) 013004},
\href{http://arxiv.org/abs/0802.0007}{{\ttfamily arXiv:0802.0007 [hep-ph]}}.

\bibitem{Fan:2018dxu}
Z.-Y. Fan, Y.-B. Yang, A.~Anthony, H.-W. Lin, and K.-F. Liu, ``{Gluon
  Quasi-Parton-Distribution Functions from Lattice QCD},''
  \href{http://dx.doi.org/10.1103/PhysRevLett.121.242001}{{\em Phys. Rev.
  Lett.} {\bfseries 121} no.~24, (2018) 242001},
  \href{http://arxiv.org/abs/1808.02077}{{\ttfamily arXiv:1808.02077
  [hep-lat]}}.

\bibitem{Balitsky:2019krf}
I.~Balitsky, W.~Morris, and A.~Radyushkin, ``{Gluon Pseudo-Distributions at
  Short Distances: Forward Case},''
  \href{http://dx.doi.org/10.1016/j.physletb.2020.135621}{{\em Phys. Lett. B}
  {\bfseries 808} (2020) 135621},
  \href{http://arxiv.org/abs/1910.13963}{{\ttfamily arXiv:1910.13963
  [hep-ph]}}.

\bibitem{Wang:2019tgg}
W.~Wang, J.-H. Zhang, S.~Zhao, and R.~Zhu, ``{Complete matching for
  quasidistribution functions in large momentum effective theory},''
  \href{http://dx.doi.org/10.1103/PhysRevD.100.074509}{{\em Phys. Rev. D}
  {\bfseries 100} no.~7, (2019) 074509},
  \href{http://arxiv.org/abs/1904.00978}{{\ttfamily arXiv:1904.00978
  [hep-ph]}}.

\bibitem{Zhang:2018diq}
J.-H. Zhang, X.~Ji, A.~Sch\"afer, W.~Wang, and S.~Zhao, ``{Accessing Gluon
  Parton Distributions in Large Momentum Effective Theory},''
  \href{http://dx.doi.org/10.1103/PhysRevLett.122.142001}{{\em Phys. Rev.
  Lett.} {\bfseries 122} no.~14, (2019) 142001},
  \href{http://arxiv.org/abs/1808.10824}{{\ttfamily arXiv:1808.10824
  [hep-ph]}}.

\bibitem{Zhang:2018nsy}
J.-H. Zhang, J.-W. Chen, L.~Jin, H.-W. Lin, A.~Sch\"afer, and Y.~Zhao, ``{First
  direct lattice-QCD calculation of the $x$-dependence of the pion parton
  distribution function},''
  \href{http://dx.doi.org/10.1103/PhysRevD.100.034505}{{\em Phys. Rev. D}
  {\bfseries 100} no.~3, (2019) 034505},
  \href{http://arxiv.org/abs/1804.01483}{{\ttfamily arXiv:1804.01483
  [hep-lat]}}.

\bibitem{Sufian:2019bol}
R.~S. Sufian, J.~Karpie, C.~Egerer, K.~Orginos, J.-W. Qiu, and D.~G. Richards,
  ``{Pion Valence Quark Distribution from Matrix Element Calculated in Lattice
  QCD},'' \href{http://dx.doi.org/10.1103/PhysRevD.99.074507}{{\em Phys. Rev.
  D} {\bfseries 99} no.~7, (2019) 074507},
  \href{http://arxiv.org/abs/1901.03921}{{\ttfamily arXiv:1901.03921
  [hep-lat]}}.

\bibitem{Izubuchi:2019lyk}
T.~Izubuchi, L.~Jin, C.~Kallidonis, N.~Karthik, S.~Mukherjee, P.~Petreczky,
  C.~Shugert, and S.~Syritsyn, ``{Valence parton distribution function of pion
  from fine lattice},''
  \href{http://dx.doi.org/10.1103/PhysRevD.100.034516}{{\em Phys. Rev. D}
  {\bfseries 100} no.~3, (2019) 034516},
  \href{http://arxiv.org/abs/1905.06349}{{\ttfamily arXiv:1905.06349
  [hep-lat]}}.

\bibitem{Joo:2019bzr}
B.~Jo\'o, J.~Karpie, K.~Orginos, A.~V. Radyushkin, D.~G. Richards, R.~S.
  Sufian, and S.~Zafeiropoulos, ``{Pion valence structure from Ioffe-time
  parton pseudodistribution functions},''
  \href{http://dx.doi.org/10.1103/PhysRevD.100.114512}{{\em Phys. Rev. D}
  {\bfseries 100} no.~11, (2019) 114512},
  \href{http://arxiv.org/abs/1909.08517}{{\ttfamily arXiv:1909.08517
  [hep-lat]}}.

\bibitem{Shugert:2020tgq}
C.~Shugert, X.~Gao, T.~Izubichi, L.~Jin, C.~Kallidonis, N.~Karthik,
  S.~Mukherjee, P.~Petreczky, S.~Syritsyn, and Y.~Zhao, ``{Pion valence quark
  PDF from lattice QCD},'' in {\em {37th International Symposium on Lattice
  Field Theory}}.
\newblock 1, 2020.
\newblock \href{http://arxiv.org/abs/2001.11650}{{\ttfamily arXiv:2001.11650
  [hep-lat]}}.

\bibitem{Chen:2020ody}
L.-B. Chen, W.~Wang, and R.~Zhu, ``{Next-to-Next-to-Leading Order Calculation
  of Quasiparton Distribution Functions},''
  \href{http://dx.doi.org/10.1103/PhysRevLett.126.072002}{{\em Phys. Rev.
  Lett.} {\bfseries 126} no.~7, (2021) 072002},
  \href{http://arxiv.org/abs/2006.14825}{{\ttfamily arXiv:2006.14825
  [hep-ph]}}.

\bibitem{Li:2020xml}
Z.-Y. Li, Y.-Q. Ma, and J.-W. Qiu, ``{Extraction of
  Next-to-Next-to-Leading-Order Parton Distribution Functions from Lattice QCD
  Calculations},'' \href{http://dx.doi.org/10.1103/PhysRevLett.126.072001}{{\em
  Phys. Rev. Lett.} {\bfseries 126} no.~7, (2021) 072001},
  \href{http://arxiv.org/abs/2006.12370}{{\ttfamily arXiv:2006.12370
  [hep-ph]}}.

\bibitem{Farrar:1975yb}
G.~R. Farrar and D.~R. Jackson, ``{Pion and Nucleon Structure Functions Near
  x=1},'' \href{http://dx.doi.org/10.1103/PhysRevLett.35.1416}{{\em Phys. Rev.
  Lett.} {\bfseries 35} (1975) 1416}.

\bibitem{Soper:1976jc}
D.~E. Soper, ``{The Parton Model and the Bethe-Salpeter Wave Function},''
  \href{http://dx.doi.org/10.1103/PhysRevD.15.1141}{{\em Phys. Rev. D}
  {\bfseries 15} (1977) 1141}.

\bibitem{Bednar:2018mtf}
K.~D. Bednar, I.~C. Clo\"et, and P.~C. Tandy, ``{Distinguishing Quarks and
  Gluons in Pion and Kaon Parton Distribution Functions},''
  \href{http://dx.doi.org/10.1103/PhysRevLett.124.042002}{{\em Phys. Rev.
  Lett.} {\bfseries 124} no.~4, (2020) 042002},
  \href{http://arxiv.org/abs/1811.12310}{{\ttfamily arXiv:1811.12310
  [nucl-th]}}.

\bibitem{Ding:2019lwe}
M.~Ding, K.~Raya, D.~Binosi, L.~Chang, C.~D. Roberts, and S.~M. Schmidt,
  ``{Symmetry, symmetry breaking, and pion parton distributions},''
  \href{http://dx.doi.org/10.1103/PhysRevD.101.054014}{{\em Phys. Rev. D}
  {\bfseries 101} no.~5, (2020) 054014},
  \href{http://arxiv.org/abs/1905.05208}{{\ttfamily arXiv:1905.05208
  [nucl-th]}}.

\bibitem{Novikov:2020snp}
I.~Novikov {\em et~al.}, ``{Parton Distribution Functions of the Charged Pion
  Within The xFitter Framework},''
  \href{http://dx.doi.org/10.1103/PhysRevD.102.014040}{{\em Phys. Rev. D}
  {\bfseries 102} no.~1, (2020) 014040},
  \href{http://arxiv.org/abs/2002.02902}{{\ttfamily arXiv:2002.02902
  [hep-ph]}}.

\bibitem{Courtoy:2020fex}
A.~Courtoy and P.~M. Nadolsky, ``{Testing momentum dependence of the
  nonperturbative hadron structure in a global QCD analysis},''
  \href{http://dx.doi.org/10.1103/PhysRevD.103.054029}{{\em Phys. Rev. D}
  {\bfseries 103} no.~5, (2021) 054029},
  \href{http://arxiv.org/abs/2011.10078}{{\ttfamily arXiv:2011.10078
  [hep-ph]}}.

\bibitem{Alexandrou:2021mmi}
{\bfseries ETM} Collaboration, C.~Alexandrou, S.~Bacchio, I.~Clo\"et,
  M.~Constantinou, K.~Hadjiyiannakou, G.~Koutsou, and C.~Lauer, ``{Pion and
  kaon \ensuremath{\langle}x3\ensuremath{\rangle} from lattice QCD and PDF
  reconstruction from Mellin moments},''
  \href{http://dx.doi.org/10.1103/PhysRevD.104.054504}{{\em Phys. Rev. D}
  {\bfseries 104} no.~5, (2021) 054504},
  \href{http://arxiv.org/abs/2104.02247}{{\ttfamily arXiv:2104.02247
  [hep-lat]}}.

\bibitem{Salas-Chavira:2021wui}
A.~Salas-Chavira, Z.~Fan, and H.-W. Lin, ``{First Glimpse into the Kaon Gluon
  Parton Distribution Using Lattice QCD},''
  \href{http://arxiv.org/abs/2112.03124}{{\ttfamily arXiv:2112.03124
  [hep-lat]}}.

\bibitem{Zhang:2017bzy}
J.-H. Zhang, J.-W. Chen, X.~Ji, L.~Jin, and H.-W. Lin, ``{Pion Distribution
  Amplitude from Lattice QCD},''
  \href{http://dx.doi.org/10.1103/PhysRevD.95.094514}{{\em Phys. Rev. D}
  {\bfseries 95} no.~9, (2017) 094514},
  \href{http://arxiv.org/abs/1702.00008}{{\ttfamily arXiv:1702.00008
  [hep-lat]}}.

\bibitem{Zhang:2017zfe}
{\bfseries LP3} Collaboration, J.-H. Zhang, L.~Jin, H.-W. Lin, A.~Sch\"afer,
  P.~Sun, Y.-B. Yang, R.~Zhang, Y.~Zhao, and J.-W. Chen, ``{Kaon Distribution
  Amplitude from Lattice QCD and the Flavor SU(3) Symmetry},''
  \href{http://dx.doi.org/10.1016/j.nuclphysb.2018.12.020}{{\em Nucl. Phys. B}
  {\bfseries 939} (2019) 429--446},
  \href{http://arxiv.org/abs/1712.10025}{{\ttfamily arXiv:1712.10025
  [hep-ph]}}.

\bibitem{Bali:2017gfr}
G.~S. Bali {\em et~al.}, ``{Pion distribution amplitude from Euclidean
  correlation functions},''
  \href{http://dx.doi.org/10.1140/epjc/s10052-018-5700-9}{{\em Eur. Phys. J. C}
  {\bfseries 78} no.~3, (2018) 217},
  \href{http://arxiv.org/abs/1709.04325}{{\ttfamily arXiv:1709.04325
  [hep-lat]}}.

\bibitem{RQCD:2019osh}
{\bfseries RQCD} Collaboration, G.~S. Bali, V.~M. Braun, S.~B\"urger,
  M.~G\"ockeler, M.~Gruber, F.~Hutzler, P.~Korcyl, A.~Sch\"afer, A.~Sternbeck,
  and P.~Wein, ``{Light-cone distribution amplitudes of pseudoscalar mesons
  from lattice QCD},'' \href{http://dx.doi.org/10.1007/JHEP08(2019)065}{{\em
  JHEP} {\bfseries 08} (2019) 065},
  \href{http://arxiv.org/abs/1903.08038}{{\ttfamily arXiv:1903.08038
  [hep-lat]}}. [Addendum: JHEP 11, 037 (2020)].

\bibitem{Hua:2020gnw}
{\bfseries Lattice Parton} Collaboration, J.~Hua, M.-H. Chu, P.~Sun, W.~Wang,
  J.~Xu, Y.-B. Yang, J.-H. Zhang, and Q.-A. Zhang, ``{Distribution Amplitudes
  of K* and \ensuremath{\phi} at the Physical Pion Mass from Lattice QCD},''
  \href{http://dx.doi.org/10.1103/PhysRevLett.127.062002}{{\em Phys. Rev.
  Lett.} {\bfseries 127} no.~6, (2021) 062002},
  \href{http://arxiv.org/abs/2011.09788}{{\ttfamily arXiv:2011.09788
  [hep-lat]}}.

\bibitem{Zhang:2020gaj}
R.~Zhang, C.~Honkala, H.-W. Lin, and J.-W. Chen, ``{Pion and kaon distribution
  amplitudes in the continuum limit},''
  \href{http://dx.doi.org/10.1103/PhysRevD.102.094519}{{\em Phys. Rev. D}
  {\bfseries 102} no.~9, (2020) 094519},
  \href{http://arxiv.org/abs/2005.13955}{{\ttfamily arXiv:2005.13955
  [hep-lat]}}.

\bibitem{Detmold:2021qln}
{\bfseries HOPE} Collaboration, W.~Detmold, A.~V. Grebe, I.~Kanamori, C.~J.~D.
  Lin, S.~Mondal, R.~J. Perry, and Y.~Zhao, ``{Parton physics from a
  heavy-quark operator product expansion: Lattice QCD calculation of the second
  moment of the pion distribution amplitude},''
  \href{http://dx.doi.org/10.1103/PhysRevD.105.034506}{{\em Phys. Rev. D}
  {\bfseries 105} no.~3, (2022) 034506},
  \href{http://arxiv.org/abs/2109.15241}{{\ttfamily arXiv:2109.15241
  [hep-lat]}}.

\bibitem{Juliano:2021hys}
N.~Juliano, R.~Zhang, C.~Honkala, and H.-W. Lin, ``{Pion Distribution
  Amplitudes in the Continuum Limit},''
  \href{http://dx.doi.org/10.22323/1.396.0436}{{\em PoS} {\bfseries
  LATTICE2021} (2022) 436}, \href{http://arxiv.org/abs/2108.04326}{{\ttfamily
  arXiv:2108.04326 [hep-lat]}}.

\bibitem{Hua:2022kcm}
J.~Hua {\em et~al.}, ``{Pion and Kaon Distribution Amplitudes from Lattice
  QCD},'' \href{http://arxiv.org/abs/2201.09173}{{\ttfamily arXiv:2201.09173
  [hep-lat]}}.

\bibitem{Gao:2022vyh}
X.~Gao, A.~D. Hanlon, N.~Karthik, S.~Mukherjee, P.~Petreczky, P.~Scior,
  S.~Syritsyn, and Y.~Zhao, ``{Pion distribution amplitude at the physical
  point using the leading-twist expansion of the quasi-DA matrix element},''
  \href{http://arxiv.org/abs/2206.04084}{{\ttfamily arXiv:2206.04084
  [hep-lat]}}.

\bibitem{Stewart:2003gt}
I.~W. Stewart, ``{Theoretical introduction to B decays and the soft collinear
  effective theory},'' in {\em {38th Rencontres de Moriond on QCD and
  High-Energy Hadronic Interactions}}.
\newblock 8, 2003.
\newblock \href{http://arxiv.org/abs/hep-ph/0308185}{{\ttfamily
  arXiv:hep-ph/0308185}}.

\bibitem{Chen:2019lcm}
J.-W. Chen, H.-W. Lin, and J.-H. Zhang, ``{Pion generalized parton distribution
  from lattice QCD},''
  \href{http://dx.doi.org/10.1016/j.nuclphysb.2020.114940}{{\em Nucl. Phys. B}
  {\bfseries 952} (2020) 114940},
  \href{http://arxiv.org/abs/1904.12376}{{\ttfamily arXiv:1904.12376
  [hep-lat]}}.

\bibitem{Alexandrou:2020zbe}
C.~Alexandrou, K.~Cichy, M.~Constantinou, K.~Hadjiyiannakou, K.~Jansen,
  A.~Scapellato, and F.~Steffens, ``{Unpolarized and helicity generalized
  parton distributions of the proton within lattice QCD},''
  \href{http://dx.doi.org/10.1103/PhysRevLett.125.262001}{{\em Phys. Rev.
  Lett.} {\bfseries 125} no.~26, (2020) 262001},
  \href{http://arxiv.org/abs/2008.10573}{{\ttfamily arXiv:2008.10573
  [hep-lat]}}.

\bibitem{Lin:2020rxa}
H.-W. Lin, ``{Nucleon Tomography and Generalized Parton Distribution at
  Physical Pion Mass from Lattice QCD},''
  \href{http://dx.doi.org/10.1103/PhysRevLett.127.182001}{{\em Phys. Rev.
  Lett.} {\bfseries 127} no.~18, (2021) 182001},
  \href{http://arxiv.org/abs/2008.12474}{{\ttfamily arXiv:2008.12474
  [hep-ph]}}.

\bibitem{Lin:2021brq}
H.-W. Lin, ``{Nucleon helicity generalized parton distribution at physical pion
  mass from lattice QCD},''
  \href{http://dx.doi.org/10.1016/j.physletb.2021.136821}{{\em Phys. Lett. B}
  {\bfseries 824} (2022) 136821},
  \href{http://arxiv.org/abs/2112.07519}{{\ttfamily arXiv:2112.07519
  [hep-lat]}}.

\bibitem{Musch:2010ka}
B.~U. Musch, P.~Hagler, J.~W. Negele, and A.~Schafer, ``{Exploring quark
  transverse momentum distributions with lattice QCD},''
  \href{http://dx.doi.org/10.1103/PhysRevD.83.094507}{{\em Phys. Rev. D}
  {\bfseries 83} (2011) 094507},
  \href{http://arxiv.org/abs/1011.1213}{{\ttfamily arXiv:1011.1213 [hep-lat]}}.

\bibitem{Musch:2011er}
B.~U. Musch, P.~Hagler, M.~Engelhardt, J.~W. Negele, and A.~Schafer, ``{Sivers
  and Boer-Mulders observables from lattice QCD},''
  \href{http://dx.doi.org/10.1103/PhysRevD.85.094510}{{\em Phys. Rev. D}
  {\bfseries 85} (2012) 094510},
  \href{http://arxiv.org/abs/1111.4249}{{\ttfamily arXiv:1111.4249 [hep-lat]}}.

\bibitem{Engelhardt:2015xja}
M.~Engelhardt, P.~H\"agler, B.~Musch, J.~Negele, and A.~Sch\"afer, ``{Lattice
  QCD study of the Boer-Mulders effect in a pion},''
  \href{http://dx.doi.org/10.1103/PhysRevD.93.054501}{{\em Phys. Rev. D}
  {\bfseries 93} no.~5, (2016) 054501},
  \href{http://arxiv.org/abs/1506.07826}{{\ttfamily arXiv:1506.07826
  [hep-lat]}}.

\bibitem{Yoon:2015ocs}
B.~Yoon, T.~Bhattacharya, M.~Engelhardt, J.~Green, R.~Gupta, P.~H\"agler,
  B.~Musch, J.~Negele, A.~Pochinsky, and S.~Syritsyn, ``{Lattice QCD
  calculations of nucleon transverse momentum-dependent parton distributions
  using clover and domain wall fermions},'' in {\em {33rd International
  Symposium on Lattice Field Theory}}.
\newblock SISSA, 11, 2015.
\newblock \href{http://arxiv.org/abs/1601.05717}{{\ttfamily arXiv:1601.05717
  [hep-lat]}}.

\bibitem{Yoon:2017qzo}
B.~Yoon, M.~Engelhardt, R.~Gupta, T.~Bhattacharya, J.~R. Green, B.~U. Musch,
  J.~W. Negele, A.~V. Pochinsky, A.~Sch\"afer, and S.~N. Syritsyn, ``{Nucleon
  Transverse Momentum-dependent Parton Distributions in Lattice QCD:
  Renormalization Patterns and Discretization Effects},''
  \href{http://dx.doi.org/10.1103/PhysRevD.96.094508}{{\em Phys. Rev. D}
  {\bfseries 96} no.~9, (2017) 094508},
  \href{http://arxiv.org/abs/1706.03406}{{\ttfamily arXiv:1706.03406
  [hep-lat]}}.

\bibitem{Shanahan:2020zxr}
P.~Shanahan, M.~Wagman, and Y.~Zhao, ``{Collins-Soper kernel for TMD evolution
  from lattice QCD},''
  \href{http://dx.doi.org/10.1103/PhysRevD.102.014511}{{\em Phys. Rev. D}
  {\bfseries 102} no.~1, (2020) 014511},
  \href{http://arxiv.org/abs/2003.06063}{{\ttfamily arXiv:2003.06063
  [hep-lat]}}.

\bibitem{LatticeParton:2020uhz}
{\bfseries Lattice Parton} Collaboration, Q.-A. Zhang {\em et~al.},
  ``{Lattice-QCD Calculations of TMD Soft Function Through Large-Momentum
  Effective Theory},'' \href{http://dx.doi.org/10.22323/1.396.0477}{{\em Phys.
  Rev. Lett.} {\bfseries 125} no.~19, (2020) 192001},
  \href{http://arxiv.org/abs/2005.14572}{{\ttfamily arXiv:2005.14572
  [hep-lat]}}.

\bibitem{Schlemmer:2021aij}
M.~Schlemmer, A.~Vladimirov, C.~Zimmermann, M.~Engelhardt, and A.~Sch\"afer,
  ``{Determination of the Collins-Soper Kernel from Lattice QCD},''
  \href{http://dx.doi.org/10.1007/JHEP08(2021)004}{{\em JHEP} {\bfseries 08}
  (2021) 004}, \href{http://arxiv.org/abs/2103.16991}{{\ttfamily
  arXiv:2103.16991 [hep-lat]}}.

\bibitem{Li:2021wvl}
Y.~Li {\em et~al.}, ``{Lattice QCD Study of Transverse-Momentum Dependent Soft
  Function},'' \href{http://dx.doi.org/10.1103/PhysRevLett.128.062002}{{\em
  Phys. Rev. Lett.} {\bfseries 128} no.~6, (2022) 062002},
  \href{http://arxiv.org/abs/2106.13027}{{\ttfamily arXiv:2106.13027
  [hep-lat]}}.

\bibitem{Zhang:2020dbb}
{\bfseries Lattice Parton} Collaboration, Q.-A. Zhang {\em et~al.},
  ``{Lattice-QCD Calculations of TMD Soft Function Through Large-Momentum
  Effective Theory},'' \href{http://dx.doi.org/10.22323/1.396.0477}{{\em Phys.
  Rev. Lett.} {\bfseries 125} no.~19, (2020) 192001},
  \href{http://arxiv.org/abs/2005.14572}{{\ttfamily arXiv:2005.14572
  [hep-lat]}}.

\bibitem{Shanahan:2021tst}
P.~Shanahan, M.~Wagman, and Y.~Zhao, ``{Lattice QCD calculation of the
  Collins-Soper kernel from quasi-TMDPDFs},''
  \href{http://dx.doi.org/10.1103/PhysRevD.104.114502}{{\em Phys. Rev. D}
  {\bfseries 104} no.~11, (2021) 114502},
  \href{http://arxiv.org/abs/2107.11930}{{\ttfamily arXiv:2107.11930
  [hep-lat]}}.

\bibitem{Yoon:2016dyh}
B.~Yoon, T.~Bhattacharya, M.~Engelhardt, J.~Green, R.~Gupta, P.~H\"agler,
  B.~Musch, J.~Negele, A.~Pochinsky, and S.~Syritsyn, ``{Lattice QCD
  calculations of nucleon transverse momentum-dependent parton distributions
  using clover and domain wall fermions},'' in {\em {33rd International
  Symposium on Lattice Field Theory}}.
\newblock SISSA, 11, 2015.
\newblock \href{http://arxiv.org/abs/1601.05717}{{\ttfamily arXiv:1601.05717
  [hep-lat]}}.

\bibitem{Fadin:1975cb}
V.~S. Fadin, E.~A. Kuraev, and L.~N. Lipatov, ``{On the Pomeranchuk Singularity
  in Asymptotically Free Theories},''
  \href{http://dx.doi.org/10.1016/0370-2693(75)90524-9}{{\em Phys. Lett. B}
  {\bfseries 60} (1975) 50--52}.

\bibitem{Kuraev:1976ge}
E.~A. Kuraev, L.~N. Lipatov, and V.~S. Fadin, ``{Multi - Reggeon Processes in
  the Yang-Mills Theory},'' {\em Sov. Phys. JETP} {\bfseries 44} (1976)
  443--450.

\bibitem{Kuraev:1977fs}
E.~A. Kuraev, L.~N. Lipatov, and V.~S. Fadin, ``{The Po\-me\-ran\-chuk
  Singularity in Nonabelian Gauge Theories},'' {\em Sov. Phys. JETP} {\bfseries
  45} (1977) 199--204.

\bibitem{Balitsky:1978ic}
I.~I. Balitsky and L.~N. Lipatov, ``{The Pomeranchuk Singularity in Quantum
  Chromodynamics},'' {\em Sov. J. Nucl. Phys.} {\bfseries 28} (1978) 822--829.

\bibitem{Boonekamp:2011ky}
M.~Boonekamp, A.~Dechambre, V.~Juranek, O.~Kepka, M.~Rangel, C.~Royon, and
  R.~Staszewski, ``{FPMC: A Generator for forward physics},''
  \href{http://arxiv.org/abs/1102.2531}{{\ttfamily arXiv:1102.2531 [hep-ph]}}.

\bibitem{Jung:2009eq}
H.~Jung and A.~de~Roeck, eds., {\em {Proceedings, HERA and the LHC Workshop
  Series on the implications of HERA for LHC physics: 2006-2008}}.
\newblock DESY, Hamburg, Germany, 3, 2009.
\newblock \href{http://arxiv.org/abs/0903.3861}{{\ttfamily arXiv:0903.3861
  [hep-ph]}}.

\bibitem{Marquet:2013rja}
C.~Marquet, C.~Royon, M.~Saimpert, and D.~Werder, ``{Probing the Pomeron
  structure using dijets and $\gamma$+jet events at the LHC},''
  \href{http://dx.doi.org/10.1103/PhysRevD.88.074029}{{\em Phys. Rev. D}
  {\bfseries 88} no.~7, (2013) 074029},
  \href{http://arxiv.org/abs/1306.4901}{{\ttfamily arXiv:1306.4901 [hep-ph]}}.

\bibitem{Kepka:2007nr}
O.~Kepka and C.~Royon, ``{Search for exclusive events using the dijet mass
  fraction at the Tevatron},''
  \href{http://dx.doi.org/10.1103/PhysRevD.76.034012}{{\em Phys. Rev. D}
  {\bfseries 76} (2007) 034012},
  \href{http://arxiv.org/abs/0704.1956}{{\ttfamily arXiv:0704.1956 [hep-ph]}}.

\bibitem{Marquet:2016ulz}
C.~Marquet, D.~E. Martins, A.~V. Pereira, M.~Rangel, and C.~Royon,
  ``{Diffractive di-jet production at the LHC with a Reggeon contribution},''
  \href{http://dx.doi.org/10.1016/j.physletb.2016.12.045}{{\em Phys. Lett. B}
  {\bfseries 766} (2017) 23--28},
  \href{http://arxiv.org/abs/1608.05674}{{\ttfamily arXiv:1608.05674
  [hep-ph]}}.

\bibitem{Chuinard:2015sva}
A.~Chuinard, C.~Royon, and R.~Staszewski, ``{Testing Pomeron flavour symmetry
  with diffractive W charge asymmetry},''
  \href{http://dx.doi.org/10.1007/JHEP04(2016)092}{{\em JHEP} {\bfseries 04}
  (2016) 092}, \href{http://arxiv.org/abs/1510.04218}{{\ttfamily
  arXiv:1510.04218 [hep-ph]}}.

\bibitem{Lukaszuk:1973nt}
L.~Lukaszuk and B.~Nicolescu, ``{A Possible interpretation of p p rising total
  cross-sections},'' \href{http://dx.doi.org/10.1007/BF02824484}{{\em Lett.
  Nuovo Cim.} {\bfseries 8} (1973) 405--413}.

\bibitem{Martynov:2018sga}
E.~Martynov and B.~Nicolescu, ``{Odderon effects in the differential
  cross-sections at Tevatron and LHC energies},''
  \href{http://dx.doi.org/10.1140/epjc/s10052-019-6954-6}{{\em Eur. Phys. J. C}
  {\bfseries 79} no.~6, (2019) 461},
  \href{http://arxiv.org/abs/1808.08580}{{\ttfamily arXiv:1808.08580
  [hep-ph]}}.

\bibitem{Breakstone:1985pe}
A.~Breakstone {\em et~al.}, ``{A Measurement of $\bar{p} p$ and $p p$ Elastic
  Scattering in the Dip Region at $\sqrt{s}=53$-{GeV}},''
  \href{http://dx.doi.org/10.1103/PhysRevLett.54.2180}{{\em Phys. Rev. Lett.}
  {\bfseries 54} (1985) 2180}.

\bibitem{Erhan:1984mv}
S.~Erhan {\em et~al.}, ``{Comparison of $\bar{p} p$ and $p p$ Elastic
  Scattering With $0.6-{\rm GeV}^ < t < 2.1-{\rm GeV}^2$ at the {CERN}
  {ISR}},'' \href{http://dx.doi.org/10.1016/0370-2693(85)91154-2}{{\em Phys.
  Lett. B} {\bfseries 152} (1985) 131--134}.

\bibitem{UA4:1986cgb}
{\bfseries UA4} Collaboration, D.~Bernard {\em et~al.}, ``{Large T Elastic
  Scattering at the {CERN} {SPS} Collider at $\sqrt{s}=630$-{GeV}},''
  \href{http://dx.doi.org/10.1016/0370-2693(86)91014-2}{{\em Phys. Lett. B}
  {\bfseries 171} (1986) 142--144}.

\bibitem{UA4:1985oqn}
{\bfseries UA4} Collaboration, M.~Bozzo {\em et~al.}, ``{Elastic Scattering at
  the {CERN} {SPS} Collider Up to a Four Momentum Transfer of 1.55-{GeV}**2},''
  \href{http://dx.doi.org/10.1016/0370-2693(85)90985-2}{{\em Phys. Lett. B}
  {\bfseries 155} (1985) 197--202}.

\bibitem{Nagy:1978iw}
E.~Nagy {\em et~al.}, ``{Measurements of Elastic Proton Proton Scattering at
  Large Momentum Transfer at the CERN Intersecting Storage Rings},''
  \href{http://dx.doi.org/10.1016/0550-3213(79)90301-8}{{\em Nucl. Phys. B}
  {\bfseries 150} (1979) 221--267}.

\bibitem{D0:2012erd}
{\bfseries D0} Collaboration, V.~M. Abazov {\em et~al.}, ``{Measurement of the
  differential cross section $d\sigma/dt$ in elastic $p\bar{p}$ scattering at
  $\sqrt{s}=1.96$ TeV},''
  \href{http://dx.doi.org/10.1103/PhysRevD.86.012009}{{\em Phys. Rev. D}
  {\bfseries 86} (2012) 012009},
  \href{http://arxiv.org/abs/1206.0687}{{\ttfamily arXiv:1206.0687 [hep-ex]}}.

\bibitem{TOTEM:2018psk}
{\bfseries TOTEM} Collaboration, G.~Antchev {\em et~al.}, ``{Elastic
  differential cross-section ${\mathrm{d}}\sigma /{\mathrm{d}}t$ at
  $\sqrt{s}=2.76\hbox { TeV}$ and implications on the existence of a colourless
  C-odd three-gluon compound state},''
  \href{http://dx.doi.org/10.1140/epjc/s10052-020-7654-y}{{\em Eur. Phys. J. C}
  {\bfseries 80} no.~2, (2020) 91},
  \href{http://arxiv.org/abs/1812.08610}{{\ttfamily arXiv:1812.08610
  [hep-ex]}}.

\bibitem{TOTEM:2011vxg}
{\bfseries TOTEM} Collaboration, G.~Antchev {\em et~al.}, ``{Proton-proton
  elastic scattering at the LHC energy of s** (1/2) = 7-TeV},''
  \href{http://dx.doi.org/10.1209/0295-5075/95/41001}{{\em EPL} {\bfseries 95}
  no.~4, (2011) 41001}, \href{http://arxiv.org/abs/1110.1385}{{\ttfamily
  arXiv:1110.1385 [hep-ex]}}.

\bibitem{TOTEM:2015oop}
{\bfseries TOTEM} Collaboration, G.~Antchev {\em et~al.}, ``{Evidence for
  non-exponential elastic proton\textendash{}proton differential cross-section
  at low |t| and $\sqrt{s}$=8 TeV by TOTEM},''
  \href{http://dx.doi.org/10.1016/j.nuclphysb.2015.08.010}{{\em Nucl. Phys. B}
  {\bfseries 899} (2015) 527--546},
  \href{http://arxiv.org/abs/1503.08111}{{\ttfamily arXiv:1503.08111
  [hep-ex]}}.

\bibitem{TOTEM:2018hki}
{\bfseries TOTEM} Collaboration, G.~Antchev {\em et~al.}, ``{Elastic
  differential cross-section measurement at $\sqrt{s}=13$ TeV by TOTEM},''
  \href{http://dx.doi.org/10.1140/epjc/s10052-019-7346-7}{{\em Eur. Phys. J. C}
  {\bfseries 79} no.~10, (2019) 861},
  \href{http://arxiv.org/abs/1812.08283}{{\ttfamily arXiv:1812.08283
  [hep-ex]}}.

\bibitem{TOTEM:2020zzr}
{\bfseries TOTEM, D0} Collaboration, V.~M. Abazov {\em et~al.}, ``{Odderon
  Exchange from Elastic Scattering Differences between $pp$ and $p \bar{p}$
  Data at 1.96~TeV and from pp Forward Scattering Measurements},''
  \href{http://dx.doi.org/10.1103/PhysRevLett.127.062003}{{\em Phys. Rev.
  Lett.} {\bfseries 127} no.~6, (2021) 062003},
  \href{http://arxiv.org/abs/2012.03981}{{\ttfamily arXiv:2012.03981
  [hep-ex]}}.

\bibitem{ALICE:2020mso}
{\bfseries ALICE} Collaboration, C.~Loizides, W.~Riegler, {\em et~al.},
  ``{Letter of Intent: A Forward Calorimeter (FoCal) in the ALICE
  experiment},''. \url{https://cds.cern.ch/record/2719928}.

\bibitem{Azzi:2019yne}
P.~Azzi {\em et~al.}, ``{Report from Working Group 1}: {Standard Model Physics
  at the HL-LHC and HE-LHC},''
  \href{http://dx.doi.org/10.23731/CYRM-2019-007.1}{{\em CERN Yellow Rep.
  Monogr.} {\bfseries 7} (2019) 1--220},
  \href{http://arxiv.org/abs/1902.04070}{{\ttfamily arXiv:1902.04070
  [hep-ph]}}.

\bibitem{FCC:2018vvp}
{\bfseries FCC} Collaboration, A.~Abada {\em et~al.}, ``{FCC-hh: The Hadron
  Collider}: {Future Circular Collider Conceptual Design Report Volume 3},''
  \href{http://dx.doi.org/10.1140/epjst/e2019-900087-0}{{\em Eur. Phys. J. ST}
  {\bfseries 228} no.~4, (2019) 755--1107}.

\bibitem{FCC:2018byv}
{\bfseries FCC} Collaboration, A.~Abada {\em et~al.}, ``{FCC Physics
  Opportunities}: {Future Circular Collider Conceptual Design Report Volume
  1},'' \href{http://dx.doi.org/10.1140/epjc/s10052-019-6904-3}{{\em Eur. Phys.
  J. C} {\bfseries 79} no.~6, (2019) 474}.

\bibitem{Mangano:2016jyj}
M.~L. Mangano {\em et~al.}, ``{Physics at a 100 TeV pp Collider: Standard Model
  Processes},'' \href{http://arxiv.org/abs/1607.01831}{{\ttfamily
  arXiv:1607.01831 [hep-ph]}}.

\bibitem{Rojo:2016kwu}
J.~Rojo, ``{Parton Distributions at a 100 TeV Hadron Collider},''
  \href{http://dx.doi.org/10.22323/1.265.0275}{{\em PoS} {\bfseries DIS2016}
  (2016) 275}, \href{http://arxiv.org/abs/1605.08302}{{\ttfamily
  arXiv:1605.08302 [hep-ph]}}.

\bibitem{PROSA:2015yid}
{\bfseries PROSA} Collaboration, O.~Zenaiev {\em et~al.}, ``{Impact of
  heavy-flavour production cross sections measured by the LHCb experiment on
  parton distribution functions at low x},''
  \href{http://dx.doi.org/10.1140/epjc/s10052-015-3618-z}{{\em Eur. Phys. J. C}
  {\bfseries 75} no.~8, (2015) 396},
  \href{http://arxiv.org/abs/1503.04581}{{\ttfamily arXiv:1503.04581
  [hep-ph]}}.

\bibitem{Gauld:2016kpd}
R.~Gauld and J.~Rojo, ``{Precision determination of the small-$x$ gluon from
  charm production at LHCb},''
  \href{http://dx.doi.org/10.1103/PhysRevLett.118.072001}{{\em Phys. Rev.
  Lett.} {\bfseries 118} no.~7, (2017) 072001},
  \href{http://arxiv.org/abs/1610.09373}{{\ttfamily arXiv:1610.09373
  [hep-ph]}}.

\bibitem{Zenaiev:2019ktw}
{\bfseries PROSA} Collaboration, O.~Zenaiev, M.~V. Garzelli, K.~Lipka, S.~O.
  Moch, A.~Cooper-Sarkar, F.~Olness, A.~Geiser, and G.~Sigl, ``{Improved
  constraints on parton distributions using LHCb, ALICE and HERA heavy-flavour
  measurements and implications for the predictions for prompt
  atmospheric-neutrino fluxes},''
  \href{http://dx.doi.org/10.1007/JHEP04(2020)118}{{\em JHEP} {\bfseries 04}
  (2020) 118}, \href{http://arxiv.org/abs/1911.13164}{{\ttfamily
  arXiv:1911.13164 [hep-ph]}}.

\bibitem{Garzelli:2020fmd}
M.~V. Garzelli, L.~Kemmler, S.~Moch, and O.~Zenaiev, ``{Heavy-flavor
  hadro-production with heavy-quark masses renormalized in the ${\overline{\rm
  MS}}$, MSR and on-shell schemes},''
  \href{http://dx.doi.org/10.1007/JHEP04(2021)043}{{\em JHEP} {\bfseries 04}
  (2021) 043}, \href{http://arxiv.org/abs/2009.07763}{{\ttfamily
  arXiv:2009.07763 [hep-ph]}}.

\bibitem{Brodsky:2015fna}
S.~J. Brodsky, A.~Kusina, F.~Lyonnet, I.~Schienbein, H.~Spiesberger, and
  R.~Vogt, ``{A review of the intrinsic heavy quark content of the nucleon},''
  \href{http://dx.doi.org/10.1155/2015/231547}{{\em Adv. High Energy Phys.}
  {\bfseries 2015} (2015) 231547},
  \href{http://arxiv.org/abs/1504.06287}{{\ttfamily arXiv:1504.06287
  [hep-ph]}}.

\bibitem{LHCb:2013xam}
{\bfseries LHCb} Collaboration, R.~Aaij {\em et~al.}, ``{Prompt charm
  production in pp collisions at sqrt(s)=7 TeV},''
  \href{http://dx.doi.org/10.1016/j.nuclphysb.2013.02.010}{{\em Nucl. Phys. B}
  {\bfseries 871} (2013) 1--20},
  \href{http://arxiv.org/abs/1302.2864}{{\ttfamily arXiv:1302.2864 [hep-ex]}}.

\bibitem{LHCb:2015swx}
{\bfseries LHCb} Collaboration, R.~Aaij {\em et~al.}, ``{Measurements of prompt
  charm production cross-sections in $pp$ collisions at $ \sqrt{s}=13 $ TeV},''
  \href{http://dx.doi.org/10.1007/JHEP03(2016)159}{{\em JHEP} {\bfseries 03}
  (2016) 159}, \href{http://arxiv.org/abs/1510.01707}{{\ttfamily
  arXiv:1510.01707 [hep-ex]}}. [Erratum: JHEP 09, 013 (2016), Erratum: JHEP 05,
  074 (2017)].

\bibitem{LHCb:2016ikn}
{\bfseries LHCb} Collaboration, R.~Aaij {\em et~al.}, ``{Measurements of prompt
  charm production cross-sections in pp collisions at $ \sqrt{s}=5 $ TeV},''
  \href{http://dx.doi.org/10.1007/JHEP06(2017)147}{{\em JHEP} {\bfseries 06}
  (2017) 147}, \href{http://arxiv.org/abs/1610.02230}{{\ttfamily
  arXiv:1610.02230 [hep-ex]}}.

\bibitem{LHCb:2021stx}
{\bfseries LHCb} Collaboration, R.~Aaij {\em et~al.}, ``{Study of $Z$ bosons
  produced in association with charm in the forward region},''
  \href{http://arxiv.org/abs/2109.08084}{{\ttfamily arXiv:2109.08084
  [hep-ex]}}.

\bibitem{Bertone:2018dse}
V.~Bertone, R.~Gauld, and J.~Rojo, ``{Neutrino Telescopes as QCD
  Microscopes},'' \href{http://dx.doi.org/10.1007/JHEP01(2019)217}{{\em JHEP}
  {\bfseries 01} (2019) 217}, \href{http://arxiv.org/abs/1808.02034}{{\ttfamily
  arXiv:1808.02034 [hep-ph]}}.

\bibitem{Garcia:2020jwr}
A.~Garcia, R.~Gauld, A.~Heijboer, and J.~Rojo, ``{Complete predictions for
  high-energy neutrino propagation in matter},''
  \href{http://dx.doi.org/10.1088/1475-7516/2020/09/025}{{\em JCAP} {\bfseries
  09} (2020) 025}, \href{http://arxiv.org/abs/2004.04756}{{\ttfamily
  arXiv:2004.04756 [hep-ph]}}.

\bibitem{Gauld:2015kvh}
R.~Gauld, J.~Rojo, L.~Rottoli, S.~Sarkar, and J.~Talbert, ``{The prompt
  atmospheric neutrino flux in the light of LHCb},''
  \href{http://dx.doi.org/10.1007/JHEP02(2016)130}{{\em JHEP} {\bfseries 02}
  (2016) 130}, \href{http://arxiv.org/abs/1511.06346}{{\ttfamily
  arXiv:1511.06346 [hep-ph]}}.

\bibitem{Garzelli:2016xmx}
{\bfseries PROSA} Collaboration, M.~V. Garzelli, S.~Moch, O.~Zenaiev,
  A.~Cooper-Sarkar, A.~Geiser, K.~Lipka, R.~Placakyte, and G.~Sigl, ``{Prompt
  neutrino fluxes in the atmosphere with PROSA parton distribution
  functions},'' \href{http://dx.doi.org/10.1007/JHEP05(2017)004}{{\em JHEP}
  {\bfseries 05} (2017) 004}, \href{http://arxiv.org/abs/1611.03815}{{\ttfamily
  arXiv:1611.03815 [hep-ph]}}.

\bibitem{IceCube:2020wum}
{\bfseries IceCube} Collaboration, R.~Abbasi {\em et~al.}, ``{The IceCube
  high-energy starting event sample: Description and flux characterization with
  7.5 years of data},''
  \href{http://dx.doi.org/10.1103/PhysRevD.104.022002}{{\em Phys. Rev. D}
  {\bfseries 104} (2020) 022002},
  \href{http://arxiv.org/abs/2011.03545}{{\ttfamily arXiv:2011.03545
  [astro-ph.HE]}}.

\bibitem{KM3Net:2016zxf}
{\bfseries KM3Net} Collaboration, S.~Adrian-Martinez {\em et~al.}, ``{Letter of
  intent for KM3NeT 2.0}''
  \href{http://dx.doi.org/10.1088/0954-3899/43/8/084001}{{\em J. Phys. G}
  {\bfseries 43} no.~8, (2016) 084001},
  \href{http://arxiv.org/abs/1601.07459}{{\ttfamily arXiv:1601.07459
  [astro-ph.IM]}}.

\bibitem{Ackermann:2022rqc}
M.~Ackermann {\em et~al.}, ``{High-energy and ultra-high-energy neutrinos: A
  Snowmass white paper},''
  \href{http://dx.doi.org/10.1016/j.jheap.2022.08.001}{{\em JHEAp} {\bfseries
  36} (2022) 55--110}, \href{http://arxiv.org/abs/2203.08096}{{\ttfamily
  arXiv:2203.08096 [hep-ph]}}.

\bibitem{Khalek:2021ulf}
R.~A. Khalek, J.~J. Ethier, E.~R. Nocera, and J.~Rojo, ``{Self-consistent
  determination of proton and nuclear PDFs at the Electron Ion Collider},''
  \href{http://dx.doi.org/10.1103/PhysRevD.103.096005}{{\em Phys. Rev. D}
  {\bfseries 103} no.~9, (2021) 096005},
  \href{http://arxiv.org/abs/2102.00018}{{\ttfamily arXiv:2102.00018
  [hep-ph]}}.

\bibitem{NuTeV:2005wsg}
{\bfseries NuTeV} Collaboration, M.~Tzanov {\em et~al.}, ``{Precise measurement
  of neutrino and anti-neutrino differential cross sections},''
  \href{http://dx.doi.org/10.1103/PhysRevD.74.012008}{{\em Phys. Rev. D}
  {\bfseries 74} (2006) 012008},
  \href{http://arxiv.org/abs/hep-ex/0509010}{{\ttfamily arXiv:hep-ex/0509010}}.

\bibitem{NOMAD:2013hbk}
{\bfseries NOMAD} Collaboration, O.~Samoylov {\em et~al.}, ``{A Precision
  Measurement of Charm Dimuon Production in Neutrino Interactions from the
  NOMAD Experiment},''
  \href{http://dx.doi.org/10.1016/j.nuclphysb.2013.08.021}{{\em Nucl. Phys. B}
  {\bfseries 876} (2013) 339--375},
  \href{http://arxiv.org/abs/1308.4750}{{\ttfamily arXiv:1308.4750 [hep-ex]}}.

\bibitem{NuTeV:2001dfo}
{\bfseries NuTeV} Collaboration, M.~Goncharov {\em et~al.}, ``{Precise
  Measurement of Dimuon Production Cross-Sections in $\nu_{\mu}$ Fe and
  $\bar{\nu}_{\mu}$ Fe Deep Inelastic Scattering at the Tevatron.},''
  \href{http://dx.doi.org/10.1103/PhysRevD.64.112006}{{\em Phys. Rev. D}
  {\bfseries 64} (2001) 112006},
  \href{http://arxiv.org/abs/hep-ex/0102049}{{\ttfamily arXiv:hep-ex/0102049}}.

\bibitem{CHORUS:1997wxi}
{\bfseries CHORUS} Collaboration, E.~Eskut {\em et~al.}, ``{The CHORUS
  experiment to search for muon-neutrino --\ensuremath{>} tau-neutrino
  oscillation},'' \href{http://dx.doi.org/10.1016/S0168-9002(97)00931-5}{{\em
  Nucl. Instrum. Meth. A} {\bfseries 401} (1997) 7--44}.

\bibitem{Chang:2014jba}
W.-C. Chang and J.-C. Peng, ``{Flavor Structure of the Nucleon Sea},''
  \href{http://dx.doi.org/10.1016/j.ppnp.2014.08.002}{{\em Prog. Part. Nucl.
  Phys.} {\bfseries 79} (2014) 95--135},
  \href{http://arxiv.org/abs/1406.1260}{{\ttfamily arXiv:1406.1260 [hep-ph]}}.

\bibitem{Alekhin:2014sya}
S.~Alekhin, J.~Blumlein, L.~Caminadac, K.~Lipka, K.~Lohwasser, S.~Moch,
  R.~Petti, and R.~Placakyte, ``{Determination of Strange Sea Quark
  Distributions from Fixed-target and Collider Data},''
  \href{http://dx.doi.org/10.1103/PhysRevD.91.094002}{{\em Phys. Rev. D}
  {\bfseries 91} no.~9, (2015) 094002},
  \href{http://arxiv.org/abs/1404.6469}{{\ttfamily arXiv:1404.6469 [hep-ph]}}.

\bibitem{Bevilacqua:2021ovq}
G.~Bevilacqua, M.~V. Garzelli, A.~Kardos, and L.~Toth, ``{W+charm production
  with massive c quarks in PowHel},''
  \href{http://arxiv.org/abs/2106.11261}{{\ttfamily arXiv:2106.11261
  [hep-ph]}}.

\bibitem{Collins:2011zzd}
J.~Collins, {\em {Foundations of perturbative QCD}}, vol.~32.
\newblock Cambridge University Press, 11, 2013.

\bibitem{Nefedov:2021vvy}
M.~Nefedov, ``{Sudakov resummation from the BFKL evolution},''
  \href{http://dx.doi.org/10.1103/PhysRevD.104.054039}{{\em Phys. Rev. D}
  {\bfseries 104} no.~5, (2021) 054039},
  \href{http://arxiv.org/abs/2105.13915}{{\ttfamily arXiv:2105.13915
  [hep-ph]}}.

\bibitem{Hentschinski:2021lsh}
M.~Hentschinski, ``{Transverse momentum dependent gluon distribution within
  high energy factorization at next-to-leading order},''
  \href{http://dx.doi.org/10.1103/PhysRevD.104.054014}{{\em Phys. Rev. D}
  {\bfseries 104} no.~5, (2021) 054014},
  \href{http://arxiv.org/abs/2107.06203}{{\ttfamily arXiv:2107.06203
  [hep-ph]}}.

\bibitem{Bacchetta:2017gcc}
A.~Bacchetta, F.~Delcarro, C.~Pisano, M.~Radici, and A.~Signori, ``{Extraction
  of partonic transverse momentum distributions from semi-inclusive
  deep-inelastic scattering, Drell-Yan and Z-boson production},''
  \href{http://dx.doi.org/10.1007/JHEP06(2017)081}{{\em JHEP} {\bfseries 06}
  (2017) 081}, \href{http://arxiv.org/abs/1703.10157}{{\ttfamily
  arXiv:1703.10157 [hep-ph]}}. [Erratum: JHEP 06, 051 (2019)].

\bibitem{Scimemi:2017etj}
I.~Scimemi and A.~Vladimirov, ``{Analysis of vector boson production within TMD
  factorization},''
  \href{http://dx.doi.org/10.1140/epjc/s10052-018-5557-y}{{\em Eur. Phys. J. C}
  {\bfseries 78} no.~2, (2018) 89},
  \href{http://arxiv.org/abs/1706.01473}{{\ttfamily arXiv:1706.01473
  [hep-ph]}}.

\bibitem{Scimemi:2019cmh}
I.~Scimemi and A.~Vladimirov, ``{Non-perturbative structure of semi-inclusive
  deep-inelastic and Drell-Yan scattering at small transverse momentum},''
  \href{http://dx.doi.org/10.1007/JHEP06(2020)137}{{\em JHEP} {\bfseries 06}
  (2020) 137}, \href{http://arxiv.org/abs/1912.06532}{{\ttfamily
  arXiv:1912.06532 [hep-ph]}}.

\bibitem{Bacchetta:2019sam}
A.~Bacchetta, V.~Bertone, C.~Bissolotti, G.~Bozzi, F.~Delcarro, F.~Piacenza,
  and M.~Radici, ``{Transverse-momentum-dependent parton distributions up to
  N$^{3}$LL from Drell-Yan data},''
  \href{http://dx.doi.org/10.1007/JHEP07(2020)117}{{\em JHEP} {\bfseries 07}
  (2020) 117}, \href{http://arxiv.org/abs/1912.07550}{{\ttfamily
  arXiv:1912.07550 [hep-ph]}}.

\bibitem{Mulders:2000sh}
P.~J. Mulders and J.~Rodrigues, ``{Transverse momentum dependence in gluon
  distribution and fragmentation functions},''
  \href{http://dx.doi.org/10.1103/PhysRevD.63.094021}{{\em Phys. Rev. D}
  {\bfseries 63} (2001) 094021},
  \href{http://arxiv.org/abs/hep-ph/0009343}{{\ttfamily arXiv:hep-ph/0009343}}.

\bibitem{Nadolsky:2007ba}
P.~M. Nadolsky, C.~Balazs, E.~L. Berger, and C.~P. Yuan, ``{Gluon-gluon
  contributions to the production of continuum diphoton pairs at hadron
  colliders},'' \href{http://dx.doi.org/10.1103/PhysRevD.76.013008}{{\em Phys.
  Rev. D} {\bfseries 76} (2007) 013008},
  \href{http://arxiv.org/abs/hep-ph/0702003}{{\ttfamily arXiv:hep-ph/0702003}}.

\bibitem{Catani:2010pd}
S.~Catani and M.~Grazzini, ``{QCD transverse-momentum resummation in gluon
  fusion processes},''
  \href{http://dx.doi.org/10.1016/j.nuclphysb.2010.12.007}{{\em Nucl. Phys. B}
  {\bfseries 845} (2011) 297--323},
  \href{http://arxiv.org/abs/1011.3918}{{\ttfamily arXiv:1011.3918 [hep-ph]}}.

\bibitem{Boer:2010zf}
D.~Boer, S.~J. Brodsky, P.~J. Mulders, and C.~Pisano, ``{Direct Probes of
  Linearly Polarized Gluons inside Unpolarized Hadrons},''
  \href{http://dx.doi.org/10.1103/PhysRevLett.106.132001}{{\em Phys. Rev.
  Lett.} {\bfseries 106} (2011) 132001},
  \href{http://arxiv.org/abs/1011.4225}{{\ttfamily arXiv:1011.4225 [hep-ph]}}.

\bibitem{Sun:2011iw}
P.~Sun, B.-W. Xiao, and F.~Yuan, ``{Gluon Distribution Functions and Higgs
  Boson Production at Moderate Transverse Momentum},''
  \href{http://dx.doi.org/10.1103/PhysRevD.84.094005}{{\em Phys. Rev.}
  {\bfseries D84} (2011) 094005},
  \href{http://arxiv.org/abs/1109.1354}{{\ttfamily arXiv:1109.1354 [hep-ph]}}.

\bibitem{Boer:2011kf}
D.~Boer, W.~J. den Dunnen, C.~Pisano, M.~Schlegel, and W.~Vogelsang,
  ``{Linearly Polarized Gluons and the Higgs Transverse Momentum
  Distribution},'' \href{http://dx.doi.org/10.1103/PhysRevLett.108.032002}{{\em
  Phys. Rev. Lett.} {\bfseries 108} (2012) 032002},
  \href{http://arxiv.org/abs/1109.1444}{{\ttfamily arXiv:1109.1444 [hep-ph]}}.

\bibitem{Pisano:2013cya}
C.~Pisano, D.~Boer, S.~J. Brodsky, M.~G. Buffing, and P.~J. Mulders, ``{Linear
  polarization of gluons and photons in unpolarized collider experiments},''
  \href{http://dx.doi.org/10.1007/JHEP10(2013)024}{{\em JHEP} {\bfseries 10}
  (2013) 024}, \href{http://arxiv.org/abs/1307.3417}{{\ttfamily arXiv:1307.3417
  [hep-ph]}}.

\bibitem{Dunnen:2014eta}
W.~J. den Dunnen, J.~P. Lansberg, C.~Pisano, and M.~Schlegel, ``{Accessing the
  Transverse Dynamics and Polarization of Gluons inside the Proton at the
  LHC},'' \href{http://dx.doi.org/10.1103/PhysRevLett.112.212001}{{\em Phys.
  Rev. Lett.} {\bfseries 112} (2014) 212001},
  \href{http://arxiv.org/abs/1401.7611}{{\ttfamily arXiv:1401.7611 [hep-ph]}}.

\bibitem{Lansberg:2017tlc}
J.-P. Lansberg, C.~Pisano, and M.~Schlegel, ``{Associated production of a
  dilepton and a $\Upsilon(J/\psi)$ at the LHC as a probe of gluon transverse
  momentum dependent distributions},''
  \href{http://dx.doi.org/10.1016/j.nuclphysb.2017.04.011}{{\em Nucl. Phys.}
  {\bfseries B920} (2017) 192--210},
  \href{http://arxiv.org/abs/1702.00305}{{\ttfamily arXiv:1702.00305
  [hep-ph]}}.

\bibitem{Chapon:2020heu}
E.~Chapon {\em et~al.}, ``{Prospects for quarkonium studies at the
  high-luminosity LHC},''
  \href{http://dx.doi.org/10.1016/j.ppnp.2021.103906}{{\em Prog. Part. Nucl.
  Phys.} {\bfseries 122} (2022) 103906},
  \href{http://arxiv.org/abs/2012.14161}{{\ttfamily arXiv:2012.14161
  [hep-ph]}}.

\bibitem{Celiberto:2020rxb}
F.~G. Celiberto, D.~Y. Ivanov, and A.~Papa, ``{Diffractive production of
  $\Lambda$ hyperons in the high-energy limit of strong interactions},''
  \href{http://dx.doi.org/10.1103/PhysRevD.102.094019}{{\em Phys. Rev. D}
  {\bfseries 102} no.~9, (2020) 094019},
  \href{http://arxiv.org/abs/2008.10513}{{\ttfamily arXiv:2008.10513
  [hep-ph]}}.

\bibitem{Celiberto:2021dzy}
F.~G. Celiberto, M.~Fucilla, D.~Y. Ivanov, and A.~Papa, ``{High-energy
  resummation in $\Lambda _c$ baryon production},''
  \href{http://dx.doi.org/10.1140/epjc/s10052-021-09448-3}{{\em Eur. Phys. J.
  C} {\bfseries 81} no.~8, (2021) 780},
  \href{http://arxiv.org/abs/2105.06432}{{\ttfamily arXiv:2105.06432
  [hep-ph]}}.

\bibitem{Bacchetta:2020vty}
A.~Bacchetta, F.~G. Celiberto, M.~Radici, and P.~Taels,
  ``{Transverse-momentum-dependent gluon distribution functions in a spectator
  model},'' \href{http://dx.doi.org/10.1140/epjc/s10052-020-8327-6}{{\em Eur.
  Phys. J. C} {\bfseries 80} no.~8, (2020) 733},
  \href{http://arxiv.org/abs/2005.02288}{{\ttfamily arXiv:2005.02288
  [hep-ph]}}.

\bibitem{Rogers:2010dm}
T.~C. Rogers and P.~J. Mulders, ``{No Generalized TMD-Factorization in
  Hadro-Production of High Transverse Momentum Hadrons},''
  \href{http://dx.doi.org/10.1103/PhysRevD.81.094006}{{\em Phys. Rev. D}
  {\bfseries 81} (2010) 094006},
  \href{http://arxiv.org/abs/1001.2977}{{\ttfamily arXiv:1001.2977 [hep-ph]}}.

\bibitem{Frankfurt:2022jns}
L.~Frankfurt, V.~Guzey, A.~Stasto, and M.~Strikman, ``{Selected topics in
  diffraction with protons and nuclei: past, present, and future},''
  \href{http://arxiv.org/abs/2203.12289}{{\ttfamily arXiv:2203.12289
  [hep-ph]}}.

\bibitem{Flett:2019pux}
C.~A. Flett, S.~P. Jones, A.~D. Martin, M.~G. Ryskin, and T.~Teubner, ``{How to
  include exclusive $J/\psi$ production data in global PDF analyses},''
  \href{http://dx.doi.org/10.1103/PhysRevD.101.094011}{{\em Phys. Rev. D}
  {\bfseries 101} no.~9, (2020) 094011},
  \href{http://arxiv.org/abs/1908.08398}{{\ttfamily arXiv:1908.08398
  [hep-ph]}}.

\bibitem{Flett:2020duk}
C.~A. Flett, A.~D. Martin, M.~G. Ryskin, and T.~Teubner, ``{Very low $x$ gluon
  density determined by LHCb exclusive $J/\psi$ data},''
  \href{http://dx.doi.org/10.1103/PhysRevD.102.114021}{{\em Phys. Rev. D}
  {\bfseries 102} (2020) 114021},
  \href{http://arxiv.org/abs/2006.13857}{{\ttfamily arXiv:2006.13857
  [hep-ph]}}.

\bibitem{Klein:2019qfb}
S.~R. Klein and H.~M\"antysaari, ``{Imaging the nucleus with high-energy
  photons},'' \href{http://dx.doi.org/10.1038/s42254-019-0107-6}{{\em Nature
  Rev. Phys.} {\bfseries 1} no.~11, (2019) 662--674},
  \href{http://arxiv.org/abs/1910.10858}{{\ttfamily arXiv:1910.10858
  [hep-ex]}}.

\bibitem{STAR:2017enh}
{\bfseries STAR} Collaboration, L.~Adamczyk {\em et~al.}, ``{Coherent
  diffractive photoproduction of \ensuremath{\rho}0 mesons on gold nuclei at
  200 GeV/nucleon-pair at the Relativistic Heavy Ion Collider},''
  \href{http://dx.doi.org/10.1103/PhysRevC.96.054904}{{\em Phys. Rev. C}
  {\bfseries 96} no.~5, (2017) 054904},
  \href{http://arxiv.org/abs/1702.07705}{{\ttfamily arXiv:1702.07705
  [nucl-ex]}}.

\bibitem{Ji:2003ak}
X.-d. Ji, ``{Viewing the proton through 'color' filters},''
  \href{http://dx.doi.org/10.1103/PhysRevLett.91.062001}{{\em Phys. Rev. Lett.}
  {\bfseries 91} (2003) 062001},
  \href{http://arxiv.org/abs/hep-ph/0304037}{{\ttfamily arXiv:hep-ph/0304037}}.

\bibitem{Belitsky:2003nz}
A.~V. Belitsky, X.-d. Ji, and F.~Yuan, ``{Quark imaging in the proton via
  quantum phase space distributions},''
  \href{http://dx.doi.org/10.1103/PhysRevD.69.074014}{{\em Phys. Rev. D}
  {\bfseries 69} (2004) 074014},
  \href{http://arxiv.org/abs/hep-ph/0307383}{{\ttfamily arXiv:hep-ph/0307383}}.

\bibitem{Meissner:2009ww}
S.~Meissner, A.~Metz, and M.~Schlegel, ``{Generalized parton correlation
  functions for a spin-1/2 hadron},''
  \href{http://dx.doi.org/10.1088/1126-6708/2009/08/056}{{\em JHEP} {\bfseries
  08} (2009) 056}, \href{http://arxiv.org/abs/0906.5323}{{\ttfamily
  arXiv:0906.5323 [hep-ph]}}.

\bibitem{Bhattacharya:2017bvs}
S.~Bhattacharya, A.~Metz, and J.~Zhou, ``{Generalized TMDs and the exclusive
  double Drell\textendash{}Yan process},''
  \href{http://dx.doi.org/10.1016/j.physletb.2017.05.081}{{\em Phys. Lett. B}
  {\bfseries 771} (2017) 396--400},
  \href{http://arxiv.org/abs/1702.04387}{{\ttfamily arXiv:1702.04387
  [hep-ph]}}. [Erratum: Phys.Lett.B 810, 135866 (2020)].

\bibitem{Hagiwara:2017fye}
Y.~Hagiwara, Y.~Hatta, R.~Pasechnik, M.~Tasevsky, and O.~Teryaev, ``{Accessing
  the gluon Wigner distribution in ultraperipheral $pA$ collisions},''
  \href{http://dx.doi.org/10.1103/PhysRevD.96.034009}{{\em Phys. Rev. D}
  {\bfseries 96} no.~3, (2017) 034009},
  \href{http://arxiv.org/abs/1706.01765}{{\ttfamily arXiv:1706.01765
  [hep-ph]}}.

\bibitem{Boussarie:2018zwg}
R.~Boussarie, Y.~Hatta, B.-W. Xiao, and F.~Yuan, ``{Probing the
  Weizs\"acker-Williams gluon Wigner distribution in $pp$ collisions},''
  \href{http://dx.doi.org/10.1103/PhysRevD.98.074015}{{\em Phys. Rev. D}
  {\bfseries 98} no.~7, (2018) 074015},
  \href{http://arxiv.org/abs/1807.08697}{{\ttfamily arXiv:1807.08697
  [hep-ph]}}.

\bibitem{CMS:2010ifv}
{\bfseries CMS} Collaboration, V.~Khachatryan {\em et~al.}, ``{Observation of
  Long-Range Near-Side Angular Correlations in Proton-Proton Collisions at the
  LHC},'' \href{http://dx.doi.org/10.1007/JHEP09(2010)091}{{\em JHEP}
  {\bfseries 09} (2010) 091}, \href{http://arxiv.org/abs/1009.4122}{{\ttfamily
  arXiv:1009.4122 [hep-ex]}}.

\bibitem{CMS:2012qk}
{\bfseries CMS} Collaboration, V.~Khachatryan {\em et~al.}, ``{Observation of
  Long-Range Near-Side Angular Correlations in Proton-Lead Collisions at the
  LHC},'' \href{http://dx.doi.org/10.1016/j.physletb.2012.11.025}{{\em Phys.
  Lett. B} {\bfseries 718} (2013) 795--814},
  \href{http://arxiv.org/abs/1210.5482}{{\ttfamily arXiv:1210.5482 [nucl-ex]}}.

\bibitem{ATLAS:2012cix}
{\bfseries ATLAS} Collaboration, G.~Aad {\em et~al.}, ``{Observation of
  Associated Near-Side and Away-Side Long-Range Correlations in
  $\sqrt{s_{NN}}$=5.02 TeV Proton-Lead Collisions with the ATLAS Detector},''
  \href{http://dx.doi.org/10.1103/PhysRevLett.110.182302}{{\em Phys. Rev.
  Lett.} {\bfseries 110} no.~18, (2013) 182302},
  \href{http://arxiv.org/abs/1212.5198}{{\ttfamily arXiv:1212.5198 [hep-ex]}}.

\bibitem{ALICE:2012eyl}
{\bfseries ALICE} Collaboration, B.~Abelev {\em et~al.}, ``{Long-range angular
  correlations on the near and away side in $p$-Pb collisions at
  $\sqrt{s_{NN}}=5.02$ TeV},''
  \href{http://dx.doi.org/10.1016/j.physletb.2013.01.012}{{\em Phys. Lett. B}
  {\bfseries 719} (2013) 29--41},
  \href{http://arxiv.org/abs/1212.2001}{{\ttfamily arXiv:1212.2001 [nucl-ex]}}.

\bibitem{LHCb:2015coe}
{\bfseries LHCb} Collaboration, R.~Aaij {\em et~al.}, ``{Measurements of
  long-range near-side angular correlations in $\sqrt{s_{\text{NN}}}=5$TeV
  proton-lead collisions in the forward region},''
  \href{http://dx.doi.org/10.1016/j.physletb.2016.09.064}{{\em Phys. Lett. B}
  {\bfseries 762} (2016) 473--483},
  \href{http://arxiv.org/abs/1512.00439}{{\ttfamily arXiv:1512.00439
  [nucl-ex]}}.

\bibitem{ATLAS:2017fur}
{\bfseries ATLAS} Collaboration, G.~Aad {\em et~al.}, ``{Evidence for
  light-by-light scattering in heavy-ion collisions with the ATLAS detector at
  the LHC},'' \href{http://dx.doi.org/10.1038/nphys4208}{{\em Nature Phys.}
  {\bfseries 13} no.~9, (2017) 852--858},
  \href{http://arxiv.org/abs/1702.01625}{{\ttfamily arXiv:1702.01625
  [hep-ex]}}.

\bibitem{CMS:2018erd}
{\bfseries CMS} Collaboration, A.~M. Sirunyan {\em et~al.}, ``{Evidence for
  light-by-light scattering and searches for axion-like particles in
  ultraperipheral PbPb collisions at $\sqrt{s_\mathrm{NN}} =$ 5.02 TeV},''
  \href{http://dx.doi.org/10.1016/j.physletb.2019.134826}{{\em Phys. Lett. B}
  {\bfseries 797} (2019) 134826},
  \href{http://arxiv.org/abs/1810.04602}{{\ttfamily arXiv:1810.04602
  [hep-ex]}}.

\bibitem{Dainese:2703572}
A.~Dainese, M.~Mangano, A.~B. Meyer, A.~Nisati, G.~Salam, and M.~A. Vesterinen,
  \href{http://dx.doi.org/10.23731/CYRM-2019-007}{``{Report on the Physics at
  the HL-LHC, and Perspectives for the HE-LHC},''} CERN Yellow Report
  CERN-2019-007, CERN, 2019.
\newblock \url{https://cds.cern.ch/record/2703572}.

\bibitem{ATLAS-ZDC-LHCC}
{ATLAS Collaboration}, ``{A Radiation-Hard Zero Degree Calorimeter for ATLAS in
  the HL-LHC era},'' 2021.
\newblock \url{https://cds.cern.ch/record/2781150}.

\bibitem{CMS-ZDC-TDR}
Y.~Bashan, Z.~Citron, B.~Cole, M.~Grosse~Perdekamp, A.~Hase, T.~Koeth,
  C.~Lantz, S.~Lascio, R.~Longo, D.~MacLean, A.~Mignerey, Y.~Moyal, M.~Murray,
  M.~Nickel, M.~Phipps, S.~Popescu, N.~Santiago, A.~Sickles, S.~Shenkar,
  P.~Steinberg, L.~Sudit, A.~Tate, Q.~Wang, and S.~Yang, ``{A Run 4 Zero Degree
  Calorimeter for CMS},'' tech. rep., CERN, Geneva, Nov, 2021.
\newblock \url{https://cds.cern.ch/record/2791533}.
\newblock This is a joint project with the ATLAS heavy ion group.

\bibitem{CMS:2667167}
{CMS Collaboration}, ``{A MIP Timing Detector for the CMS Phase-2 Upgrade},''
  tech. rep., CERN, Geneva, Mar, 2019.
\newblock \url{https://cds.cern.ch/record/2667167}.

\bibitem{ALICE:2803563}
C.~ALICE, ``{Letter of intent for ALICE 3: A next generation heavy-ion
  experiment at the LHC},'' tech. rep., CERN, Geneva, Mar, 2022.
\newblock \url{https://cds.cern.ch/record/2803563}.

\bibitem{PHENIX:2015siv}
{\bfseries PHENIX} Collaboration, A.~Adare {\em et~al.}, ``{An Upgrade Proposal
  from the PHENIX Collaboration},''
  \href{http://arxiv.org/abs/1501.06197}{{\ttfamily arXiv:1501.06197
  [nucl-ex]}}.

\bibitem{ATLAS:2010isq}
{\bfseries ATLAS} Collaboration, G.~Aad {\em et~al.}, ``{Observation of a
  Centrality-Dependent Dijet Asymmetry in Lead-Lead Collisions at
  $\sqrt{s_{NN}}=2.77$ TeV with the ATLAS Detector at the LHC},''
  \href{http://dx.doi.org/10.1103/PhysRevLett.105.252303}{{\em Phys. Rev.
  Lett.} {\bfseries 105} (2010) 252303},
  \href{http://arxiv.org/abs/1011.6182}{{\ttfamily arXiv:1011.6182 [hep-ex]}}.

\bibitem{CMS:2011iwn}
{\bfseries CMS} Collaboration, S.~Chatrchyan {\em et~al.}, ``{Observation and
  studies of jet quenching in PbPb collisions at nucleon-nucleon center-of-mass
  energy = 2.76 TeV},''
  \href{http://dx.doi.org/10.1103/PhysRevC.84.024906}{{\em Phys. Rev. C}
  {\bfseries 84} (2011) 024906},
  \href{http://arxiv.org/abs/1102.1957}{{\ttfamily arXiv:1102.1957 [nucl-ex]}}.

\bibitem{CMS:2012ulu}
{\bfseries CMS} Collaboration, S.~Chatrchyan {\em et~al.}, ``{Jet momentum
  dependence of jet quenching in PbPb collisions at $\sqrt{s_{NN}}=2.76$
  TeV},'' \href{http://dx.doi.org/10.1016/j.physletb.2012.04.058}{{\em Phys.
  Lett. B} {\bfseries 712} (2012) 176--197},
  \href{http://arxiv.org/abs/1202.5022}{{\ttfamily arXiv:1202.5022 [nucl-ex]}}.

\bibitem{CMS:2012ytf}
{\bfseries CMS} Collaboration, S.~Chatrchyan {\em et~al.}, ``{Studies of jet
  quenching using isolated-photon+jet correlations in PbPb and $pp$ collisions
  at $\sqrt{s_{NN}}=2.76$ TeV},''
  \href{http://dx.doi.org/10.1016/j.physletb.2012.11.003}{{\em Phys. Lett. B}
  {\bfseries 718} (2013) 773--794},
  \href{http://arxiv.org/abs/1205.0206}{{\ttfamily arXiv:1205.0206 [nucl-ex]}}.

\bibitem{CMS:2017ehl}
{\bfseries CMS} Collaboration, A.~M. Sirunyan {\em et~al.}, ``{Study of jet
  quenching with isolated-photon+jet correlations in PbPb and pp collisions at
  $\sqrt{s_{_{\mathrm{NN}}}} =$ 5.02 TeV},''
  \href{http://dx.doi.org/10.1016/j.physletb.2018.07.061}{{\em Phys. Lett. B}
  {\bfseries 785} (2018) 14--39},
  \href{http://arxiv.org/abs/1711.09738}{{\ttfamily arXiv:1711.09738
  [nucl-ex]}}.

\bibitem{ATLAS:2018dgb}
{\bfseries ATLAS} Collaboration, M.~Aaboud {\em et~al.}, ``{Measurement of
  photon\textendash{}jet transverse momentum correlations in 5.02 TeV Pb + Pb
  and $pp$ collisions with ATLAS},''
  \href{http://dx.doi.org/10.1016/j.physletb.2018.12.023}{{\em Phys. Lett. B}
  {\bfseries 789} (2019) 167--190},
  \href{http://arxiv.org/abs/1809.07280}{{\ttfamily arXiv:1809.07280
  [nucl-ex]}}.

\bibitem{CMS:2017eqd}
{\bfseries CMS} Collaboration, A.~M. Sirunyan {\em et~al.}, ``{Study of Jet
  Quenching with $Z+\text{jet}$ Correlations in Pb-Pb and $pp$ Collisions at
  ${\sqrt{s}}_{NN}=5.02\text{ }\text{ }\mathrm{TeV}$},''
  \href{http://dx.doi.org/10.1103/PhysRevLett.119.082301}{{\em Phys. Rev.
  Lett.} {\bfseries 119} no.~8, (2017) 082301},
  \href{http://arxiv.org/abs/1702.01060}{{\ttfamily arXiv:1702.01060
  [nucl-ex]}}.

\bibitem{ATLAS:2012tjt}
{\bfseries ATLAS} Collaboration, G.~Aad {\em et~al.}, ``{Measurement of the jet
  radius and transverse momentum dependence of inclusive jet suppression in
  lead-lead collisions at $\sqrt{s_{NN}}$= 2.76 TeV with the ATLAS detector},''
  \href{http://dx.doi.org/10.1016/j.physletb.2013.01.024}{{\em Phys. Lett. B}
  {\bfseries 719} (2013) 220--241},
  \href{http://arxiv.org/abs/1208.1967}{{\ttfamily arXiv:1208.1967 [hep-ex]}}.

\bibitem{ATLAS:2014ipv}
{\bfseries ATLAS} Collaboration, G.~Aad {\em et~al.}, ``{Measurements of the
  Nuclear Modification Factor for Jets in Pb+Pb Collisions at
  $\sqrt{s_{\mathrm{NN}}}=2.76$ TeV with the ATLAS Detector},''
  \href{http://dx.doi.org/10.1103/PhysRevLett.114.072302}{{\em Phys. Rev.
  Lett.} {\bfseries 114} no.~7, (2015) 072302},
  \href{http://arxiv.org/abs/1411.2357}{{\ttfamily arXiv:1411.2357 [hep-ex]}}.

\bibitem{CMS:2016uxf}
{\bfseries CMS} Collaboration, V.~Khachatryan {\em et~al.}, ``{Measurement of
  inclusive jet cross sections in $pp$ and PbPb collisions at $\sqrt{s_{NN}}=$
  2.76 TeV},'' \href{http://dx.doi.org/10.1103/PhysRevC.96.015202}{{\em Phys.
  Rev. C} {\bfseries 96} no.~1, (2017) 015202},
  \href{http://arxiv.org/abs/1609.05383}{{\ttfamily arXiv:1609.05383
  [nucl-ex]}}.

\bibitem{ATLAS:2018gwx}
{\bfseries ATLAS} Collaboration, M.~Aaboud {\em et~al.}, ``{Measurement of the
  nuclear modification factor for inclusive jets in Pb+Pb collisions at
  $\sqrt{s_\mathrm{NN}}=5.02$ TeV with the ATLAS detector},''
  \href{http://dx.doi.org/10.1016/j.physletb.2018.10.076}{{\em Phys. Lett. B}
  {\bfseries 790} (2019) 108--128},
  \href{http://arxiv.org/abs/1805.05635}{{\ttfamily arXiv:1805.05635
  [nucl-ex]}}.

\bibitem{CMS:2021vui}
{\bfseries CMS} Collaboration, A.~M. Sirunyan {\em et~al.}, ``{First
  measurement of large area jet transverse momentum spectra in heavy-ion
  collisions},'' \href{http://dx.doi.org/10.1007/JHEP05(2021)284}{{\em JHEP}
  {\bfseries 05} (2021) 284}, \href{http://arxiv.org/abs/2102.13080}{{\ttfamily
  arXiv:2102.13080 [hep-ex]}}.

\bibitem{ALICE:2015mjv}
{\bfseries ALICE} Collaboration, J.~Adam {\em et~al.}, ``{Measurement of jet
  suppression in central Pb-Pb collisions at $\sqrt{s_{\rm NN}}$ = 2.76 TeV},''
  \href{http://dx.doi.org/10.1016/j.physletb.2015.04.039}{{\em Phys. Lett. B}
  {\bfseries 746} (2015) 1--14},
  \href{http://arxiv.org/abs/1502.01689}{{\ttfamily arXiv:1502.01689
  [nucl-ex]}}.

\bibitem{CMS:2015hkr}
{\bfseries CMS} Collaboration, V.~Khachatryan {\em et~al.}, ``{Measurement of
  transverse momentum relative to dijet systems in PbPb and pp collisions at $
  \sqrt{s_{\mathrm{NN}}}=2.76 $ TeV},''
  \href{http://dx.doi.org/10.1007/JHEP01(2016)006}{{\em JHEP} {\bfseries 01}
  (2016) 006}, \href{http://arxiv.org/abs/1509.09029}{{\ttfamily
  arXiv:1509.09029 [nucl-ex]}}.

\bibitem{CMS:2013lhm}
{\bfseries CMS} Collaboration, S.~Chatrchyan {\em et~al.}, ``{Modification of
  Jet Shapes in PbPb Collisions at $\sqrt {s_{NN}} = 2.76$ TeV},''
  \href{http://dx.doi.org/10.1016/j.physletb.2014.01.042}{{\em Phys. Lett. B}
  {\bfseries 730} (2014) 243--263},
  \href{http://arxiv.org/abs/1310.0878}{{\ttfamily arXiv:1310.0878 [nucl-ex]}}.

\bibitem{CMS:2014jjt}
{\bfseries CMS} Collaboration, S.~Chatrchyan {\em et~al.}, ``{Measurement of
  Jet Fragmentation in PbPb and pp Collisions at $\sqrt{s_{NN}}= 2.76$ TeV},''
  \href{http://dx.doi.org/10.1103/PhysRevC.90.024908}{{\em Phys. Rev. C}
  {\bfseries 90} no.~2, (2014) 024908},
  \href{http://arxiv.org/abs/1406.0932}{{\ttfamily arXiv:1406.0932 [nucl-ex]}}.

\bibitem{ATLAS:2014dtd}
{\bfseries ATLAS} Collaboration, G.~Aad {\em et~al.}, ``{Measurement of
  inclusive jet charged-particle fragmentation functions in Pb+Pb collisions at
  $\sqrt{s_{NN}}=2.76$ TeV with the ATLAS detector},''
  \href{http://dx.doi.org/10.1016/j.physletb.2014.10.065}{{\em Phys. Lett. B}
  {\bfseries 739} (2014) 320--342},
  \href{http://arxiv.org/abs/1406.2979}{{\ttfamily arXiv:1406.2979 [hep-ex]}}.

\bibitem{CMS:2017qlm}
{\bfseries CMS} Collaboration, A.~M. Sirunyan {\em et~al.}, ``{Measurement of
  the Splitting Function in $pp$ and Pb-Pb Collisions at
  $\sqrt{s_{_{\mathrm{NN}}}} =$ 5.02 TeV},''
  \href{http://dx.doi.org/10.1103/PhysRevLett.120.142302}{{\em Phys. Rev.
  Lett.} {\bfseries 120} no.~14, (2018) 142302},
  \href{http://arxiv.org/abs/1708.09429}{{\ttfamily arXiv:1708.09429
  [nucl-ex]}}.

\bibitem{CMS:2018jco}
{\bfseries CMS} Collaboration, A.~M. Sirunyan {\em et~al.}, ``{Jet Shapes of
  Isolated Photon-Tagged Jets in Pb-Pb and pp Collisions at
  $\sqrt{s_\mathrm{NN}} =$ 5.02 TeV},''
  \href{http://dx.doi.org/10.1103/PhysRevLett.122.152001}{{\em Phys. Rev.
  Lett.} {\bfseries 122} no.~15, (2019) 152001},
  \href{http://arxiv.org/abs/1809.08602}{{\ttfamily arXiv:1809.08602
  [hep-ex]}}.

\bibitem{ATLAS:2020wmg}
{\bfseries ATLAS} Collaboration, G.~Aad {\em et~al.}, ``{Medium-Induced
  Modification of $Z$-Tagged Charged Particle Yields in $Pb+Pb$ Collisions at
  5.02 TeV with the ATLAS Detector},''
  \href{http://dx.doi.org/10.1103/PhysRevLett.126.072301}{{\em Phys. Rev.
  Lett.} {\bfseries 126} no.~7, (2021) 072301},
  \href{http://arxiv.org/abs/2008.09811}{{\ttfamily arXiv:2008.09811
  [nucl-ex]}}.

\bibitem{ATLAS:2018bvp}
{\bfseries ATLAS} Collaboration, M.~Aaboud {\em et~al.}, ``{Measurement of jet
  fragmentation in Pb+Pb and $pp$ collisions at $\sqrt{s_{NN}} = 5.02$ TeV with
  the ATLAS detector},''
  \href{http://dx.doi.org/10.1103/PhysRevC.98.024908}{{\em Phys. Rev. C}
  {\bfseries 98} no.~2, (2018) 024908},
  \href{http://arxiv.org/abs/1805.05424}{{\ttfamily arXiv:1805.05424
  [nucl-ex]}}.

\bibitem{CMS:2018mqn}
{\bfseries CMS} Collaboration, A.~M. Sirunyan {\em et~al.}, ``{Observation of
  Medium-Induced Modifications of Jet Fragmentation in Pb-Pb Collisions at
  $\sqrt{s_{NN}}=$ 5.02 TeV Using Isolated Photon-Tagged Jets},''
  \href{http://dx.doi.org/10.1103/PhysRevLett.121.242301}{{\em Phys. Rev.
  Lett.} {\bfseries 121} no.~24, (2018) 242301},
  \href{http://arxiv.org/abs/1801.04895}{{\ttfamily arXiv:1801.04895
  [hep-ex]}}.

\bibitem{CMS:2018fof}
{\bfseries CMS} Collaboration, A.~M. Sirunyan {\em et~al.}, ``{Measurement of
  the groomed jet mass in PbPb and pp collisions at $
  \sqrt{s_{\mathrm{NN}}}=5.02 $ TeV},''
  \href{http://dx.doi.org/10.1007/JHEP10(2018)161}{{\em JHEP} {\bfseries 10}
  (2018) 161}, \href{http://arxiv.org/abs/1805.05145}{{\ttfamily
  arXiv:1805.05145 [hep-ex]}}.

\bibitem{ALICE:2019ykw}
{\bfseries ALICE} Collaboration, S.~Acharya {\em et~al.}, ``{Exploration of jet
  substructure using iterative declustering in pp and Pb\textendash{}Pb
  collisions at LHC energies},''
  \href{http://dx.doi.org/10.1016/j.physletb.2020.135227}{{\em Phys. Lett. B}
  {\bfseries 802} (2020) 135227},
  \href{http://arxiv.org/abs/1905.02512}{{\ttfamily arXiv:1905.02512
  [nucl-ex]}}.

\bibitem{ALICE:2018dxf}
{\bfseries ALICE} Collaboration, S.~Acharya {\em et~al.}, ``{Medium
  modification of the shape of small-radius jets in central Pb-Pb collisions at
  $\sqrt{s_{\mathrm {NN}}} = 2.76\,\rm{TeV}$},''
  \href{http://dx.doi.org/10.1007/JHEP10(2018)139}{{\em JHEP} {\bfseries 10}
  (2018) 139}, \href{http://arxiv.org/abs/1807.06854}{{\ttfamily
  arXiv:1807.06854 [nucl-ex]}}.

\bibitem{CMS:2020plq}
{\bfseries CMS} Collaboration, A.~M. Sirunyan {\em et~al.}, ``{Measurement of
  quark- and gluon-like jet fractions using jet charge in PbPb and pp
  collisions at 5.02 TeV},''
  \href{http://dx.doi.org/10.1007/JHEP07(2020)115}{{\em JHEP} {\bfseries 07}
  (2020) 115}, \href{http://arxiv.org/abs/2004.00602}{{\ttfamily
  arXiv:2004.00602 [hep-ex]}}.

\bibitem{CMS:2021otx}
{\bfseries CMS} Collaboration, A.~M. Sirunyan {\em et~al.}, ``{Using Z Boson
  Events to Study Parton-Medium Interactions in Pb-Pb Collisions},''
  \href{http://dx.doi.org/10.1103/PhysRevLett.128.122301}{{\em Phys. Rev.
  Lett.} {\bfseries 128} no.~12, (2022) 122301},
  \href{http://arxiv.org/abs/2103.04377}{{\ttfamily arXiv:2103.04377
  [hep-ex]}}.

\bibitem{FTR-18-025}
{CMS Collaboration}, ``{Performance of jet quenching measurements in pp and
  PbPb collisions with CMS at the HL-LHC},'' CMS Physics Analysis Summary
  CMS-PAS-FTR-18-025, CERN, 2018.
\newblock \url{http://cds.cern.ch/record/2651892}.

\bibitem{ATL-PHYS-PUB-2018-019}
{ATLAS Collaboration}, ``{Projections for ATLAS Measurements of Jet
  Modifications in Pb+Pb Collisions in LHC Runs 3 and 4}.''
  {ATL-PHYS-PUB-2018-019}, 2018.
\newblock \url{https://cds.cern.ch/record/2644406}.

\bibitem{sPHENIX:2017lqb}
{\bfseries sPHENIX} Collaboration, C.~A. Aidala {\em et~al.}, ``{Design and
  Beam Test Results for the sPHENIX Electromagnetic and Hadronic Calorimeter
  Prototypes},'' \href{http://dx.doi.org/10.1109/TNS.2018.2879047}{{\em IEEE
  Trans. Nucl. Sci.} {\bfseries 65} no.~12, (2018) 2901--2919},
  \href{http://arxiv.org/abs/1704.01461}{{\ttfamily arXiv:1704.01461
  [physics.ins-det]}}.

\bibitem{Kopeliovich:2014una}
B.~Z. Kopeliovich, I.~K. Potashnikova, I.~Schmidt, and M.~Siddikov, ``{Survival
  of charmonia in a hot environment},''
  \href{http://dx.doi.org/10.1103/PhysRevC.91.024911}{{\em Phys. Rev. C}
  {\bfseries 91} no.~2, (2015) 024911},
  \href{http://arxiv.org/abs/1409.5147}{{\ttfamily arXiv:1409.5147 [hep-ph]}}.

\bibitem{Aronson:2017ymv}
S.~Aronson, E.~Borras, B.~Odegard, R.~Sharma, and I.~Vitev, ``{Collisional and
  thermal dissociation of $J/\psi$ and $\Upsilon$ states at the LHC},''
  \href{http://dx.doi.org/10.1016/j.physletb.2018.01.038}{{\em Phys. Lett. B}
  {\bfseries 778} (2018) 384--391},
  \href{http://arxiv.org/abs/1709.02372}{{\ttfamily arXiv:1709.02372
  [hep-ph]}}.

\bibitem{Du:2017qkv}
X.~Du, R.~Rapp, and M.~He, ``{Color Screening and Regeneration of Bottomonia in
  High-Energy Heavy-Ion Collisions},''
  \href{http://dx.doi.org/10.1103/PhysRevC.96.054901}{{\em Phys. Rev. C}
  {\bfseries 96} no.~5, (2017) 054901},
  \href{http://arxiv.org/abs/1706.08670}{{\ttfamily arXiv:1706.08670
  [hep-ph]}}.

\bibitem{CMS:2017uoy}
{\bfseries CMS} Collaboration, A.~M. Sirunyan {\em et~al.}, ``{Measurement of
  the ${B}^{\pm}$ Meson Nuclear Modification Factor in Pb-Pb Collisions at
  $\sqrt{{s}_{NN}}=5.02\text{ }\text{ }\mathrm{TeV}$},''
  \href{http://dx.doi.org/10.1103/PhysRevLett.119.152301}{{\em Phys. Rev.
  Lett.} {\bfseries 119} no.~15, (2017) 152301},
  \href{http://arxiv.org/abs/1705.04727}{{\ttfamily arXiv:1705.04727
  [hep-ex]}}.

\bibitem{CMS:2018eso}
{\bfseries CMS} Collaboration, A.~M. Sirunyan {\em et~al.}, ``{Measurement of
  B$^0_\mathrm{s}$ meson production in pp and PbPb collisions at
  $\sqrt{s_\mathrm{NN}} =$ 5.02 TeV},''
  \href{http://dx.doi.org/10.1016/j.physletb.2019.07.014}{{\em Phys. Lett. B}
  {\bfseries 796} (2019) 168--190},
  \href{http://arxiv.org/abs/1810.03022}{{\ttfamily arXiv:1810.03022
  [hep-ex]}}.

\bibitem{ALICE:2018lyv}
{\bfseries ALICE} Collaboration, S.~Acharya {\em et~al.}, ``{Measurement of
  D$^{0}$, D$^{+}$, D$^{*+}$ and D$_{s}^{+}$ production in Pb-Pb collisions at
  $ \sqrt{{\mathrm{s}}_{\mathrm{NN}}}=5.02 $ TeV},''
  \href{http://dx.doi.org/10.1007/JHEP10(2018)174}{{\em JHEP} {\bfseries 10}
  (2018) 174}, \href{http://arxiv.org/abs/1804.09083}{{\ttfamily
  arXiv:1804.09083 [nucl-ex]}}.

\bibitem{CMS:2017qjw}
{\bfseries CMS} Collaboration, A.~M. Sirunyan {\em et~al.}, ``{Nuclear
  modification factor of D$^0$ mesons in PbPb collisions at
  $\sqrt{s_\mathrm{NN}} = 5.02$ TeV},''
  \href{http://dx.doi.org/10.1016/j.physletb.2018.05.074}{{\em Phys. Lett. B}
  {\bfseries 782} (2018) 474--496},
  \href{http://arxiv.org/abs/1708.04962}{{\ttfamily arXiv:1708.04962
  [nucl-ex]}}.

\bibitem{ALICE:2015vxz}
{\bfseries ALICE} Collaboration, J.~Adam {\em et~al.}, ``{Transverse momentum
  dependence of D-meson production in Pb-Pb collisions at $
  \sqrt{{\mathrm{s}}_{\mathrm{NN}}}=$ 2.76 TeV},''
  \href{http://dx.doi.org/10.1007/JHEP03(2016)081}{{\em JHEP} {\bfseries 03}
  (2016) 081}, \href{http://arxiv.org/abs/1509.06888}{{\ttfamily
  arXiv:1509.06888 [nucl-ex]}}.

\bibitem{ALICE:2021mgk}
{\bfseries ALICE} Collaboration, S.~Acharya {\em et~al.}, ``{Measurement of
  beauty and charm production in pp collisions at $ \sqrt{s} $ = 5.02 TeV via
  non-prompt and prompt D mesons},''
  \href{http://dx.doi.org/10.1007/JHEP05(2021)220}{{\em JHEP} {\bfseries 05}
  (2021) 220}, \href{http://arxiv.org/abs/2102.13601}{{\ttfamily
  arXiv:2102.13601 [nucl-ex]}}.

\bibitem{CMS:2018bwt}
{\bfseries CMS} Collaboration, A.~M. Sirunyan {\em et~al.}, ``{Studies of
  Beauty Suppression via Nonprompt $D^0$ Mesons in Pb-Pb Collisions at $Q^2 =
  4$ $\rm GeV^2$},''
  \href{http://dx.doi.org/10.1103/PhysRevLett.123.022001}{{\em Phys. Rev.
  Lett.} {\bfseries 123} no.~2, (2019) 022001},
  \href{http://arxiv.org/abs/1810.11102}{{\ttfamily arXiv:1810.11102
  [hep-ex]}}.

\bibitem{CMS:2017exb}
{\bfseries CMS} Collaboration, A.~M. Sirunyan {\em et~al.}, ``{Measurement of
  prompt and nonprompt $\mathrm{J}/{\psi }$ production in $\mathrm {p}\mathrm
  {p}$ and $\mathrm {p}\mathrm {Pb}$ collisions at $\sqrt{s_{\mathrm {NN}}}
  =5.02\,\text {TeV} $},''
  \href{http://dx.doi.org/10.1140/epjc/s10052-017-4828-3}{{\em Eur. Phys. J. C}
  {\bfseries 77} no.~4, (2017) 269},
  \href{http://arxiv.org/abs/1702.01462}{{\ttfamily arXiv:1702.01462
  [nucl-ex]}}.

\bibitem{ATLAS:2018hqe}
{\bfseries ATLAS} Collaboration, M.~Aaboud {\em et~al.}, ``{Prompt and
  non-prompt $J/\psi $ and $\psi (2\mathrm {S})$ suppression at high transverse
  momentum in $5.02~\mathrm {TeV}$ Pb+Pb collisions with the ATLAS
  experiment},'' \href{http://dx.doi.org/10.1140/epjc/s10052-018-6219-9}{{\em
  Eur. Phys. J. C} {\bfseries 78} no.~9, (2018) 762},
  \href{http://arxiv.org/abs/1805.04077}{{\ttfamily arXiv:1805.04077
  [nucl-ex]}}.

\bibitem{CMS:2011all}
{\bfseries CMS} Collaboration, S.~Chatrchyan {\em et~al.}, ``{Indications of
  suppression of excited $\Upsilon$ states in PbPb collisions at
  $\sqrt{S_{NN}}$ = 2.76 TeV},''
  \href{http://dx.doi.org/10.1103/PhysRevLett.107.052302}{{\em Phys. Rev.
  Lett.} {\bfseries 107} (2011) 052302},
  \href{http://arxiv.org/abs/1105.4894}{{\ttfamily arXiv:1105.4894 [nucl-ex]}}.

\bibitem{CMS:2012bms}
{\bfseries CMS} Collaboration, S.~Chatrchyan {\em et~al.}, ``{Suppression of
  non-prompt $J/\psi$, prompt $J/\psi$, and Y(1S) in PbPb collisions at
  $\sqrt{s_{NN}}=2.76$ TeV},''
  \href{http://dx.doi.org/10.1007/JHEP05(2012)063}{{\em JHEP} {\bfseries 05}
  (2012) 063}, \href{http://arxiv.org/abs/1201.5069}{{\ttfamily arXiv:1201.5069
  [nucl-ex]}}.

\bibitem{CMS:2012gvv}
{\bfseries CMS} Collaboration, S.~Chatrchyan {\em et~al.}, ``{Observation of
  Sequential Upsilon Suppression in PbPb Collisions},''
  \href{http://dx.doi.org/10.1103/PhysRevLett.109.222301}{{\em Phys. Rev.
  Lett.} {\bfseries 109} (2012) 222301},
  \href{http://arxiv.org/abs/1208.2826}{{\ttfamily arXiv:1208.2826 [nucl-ex]}}.
  [Erratum: Phys.Rev.Lett. 120, 199903 (2018)].

\bibitem{CMS:2016rpc}
{\bfseries CMS} Collaboration, V.~Khachatryan {\em et~al.}, ``{Suppression of
  $\Upsilon(1S), \Upsilon(2S)$ and $\Upsilon(3S)$ production in PbPb collisions
  at $\sqrt{s_{\rm NN}}$ = 2.76 TeV},''
  \href{http://dx.doi.org/10.1016/j.physletb.2017.04.031}{{\em Phys. Lett. B}
  {\bfseries 770} (2017) 357--379},
  \href{http://arxiv.org/abs/1611.01510}{{\ttfamily arXiv:1611.01510
  [nucl-ex]}}.

\bibitem{ATLAS:2010xzb}
{\bfseries ATLAS} Collaboration, G.~Aad {\em et~al.}, ``{Measurement of the
  centrality dependence of J/\ensuremath{\psi} yields and observation of Z
  production in lead\textendash{}lead collisions with the ATLAS detector at the
  LHC},'' \href{http://dx.doi.org/10.1016/j.physletb.2011.02.006}{{\em Phys.
  Lett. B} {\bfseries 697} (2011) 294--312},
  \href{http://arxiv.org/abs/1012.5419}{{\ttfamily arXiv:1012.5419 [hep-ex]}}.

\bibitem{ALICE:2012jsl}
{\bfseries ALICE} Collaboration, B.~Abelev {\em et~al.}, ``{$J/\psi$
  suppression at forward rapidity in Pb-Pb collisions at $\sqrt{s_{NN}}=2.76$
  TeV},'' \href{http://dx.doi.org/10.1103/PhysRevLett.109.072301}{{\em Phys.
  Rev. Lett.} {\bfseries 109} (2012) 072301},
  \href{http://arxiv.org/abs/1202.1383}{{\ttfamily arXiv:1202.1383 [hep-ex]}}.

\bibitem{CMS:2014vjg}
{\bfseries CMS} Collaboration, V.~Khachatryan {\em et~al.}, ``{Measurement of
  Prompt $\psi(2S) \to J/\psi$ Yield Ratios in Pb-Pb and $p-p$ Collisions at
  $\sqrt {s_{NN}}=$ 2.76 TeV},''
  \href{http://dx.doi.org/10.1103/PhysRevLett.113.262301}{{\em Phys. Rev.
  Lett.} {\bfseries 113} no.~26, (2014) 262301},
  \href{http://arxiv.org/abs/1410.1804}{{\ttfamily arXiv:1410.1804 [nucl-ex]}}.

\bibitem{ALICE:2015jrl}
{\bfseries ALICE} Collaboration, J.~Adam {\em et~al.}, ``{Differential studies
  of inclusive J/\ensuremath{\psi} and \ensuremath{\psi}(2S) production at
  forward rapidity in Pb-Pb collisions at $ \sqrt{s_{\mathrm{NN}}}=2.76 $
  TeV},'' \href{http://dx.doi.org/10.1007/JHEP05(2016)179}{{\em JHEP}
  {\bfseries 05} (2016) 179}, \href{http://arxiv.org/abs/1506.08804}{{\ttfamily
  arXiv:1506.08804 [nucl-ex]}}.

\bibitem{Reidt:2021tvq}
{\bfseries ALICE} Collaboration, F.~Reidt, ``{Upgrade of the ALICE ITS
  detector},'' \href{http://dx.doi.org/10.1016/j.nima.2022.166632}{{\em Nucl.
  Instrum. Meth. A} {\bfseries 1032} (2022) 166632},
  \href{http://arxiv.org/abs/2111.08301}{{\ttfamily arXiv:2111.08301
  [physics.ins-det]}}.

\bibitem{Colella:2021stb}
{\bfseries ALICE} Collaboration, D.~Colella, ``{ALICE ITS 3: the first truly
  cylindrical inner tracker},'' in {\em {12th International Conference on
  Position Sensitive Detectors}}.
\newblock 11, 2021.
\newblock \href{http://arxiv.org/abs/2111.09689}{{\ttfamily arXiv:2111.09689
  [physics.ins-det]}}.

\bibitem{FTR-18-024}
{CMS Collaboration}, ``{Open heavy flavor and quarkonia in heavy ion collisions
  at HL-LHC},'' CMS Physics Analysis Summary CMS-PAS-FTR-18-024, CERN, 2018.
\newblock \url{http://cds.cern.ch/record/2650897}.

\bibitem{FTR-18-026}
{CMS Collaboration}, ``{Predictions on the precision achievable for small
  system flow observables in the context of HL-LHC},'' CMS Physics Analysis
  Summary CMS-PAS-FTR-18-026, CERN, 2018.
\newblock \url{http://cds.cern.ch/record/2650773}.

\bibitem{ATL-PHYS-PUB-2018-020}
{\bfseries ATLAS} Collaboration, {ATLAS collaboration}, ``{Projections for
  ATLAS Measurements of Bulk Properties of Pb+Pb, $p$+Pb, and $pp$ Collisions
  in LHC Runs 3 and 4},'' tech. rep., CERN, Geneva, Oct, 2018.
\newblock \url{http://cds.cern.ch/record/2644407}.

\bibitem{LHCb:2020sey}
{\bfseries LHCb} Collaboration, R.~Aaij {\em et~al.}, ``{Observation of
  Multiplicity Dependent Prompt $\chi_{c1}(3872)$ and $\psi(2S)$ Production in
  $pp$ Collisions},''
  \href{http://dx.doi.org/10.1103/PhysRevLett.126.092001}{{\em Phys. Rev.
  Lett.} {\bfseries 126} no.~9, (2021) 092001},
  \href{http://arxiv.org/abs/2009.06619}{{\ttfamily arXiv:2009.06619
  [hep-ex]}}.

\bibitem{CMS:2021znk}
{\bfseries CMS} Collaboration, A.~M. Sirunyan {\em et~al.}, ``{Evidence for
  X(3872) in Pb-Pb Collisions and Studies of its Prompt Production at $\sqrt
  {s_{NN}}$=5.02\,\,TeV},''
  \href{http://dx.doi.org/10.1103/PhysRevLett.128.032001}{{\em Phys. Rev.
  Lett.} {\bfseries 128} no.~3, (2022) 032001},
  \href{http://arxiv.org/abs/2102.13048}{{\ttfamily arXiv:2102.13048
  [hep-ex]}}.

\bibitem{ALICE:2016kpq}
{\bfseries ALICE} Collaboration, {ALICE Collaboration}, ``{Correlated
  event-by-event fluctuations of flow harmonics in Pb-Pb collisions at
  $\sqrt{s_{_{\mathrm{NN}}}}=2.76$ TeV},''
  \href{http://dx.doi.org/10.1103/PhysRevLett.117.182301}{{\em Phys. Rev.
  Lett.} {\bfseries 117} (2016) 182301},
  \href{http://arxiv.org/abs/1604.07663}{{\ttfamily arXiv:1604.07663
  [nucl-ex]}}.

\bibitem{CMS-QCD-10-002}
{CMS Collaboration}, ``{Observation of long-range, near-side angular
  correlations in proton--proton collisions at the LHC},''
  \href{http://dx.doi.org/10.1007/JHEP09(2010)091}{{\em JHEP} {\bfseries 09}
  (2010) 091}, \href{http://arxiv.org/abs/1009.4122}{{\ttfamily arXiv:1009.4122
  [hep-ex]}}.

\bibitem{CMS-HIN-12-005}
{CMS Collaboration}, ``{Observation of long-range, near-side angular
  correlations in \(p\)Pb collisions at the LHC},''
  \href{http://dx.doi.org/10.1016/j.physletb.2012.11.025}{{\em Phys. Lett. B}
  {\bfseries 718} (2013) 795}, \href{http://arxiv.org/abs/1210.5482}{{\ttfamily
  arXiv:1210.5482 [hep-ex]}}.

\bibitem{HION-2012-13}
{ATLAS Collaboration}, ``{Observation of Associated Near-Side and Away-Side
  Long-Range Correlations in \(\sqrt{s_{\text{NN}}} = 5.02\,\text{TeV}\)
  Proton--Lead Collisions with the ATLAS Detector},''
  \href{http://dx.doi.org/10.1103/PhysRevLett.110.182302}{{\em Phys. Rev.
  Lett.} {\bfseries 110} (2013) 182302},
  \href{http://arxiv.org/abs/1212.5198}{{\ttfamily arXiv:1212.5198 [hep-ex]}}.

\bibitem{Gardim:2012im}
F.~G. Gardim, F.~Grassi, M.~Luzum, and J.-Y. Ollitrault, ``{Breaking of
  factorization of two-particle correlations in hydrodynamics},''
  \href{http://dx.doi.org/10.1103/PhysRevC.87.031901}{{\em Phys. Rev. C}
  {\bfseries 87} no.~3, (2013) 031901},
  \href{http://arxiv.org/abs/1211.0989}{{\ttfamily arXiv:1211.0989 [nucl-th]}}.

\bibitem{Bertulani:2005ru}
C.~A. Bertulani, S.~R. Klein, and J.~Nystrand, ``{Physics of ultra-peripheral
  nuclear collisions},''
  \href{http://dx.doi.org/10.1146/annurev.nucl.55.090704.151526}{{\em Ann. Rev.
  Nucl. Part. Sci.} {\bfseries 55} (2005) 271--310},
  \href{http://arxiv.org/abs/nucl-ex/0502005}{{\ttfamily
  arXiv:nucl-ex/0502005}}.

\bibitem{Contreras:2015dqa}
J.~G. Contreras and J.~D. Tapia~Takaki, ``{Ultra-peripheral heavy-ion
  collisions at the LHC},''
  \href{http://dx.doi.org/10.1142/S0217751X15420129}{{\em Int. J. Mod. Phys. A}
  {\bfseries 30} (2015) 1542012}.

\bibitem{Klein:2020fmr}
S.~Klein and P.~Steinberg, ``{Photonuclear and Two-photon Interactions at
  High-Energy Nuclear Colliders},''
  \href{http://dx.doi.org/10.1146/annurev-nucl-030320-033923}{{\em Ann. Rev.
  Nucl. Part. Sci.} {\bfseries 70} (2020) 323--354},
  \href{http://arxiv.org/abs/2005.01872}{{\ttfamily arXiv:2005.01872
  [nucl-ex]}}.

\bibitem{Baltz:2007kq}
A.~J. Baltz, ``{The Physics of Ultraperipheral Collisions at the LHC},''
  \href{http://dx.doi.org/10.1016/j.physrep.2007.12.001}{{\em Phys. Rept.}
  {\bfseries 458} (2008) 1--171},
  \href{http://arxiv.org/abs/0706.3356}{{\ttfamily arXiv:0706.3356 [nucl-ex]}}.

\bibitem{Citron:2018lsq}
Z.~Citron {\em et~al.}, ``{Report from Working Group 5}: {Future physics
  opportunities for high-density QCD at the LHC with heavy-ion and proton
  beams},'' \href{http://dx.doi.org/10.23731/CYRM-2019-007.1159}{{\em CERN
  Yellow Rep. Monogr.} {\bfseries 7} (2019) 1159--1410},
  \href{http://arxiv.org/abs/1812.06772}{{\ttfamily arXiv:1812.06772
  [hep-ph]}}.

\bibitem{Burmasov:2020doi}
N.~Burmasov, ``{Central Diffraction and Ultra-Peripheral Collisions in ALICE in
  Run 3 and 4},'' \href{http://dx.doi.org/10.1134/S106377962202023X}{{\em Phys.
  Part. Nucl.} {\bfseries 53} no.~2, (2022) 297--302},
  \href{http://arxiv.org/abs/2010.09752}{{\ttfamily arXiv:2010.09752
  [hep-ex]}}.

\bibitem{Klein:2017vua}
S.~R. Klein, ``{Ultra-peripheral collisions and hadronic structure},''
  \href{http://dx.doi.org/10.1016/j.nuclphysa.2017.05.098}{{\em Nucl. Phys. A}
  {\bfseries 967} (2017) 249--256},
  \href{http://arxiv.org/abs/1704.04715}{{\ttfamily arXiv:1704.04715
  [nucl-ex]}}.

\bibitem{Klein:2000dk}
S.~R. Klein, J.~Nystrand, and R.~Vogt, ``{Photoproduction of top in peripheral
  heavy ion collisions},'' \href{http://dx.doi.org/10.1007/s100520100739}{{\em
  Eur. Phys. J. C} {\bfseries 21} (2001) 563--566},
  \href{http://arxiv.org/abs/hep-ph/0005157}{{\ttfamily arXiv:hep-ph/0005157}}.

\bibitem{Klein:2002wm}
S.~R. Klein, J.~Nystrand, and R.~Vogt, ``{Heavy quark photoproduction in
  ultraperipheral heavy ion collisions},''
  \href{http://dx.doi.org/10.1103/PhysRevC.66.044906}{{\em Phys. Rev. C}
  {\bfseries 66} (2002) 044906},
  \href{http://arxiv.org/abs/hep-ph/0206220}{{\ttfamily arXiv:hep-ph/0206220}}.

\bibitem{Goncalves:2006xi}
V.~P. Goncalves and M.~V.~T. Machado, ``{Diffractive photoproduction of heavy
  quarks in hadronic collisions},''
  \href{http://dx.doi.org/10.1103/PhysRevD.75.031502}{{\em Phys. Rev. D}
  {\bfseries 75} (2007) 031502},
  \href{http://arxiv.org/abs/hep-ph/0612265}{{\ttfamily arXiv:hep-ph/0612265}}.

\bibitem{Adeluyi:2012sw}
A.~Adeluyi and T.~Nguyen, ``{Photoproduction of heavy quarks in ultraperipheral
  pp, pA, and AA collisions at the CERN Large Hadron Collider},''
  \href{http://arxiv.org/abs/1210.3327}{{\ttfamily arXiv:1210.3327 [nucl-th]}}.

\bibitem{Goncalves:2017zdx}
V.~P. Gon\c{c}alves, G.~Sampaio~dos Santos, and C.~R. Sena, ``{Inclusive heavy
  quark photoproduction in $pp$, $pPb$ and $PbPb$ collisions at Run 2 LHC
  energies},'' \href{http://dx.doi.org/10.1016/j.nuclphysa.2018.05.002}{{\em
  Nucl. Phys. A} {\bfseries 976} (2018) 33--45},
  \href{http://arxiv.org/abs/1711.04497}{{\ttfamily arXiv:1711.04497
  [hep-ph]}}.

\bibitem{Frankfurt:2008vi}
L.~Frankfurt, M.~Strikman, D.~Treleani, and C.~Weiss, ``{Evidence for color
  fluctuations in the nucleon in high-energy scattering},''
  \href{http://dx.doi.org/10.1103/PhysRevLett.101.202003}{{\em Phys. Rev.
  Lett.} {\bfseries 101} (2008) 202003},
  \href{http://arxiv.org/abs/0808.0182}{{\ttfamily arXiv:0808.0182 [hep-ph]}}.

\bibitem{Fichet:2014uka}
S.~Fichet, G.~von Gersdorff, B.~Lenzi, C.~Royon, and M.~Saimpert,
  ``{Light-by-light scattering with intact protons at the LHC: from Standard
  Model to New Physics},''
  \href{http://dx.doi.org/10.1007/JHEP02(2015)165}{{\em JHEP} {\bfseries 02}
  (2015) 165}, \href{http://arxiv.org/abs/1411.6629}{{\ttfamily arXiv:1411.6629
  [hep-ph]}}.

\bibitem{Fichet:2013gsa}
S.~Fichet, G.~von Gersdorff, O.~Kepka, B.~Lenzi, C.~Royon, and M.~Saimpert,
  ``{Probing new physics in diphoton production with proton tagging at the
  Large Hadron Collider},''
  \href{http://dx.doi.org/10.1103/PhysRevD.89.114004}{{\em Phys. Rev. D}
  {\bfseries 89} (2014) 114004},
  \href{http://arxiv.org/abs/1312.5153}{{\ttfamily arXiv:1312.5153 [hep-ph]}}.

\bibitem{ATL-PHYS-PUB-2018-018}
{ATLAS Collaboration}, ``{Prospects for Measurements of Photon-Induced
  Processes in Ultra-Peripheral Collisions of Heavy Ions with the ATLAS
  Detector in the LHC Runs~3 and 4}.'' {ATL-PHYS-PUB-2018-018}, 2018.
\newblock \url{https://cds.cern.ch/record/2641655}.

\bibitem{dEnterria:2013zqi}
D.~d'Enterria and G.~G. da~Silveira, ``{Observing light-by-light scattering at
  the Large Hadron Collider},''
  \href{http://dx.doi.org/10.1103/PhysRevLett.111.080405}{{\em Phys. Rev.
  Lett.} {\bfseries 111} (2013) 080405},
  \href{http://arxiv.org/abs/1305.7142}{{\ttfamily arXiv:1305.7142 [hep-ph]}}.
  [Erratum: Phys.Rev.Lett. 116, 129901 (2016)].

\bibitem{Knapen:2016moh}
S.~Knapen, T.~Lin, H.~K. Lou, and T.~Melia, ``{Searching for Axionlike
  Particles with Ultraperipheral Heavy-Ion Collisions},''
  \href{http://dx.doi.org/10.1103/PhysRevLett.118.171801}{{\em Phys. Rev.
  Lett.} {\bfseries 118} no.~17, (2017) 171801},
  \href{http://arxiv.org/abs/1607.06083}{{\ttfamily arXiv:1607.06083
  [hep-ph]}}.

\bibitem{ATLAS-TDR-24}
{ATLAS Collaboration}, ``{ATLAS Forward Proton Phase-I Upgrade: Technical
  Design Report},'' 2015.
\newblock \url{https://cds.cern.ch/record/2017378}.

\bibitem{CMS-TDR-13}
{CMS-TOTEM Collaboration}, ``{CMS-TOTEM Precision Proton Spectrometer},'' tech.
  rep., CERN, Sep, 2014.
\newblock \url{https://cds.cern.ch/record/1753795}.

\bibitem{Bauer:2017ris}
M.~Bauer, M.~Neubert, and A.~Thamm, ``{Collider Probes of Axion-Like
  Particles},'' \href{http://dx.doi.org/10.1007/JHEP12(2017)044}{{\em JHEP}
  {\bfseries 12} (2017) 044}, \href{http://arxiv.org/abs/1708.00443}{{\ttfamily
  arXiv:1708.00443 [hep-ph]}}.

\bibitem{CMS-FSQ-16-012}
{CMS Collaboration}, ``{Evidence for light-by-light scattering and searches for
  axion-like particles in ultraperipheral PbPb collisions at
  \(\sqrt{s_{\text{NN}}} = 5.02\,\text{TeV}\)},''
  \href{http://dx.doi.org/10.1016/j.physletb.2019.134826}{{\em Phys. Lett. B}
  {\bfseries 797} (2019) 134826},
  \href{http://arxiv.org/abs/1810.04602}{{\ttfamily arXiv:1810.04602
  [hep-ex]}}.

\bibitem{Adamova:2019vkf}
D.~Adamov\'a {\em et~al.}, ``{A next-generation LHC heavy-ion experiment},''
  \href{http://arxiv.org/abs/1902.01211}{{\ttfamily arXiv:1902.01211
  [physics.ins-det]}}.

\bibitem{Baldenegro:2018hng}
C.~Baldenegro, S.~Fichet, G.~von Gersdorff, and C.~Royon, ``{Searching for
  axion-like particles with proton tagging at the LHC},''
  \href{http://dx.doi.org/10.1007/JHEP06(2018)131}{{\em JHEP} {\bfseries 06}
  (2018) 131}, \href{http://arxiv.org/abs/1803.10835}{{\ttfamily
  arXiv:1803.10835 [hep-ph]}}.

\bibitem{Baldenegro:2019whq}
C.~Baldenegro, S.~Hassani, C.~Royon, and L.~Schoeffel, ``{Extending the
  constraint for axion-like particles as resonances at the LHC and laser beam
  experiments},'' \href{http://dx.doi.org/10.1016/j.physletb.2019.06.029}{{\em
  Phys. Lett. B} {\bfseries 795} (2019) 339--345},
  \href{http://arxiv.org/abs/1903.04151}{{\ttfamily arXiv:1903.04151
  [hep-ph]}}.

\bibitem{deFavereaudeJeneret:2009db}
J.~de~Favereau~de Jeneret, V.~Lemaitre, Y.~Liu, S.~Ovyn, T.~Pierzchala,
  K.~Piotrzkowski, X.~Rouby, N.~Schul, and M.~Vander~Donckt, ``{High energy
  photon interactions at the LHC},''
  \href{http://arxiv.org/abs/0908.2020}{{\ttfamily arXiv:0908.2020 [hep-ph]}}.

\bibitem{Pierzchala:2008xc}
T.~Pierzchala and K.~Piotrzkowski, ``{Sensitivity to anomalous quartic gauge
  couplings in photon-photon interactions at the LHC},''
  \href{http://dx.doi.org/10.1016/j.nuclphysbps.2008.07.032}{{\em Nucl. Phys. B
  Proc. Suppl.} {\bfseries 179-180} (2008) 257--264},
  \href{http://arxiv.org/abs/0807.1121}{{\ttfamily arXiv:0807.1121 [hep-ph]}}.

\bibitem{Chapon:2009hh}
E.~Chapon, C.~Royon, and O.~Kepka, ``{Anomalous quartic W W gamma gamma, Z Z
  gamma gamma, and trilinear WW gamma couplings in two-photon processes at high
  luminosity at the LHC},''
  \href{http://dx.doi.org/10.1103/PhysRevD.81.074003}{{\em Phys. Rev. D}
  {\bfseries 81} (2010) 074003},
  \href{http://arxiv.org/abs/0912.5161}{{\ttfamily arXiv:0912.5161 [hep-ph]}}.

\bibitem{STDM-2017-21}
{ATLAS Collaboration}, ``{Observation of photon-induced \(W^+W^-\) production
  in \(pp\) collisions at \(\sqrt{s} = 13\,\text{TeV}\) using the ATLAS
  detector},'' \href{http://dx.doi.org/10.1016/j.physletb.2021.136190}{{\em
  Phys. Lett. B} {\bfseries 816} (2021) 136190},
  \href{http://arxiv.org/abs/2010.04019}{{\ttfamily arXiv:2010.04019
  [hep-ex]}}.

\bibitem{ATL-PHYS-PUB-2021-026}
{ATLAS Collaboration}, ``{Sensitivity to exclusive \(WW\) production in photon
  scattering at the High Luminosity LHC}.'' {ATL-PHYS-PUB-2021-026}, 2021.
\newblock \url{https://cds.cern.ch/record/2776764}.

\bibitem{Kepka:2008yx}
O.~Kepka and C.~Royon, ``{Anomalous $W W \gamma$ coupling in photon-induced
  processes using forward detectors at the LHC},''
  \href{http://dx.doi.org/10.1103/PhysRevD.78.073005}{{\em Phys. Rev. D}
  {\bfseries 78} (2008) 073005},
  \href{http://arxiv.org/abs/0808.0322}{{\ttfamily arXiv:0808.0322 [hep-ph]}}.

\bibitem{Fichet:2015vvy}
S.~Fichet, G.~von Gersdorff, and C.~Royon, ``{Scattering light by light at 750
  GeV at the LHC},'' \href{http://dx.doi.org/10.1103/PhysRevD.93.075031}{{\em
  Phys. Rev. D} {\bfseries 93} no.~7, (2016) 075031},
  \href{http://arxiv.org/abs/1512.05751}{{\ttfamily arXiv:1512.05751
  [hep-ph]}}.

\bibitem{Fichet:2016pvq}
S.~Fichet, G.~von Gersdorff, and C.~Royon, ``{Measuring the Diphoton Coupling
  of a 750 GeV Resonance},''
  \href{http://dx.doi.org/10.1103/PhysRevLett.116.231801}{{\em Phys. Rev.
  Lett.} {\bfseries 116} no.~23, (2016) 231801},
  \href{http://arxiv.org/abs/1601.01712}{{\ttfamily arXiv:1601.01712
  [hep-ph]}}.

\bibitem{Baldenegro:2017aen}
C.~Baldenegro, S.~Fichet, G.~von Gersdorff, and C.~Royon, ``{Probing the
  anomalous \ensuremath{\gamma}\ensuremath{\gamma}\ensuremath{\gamma}Z coupling
  at the LHC with proton tagging},''
  \href{http://dx.doi.org/10.1007/JHEP06(2017)142}{{\em JHEP} {\bfseries 06}
  (2017) 142}, \href{http://arxiv.org/abs/1703.10600}{{\ttfamily
  arXiv:1703.10600 [hep-ph]}}.

\bibitem{Baldenegro:2020qut}
C.~Baldenegro, G.~Biagi, G.~Legras, and C.~Royon, ``{Central exclusive
  production of $W$ boson pairs in $pp$ collisions at the LHC in hadronic and
  semi-leptonic final states},''
  \href{http://dx.doi.org/10.1007/JHEP12(2020)165}{{\em JHEP} {\bfseries 12}
  (2020) 165}, \href{http://arxiv.org/abs/2009.08331}{{\ttfamily
  arXiv:2009.08331 [hep-ph]}}.

\bibitem{Baldenegro:2022kaa}
C.~Baldenegro, A.~Bellora, S.~Fichet, G.~von Gersdorff, M.~Pitt, and C.~Royon,
  ``{Searching for anomalous top quark interactions with proton tagging and
  timing detectors at the LHC},''
  \href{http://arxiv.org/abs/2205.01173}{{\ttfamily arXiv:2205.01173
  [hep-ph]}}.

\bibitem{Brodsky:1998kn}
S.~J. Brodsky, V.~S. Fadin, V.~T. Kim, L.~N. Lipatov, and G.~B. Pivovarov,
  ``{The QCD pomeron with optimal renormalization},''
  \href{http://dx.doi.org/10.1134/1.568145}{{\em JETP Lett.} {\bfseries 70}
  (1999) 155--160}, \href{http://arxiv.org/abs/hep-ph/9901229}{{\ttfamily
  arXiv:hep-ph/9901229}}.

\bibitem{Brodsky:2002ka}
S.~J. Brodsky, V.~S. Fadin, V.~T. Kim, L.~N. Lipatov, and G.~B. Pivovarov,
  ``{High-energy QCD asymptotics of photon-photon collisions},''
  \href{http://dx.doi.org/10.1134/1.1520615}{{\em JETP Lett.} {\bfseries 76}
  (2002) 249--252}, \href{http://arxiv.org/abs/hep-ph/0207297}{{\ttfamily
  arXiv:hep-ph/0207297}}.

\bibitem{Caporale:2008is}
F.~Caporale, D.~Y. Ivanov, and A.~Papa, ``{BFKL resummation effects in the
  gamma* gamma* total hadronic cross section},''
  \href{http://dx.doi.org/10.1140/epjc/s10052-008-0732-1}{{\em Eur. Phys. J. C}
  {\bfseries 58} (2008) 1--7}, \href{http://arxiv.org/abs/0807.3231}{{\ttfamily
  arXiv:0807.3231 [hep-ph]}}.

\bibitem{Zheng:2013uja}
X.-C. Zheng, X.-G. Wu, S.-Q. Wang, J.-M. Shen, and Q.-L. Zhang, ``{Reanalysis
  of the BFKL Pomeron at the next-to-leading logarithmic accuracy},''
  \href{http://dx.doi.org/10.1007/JHEP10(2013)117}{{\em JHEP} {\bfseries 10}
  (2013) 117}, \href{http://arxiv.org/abs/1308.2381}{{\ttfamily arXiv:1308.2381
  [hep-ph]}}.

\bibitem{Chirilli:2014dcb}
G.~A. Chirilli and Y.~V. Kovchegov, ``{$\gamma^* \gamma^*$ Cross Section at NLO
  and Properties of the BFKL Evolution at Higher Orders},''
  \href{http://dx.doi.org/10.1007/JHEP05(2014)099}{{\em JHEP} {\bfseries 05}
  (2014) 099}, \href{http://arxiv.org/abs/1403.3384}{{\ttfamily arXiv:1403.3384
  [hep-ph]}}. [Erratum: JHEP 08, 075 (2015)].

\bibitem{Ivanov:2014hpa}
D.~Y. Ivanov, B.~Murdaca, and A.~Papa, ``{The $\gamma^* \gamma^*$ total cross
  section in next-to-leading order BFKL and LEP2 data},''
  \href{http://dx.doi.org/10.1007/JHEP10(2014)058}{{\em JHEP} {\bfseries 10}
  (2014) 058}, \href{http://arxiv.org/abs/1407.8447}{{\ttfamily arXiv:1407.8447
  [hep-ph]}}.

\bibitem{Ivanov:2005gn}
D.~{\relax Yu}. Ivanov and A.~Papa, ``{Electroproduction of two light vector
  mesons in the next-to-leading approximation},''
  \href{http://dx.doi.org/10.1016/j.nuclphysb.2005.10.028}{{\em Nucl. Phys. B}
  {\bfseries 732} (2006) 183--199},
  \href{http://arxiv.org/abs/hep-ph/0508162}{{\ttfamily arXiv:hep-ph/0508162}}.

\bibitem{Ivanov:2006gt}
D.~{\relax Yu}. Ivanov and A.~Papa, ``{Electroproduction of two light vector
  mesons in next-to-leading BFKL: Study of systematic effects},''
  \href{http://dx.doi.org/10.1140/epjc/s10052-006-0180-8}{{\em Eur. Phys. J. C}
  {\bfseries 49} (2007) 947--955},
  \href{http://arxiv.org/abs/hep-ph/0610042}{{\ttfamily arXiv:hep-ph/0610042}}.

\bibitem{Enberg:2005eq}
R.~Enberg, B.~Pire, L.~Szymanowski, and S.~Wallon, ``{BFKL resummation effects
  in gamma* gamma* ---\ensuremath{>} rho rho},''
  \href{http://dx.doi.org/10.1140/epjc/s10052-007-0375-7}{{\em Eur. Phys. J. C}
  {\bfseries 45} (2006) 759--769},
  \href{http://arxiv.org/abs/hep-ph/0508134}{{\ttfamily arXiv:hep-ph/0508134}}.
  [Erratum: Eur.Phys.J.C 51, 1015 (2007)].

\bibitem{Kwiecinski:1998sa}
J.~Kwiecinski and L.~Motyka, ``{Diffractive J / psi production in high-energy
  gamma gamma collisions as a probe of the QCD pomeron},''
  \href{http://dx.doi.org/10.1016/S0370-2693(98)00958-7}{{\em Phys. Lett. B}
  {\bfseries 438} (1998) 203--210},
  \href{http://arxiv.org/abs/hep-ph/9806260}{{\ttfamily arXiv:hep-ph/9806260}}.

\bibitem{Celiberto:2017nyx}
F.~G. Celiberto, D.~{\relax Yu}. Ivanov, B.~Murdaca, and A.~Papa,
  ``{High-energy resummation in heavy-quark pair photoproduction},''
  \href{http://dx.doi.org/10.1016/j.physletb.2017.12.020}{{\em Phys. Lett. B}
  {\bfseries 777} (2018) 141--150},
  \href{http://arxiv.org/abs/1709.10032}{{\ttfamily arXiv:1709.10032
  [hep-ph]}}.

\bibitem{TFReport}
N.~Craig, C.~Csaki, and A.~El-Khadra, ``{Theory Frontier Summary Report},''.
  {Snowmass 2021 Community Study}.

\bibitem{Huss:2022ful}
A.~Huss, J.~Huston, S.~Jones, and M.~Pellen, ``{Les Houches 2021: Physics at
  TeV Colliders: Report on the Standard Model Precision Wishlist},''
  \href{http://arxiv.org/abs/2207.02122}{{\ttfamily arXiv:2207.02122
  [hep-ph]}}.

\bibitem{Heinrich:2020ybq}
G.~Heinrich, ``{Collider Physics at the Precision Frontier},''
  \href{http://dx.doi.org/10.1016/j.physrep.2021.03.006}{{\em Phys. Rept.}
  {\bfseries 922} (2021) 1--69},
  \href{http://arxiv.org/abs/2009.00516}{{\ttfamily arXiv:2009.00516
  [hep-ph]}}.

\bibitem{Cordero:2022gsh}
F.~Febres~Cordero, A.~von Manteuffel, and T.~Neumann, ``{Computational
  challenges for multi-loop collider phenomenology},'' in {\em {2022 Snowmass
  Summer Study}}.
\newblock 4, 2022.
\newblock \href{http://arxiv.org/abs/2204.04200}{{\ttfamily arXiv:2204.04200
  [hep-ph]}}.

\bibitem{Travaglini:2022uwo}
G.~Travaglini {\em et~al.}, ``{The SAGEX Review on Scattering Amplitudes},''
  \href{http://arxiv.org/abs/2203.13011}{{\ttfamily arXiv:2203.13011
  [hep-th]}}.

\bibitem{Chen:2019wyb}
L.~Chen, ``{A prescription for projectors to compute helicity amplitudes in D
  dimensions},'' \href{http://dx.doi.org/10.1140/epjc/s10052-021-09210-9}{{\em
  Eur. Phys. J. C} {\bfseries 81} no.~5, (2021) 417},
  \href{http://arxiv.org/abs/1904.00705}{{\ttfamily arXiv:1904.00705
  [hep-ph]}}.

\bibitem{Peraro:2019cjj}
T.~Peraro and L.~Tancredi, ``{Physical projectors for multi-leg helicity
  amplitudes},'' \href{http://dx.doi.org/10.1007/JHEP07(2019)114}{{\em JHEP}
  {\bfseries 07} (2019) 114}, \href{http://arxiv.org/abs/1906.03298}{{\ttfamily
  arXiv:1906.03298 [hep-ph]}}.

\bibitem{Peraro:2020sfm}
T.~Peraro and L.~Tancredi, ``{Tensor decomposition for bosonic and fermionic
  scattering amplitudes},''
  \href{http://dx.doi.org/10.1103/PhysRevD.103.054042}{{\em Phys. Rev. D}
  {\bfseries 103} no.~5, (2021) 054042},
  \href{http://arxiv.org/abs/2012.00820}{{\ttfamily arXiv:2012.00820
  [hep-ph]}}.

\bibitem{Tkachov:1981wb}
F.~V. Tkachov, ``{A Theorem on Analytical Calculability of Four Loop
  Renormalization Group Functions},''
  \href{http://dx.doi.org/10.1016/0370-2693(81)90288-4}{{\em Phys. Lett. B}
  {\bfseries 100} (1981) 65--68}.

\bibitem{Chetyrkin:1981qh}
K.~G. Chetyrkin and F.~V. Tkachov, ``{Integration by Parts: The Algorithm to
  Calculate beta Functions in 4 Loops},''
  \href{http://dx.doi.org/10.1016/0550-3213(81)90199-1}{{\em Nucl. Phys. B}
  {\bfseries 192} (1981) 159--204}.

\bibitem{Gehrmann:1999as}
T.~Gehrmann and E.~Remiddi, ``{Differential equations for two loop four point
  functions},'' \href{http://dx.doi.org/10.1016/S0550-3213(00)00223-6}{{\em
  Nucl. Phys. B} {\bfseries 580} (2000) 485--518},
  \href{http://arxiv.org/abs/hep-ph/9912329}{{\ttfamily arXiv:hep-ph/9912329}}.

\bibitem{Laporta:2000dsw}
S.~Laporta, ``{High precision calculation of multiloop Feynman integrals by
  difference equations},''
  \href{http://dx.doi.org/10.1142/S0217751X00002159}{{\em Int. J. Mod. Phys. A}
  {\bfseries 15} (2000) 5087--5159},
  \href{http://arxiv.org/abs/hep-ph/0102033}{{\ttfamily arXiv:hep-ph/0102033}}.

\bibitem{Kant:2013vta}
P.~Kant, ``{Finding Linear Dependencies in Integration-By-Parts Equations: A
  Monte Carlo Approach},''
  \href{http://dx.doi.org/10.1016/j.cpc.2014.01.017}{{\em Comput. Phys.
  Commun.} {\bfseries 185} (2014) 1473--1476},
  \href{http://arxiv.org/abs/1309.7287}{{\ttfamily arXiv:1309.7287 [hep-ph]}}.

\bibitem{vonManteuffel:2014ixa}
A.~von Manteuffel and R.~M. Schabinger, ``{A novel approach to integration by
  parts reduction},''
  \href{http://dx.doi.org/10.1016/j.physletb.2015.03.029}{{\em Phys. Lett. B}
  {\bfseries 744} (2015) 101--104},
  \href{http://arxiv.org/abs/1406.4513}{{\ttfamily arXiv:1406.4513 [hep-ph]}}.

\bibitem{Peraro:2016wsq}
T.~Peraro, ``{Scattering amplitudes over finite fields and multivariate
  functional reconstruction},''
  \href{http://dx.doi.org/10.1007/JHEP12(2016)030}{{\em JHEP} {\bfseries 12}
  (2016) 030}, \href{http://arxiv.org/abs/1608.01902}{{\ttfamily
  arXiv:1608.01902 [hep-ph]}}.

\bibitem{Weinzierl:2022eaz}
S.~Weinzierl, ``{Feynman Integrals},''
  \href{http://arxiv.org/abs/2201.03593}{{\ttfamily arXiv:2201.03593
  [hep-th]}}.

\bibitem{Abreu:2022mfk}
S.~Abreu, R.~Britto, and C.~Duhr, ``{The SAGEX Review on Scattering Amplitudes,
  Chapter 3: Mathematical structures in Feynman integrals},''
  \href{http://arxiv.org/abs/2203.13014}{{\ttfamily arXiv:2203.13014
  [hep-th]}}.

\bibitem{Blumlein:2022zkr}
J.~Bl\"umlein and C.~Schneider, ``{The SAGEX Review on Scattering Amplitudes,
  Chapter 4: Multi-loop Feynman Integrals},''
  \href{http://arxiv.org/abs/2203.13015}{{\ttfamily arXiv:2203.13015
  [hep-th]}}.

\bibitem{Abreu:2017xsl}
S.~Abreu, F.~Febres~Cordero, H.~Ita, M.~Jaquier, B.~Page, and M.~Zeng,
  ``{Two-Loop Four-Gluon Amplitudes from Numerical Unitarity},''
  \href{http://dx.doi.org/10.1103/PhysRevLett.119.142001}{{\em Phys. Rev.
  Lett.} {\bfseries 119} no.~14, (2017) 142001},
  \href{http://arxiv.org/abs/1703.05273}{{\ttfamily arXiv:1703.05273
  [hep-ph]}}.

\bibitem{Abreu:2017hqn}
S.~Abreu, F.~Febres~Cordero, H.~Ita, B.~Page, and M.~Zeng, ``{Planar Two-Loop
  Five-Gluon Amplitudes from Numerical Unitarity},''
  \href{http://dx.doi.org/10.1103/PhysRevD.97.116014}{{\em Phys. Rev. D}
  {\bfseries 97} no.~11, (2018) 116014},
  \href{http://arxiv.org/abs/1712.03946}{{\ttfamily arXiv:1712.03946
  [hep-ph]}}.

\bibitem{Abreu:2019odu}
S.~Abreu, J.~Dormans, F.~Febres~Cordero, H.~Ita, B.~Page, and V.~Sotnikov,
  ``{Analytic Form of the Planar Two-Loop Five-Parton Scattering Amplitudes in
  QCD},'' \href{http://dx.doi.org/10.1007/JHEP05(2019)084}{{\em JHEP}
  {\bfseries 05} (2019) 084}, \href{http://arxiv.org/abs/1904.00945}{{\ttfamily
  arXiv:1904.00945 [hep-ph]}}.

\bibitem{Kotikov:1990kg}
A.~V. Kotikov, ``{Differential equations method: New technique for massive
  Feynman diagrams calculation},''
  \href{http://dx.doi.org/10.1016/0370-2693(91)90413-K}{{\em Phys. Lett. B}
  {\bfseries 254} (1991) 158--164}.

\bibitem{Henn:2013pwa}
J.~M. Henn, ``{Multiloop integrals in dimensional regularization made
  simple},'' \href{http://dx.doi.org/10.1103/PhysRevLett.110.251601}{{\em Phys.
  Rev. Lett.} {\bfseries 110} (2013) 251601},
  \href{http://arxiv.org/abs/1304.1806}{{\ttfamily arXiv:1304.1806 [hep-th]}}.

\bibitem{Abreu:2020jxa}
S.~Abreu, H.~Ita, F.~Moriello, B.~Page, W.~Tschernow, and M.~Zeng, ``{Two-Loop
  Integrals for Planar Five-Point One-Mass Processes},''
  \href{http://dx.doi.org/10.1007/JHEP11(2020)117}{{\em JHEP} {\bfseries 11}
  (2020) 117}, \href{http://arxiv.org/abs/2005.04195}{{\ttfamily
  arXiv:2005.04195 [hep-ph]}}.

\bibitem{Frellesvig:2021hkr}
H.~Frellesvig, ``{On epsilon factorized differential equations for elliptic
  Feynman integrals},'' \href{http://dx.doi.org/10.1007/JHEP03(2022)079}{{\em
  JHEP} {\bfseries 03} (2022) 079},
  \href{http://arxiv.org/abs/2110.07968}{{\ttfamily arXiv:2110.07968
  [hep-th]}}.

\bibitem{Dlapa:2021qsl}
C.~Dlapa, X.~Li, and Y.~Zhang, ``{Leading singularities in Baikov
  representation and Feynman integrals with uniform transcendental weight},''
  \href{http://dx.doi.org/10.1007/JHEP07(2021)227}{{\em JHEP} {\bfseries 07}
  (2021) 227}, \href{http://arxiv.org/abs/2103.04638}{{\ttfamily
  arXiv:2103.04638 [hep-th]}}.

\bibitem{Syrrakos:2020kba}
N.~Syrrakos, ``{Pentagon integrals to arbitrary order in the dimensional
  regulator},'' \href{http://dx.doi.org/10.1007/JHEP06(2021)037}{{\em JHEP}
  {\bfseries 06} (2021) 037}, \href{http://arxiv.org/abs/2012.10635}{{\ttfamily
  arXiv:2012.10635 [hep-ph]}}.

\bibitem{Kardos:2022tpo}
A.~Kardos, C.~G. Papadopoulos, A.~V. Smirnov, N.~Syrrakos, and C.~Wever,
  ``{Two-loop non-planar hexa-box integrals with one massive leg},''
  \href{http://dx.doi.org/10.1007/JHEP05(2022)033}{{\em JHEP} {\bfseries 05}
  (2022) 033}, \href{http://arxiv.org/abs/2201.07509}{{\ttfamily
  arXiv:2201.07509 [hep-ph]}}.

\bibitem{Henn:2021cyv}
J.~Henn, T.~Peraro, Y.~Xu, and Y.~Zhang, ``{A first look at the function space
  for planar two-loop six-particle Feynman integrals},''
  \href{http://dx.doi.org/10.1007/JHEP03(2022)056}{{\em JHEP} {\bfseries 03}
  (2022) 056}, \href{http://arxiv.org/abs/2112.10605}{{\ttfamily
  arXiv:2112.10605 [hep-th]}}.

\bibitem{Abreu:2021smk}
S.~Abreu, H.~Ita, B.~Page, and W.~Tschernow, ``{Two-loop hexa-box integrals for
  non-planar five-point one-mass processes},''
  \href{http://dx.doi.org/10.1007/JHEP03(2022)182}{{\em JHEP} {\bfseries 03}
  (2022) 182}, \href{http://arxiv.org/abs/2107.14180}{{\ttfamily
  arXiv:2107.14180 [hep-ph]}}.

\bibitem{Papadopoulos:2014lla}
C.~G. Papadopoulos, ``{Simplified differential equations approach for Master
  Integrals},'' \href{http://dx.doi.org/10.1007/JHEP07(2014)088}{{\em JHEP}
  {\bfseries 07} (2014) 088}, \href{http://arxiv.org/abs/1401.6057}{{\ttfamily
  arXiv:1401.6057 [hep-ph]}}.

\bibitem{Canko:2020ylt}
D.~D. Canko, C.~G. Papadopoulos, and N.~Syrrakos, ``{Analytic representation of
  all planar two-loop five-point Master Integrals with one off-shell leg},''
  \href{http://dx.doi.org/10.1007/JHEP01(2021)199}{{\em JHEP} {\bfseries 01}
  (2021) 199}, \href{http://arxiv.org/abs/2009.13917}{{\ttfamily
  arXiv:2009.13917 [hep-ph]}}.

\bibitem{Argeri:2007up}
M.~Argeri and P.~Mastrolia, ``{Feynman Diagrams and Differential Equations},''
  \href{http://dx.doi.org/10.1142/S0217751X07037147}{{\em Int. J. Mod. Phys. A}
  {\bfseries 22} (2007) 4375--4436},
  \href{http://arxiv.org/abs/0707.4037}{{\ttfamily arXiv:0707.4037 [hep-ph]}}.

\bibitem{Henn:2014qga}
J.~M. Henn, ``{Lectures on differential equations for Feynman integrals},''
  \href{http://dx.doi.org/10.1088/1751-8113/48/15/153001}{{\em J. Phys. A}
  {\bfseries 48} (2015) 153001},
  \href{http://arxiv.org/abs/1412.2296}{{\ttfamily arXiv:1412.2296 [hep-ph]}}.

\bibitem{Gehrmann-DeRidder:2005btv}
A.~Gehrmann-De~Ridder, T.~Gehrmann, and E.~W.~N. Glover, ``{Antenna subtraction
  at NNLO},'' \href{http://dx.doi.org/10.1088/1126-6708/2005/09/056}{{\em JHEP}
  {\bfseries 09} (2005) 056},
  \href{http://arxiv.org/abs/hep-ph/0505111}{{\ttfamily arXiv:hep-ph/0505111}}.

\bibitem{Currie:2013vh}
J.~Currie, E.~W.~N. Glover, and S.~Wells, ``{Infrared Structure at NNLO Using
  Antenna Subtraction},'' \href{http://dx.doi.org/10.1007/JHEP04(2013)066}{{\em
  JHEP} {\bfseries 04} (2013) 066},
  \href{http://arxiv.org/abs/1301.4693}{{\ttfamily arXiv:1301.4693 [hep-ph]}}.

\bibitem{Gauld:2019yng}
R.~Gauld, A.~Gehrmann-De~Ridder, E.~W.~N. Glover, A.~Huss, and I.~Majer,
  ``{Associated production of a Higgs boson decaying into bottom quarks and a
  weak vector boson decaying leptonically at NNLO in QCD},''
  \href{http://dx.doi.org/10.1007/JHEP10(2019)002}{{\em JHEP} {\bfseries 10}
  (2019) 002}, \href{http://arxiv.org/abs/1907.05836}{{\ttfamily
  arXiv:1907.05836 [hep-ph]}}.

\bibitem{Gauld:2020deh}
R.~Gauld, A.~Gehrmann-De~Ridder, E.~W.~N. Glover, A.~Huss, and I.~Majer,
  ``{Predictions for $Z$ -Boson Production in Association with a $b$-Jet at
  $\mathcal {O}(\alpha_s^3)$},''
  \href{http://dx.doi.org/10.1103/PhysRevLett.125.222002}{{\em Phys. Rev.
  Lett.} {\bfseries 125} no.~22, (2020) 222002},
  \href{http://arxiv.org/abs/2005.03016}{{\ttfamily arXiv:2005.03016
  [hep-ph]}}.

\bibitem{Gehrmann:2022cih}
T.~Gehrmann and R.~Sch\"urmann, ``{Photon fragmentation in the antenna
  subtraction formalism},''
  \href{http://dx.doi.org/10.1007/JHEP04(2022)031}{{\em JHEP} {\bfseries 04}
  (2022) 031}, \href{http://arxiv.org/abs/2201.06982}{{\ttfamily
  arXiv:2201.06982 [hep-ph]}}.

\bibitem{Chen:2022gpk}
X.~Chen, T.~Gehrmann, E.~W.~N. Glover, M.~H\"ofer, A.~Huss, and R.~Sch\"urmann,
  ``{Single Photon Production at Hadron Colliders at NNLO QCD with Realistic
  Photon Isolation},'' \href{http://arxiv.org/abs/2205.01516}{{\ttfamily
  arXiv:2205.01516 [hep-ph]}}.

\bibitem{Czakon:2010td}
M.~Czakon, ``{A novel subtraction scheme for double-real radiation at NNLO},''
  \href{http://dx.doi.org/10.1016/j.physletb.2010.08.036}{{\em Phys. Lett. B}
  {\bfseries 693} (2010) 259--268},
  \href{http://arxiv.org/abs/1005.0274}{{\ttfamily arXiv:1005.0274 [hep-ph]}}.

\bibitem{Czakon:2011ve}
M.~Czakon, ``{Double-real radiation in hadronic top quark pair production as a
  proof of a certain concept},''
  \href{http://dx.doi.org/10.1016/j.nuclphysb.2011.03.020}{{\em Nucl. Phys. B}
  {\bfseries 849} (2011) 250--295},
  \href{http://arxiv.org/abs/1101.0642}{{\ttfamily arXiv:1101.0642 [hep-ph]}}.

\bibitem{Boughezal:2011jf}
R.~Boughezal, K.~Melnikov, and F.~Petriello, ``{A subtraction scheme for NNLO
  computations},'' \href{http://dx.doi.org/10.1103/PhysRevD.85.034025}{{\em
  Phys. Rev. D} {\bfseries 85} (2012) 034025},
  \href{http://arxiv.org/abs/1111.7041}{{\ttfamily arXiv:1111.7041 [hep-ph]}}.

\bibitem{Frixione:1995ms}
S.~Frixione, Z.~Kunszt, and A.~Signer, ``{Three jet cross-sections to
  next-to-leading order},''
  \href{http://dx.doi.org/10.1016/0550-3213(96)00110-1}{{\em Nucl. Phys. B}
  {\bfseries 467} (1996) 399--442},
  \href{http://arxiv.org/abs/hep-ph/9512328}{{\ttfamily arXiv:hep-ph/9512328}}.

\bibitem{Frederix:2009yq}
R.~Frederix, S.~Frixione, F.~Maltoni, and T.~Stelzer, ``{Automation of
  next-to-leading order computations in QCD: The FKS subtraction},''
  \href{http://dx.doi.org/10.1088/1126-6708/2009/10/003}{{\em JHEP} {\bfseries
  10} (2009) 003}, \href{http://arxiv.org/abs/0908.4272}{{\ttfamily
  arXiv:0908.4272 [hep-ph]}}.

\bibitem{Binoth:2000ps}
T.~Binoth and G.~Heinrich, ``{An automatized algorithm to compute infrared
  divergent multiloop integrals},''
  \href{http://dx.doi.org/10.1016/S0550-3213(00)00429-6}{{\em Nucl. Phys. B}
  {\bfseries 585} (2000) 741--759},
  \href{http://arxiv.org/abs/hep-ph/0004013}{{\ttfamily arXiv:hep-ph/0004013}}.

\bibitem{Heinrich:2002rc}
G.~Heinrich, ``{A numerical method for NNLO calculations},''
  \href{http://dx.doi.org/10.1016/S0920-5632(03)80201-3}{{\em Nucl. Phys. B
  Proc. Suppl.} {\bfseries 116} (2003) 368--372},
  \href{http://arxiv.org/abs/hep-ph/0211144}{{\ttfamily arXiv:hep-ph/0211144}}.

\bibitem{Anastasiou:2003gr}
C.~Anastasiou, K.~Melnikov, and F.~Petriello, ``{A new method for real
  radiation at NNLO},''
  \href{http://dx.doi.org/10.1103/PhysRevD.69.076010}{{\em Phys. Rev. D}
  {\bfseries 69} (2004) 076010},
  \href{http://arxiv.org/abs/hep-ph/0311311}{{\ttfamily arXiv:hep-ph/0311311}}.

\bibitem{Binoth:2004jv}
T.~Binoth and G.~Heinrich, ``{Numerical evaluation of phase space integrals by
  sector decomposition},''
  \href{http://dx.doi.org/10.1016/j.nuclphysb.2004.06.005}{{\em Nucl. Phys. B}
  {\bfseries 693} (2004) 134--148},
  \href{http://arxiv.org/abs/hep-ph/0402265}{{\ttfamily arXiv:hep-ph/0402265}}.

\bibitem{Czakon:2020coa}
M.~Czakon, A.~Mitov, M.~Pellen, and R.~Poncelet, ``{NNLO QCD predictions for
  W+c-jet production at the LHC},''
  \href{http://dx.doi.org/10.1007/JHEP06(2021)100}{{\em JHEP} {\bfseries 06}
  (2021) 100}, \href{http://arxiv.org/abs/2011.01011}{{\ttfamily
  arXiv:2011.01011 [hep-ph]}}.

\bibitem{Czakon:2021ohs}
M.~L. Czakon, T.~Generet, A.~Mitov, and R.~Poncelet, ``{B-hadron production in
  NNLO QCD: application to LHC t$ \overline{t} $ events with leptonic
  decays},'' \href{http://dx.doi.org/10.1007/JHEP10(2021)216}{{\em JHEP}
  {\bfseries 10} (2021) 216}, \href{http://arxiv.org/abs/2102.08267}{{\ttfamily
  arXiv:2102.08267 [hep-ph]}}.

\bibitem{Catani:2007vq}
S.~Catani and M.~Grazzini, ``{An NNLO subtraction formalism in hadron
  collisions and its application to Higgs boson production at the LHC},''
  \href{http://dx.doi.org/10.1103/PhysRevLett.98.222002}{{\em Phys. Rev. Lett.}
  {\bfseries 98} (2007) 222002},
  \href{http://arxiv.org/abs/hep-ph/0703012}{{\ttfamily arXiv:hep-ph/0703012}}.

\bibitem{Grazzini:2017mhc}
M.~Grazzini, S.~Kallweit, and M.~Wiesemann, ``{Fully differential NNLO
  computations with MATRIX},''
  \href{http://dx.doi.org/10.1140/epjc/s10052-018-5771-7}{{\em Eur. Phys. J. C}
  {\bfseries 78} no.~7, (2018) 537},
  \href{http://arxiv.org/abs/1711.06631}{{\ttfamily arXiv:1711.06631
  [hep-ph]}}.

\bibitem{Campbell:2022gdq}
J.~M. Campbell, R.~K. Ellis, and S.~Seth, ``{Non-local slicing approaches for
  NNLO QCD in MCFM},'' \href{http://dx.doi.org/10.1007/JHEP06(2022)002}{{\em
  JHEP} {\bfseries 06} (2022) 002},
  \href{http://arxiv.org/abs/2202.07738}{{\ttfamily arXiv:2202.07738
  [hep-ph]}}.

\bibitem{Bonciani:2015sha}
R.~Bonciani, S.~Catani, M.~Grazzini, H.~Sargsyan, and A.~Torre, ``{The $q_T$
  subtraction method for top quark production at hadron colliders},''
  \href{http://dx.doi.org/10.1140/epjc/s10052-015-3793-y}{{\em Eur. Phys. J. C}
  {\bfseries 75} no.~12, (2015) 581},
  \href{http://arxiv.org/abs/1508.03585}{{\ttfamily arXiv:1508.03585
  [hep-ph]}}.

\bibitem{Angeles-Martinez:2018mqh}
R.~Angeles-Martinez, M.~Czakon, and S.~Sapeta, ``{NNLO soft function for top
  quark pair production at small transverse momentum},''
  \href{http://dx.doi.org/10.1007/JHEP10(2018)201}{{\em JHEP} {\bfseries 10}
  (2018) 201}, \href{http://arxiv.org/abs/1809.01459}{{\ttfamily
  arXiv:1809.01459 [hep-ph]}}.

\bibitem{Catani:2019iny}
S.~Catani, S.~Devoto, M.~Grazzini, S.~Kallweit, J.~Mazzitelli, and H.~Sargsyan,
  ``{Top-quark pair hadroproduction at next-to-next-to-leading order in QCD},''
  \href{http://dx.doi.org/10.1103/PhysRevD.99.051501}{{\em Phys. Rev. D}
  {\bfseries 99} no.~5, (2019) 051501},
  \href{http://arxiv.org/abs/1901.04005}{{\ttfamily arXiv:1901.04005
  [hep-ph]}}.

\bibitem{Catani:2019hip}
S.~Catani, S.~Devoto, M.~Grazzini, S.~Kallweit, and J.~Mazzitelli, ``{Top-quark
  pair production at the LHC: Fully differential QCD predictions at NNLO},''
  \href{http://dx.doi.org/10.1007/JHEP07(2019)100}{{\em JHEP} {\bfseries 07}
  (2019) 100}, \href{http://arxiv.org/abs/1906.06535}{{\ttfamily
  arXiv:1906.06535 [hep-ph]}}.

\bibitem{Catani:2020kkl}
S.~Catani, S.~Devoto, M.~Grazzini, S.~Kallweit, and J.~Mazzitelli,
  ``{Bottom-quark production at hadron colliders: fully differential
  predictions in NNLO QCD},''
  \href{http://dx.doi.org/10.1007/JHEP03(2021)029}{{\em JHEP} {\bfseries 03}
  (2021) 029}, \href{http://arxiv.org/abs/2010.11906}{{\ttfamily
  arXiv:2010.11906 [hep-ph]}}.

\bibitem{Boughezal:2015eha}
R.~Boughezal, X.~Liu, and F.~Petriello, ``{$N$-jettiness soft function at
  next-to-next-to-leading order},''
  \href{http://dx.doi.org/10.1103/PhysRevD.91.094035}{{\em Phys. Rev. D}
  {\bfseries 91} no.~9, (2015) 094035},
  \href{http://arxiv.org/abs/1504.02540}{{\ttfamily arXiv:1504.02540
  [hep-ph]}}.

\bibitem{Boughezal:2015dva}
R.~Boughezal, C.~Focke, X.~Liu, and F.~Petriello, ``{$W$-boson production in
  association with a jet at next-to-next-to-leading order in perturbative
  QCD},'' \href{http://dx.doi.org/10.1103/PhysRevLett.115.062002}{{\em Phys.
  Rev. Lett.} {\bfseries 115} no.~6, (2015) 062002},
  \href{http://arxiv.org/abs/1504.02131}{{\ttfamily arXiv:1504.02131
  [hep-ph]}}.

\bibitem{Gaunt:2015pea}
J.~Gaunt, M.~Stahlhofen, F.~J. Tackmann, and J.~R. Walsh, ``{N-jettiness
  Subtractions for NNLO QCD Calculations},''
  \href{http://dx.doi.org/10.1007/JHEP09(2015)058}{{\em JHEP} {\bfseries 09}
  (2015) 058}, \href{http://arxiv.org/abs/1505.04794}{{\ttfamily
  arXiv:1505.04794 [hep-ph]}}.

\bibitem{Boughezal:2016wmq}
R.~Boughezal, J.~M. Campbell, R.~K. Ellis, C.~Focke, W.~Giele, X.~Liu,
  F.~Petriello, and C.~Williams, ``{Color singlet production at NNLO in
  MCFM},'' \href{http://dx.doi.org/10.1140/epjc/s10052-016-4558-y}{{\em Eur.
  Phys. J. C} {\bfseries 77} no.~1, (2017) 7},
  \href{http://arxiv.org/abs/1605.08011}{{\ttfamily arXiv:1605.08011
  [hep-ph]}}.

\bibitem{Campbell:2019dru}
J.~Campbell and T.~Neumann, ``{Precision Phenomenology with MCFM},''
  \href{http://dx.doi.org/10.1007/JHEP12(2019)034}{{\em JHEP} {\bfseries 12}
  (2019) 034}, \href{http://arxiv.org/abs/1909.09117}{{\ttfamily
  arXiv:1909.09117 [hep-ph]}}.

\bibitem{DelDuca:2015zqa}
V.~Del~Duca, C.~Duhr, G.~Somogyi, F.~Tramontano, and Z.~Tr\'ocs\'anyi, ``{Higgs
  boson decay into b-quarks at NNLO accuracy},''
  \href{http://dx.doi.org/10.1007/JHEP04(2015)036}{{\em JHEP} {\bfseries 04}
  (2015) 036}, \href{http://arxiv.org/abs/1501.07226}{{\ttfamily
  arXiv:1501.07226 [hep-ph]}}.

\bibitem{Catani:1996vz}
S.~Catani and M.~H. Seymour, ``{A General algorithm for calculating jet
  cross-sections in NLO QCD},''
  \href{http://dx.doi.org/10.1016/S0550-3213(96)00589-5}{{\em Nucl. Phys. B}
  {\bfseries 485} (1997) 291--419},
  \href{http://arxiv.org/abs/hep-ph/9605323}{{\ttfamily arXiv:hep-ph/9605323}}.
  [Erratum: Nucl.Phys.B 510, 503--504 (1998)].

\bibitem{Caola:2017dug}
F.~Caola, K.~Melnikov, and R.~R\"ontsch, ``{Nested soft-collinear subtractions
  in NNLO QCD computations},''
  \href{http://dx.doi.org/10.1140/epjc/s10052-017-4774-0}{{\em Eur. Phys. J. C}
  {\bfseries 77} no.~4, (2017) 248},
  \href{http://arxiv.org/abs/1702.01352}{{\ttfamily arXiv:1702.01352
  [hep-ph]}}.

\bibitem{Caola:2018pxp}
F.~Caola, M.~Delto, H.~Frellesvig, and K.~Melnikov, ``{The double-soft integral
  for an arbitrary angle between hard radiators},''
  \href{http://dx.doi.org/10.1140/epjc/s10052-018-6180-7}{{\em Eur. Phys. J. C}
  {\bfseries 78} no.~8, (2018) 687},
  \href{http://arxiv.org/abs/1807.05835}{{\ttfamily arXiv:1807.05835
  [hep-ph]}}.

\bibitem{Delto:2019asp}
M.~Delto and K.~Melnikov, ``{Integrated triple-collinear counter-terms for the
  nested soft-collinear subtraction scheme},''
  \href{http://dx.doi.org/10.1007/JHEP05(2019)148}{{\em JHEP} {\bfseries 05}
  (2019) 148}, \href{http://arxiv.org/abs/1901.05213}{{\ttfamily
  arXiv:1901.05213 [hep-ph]}}.

\bibitem{Magnea:2018hab}
L.~Magnea, E.~Maina, G.~Pelliccioli, C.~Signorile-Signorile, P.~Torrielli, and
  S.~Uccirati, ``{Local analytic sector subtraction at NNLO},''
  \href{http://dx.doi.org/10.1007/JHEP12(2018)107}{{\em JHEP} {\bfseries 12}
  (2018) 107}, \href{http://arxiv.org/abs/1806.09570}{{\ttfamily
  arXiv:1806.09570 [hep-ph]}}. [Erratum: JHEP 06, 013 (2019)].

\bibitem{Magnea:2018ebr}
L.~Magnea, E.~Maina, G.~Pelliccioli, C.~Signorile-Signorile, P.~Torrielli, and
  S.~Uccirati, ``{Factorisation and Subtraction beyond NLO},''
  \href{http://dx.doi.org/10.1007/JHEP12(2018)062}{{\em JHEP} {\bfseries 12}
  (2018) 062}, \href{http://arxiv.org/abs/1809.05444}{{\ttfamily
  arXiv:1809.05444 [hep-ph]}}.

\bibitem{Cacciari:2015jma}
M.~Cacciari, F.~A. Dreyer, A.~Karlberg, G.~P. Salam, and G.~Zanderighi,
  ``{Fully Differential Vector-Boson-Fusion Higgs Production at
  Next-to-Next-to-Leading Order},''
  \href{http://dx.doi.org/10.1103/PhysRevLett.115.082002}{{\em Phys. Rev.
  Lett.} {\bfseries 115} no.~8, (2015) 082002},
  \href{http://arxiv.org/abs/1506.02660}{{\ttfamily arXiv:1506.02660
  [hep-ph]}}. [Erratum: Phys.Rev.Lett. 120, 139901 (2018)].

\bibitem{Carli:2010rw}
T.~Carli, D.~Clements, A.~Cooper-Sarkar, C.~Gwenlan, G.~P. Salam, F.~Siegert,
  P.~Starovoitov, and M.~Sutton, ``{A posteriori inclusion of parton density
  functions in NLO QCD final-state calculations at hadron colliders: The
  APPLGRID Project},''
  \href{http://dx.doi.org/10.1140/epjc/s10052-010-1255-0}{{\em Eur. Phys. J. C}
  {\bfseries 66} (2010) 503--524},
  \href{http://arxiv.org/abs/0911.2985}{{\ttfamily arXiv:0911.2985 [hep-ph]}}.

\bibitem{Kluge:2006xs}
T.~Kluge, K.~Rabbertz, and M.~Wobisch,
  \href{http://dx.doi.org/10.1142/9789812706706_0110}{``{FastNLO: Fast pQCD
  calculations for PDF fits},''} in {\em {14th International Workshop on Deep
  Inelastic Scattering}}, pp.~483--486.
\newblock 9, 2006.
\newblock \href{http://arxiv.org/abs/hep-ph/0609285}{{\ttfamily
  hep-ph/0609285}}.

\bibitem{Carrazza:2020gss}
S.~Carrazza, E.~R. Nocera, C.~Schwan, and M.~Zaro, ``{PineAPPL: combining EW
  and QCD corrections for fast evaluation of LHC processes},''
  \href{http://dx.doi.org/10.1007/JHEP12(2020)108}{{\em JHEP} {\bfseries 12}
  (2020) 108}, \href{http://arxiv.org/abs/2008.12789}{{\ttfamily
  arXiv:2008.12789 [hep-ph]}}.

\bibitem{Abdesselam:2010pt}
A.~Abdesselam {\em et~al.}, ``{Boosted objects: a probe of beyond the standard
  model physics},''
  \href{http://dx.doi.org/10.1140/epjc/s10052-011-1661-y}{{\em
  EPHJA,C71,1661.2011} {\bfseries C71} (2011) 1661},
  \href{http://arxiv.org/abs/1012.5412}{{\ttfamily arXiv:1012.5412 [hep-ph]}}.

\bibitem{Altheimer:2012mn}
A.~Altheimer {\em et~al.}, ``{Jet Substructure at the Tevatron and LHC: New
  results, new tools, new benchmarks},''
  \href{http://arxiv.org/abs/1201.0008}{{\ttfamily arXiv:1201.0008 [hep-ph]}}.
Long author list - awaiting processing.

\bibitem{Altheimer:2013yza}
A.~Altheimer {\em et~al.}, ``{Boosted Objects and Jet Substructure at the LHC.
  Report of BOOST2012, held at IFIC Valencia, 23rd-27th of July 2012},''
  \href{http://dx.doi.org/10.1140/epjc/s10052-014-2792-8}{{\em Eur. Phys. J. C}
  {\bfseries 74} no.~3, (2014) 2792},
  \href{http://arxiv.org/abs/1311.2708}{{\ttfamily arXiv:1311.2708 [hep-ex]}}.

\bibitem{Adams:2015hiv}
D.~Adams {\em et~al.}, ``{Towards an Understanding of the Correlations in Jet
  Substructure},'' \href{http://dx.doi.org/10.1140/epjc/s10052-015-3587-2}{{\em
  Eur. Phys. J. C} {\bfseries 75} no.~9, (2015) 409},
  \href{http://arxiv.org/abs/1504.00679}{{\ttfamily arXiv:1504.00679
  [hep-ph]}}.

\bibitem{Kogler:2018hem}
R.~Kogler, B.~Nachman, A.~Schmidt~(editors), {\em et~al.}, ``{Jet Substructure
  at the Large Hadron Collider: Experimental Review},''
  \href{http://dx.doi.org/10.1103/RevModPhys.91.045003}{{\em Rev. Mod. Phys.}
  {\bfseries 91} no.~4, (2019) 045003},
  \href{http://arxiv.org/abs/1803.06991}{{\ttfamily arXiv:1803.06991
  [hep-ex]}}.

\bibitem{Kogler:2021kkw}
R.~Kogler, \href{http://dx.doi.org/10.1007/978-3-030-72858-8}{{\em {Advances in
  Jet Substructure at the LHC: Algorithms, Measurements and Searches for New
  Physical Phenomena}}}, vol.~284 of {\em Springer Tracts Mod. Phys.}
\newblock Springer, 2021.

\bibitem{Yeh:2019xbj}
C.~H. Yeh, S.~V. Chekanov, A.~V. Kotwal, J.~Proudfoot, S.~Sen, N.~V. Tran, and
  S.~S. Yu, ``{Studies of granularity of a hadronic calorimeter for tens-of-TeV
  jets at a 100 TeV $pp$ collider},''
  \href{http://dx.doi.org/10.1088/1748-0221/14/05/P05008}{{\em JINST}
  {\bfseries 14} no.~05, (2019) P05008},
  \href{http://arxiv.org/abs/1901.11146}{{\ttfamily arXiv:1901.11146
  [physics.ins-det]}}.

\bibitem{Coleman:2017fiq}
E.~Coleman, M.~Freytsis, A.~Hinzmann, M.~Narain, J.~Thaler, N.~Tran, and
  C.~Vernieri, ``{The importance of calorimetry for highly-boosted jet
  substructure},'' \href{http://dx.doi.org/10.1088/1748-0221/13/01/T01003}{{\em
  JINST} {\bfseries 13} no.~01, (2018) T01003},
  \href{http://arxiv.org/abs/1709.08705}{{\ttfamily arXiv:1709.08705
  [hep-ph]}}.

\bibitem{Proceedings:2018jsb}
J.~R. Andersen {\em et~al.}, ``{Les Houches 2017: Physics at TeV Colliders
  Standard Model Working Group Report},'' 2018.

\bibitem{CMS:2017wtu}
{\bfseries CMS} Collaboration, A.~M. Sirunyan {\em et~al.}, ``{Identification
  of heavy-flavour jets with the CMS detector in pp collisions at 13 TeV},''
  \href{http://dx.doi.org/10.1088/1748-0221/13/05/P05011}{{\em JINST}
  {\bfseries 13} no.~05, (2018) P05011},
  \href{http://arxiv.org/abs/1712.07158}{{\ttfamily arXiv:1712.07158
  [physics.ins-det]}}.

\bibitem{ATLAS:2019bwq}
{\bfseries ATLAS} Collaboration, G.~Aad {\em et~al.}, ``{ATLAS b-jet
  identification performance and efficiency measurement with $t{\bar{t}}$
  events in pp collisions at $\sqrt{s}=13$ TeV},''
  \href{http://dx.doi.org/10.1140/epjc/s10052-019-7450-8}{{\em Eur. Phys. J. C}
  {\bfseries 79} no.~11, (2019) 970},
  \href{http://arxiv.org/abs/1907.05120}{{\ttfamily arXiv:1907.05120
  [hep-ex]}}.

\bibitem{CMS:2021scf}
{\bfseries CMS} Collaboration, ``{A new calibration method for charm jet
  identification validated with proton-proton collision events at $\sqrt{s}$
  =13 TeV},'' 11, 2021.

\bibitem{ATLAS:2021cxe}
{\bfseries ATLAS} Collaboration, G.~Aad {\em et~al.}, ``{Measurement of the
  c-jet mistagging efficiency in $t\bar{t}$~events using pp collision data at
  $\sqrt{s}=13$~$\text {TeV}$ collected with the ATLAS detector},''
  \href{http://dx.doi.org/10.1140/epjc/s10052-021-09843-w}{{\em Eur. Phys. J.
  C} {\bfseries 82} no.~1, (2022) 95},
  \href{http://arxiv.org/abs/2109.10627}{{\ttfamily arXiv:2109.10627
  [hep-ex]}}.

\bibitem{Erdmann:2020ovh}
J.~Erdmann, O.~Nackenhorst, and S.~V. Zei\ss{}ner, ``{Maximum performance of
  strange-jet tagging at hadron colliders},''
  \href{http://dx.doi.org/10.1088/1748-0221/16/08/P08039}{{\em JINST}
  {\bfseries 16} no.~08, (2021) P08039},
  \href{http://arxiv.org/abs/2011.10736}{{\ttfamily arXiv:2011.10736
  [hep-ex]}}.

\bibitem{Nakai:2020kuu}
Y.~Nakai, D.~Shih, and S.~Thomas, ``{Strange Jet Tagging},''
  \href{http://arxiv.org/abs/2003.09517}{{\ttfamily arXiv:2003.09517
  [hep-ph]}}.

\bibitem{Erdmann:2019blf}
J.~Erdmann, ``{A tagger for strange jets based on tracking information using
  long short-term memory},''
  \href{http://dx.doi.org/10.1088/1748-0221/15/01/P01021}{{\em JINST}
  {\bfseries 15} no.~01, (2020) P01021},
  \href{http://arxiv.org/abs/1907.07505}{{\ttfamily arXiv:1907.07505
  [physics.ins-det]}}.

\bibitem{Chiu:2021sgs}
W.~H. Chiu, Z.~Liu, M.~Low, and L.-T. Wang, ``{Jet timing},''
  \href{http://dx.doi.org/10.1007/JHEP01(2022)014}{{\em JHEP} {\bfseries 01}
  (2022) 014}, \href{http://arxiv.org/abs/2109.01682}{{\ttfamily
  arXiv:2109.01682 [hep-ph]}}.

\bibitem{Chatterjee:2019brg}
S.~Chatterjee, R.~Godbole, and T.~S. Roy, ``{Jets with electrons from boosted
  top quarks},'' \href{http://dx.doi.org/10.1007/JHEP01(2020)170}{{\em JHEP}
  {\bfseries 01} (2020) 170}, \href{http://arxiv.org/abs/1909.11041}{{\ttfamily
  arXiv:1909.11041 [hep-ph]}}.

\bibitem{Mitra:2016kov}
M.~Mitra, R.~Ruiz, D.~J. Scott, and M.~Spannowsky, ``{Neutrino Jets from
  High-Mass $W_R$ Gauge Bosons in TeV-Scale Left-Right Symmetric Models},''
  \href{http://dx.doi.org/10.1103/PhysRevD.94.095016}{{\em Phys. Rev. D}
  {\bfseries 94} no.~9, (2016) 095016},
  \href{http://arxiv.org/abs/1607.03504}{{\ttfamily arXiv:1607.03504
  [hep-ph]}}.

\bibitem{Nemevsek:2018bbt}
M.~Nemev\v{s}ek, F.~Nesti, and G.~Popara, ``{Keung-Senjanovi\'c process at the
  LHC: From lepton number violation to displaced vertices to invisible
  decays},'' \href{http://dx.doi.org/10.1103/PhysRevD.97.115018}{{\em Phys.
  Rev. D} {\bfseries 97} no.~11, (2018) 115018},
  \href{http://arxiv.org/abs/1801.05813}{{\ttfamily arXiv:1801.05813
  [hep-ph]}}.

\bibitem{duPlessis:2021xuc}
K.~du~Plessis, M.~M. Flores, D.~Kar, S.~Sinha, and H.~van~der Schyf, ``{Hitting
  two BSM particles with one lepton-jet: search for a top partner decaying to a
  dark photon, resulting in a lepton-jet},''
  \href{http://arxiv.org/abs/2112.08425}{{\ttfamily arXiv:2112.08425
  [hep-ph]}}.

\bibitem{Dube:2017jgo}
S.~Dube, D.~Gadkari, and A.~M. Thalapillil, ``{Lepton-Jets and Low-Mass Sterile
  Neutrinos at Hadron Colliders},''
  \href{http://dx.doi.org/10.1103/PhysRevD.96.055031}{{\em Phys. Rev. D}
  {\bfseries 96} no.~5, (2017) 055031},
  \href{http://arxiv.org/abs/1707.00008}{{\ttfamily arXiv:1707.00008
  [hep-ph]}}.

\bibitem{Wang:2021uyb}
D.~Wang, L.~Wu, J.~M. Yang, and M.~Zhang, ``{Photon-jet events as a probe of
  axionlike particles at the LHC},''
  \href{http://dx.doi.org/10.1103/PhysRevD.104.095016}{{\em Phys. Rev. D}
  {\bfseries 104} no.~9, (2021) 095016},
  \href{http://arxiv.org/abs/2102.01532}{{\ttfamily arXiv:2102.01532
  [hep-ph]}}.

\bibitem{Sheff:2020jyw}
B.~Sheff, N.~Steinberg, and J.~D. Wells, ``{Higgs boson decays into narrow
  diphoton jets and their search strategies at the Large Hadron Collider},''
  \href{http://dx.doi.org/10.1103/PhysRevD.104.036009}{{\em Phys. Rev. D}
  {\bfseries 104} no.~3, (2021) 036009},
  \href{http://arxiv.org/abs/2008.10568}{{\ttfamily arXiv:2008.10568
  [hep-ph]}}.

\bibitem{Kar:2020bws}
D.~Kar and S.~Sinha, ``{Exploring jet substructure in semi-visible jets},''
  \href{http://dx.doi.org/10.21468/SciPostPhys.10.4.084}{{\em SciPost Phys.}
  {\bfseries 10} no.~4, (2021) 084},
  \href{http://arxiv.org/abs/2007.11597}{{\ttfamily arXiv:2007.11597
  [hep-ph]}}.

\bibitem{Canelli:2021aps}
F.~Canelli, A.~de~Cosa, L.~L. Pottier, J.~Niedziela, K.~Pedro, and M.~Pierini,
  ``{Autoencoders for Semivisible Jet Detection},''
  \href{http://arxiv.org/abs/2112.02864}{{\ttfamily arXiv:2112.02864
  [hep-ph]}}.

\bibitem{CMS:2021dzb}
{\bfseries CMS} Collaboration, A.~Tumasyan {\em et~al.}, ``{Search for a
  right-handed W boson and a heavy neutrino in proton-proton collisions at
  $\sqrt{s}$ = 13 TeV},'' \href{http://arxiv.org/abs/2112.03949}{{\ttfamily
  arXiv:2112.03949 [hep-ex]}}.

\bibitem{ATLAS:2019isd}
{\bfseries ATLAS} Collaboration, M.~Aaboud {\em et~al.}, ``{Search for a
  right-handed gauge boson decaying into a high-momentum heavy neutrino and a
  charged lepton in $pp$ collisions with the ATLAS detector at $\sqrt{s}=13$
  TeV},'' \href{http://dx.doi.org/10.1016/j.physletb.2019.134942}{{\em Phys.
  Lett. B} {\bfseries 798} (2019) 134942},
  \href{http://arxiv.org/abs/1904.12679}{{\ttfamily arXiv:1904.12679
  [hep-ex]}}.

\bibitem{CMS:2021dzg}
{\bfseries CMS} Collaboration, A.~Tumasyan {\em et~al.}, ``{Search for resonant
  production of strongly coupled dark matter in proton-proton collisions at 13
  TeV},'' \href{http://arxiv.org/abs/2112.11125}{{\ttfamily arXiv:2112.11125
  [hep-ex]}}.

\bibitem{ATLAS:2019tkk}
{\bfseries ATLAS} Collaboration, G.~Aad {\em et~al.}, ``{Search for light
  long-lived neutral particles produced in $pp$ collisions at $\sqrt{s} =$ 13
  TeV and decaying into collimated leptons or light hadrons with the ATLAS
  detector},'' \href{http://dx.doi.org/10.1140/epjc/s10052-020-7997-4}{{\em
  Eur. Phys. J. C} {\bfseries 80} no.~5, (2020) 450},
  \href{http://arxiv.org/abs/1909.01246}{{\ttfamily arXiv:1909.01246
  [hep-ex]}}.

\bibitem{CMS:2019qjk}
{\bfseries CMS} Collaboration, A.~M. Sirunyan {\em et~al.}, ``{Search for
  long-lived particles using nonprompt jets and missing transverse momentum
  with proton-proton collisions at $\sqrt{s} =$ 13 TeV},''
  \href{http://dx.doi.org/10.1016/j.physletb.2019.134876}{{\em Phys. Lett. B}
  {\bfseries 797} (2019) 134876},
  \href{http://arxiv.org/abs/1906.06441}{{\ttfamily arXiv:1906.06441
  [hep-ex]}}.

\bibitem{Chen:2022jhb}
H.~Chen, I.~Moult, J.~Sandor, and H.~X. Zhu, ``{Celestial Blocks and Transverse
  Spin in the Three-Point Energy Correlator},''
  \href{http://arxiv.org/abs/2202.04085}{{\ttfamily arXiv:2202.04085
  [hep-ph]}}.

\bibitem{Chen:2021gdk}
H.~Chen, I.~Moult, and H.~X. Zhu, ``{Spinning Gluons from the QCD Light-Ray
  OPE},'' \href{http://arxiv.org/abs/2104.00009}{{\ttfamily arXiv:2104.00009
  [hep-ph]}}.

\bibitem{Chen:2020adz}
H.~Chen, I.~Moult, and H.~X. Zhu, ``{Quantum Interference in Jet Substructure
  from Spinning Gluons},''
  \href{http://dx.doi.org/10.1103/PhysRevLett.126.112003}{{\em Phys. Rev.
  Lett.} {\bfseries 126} no.~11, (2021) 112003},
  \href{http://arxiv.org/abs/2011.02492}{{\ttfamily arXiv:2011.02492
  [hep-ph]}}.

\bibitem{Chen:2019bpb}
H.~Chen, M.-X. Luo, I.~Moult, T.-Z. Yang, X.~Zhang, and H.~X. Zhu, ``{Three
  point energy correlators in the collinear limit: symmetries, dualities and
  analytic results},'' \href{http://dx.doi.org/10.1007/JHEP08(2020)028}{{\em
  JHEP} {\bfseries 08} no.~08, (2020) 028},
  \href{http://arxiv.org/abs/1912.11050}{{\ttfamily arXiv:1912.11050
  [hep-ph]}}.

\bibitem{Chen:2020vvp}
H.~Chen, I.~Moult, X.~Zhang, and H.~X. Zhu, ``{Rethinking jets with energy
  correlators: Tracks, resummation, and analytic continuation},''
  \href{http://dx.doi.org/10.1103/PhysRevD.102.054012}{{\em Phys. Rev. D}
  {\bfseries 102} no.~5, (2020) 054012},
  \href{http://arxiv.org/abs/2004.11381}{{\ttfamily arXiv:2004.11381
  [hep-ph]}}.

\bibitem{Holguin:2022epo}
J.~Holguin, I.~Moult, A.~Pathak, and M.~Procura, ``{A New Paradigm for
  Precision Top Physics: Weighing the Top with Energy Correlators},''
  \href{http://arxiv.org/abs/2201.08393}{{\ttfamily arXiv:2201.08393
  [hep-ph]}}.

\bibitem{Komiske:2022enw}
P.~T. Komiske, I.~Moult, J.~Thaler, and H.~X. Zhu, ``{Analyzing N-point Energy
  Correlators Inside Jets with CMS Open Data},''
  \href{http://arxiv.org/abs/2201.07800}{{\ttfamily arXiv:2201.07800
  [hep-ph]}}.

\bibitem{Kologlu:2019mfz}
M.~Kologlu, P.~Kravchuk, D.~Simmons-Duffin, and A.~Zhiboedov, ``{The light-ray
  OPE and conformal colliders},''
  \href{http://dx.doi.org/10.1007/JHEP01(2021)128}{{\em JHEP} {\bfseries 01}
  (2021) 128}, \href{http://arxiv.org/abs/1905.01311}{{\ttfamily
  arXiv:1905.01311 [hep-th]}}.

\bibitem{Dasgupta:2013ihk}
M.~Dasgupta, A.~Fregoso, S.~Marzani, and G.~P. Salam, ``{Towards an
  understanding of jet substructure},''
  \href{http://dx.doi.org/10.1007/JHEP09(2013)029}{{\em JHEP} {\bfseries 1309}
  (2013) 029},
\href{http://arxiv.org/abs/1307.0007}{{\ttfamily arXiv:1307.0007 [hep-ph]}}.

\bibitem{Larkoski:2014wba}
A.~J. Larkoski, S.~Marzani, G.~Soyez, and J.~Thaler, ``{Soft Drop},''
  \href{http://dx.doi.org/10.1007/JHEP05(2014)146}{{\em JHEP} {\bfseries 1405}
  (2014) 146},
\href{http://arxiv.org/abs/1402.2657}{{\ttfamily arXiv:1402.2657 [hep-ph]}}.

\bibitem{Luisoni:2020efy}
G.~Luisoni, P.~F. Monni, and G.~P. Salam, ``{$C$-parameter hadronisation in the
  symmetric 3-jet limit and impact on $\alpha_s$ fits},''
  \href{http://dx.doi.org/10.1140/epjc/s10052-021-08941-z}{{\em Eur. Phys. J.
  C} {\bfseries 81} no.~2, (2021) 158},
  \href{http://arxiv.org/abs/2012.00622}{{\ttfamily arXiv:2012.00622
  [hep-ph]}}.

\bibitem{Caola:2021kzt}
F.~Caola, S.~Ferrario~Ravasio, G.~Limatola, K.~Melnikov, and P.~Nason, ``{On
  linear power corrections in certain collider observables},''
  \href{http://dx.doi.org/10.1007/JHEP01(2022)093}{{\em JHEP} {\bfseries 01}
  (2022) 093}, \href{http://arxiv.org/abs/2108.08897}{{\ttfamily
  arXiv:2108.08897 [hep-ph]}}.

\bibitem{Caola:2022vea}
F.~Caola, S.~Ferrario~Ravasio, G.~Limatola, K.~Melnikov, P.~Nason, and M.~A.
  Ozcelik, ``{Linear power corrections to $e^+e^-$ shape variables in the
  three-jet region},'' \href{http://arxiv.org/abs/2204.02247}{{\ttfamily
  arXiv:2204.02247 [hep-ph]}}.

\bibitem{Hoang:2019ceu}
A.~H. Hoang, S.~Mantry, A.~Pathak, and I.~W. Stewart, ``{Nonperturbative
  Corrections to Soft Drop Jet Mass},''
  \href{http://dx.doi.org/10.1007/JHEP12(2019)002}{{\em JHEP} {\bfseries 12}
  (2019) 002}, \href{http://arxiv.org/abs/1906.11843}{{\ttfamily
  arXiv:1906.11843 [hep-ph]}}.

\bibitem{Dixon:2019uzg}
L.~J. Dixon, I.~Moult, and H.~X. Zhu, ``{Collinear limit of the energy-energy
  correlator},'' \href{http://dx.doi.org/10.1103/PhysRevD.100.014009}{{\em
  Phys. Rev. D} {\bfseries 100} no.~1, (2019) 014009},
  \href{http://arxiv.org/abs/1905.01310}{{\ttfamily arXiv:1905.01310
  [hep-ph]}}.

\bibitem{Dixon:2018qgp}
L.~J. Dixon, M.-X. Luo, V.~Shtabovenko, T.-Z. Yang, and H.~X. Zhu,
  ``{Analytical Computation of Energy-Energy Correlation at Next-to-Leading
  Order in QCD},'' \href{http://dx.doi.org/10.1103/PhysRevLett.120.102001}{{\em
  Phys. Rev. Lett.} {\bfseries 120} no.~10, (2018) 102001},
  \href{http://arxiv.org/abs/1801.03219}{{\ttfamily arXiv:1801.03219
  [hep-ph]}}.

\bibitem{Luo:2019nig}
M.-X. Luo, V.~Shtabovenko, T.-Z. Yang, and H.~X. Zhu, ``{Analytic
  Next-To-Leading Order Calculation of Energy-Energy Correlation in
  Gluon-Initiated Higgs Decays},''
  \href{http://dx.doi.org/10.1007/JHEP06(2019)037}{{\em JHEP} {\bfseries 06}
  (2019) 037}, \href{http://arxiv.org/abs/1903.07277}{{\ttfamily
  arXiv:1903.07277 [hep-ph]}}.

\bibitem{Ebert:2020sfi}
M.~A. Ebert, B.~Mistlberger, and G.~Vita, ``{The Energy-Energy Correlation in
  the back-to-back limit at N$^{3}$LO and N$^{3}$LL'},''
  \href{http://dx.doi.org/10.1007/JHEP08(2021)022}{{\em JHEP} {\bfseries 08}
  (2021) 022}, \href{http://arxiv.org/abs/2012.07859}{{\ttfamily
  arXiv:2012.07859 [hep-ph]}}.

\bibitem{Kardos:2020gty}
A.~Kardos, A.~J. Larkoski, and Z.~Tr\'ocs\'anyi, ``{Groomed jet mass at high
  precision},'' \href{http://dx.doi.org/10.1016/j.physletb.2020.135704}{{\em
  Phys. Lett. B} {\bfseries 809} (2020) 135704},
  \href{http://arxiv.org/abs/2002.00942}{{\ttfamily arXiv:2002.00942
  [hep-ph]}}.

\bibitem{Bern:2013zja}
Z.~Bern, L.~J. Dixon, F.~Febres~Cordero, S.~H\"oche, H.~Ita, D.~A. Kosower, and
  D.~Maitre, ``{Ntuples for NLO Events at Hadron Colliders},''
  \href{http://dx.doi.org/10.1016/j.cpc.2014.01.011}{{\em Comput. Phys.
  Commun.} {\bfseries 185} (2014) 1443--1460},
  \href{http://arxiv.org/abs/1310.7439}{{\ttfamily arXiv:1310.7439 [hep-ph]}}.

\bibitem{Cranmer:2021urp}
K.~Cranmer {\em et~al.}, ``{Publishing statistical models: Getting the most out
  of particle physics experiments},''
  \href{http://dx.doi.org/10.21468/SciPostPhys.12.1.037}{{\em SciPost Phys.}
  {\bfseries 12} (2022) 037}, \href{http://arxiv.org/abs/2109.04981}{{\ttfamily
  arXiv:2109.04981 [hep-ph]}}.

\bibitem{MengXL:2018}
X.-L. Meng, ``\protect{Statistical paradises and paradoxes in big data (I): Law
  of large populations, big data paradox, and the 2016 US presidential
  election},'' \href{http://dx.doi.org/10.1214/18-AOAS1161SF}{{\em The Annals
  of Applied Statistics} {\bfseries 12} no.~2, (2018) 685}.

\bibitem{hepmllivingreview}
{HEP ML Community}, ``{A Living Review of Machine Learning for Particle
  Physics}.''
\newblock \url{https://iml-wg.github.io/HEPML-LivingReview/}.

\bibitem{Butter:2022rso}
S.~Badger {\em et~al.}, ``{Machine Learning and LHC Event Generation},''
  \href{http://arxiv.org/abs/2203.07460}{{\ttfamily arXiv:2203.07460
  [hep-ph]}}.

\bibitem{Boyda:2022nmh}
D.~Boyda {\em et~al.}, ``{Applications of Machine Learning to Lattice Quantum
  Field Theory},'' in {\em {2022 Snowmass Summer Study}}.
\newblock 2, 2022.
\newblock \href{http://arxiv.org/abs/2202.05838}{{\ttfamily arXiv:2202.05838
  [hep-lat]}}.

\bibitem{Shanahan:2022ifi}
P.~Shanahan {\em et~al.}, ``{Snowmass 2021 Computational Frontier CompF03
  Topical Group Report: Machine Learning},''
  \href{http://arxiv.org/abs/2209.07559}{{\ttfamily arXiv:2209.07559
  [physics.comp-ph]}}.

\bibitem{Manohar:1996cq}
A.~V. Manohar, ``{Effective field theories},''
  \href{http://dx.doi.org/10.1007/BFb0104294}{{\em Lect. Notes Phys.}
  {\bfseries 479} (1997) 311--362},
  \href{http://arxiv.org/abs/hep-ph/9606222}{{\ttfamily arXiv:hep-ph/9606222}}.

\bibitem{Brivio:2017vri}
I.~Brivio and M.~Trott, ``{The Standard Model as an Effective Field Theory},''
  \href{http://dx.doi.org/10.1016/j.physrep.2018.11.002}{{\em Phys. Rept.}
  {\bfseries 793} (2019) 1--98},
  \href{http://arxiv.org/abs/1706.08945}{{\ttfamily arXiv:1706.08945
  [hep-ph]}}.

\bibitem{Brivio:2019ius}
I.~Brivio, S.~Bruggisser, F.~Maltoni, R.~Moutafis, T.~Plehn, E.~Vryonidou,
  S.~Westhoff, and C.~Zhang, ``{O new physics, where art thou? A global search
  in the top sector},'' \href{http://dx.doi.org/10.1007/JHEP02(2020)131}{{\em
  JHEP} {\bfseries 02} (2020) 131},
  \href{http://arxiv.org/abs/1910.03606}{{\ttfamily arXiv:1910.03606
  [hep-ph]}}.

\bibitem{Ellis:2020unq}
J.~Ellis, M.~Madigan, K.~Mimasu, V.~Sanz, and T.~You, ``{Top, Higgs, Diboson
  and Electroweak Fit to the Standard Model Effective Field Theory},''
  \href{http://dx.doi.org/10.1007/JHEP04(2021)279}{{\em JHEP} {\bfseries 04}
  (2021) 279}, \href{http://arxiv.org/abs/2012.02779}{{\ttfamily
  arXiv:2012.02779 [hep-ph]}}.

\bibitem{Ethier:2021bye}
{\bfseries SMEFiT} Collaboration, J.~J. Ethier, G.~Magni, F.~Maltoni,
  L.~Mantani, E.~R. Nocera, J.~Rojo, E.~Slade, E.~Vryonidou, and C.~Zhang,
  ``{Combined SMEFT interpretation of Higgs, diboson, and top quark data from
  the LHC},'' \href{http://dx.doi.org/10.1007/JHEP11(2021)089}{{\em JHEP}
  {\bfseries 11} (2021) 089}, \href{http://arxiv.org/abs/2105.00006}{{\ttfamily
  arXiv:2105.00006 [hep-ph]}}.

\bibitem{Corbett:2021eux}
T.~Corbett, A.~Helset, A.~Martin, and M.~Trott, ``{EWPD in the SMEFT to
  dimension eight},'' \href{http://dx.doi.org/10.1007/JHEP06(2021)076}{{\em
  JHEP} {\bfseries 06} (2021) 076},
  \href{http://arxiv.org/abs/2102.02819}{{\ttfamily arXiv:2102.02819
  [hep-ph]}}.

\bibitem{Boughezal:2021tih}
R.~Boughezal, E.~Mereghetti, and F.~Petriello, ``{Dilepton production in the
  SMEFT at O(1/\ensuremath{\Lambda}4)},''
  \href{http://dx.doi.org/10.1103/PhysRevD.104.095022}{{\em Phys. Rev. D}
  {\bfseries 104} no.~9, (2021) 095022},
  \href{http://arxiv.org/abs/2106.05337}{{\ttfamily arXiv:2106.05337
  [hep-ph]}}.

\bibitem{Ethier:2021ydt}
J.~J. Ethier, R.~Gomez-Ambrosio, G.~Magni, and J.~Rojo, ``{SMEFT analysis of
  vector boson scattering and diboson data from the LHC Run II},''
  \href{http://dx.doi.org/10.1140/epjc/s10052-021-09347-7}{{\em Eur. Phys. J.
  C} {\bfseries 81} no.~6, (2021) 560},
  \href{http://arxiv.org/abs/2101.03180}{{\ttfamily arXiv:2101.03180
  [hep-ph]}}.

\bibitem{Miralles:2021dyw}
V.~Miralles, M.~M. L\'opez, M.~M. Ll\'acer, A.~Pe\~nuelas, M.~Perell\'o, and
  M.~Vos, ``{The top quark electro-weak couplings after LHC Run 2},''
  \href{http://dx.doi.org/10.1007/JHEP02(2022)032}{{\em JHEP} {\bfseries 02}
  (2022) 032}, \href{http://arxiv.org/abs/2107.13917}{{\ttfamily
  arXiv:2107.13917 [hep-ph]}}.

\bibitem{Durieux:2022cvf}
G.~Durieux, A.~G. Camacho, L.~Mantani, V.~Miralles, M.~M. L\'opez,
  M.~Ll\'acer~Moreno, R.~Poncelet, E.~Vryonidou, and M.~Vos, ``{Snowmass White
  Paper: prospects for the measurement of top-quark couplings},'' in {\em {2022
  Snowmass Summer Study}}.
\newblock 5, 2022.
\newblock \href{http://arxiv.org/abs/2205.02140}{{\ttfamily arXiv:2205.02140
  [hep-ph]}}.

\bibitem{Greljo:2021kvv}
A.~Greljo, S.~Iranipour, Z.~Kassabov, M.~Madigan, J.~Moore, J.~Rojo, M.~Ubiali,
  and C.~Voisey, ``{Parton distributions in the SMEFT from high-energy
  Drell-Yan tails},'' \href{http://dx.doi.org/10.1007/JHEP07(2021)122}{{\em
  JHEP} {\bfseries 07} (2021) 122},
  \href{http://arxiv.org/abs/2104.02723}{{\ttfamily arXiv:2104.02723
  [hep-ph]}}.

\bibitem{Iranipour:2022iak}
S.~Iranipour and M.~Ubiali, ``{A new generation of simultaneous fits to LHC
  data using deep learning},''
  \href{http://dx.doi.org/10.1007/JHEP05(2022)032}{{\em JHEP} {\bfseries 05}
  (2022) 032}, \href{http://arxiv.org/abs/2201.07240}{{\ttfamily
  arXiv:2201.07240 [hep-ph]}}.

\bibitem{Liu:2022plj}
D.~Liu, C.~Sun, and J.~Gao, ``{Machine learning of log-likelihood functions in
  global analysis of parton distributions},''
  \href{http://arxiv.org/abs/2201.06586}{{\ttfamily arXiv:2201.06586
  [hep-ph]}}.

\bibitem{Carrazza:2019sec}
S.~Carrazza, C.~Degrande, S.~Iranipour, J.~Rojo, and M.~Ubiali, ``{Can New
  Physics hide inside the proton?},''
  \href{http://dx.doi.org/10.1103/PhysRevLett.123.132001}{{\em Phys. Rev.
  Lett.} {\bfseries 123} no.~13, (2019) 132001},
  \href{http://arxiv.org/abs/1905.05215}{{\ttfamily arXiv:1905.05215
  [hep-ph]}}.

\bibitem{ZEUS:2019cou}
{\bfseries ZEUS} Collaboration, H.~Abramowicz {\em et~al.}, ``{Limits on
  contact interactions and leptoquarks at HERA},''
  \href{http://dx.doi.org/10.1103/PhysRevD.99.092006}{{\em Phys. Rev. D}
  {\bfseries 99} no.~9, (2019) 092006},
  \href{http://arxiv.org/abs/1902.03048}{{\ttfamily arXiv:1902.03048
  [hep-ex]}}.

\bibitem{CMS:2021yzl}
{\bfseries CMS} Collaboration, A.~Tumasyan {\em et~al.}, ``{Measurement and QCD
  analysis of double-differential inclusive jet cross sections in proton-proton
  collisions at $ \sqrt{s} $ = 13 TeV},''
  \href{http://dx.doi.org/10.1007/JHEP02(2022)142}{{\em JHEP} {\bfseries 02}
  (2022) 142}, \href{http://arxiv.org/abs/2111.10431}{{\ttfamily
  arXiv:2111.10431 [hep-ex]}}.

\bibitem{Boughezal:2020uwq}
R.~Boughezal, F.~Petriello, and D.~Wiegand, ``{Removing flat directions in
  standard model EFT fits: How polarized electron-ion collider data can
  complement the LHC},''
  \href{http://dx.doi.org/10.1103/PhysRevD.101.116002}{{\em Phys. Rev. D}
  {\bfseries 101} no.~11, (2020) 116002},
  \href{http://arxiv.org/abs/2004.00748}{{\ttfamily arXiv:2004.00748
  [hep-ph]}}.

\bibitem{Boughezal:2021kla}
R.~Boughezal, F.~Petriello, and D.~Wiegand, ``{Disentangling Standard Model EFT
  operators with future low-energy parity-violating electron scattering
  experiments},'' \href{http://dx.doi.org/10.1103/PhysRevD.104.016005}{{\em
  Phys. Rev. D} {\bfseries 104} no.~1, (2021) 016005},
  \href{http://arxiv.org/abs/2104.03979}{{\ttfamily arXiv:2104.03979
  [hep-ph]}}.

\bibitem{GEANT4:2002zbu}
{\bfseries GEANT4} Collaboration, S.~Agostinelli {\em et~al.}, ``{GEANT4--a
  simulation toolkit},''
  \href{http://dx.doi.org/10.1016/S0168-9002(03)01368-8}{{\em Nucl. Instrum.
  Meth. A} {\bfseries 506} (2003) 250--303}.

\bibitem{CALICE:2019vza}
{\bfseries CALICE} Collaboration, G.~Eigen {\em et~al.}, ``{Characterisation of
  different stages of hadronic showers using the CALICE Si-W ECAL physics
  prototype},'' \href{http://dx.doi.org/10.1016/j.nima.2019.04.111}{{\em Nucl.
  Instrum. Meth. A} {\bfseries 937} (2019) 41},
  \href{http://arxiv.org/abs/1902.06161}{{\ttfamily arXiv:1902.06161
  [physics.ins-det]}}.

\bibitem{Han:2020uid}
T.~Han, Y.~Ma, and K.~Xie, ``{High energy leptonic collisions and electroweak
  parton distribution functions},''
  \href{http://dx.doi.org/10.1103/PhysRevD.103.L031301}{{\em Phys. Rev. D}
  {\bfseries 103} no.~3, (2021) L031301},
  \href{http://arxiv.org/abs/2007.14300}{{\ttfamily arXiv:2007.14300
  [hep-ph]}}.

\end{thebibliography}

\providecommand{\href}[2]{#2}\begingroup\raggedright\endgroup

\end{document}